\documentclass[a4paper,11pt]{article}
\usepackage{natbib}
\usepackage{authblk}
\usepackage[english]{babel}
\bibpunct{(}{)}{;}{a}{}{,}
\usepackage{epsfig}
\usepackage{fancyhdr,amssymb,a4wide,hyperref}
\hypersetup{
    colorlinks=true, 
    linktoc=all,     
    allcolors=blue
}
\usepackage{amsmath}
\usepackage{graphicx}
\usepackage{tabularx}
\usepackage{multicol}
\usepackage{appendix}
\usepackage[nameinlink,capitalise]{cleveref}

\usepackage[table,svgnames,dvipsnames]{xcolor}
\usepackage{subcaption}

\usepackage[nopostdot, nonumberlist, acronym]{glossaries}

\glsdisablehyper

\def\feh{\mathrm{[Fe/H]}}

\newcommand{\lya}{\ensuremath{\mathrm{Ly}-\alpha}}

\newcommand{\hii}{H\textsc{ii}}
\newcommand{\civ}{\mbox{C\,{\sc iv}}}
\newcommand{\mgii}{\mbox{Mg\,{\sc ii}}}

\newcommand{\kms}{\,km\,s$^{-1}$} 

\newcommand{\wst}{\textit{WST}} 
\newcommand{\gaia}{\textit{Gaia}} 
\newcommand{\lsst}{\textit{Rubin LSST}} 
 
\newcommand{\euclid}{\textit{Euclid}}
\newcommand{\nancy}{\textit{Roman Space Telescope}}
\newcommand{\moons}{\textit{MOONS}} 
\newcommand{\pfs}{\textit{PFS}} 
\newcommand{\fourmost}{\textit{4MOST}}
\newcommand{\weave}{\textit{WEAVE}} 
\newcommand{\desi}{\textit{DESI}} 
\newcommand{\jwst}{\textit{JWST}} 
\newcommand{\muse}{\textit{MUSE}} 
\newcommand{\kcwi}{\textit{KCWI}} 
\newcommand{\mosaic}{\textit{MOSAIC}} 
\newcommand{\elt}{\textit{ELT}} 
\newcommand{\vlt}{\textit{VLT}} 
\newcommand{\athena}{\textit{ATHENA}} 
\newcommand{\mse}{\textit{MSE}} 
\newcommand{\megamapper}{\textit{MegaMapper}}

\newcommand{\AB}[1]{\ensuremath{#1_\mathrm{AB}}}

\newcommand{\fov}{FoV}

\begin{document}
\title{The Wide-field Spectroscopic Telescope (WST) \\ Science White Paper \\v1 }
\author{}

\newacronym{2dFGRS}{2dFGRS}{2-degree Field Galaxy Redshift Survey}
\newacronym[longplural={active galactic nuclei}]{AGN}{AGN}{active galactic nucleus}
\newacronym{WST}{WST}{Wide-field Spectroscopic Telescope}
\newacronym{FoV}{FoV}{field of view}
\newacronym{ToO}{ToO}{target of opportunity}
\newacronym{IFS}{IFS}{integral field spectrograph}
\newacronym{MOS}{MOS}{multi-object spectrograph}
\newacronym{ngVLA}{ngVLA}{Next Generation Very Large Array}
\newacronym{SKA}{SKA}{Square Kilometer Array}
\newacronym{BOSS}{BOSS}{Baryon Oscillation Spectroscopic Survey}
\newacronym{DESI}{DESI}{Dark Energy Spectroscopic Instrument experiment}
\newacronym{ELT}{ELT}{Extremely Large Telescope}
\newacronym{HST}{HST}{Hubble Space Telescope}
\newacronym{JWST}{JWST}{James Webb Space Telescope}
\newacronym{LF}{LF}{luminosity function}
\newacronym{SMBH}{SMBH}{supermassive black hole}
\newacronym{VLT}{VLT}{Very Large Telescope}
\newacronym{ISM}{ISM}{interstellar medium}
\newacronym{MW}{MW}{Milky Way}

\maketitle

\begin{figure}[h!]
  \centering
  \includegraphics[width=\textwidth]{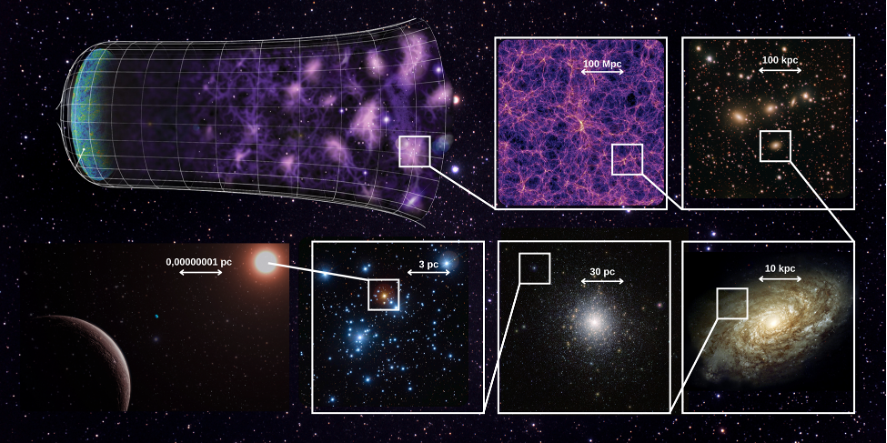}
\end{figure}

 \newpage
 
{\small \tableofcontents}

\pagebreak

\clearpage

\section*{Preface}
\addcontentsline{toc}{section}{\protect\numberline{}Preface}

For several years now, the astronomical community has expressed a strong and widespread demand for a 10-meter class telescope devoted to spectroscopic surveys, a need that is prominently highlighted in numerous strategic science plans. The latest ESO pool among its users showed that $75\%$ of that community identified a "dedicated optical/near-infrared spectroscopic facility at a 10-m class telescope" as the most crucial one for the future \citep{Merand+2021}.

\medskip

Motivated by this widespread scientific interest, a project named SpecTel has been studied in the period 2016-2020. First, a group of ESO scientists and engineers led by Luca Pasquini and Bernard Delabre proposed two innovative designs for a wide-field 10-meter  spectroscopic telescope \citep{Pasquini2016}. At the same time, ESO established a Working Group, chaired by Richard Ellis, with the mandate of developing the scientific case for a large aperture optical spectroscopic survey telescope. The scientific requirements that emerged from the Working Group report \citep{Ellis+2017} identified the need of the largest possible field of view in a 10-meter class telescope and the use of optical fibres feeding intermediate and high resolution spectrographs. In addition, the availability of a large \gls{IFS} at a separate focal station was considered a valuable additional option. A design concept for a facility full-filling such requirements was presented in \citealt{Pasquini2018}. 

\medskip

In 2021, a large European and Australian consortium was formed under the leadership of Roland Bacon to build upon those initial ideas and study a novel concept that could lead to the realization of the world-wide request for a 10-meter class telescope devoted to spectroscopic surveys. The new facility was named \textit{\gls{WST}}\footnote{https://www.wstelescope.com/}. The revised project inherits from the SpecTel studies, with one major evolution, namely the ambition to have simultaneous operation of a large field-of-view, high multiplex \gls{MOS}, including both low- and high-resolution modes, and a giant panoramic central IFS.  Current instrumentation at 10-meter class telescopes has demonstrated the great complementarity between \gls{MOS} and IFS instruments, therefore their simultaneous operation at the \textit{\gls{WST}} will greatly increase the scientific return of this facility. 

\medskip

As shown in this document, \textit{\gls{WST}} will have an impact on an extremely broad variety of  astrophysics research areas. The Science Team currently involves more than 500 scientists distributed in 32 countries over five continents (see Appendix \ref{appendix}), witnessing the widespread scientific interest for \textit{\gls{WST}}. A major international conference was hold in Vienna (May 23-26, 2023)\footnote{https://www.wstelescope.com/meetings/symposium} where the technical concept of the facility was presented and a wide set of scientific ideas was discussed. Following this successful meeting, the Science Team has been working to prepare a "\textit{\gls{WST}} Science White Paper" to describe the range of science that this new facility would allow to tackle.

\medskip

At the time of writing, the \textit{\gls{WST}} team is preparing a proposal for HORIZON funding to perform a three years (2025-2027) concept study. During the study, we will be able to develop the technical and operational concept of \textit{\gls{WST}} as well as plans for data reduction, analysis, and archiving; at the same time we will further develop, expand and refine the science drivers. Membership to the Science Team will continue to be open to all scientists in the research community who are interested to contribute\footnote{https://www.wstelescope.com/for-scientists/participate}. We plan at least two international conferences during this period and a comprehensive new version of the "\textit{\gls{WST}} Science White Paper" at the end of the study.

\medskip
The Science Team is led by the project scientist, Vincenzo Mainieri, and it is currently organised in five working groups:
\begin{itemize}
    \item Time domain (led by Richard I. Anderson, Cyrielle Opitom, Paula S\'anchez S\'aez)
    \item Galactic (led by Vanessa Hill, Rodolfo Smiljanic, Eline Tolstoy)
    \item Resolved stellar populations beyond the Milky Way (led by Anna McLeod, Martin M. Roth)
    \item Extra-galactic (led by Jarle Brinchmann, Richard Ellis)
    \item Cosmology (led by Andrea Cimatti, Jean-Paul Kneib)
\end{itemize}

\clearpage


\clearpage

\section*{Scientific summary}
\addcontentsline{toc}{section}{\protect\numberline{}Scientific summary}

The \textit{\gls{WST}} is proposed as a new facility dedicated to the efficient delivery of spectroscopic surveys and this white paper summarises the initial concept as well as the corresponding science cases. It will provide an end-to-end service from the preparation of surveys to the delivery of high-quality calibrated science data product. Uniquely, it will have the ability to simultaneously deliver fibre-fed multi-object spectroscopy over a large field of view (MOS) and integral field imaging spectroscopy of a central field of $3\times3$ arcmin$^2$ (IFS). \textit{\gls{WST}} will fill a crucial gap in the astronomical landscape and will allow full exploitation of major current and upcoming ground and space-based facilities. \textit{\gls{WST}} will address outstanding scientific questions in the areas of cosmology; galaxy assembly, evolution, and enrichment, including our own Milky Way; origin of stars and planets; time domain and multi-messenger astrophysics.

\textit{\gls{WST}} features a 12 meter aperture telescope that covers the UV to the H band and operates in seeing limited conditions. Its \gls{FoV} of 3.1 deg$^2$ will be the largest of any telescope equipped with a MOS. In particular, its \gls{FoV} is 2.5 times larger than PFS@Subaru and a remarkable 22 times larger than MOONS@VLT. In addition, the \textit{\gls{WST}} has a substantial aperture, twice as large (100 square metres) as its 8 meter class competitors. A central circular area with a diameter of 13 arcmin is available for simultaneous IFS observations. Within this area, the IFS with a \gls{FoV} of $3\times3$ arcmin$^2$ (9 times larger than MUSE@VLT) can be utilised simultaneously with the MOS. In addition, the IFS can be used to mosaic a larger area within the 13 arcmin circle (an area larger than the entire \gls{FoV} of the ELT) while the MOS positioner can perform multiple configurations. The \textit{\gls{WST}} operational model will also implement \gls{ToO} operations at both the telescope and MOS fibre levels to support the time domain science cases.

The demand for a facility dedicated to spectroscopic surveys is an interest shared worldwide and figures explicitly in many strategic science plans. The \textit{\gls{WST}} project as presented here is conceived to address this demand. This first version of the Science White Paper aims to highlight the transformational science that \textit{\gls{WST}} will achieve on its own, but at the same time show how \textit{\gls{WST}} naturally complements many other major upcoming facilities and ultimately will contribute to a major progress in science.

\bigskip

{\bf Time-domain and multi-messenger astronomy} are key priorities identified by major planning exercises, such as the Astro2020 decadal survey, and they are tightly connected since multi-messenger events are typically of a transient nature. The upcoming investment in deep survey imaging with dedicated facilities promises many unforeseen scientific discoveries in this area: e.g., the \lsst\ facility will undertake a nightly scan leading to a ‘real time video’ of the southern sky. However, spectroscopic follow-up is key to delivering this extraordinary scientific bounty. Recognising this, upcoming 4-meter telescopes such as \fourmost\ and \weave\ are poised to follow-up time-variable sources. However, all current and upcoming facilities together can follow up only a few percent of the expected transient alerts, \textit{\gls{WST}} will take over the mantle and supersede these early spectroscopic campaigns. \textit{\gls{WST}} is being conceived with a time-domain mindset, to enable maximum operational flexibility and rapid data processing in order to serve the needs of this exciting and diverse field full of serendipitous discovery potential, with science cases spanning all physical scales from Solar system objects to Cosmology and all-time scales from hours to decades.
In addition to finding hitherto unknown categories of variable events thereby revealing new astrophysical processes, \textit{\gls{WST}} can make huge progress in the revolution of exploiting {\bf sources of gravitational waves (GWs)}. Whereas the current rate of LIGO events is modest, when \textit{\gls{WST}} is operational we can expect a considerable increase necessitating dedicated campaigns to pinpoint their hosts through deep extragalactic surveys to redshifts z$>3$. The potential for using both neutron star mergers with electromagnetic counterparts and dark sirens from black hole mergers as distance indicators is exciting both for cosmology and understanding their demographics. {\bf Active galactic nuclei} (AGN) are likewise extragalactic variables but the physics of their variability, both on long timescales and via occasional short-term flares, is poorly understood. Reverberation Mapping, which a dedicated facility such as \textit{\gls{WST}} can undertake though high signal/noise monitoring campaigns, offers the prospect of unveiling spatial structures in the accretion disk surrounding the black hole to redshifts beyond the peak of AGN activity. {\bf Supernovae} were originally classified into two broad categories, largely on the basis of their light curves and spectra. However, it is now recognised there is a much wider diversity related to variations in the nature of the explosion mechanism and their local environment. \textit{\gls{WST}} has the capability to secure high s/n spectra shortly after the initial explosion, thereby probing interactions with stellar companions, as well as trends related to the nature of their host galaxy. Rarer phenomena such a {\bf gamma ray bursts} and {\bf tidal disruption events} can also be analysed spectroscopically, to respectively study the ionisation state of the early intergalactic medium and provide black holes masses independent of other, less direct, measures. Finally, {\bf classical variable stars} in the Milky Way and nearby galaxies remain unique laboratories for detailed studies. Their pulsation velocities, line profiles and chemical abundances are crucial to understand their role in establishing the cosmological distance scale.

Our own {\bf solar system} is the only planetary system for which we can study a range of bodies resulting from the planetary formation process in detail and sometimes even in-situ. Understanding how our own solar system formed and evolved is crucial to understand the formation and evolution of planetary systems. Small bodies of the solar system, like {\bf asteroids and comets}, are remnants from the planetary formation process. They are thus essential tools to understand the history of our solar system. When a comet approaches the Sun, the ices contained in its nucleus sublimate to form an atmosphere of gas and dust around the nucleus called the coma. The \textit{\gls{WST}} IFS will provide an incredible opportunity to detect and map the emission of different radicals in the coma, in a range of comets and interstellar objects permitting to significantly increase our understanding of cometary activity and the release of species in the coma. At the same time the IFS will allow to observe multiple asteroids at the same time and quickly build a representative sample of asteroid surface spectra for different types of objects.

Although \textit{\gls{WST}} is a massively-multiplexed spectroscopic facility, with careful planning through target-of-opportunity and long-term monitoring campaigns, it can fully exploit the remarkable revolution we are witnessing in multi-messenger astronomy. It will benefit from the experience of 4-meter facilities which have recognised this opportunity and will continue the journey of exploring this exciting frontier in astrophysics. 

\bigskip

We are living in a golden era for {\bf Galactic, stellar, and exoplanet astrophysics}, thanks to the ESA \gaia\ mission, along with the numerous stellar surveys undertaken in the last decade. \textit{\gls{WST}} will allow us to further push the boundaries of our understanding of the \gls{MW} and its component stars and populations. 
The final data release of \gaia\ is expected around 2030; in the coming years the \lsst\ will provide photometry and astrometry extending samples to $\sim 4$ magnitudes fainter than \gaia, while the \nancy\ will perform a deep infrared survey of the entire MW plane and the Bulge. At the same time, future space missions, including \nancy\ itself, will detect tens of thousands of new exoplanets, also orbiting faint stars. These exquisite new datasets have an enormous potential for breakthrough science but will be best exploited when complemented by spectroscopic information. \textit{\gls{WST}} will indeed crucially allow measurements of precise radial velocities, metallicities, elemental abundances and other stellar properties, reaching distant and faint populations.  The combination of the quality of a high-resolution sample and the depth of a low resolution one, along with IFS observations of denser regions and stellar systems like star clusters, will significantly enhance our comprehension of the MW’s stellar populations and its system of neighboring galaxies. 

Precise abundances of several chemical elements, covering the different nucleosynthesis channels, for a few million stars will allow a more comprehensive understanding of {\bf the origin of the elements} which is essential for a variety of other astrophysical issues. \textit{\gls{WST}} will also greatly enhance our ability to measure abundance ratios that are sensitive to age (the so-called “chemical clocks”) for statistically significant samples of stars and to produce {\bf extensive age maps of the Galactic disk}. The combination of chemistry, kinematics and ages will maximize our ability to accurately identify related groups of stars and reconstruct the {\bf star-formation history of the Milky Way disk} in extremely fine detail.

Low and high resolution MOS observations, as well as IFS spectra in the densest parts, will allow to {\bf disentangle co-spatial stellar populations} in the {\bf MW Bulge} and promise to be transformative; they will indeed open a window into this critical component of our MW that is seen in the formation stages in other galaxies at high redshift with \textit{\gls{JWST}}. The complementarity will be extremely powerful.

\textit{\gls{WST}} will give a substantial contribution to identify and characterize the assembly and accretion history of the MW, in particular by the {\bf chemodynamical identifications of past accretions}, through large samples that extend to fainter magnitudes to probe further out in the halo ($>10$ kpc). Reconstruction of the full assembly/accretion history of the MW, discriminating between stars formed in situ and those formed in progenitor galaxies of different masses and star formation efficiencies, will in turn deepen {\bf our understanding of the Milky Way’s role in the broader cosmological context} and how our Galaxy can help us to understand the properties of other galaxies across cosmic time.

Our comprehension of {\bf star and planet formation} is still incomplete; in particular, the traditional view that high-density clusters are the primary sites of star formation (SF) has been challenged by recent results.  Also, whilst models predict that mass accretion and outflow play a fundamental role in shaping the evolution of the star$+$circumstellar disk systems and, hence, the planet formation process, clear observational tests are still missing. \textit{\gls{WST}} can play a transformational role in this area. Specifically, {\bf the IFS} will allow the study with unprecedented detail of the {\bf SF processes in massive and dense environments; the low-resolution MOS} is ideal to investigate {\bf the properties of dispersed populations, while the high-resolution MOS} would be essential to measure {\bf abundances and infer stellar activity in large samples of planet-hosting stars}, thus unveiling the role of host-star chemical composition and magnetic fields in shaping planetary systems. 

\textit{\gls{WST}} will be able to probe the {\bf interstellar gas} by means of absorption lines and will also allow detailed studies of Diffuse Interstellar Bands; the IFS will be the best instrument to study molecular clouds and perform extended source spectroscopy, {\bf mapping the distribution of low-energy cosmic rays in the Galaxy}, identifying the sources of cosmic rays, and constraining the propagation models. 

Finally, by delivering high quality MOS and IFS spectra of  {\bf large samples of stars at all stages of evolution}, including binary systems, white dwarfs, and ionized nebulae,  \textit{\gls{WST}} will be key to address {\bf many unresolved questions in the field of stellar evolution} that impact all areas of astronomy focusing on stellar populations, as stellar evolution models are widely used to interpret the properties of individual stars and the integrated properties of galaxies anywhere in the Universe. 

\bigskip

The \textit{\gls{WST}} IFS has the perfect combination of \gls{FoV} and wavelength coverage to enable detailed investigations of {\bf resolved stellar populations} in densely populated regions such as the central regions of dense stellar systems from the Milky Way to nearby Local Volume galaxies. The {\bf Magellanic Clouds} are ideal cosmic laboratories to perform detailed studies of star formation in low-mass galaxies and as a function of metallicity. The \textit{\gls{WST}} IFS has the combination of \gls{FoV} and collecting area to probe age sensitive regions of the colour-magnitude diagram, such as the sub-giant branch, for even the oldest stellar populations in the Magellanic Clouds. \textit{\gls{WST}} will be able to unveil many of the secrets still hidden by the clouds: e.g., their history of interactions, both mutually and with the Milky Way, the existence of central massive black holes, or the chemical enrichment with time. \textit{\gls{WST}} will be targeting both young massive clusters, which are abundant in the Magellanic Clouds and other nearby star-forming galaxies, and old globular clusters in the Milky Way, to understand how binary stars impact the evolution of massive clusters and vice versa. Extending the scope from the Milky Way to other nearby galaxies is important in order to encompass different metallicity environments.

The formation and evolution of {\bf massive stars} are far from being understood. Massive stars (M$>8\,$M$_\odot$) can have a key role in shaping chemically and dynamically galaxies: as principal sources of heavy elements and UV radiation, the most massive stars, despite their brief lives, play a fundamental role in the composition and ionization of the Universe;  their mighty ends, as supernova explosion, make their impact in the interstellar medium more acute, and label massive stars as the most plausible progenitors of long $\gamma$-ray bursts. In a Universe of ever-increasing chemical complexity, understanding how the physics of massive stars depends on metallicity (Z), particularly in metal-poor environments (low-Z) is crucial to assess their impact on their host galaxies along cosmic history. Massive stars in the Small Magellanic Cloud (SMC, 1/5 Z$_\odot$) have so far set the standard for large samples in the low-Z regime. However, their metallicity is representative of the relatively late Universe: the lessons learned from SMC massive stars do not apply to important epochs such as the peak of the SFR of the Universe. The WST allows to supersede the limitations of current observations at lower metallicities in more distant galaxies, and to systematically study and resolve stellar populations in a range of different low-Z environments.

{\bf Dwarf galaxies}, i.e. systems with stellar mass at least one order of magnitude smaller than that of the \gls{MW}, are the most common type of galaxies found in the Universe today, therefore learning about their formation and evolution implies learning about the most common mode of galaxy formation. They are also widely considered as some of the best systems from which we can gather constraints on the nature of dark matter (DM) and on the effect that baryonic processes can have on modifying the inner structure of DM haloes. Expanding studies of radial velocities, metallicities, and chemical abundances of individual stars to fainter limits with \textit{\gls{WST}} (MOS and IFS) will be especially powerful for advancing the study of the ultra faint dwarf galaxy population in the Local Group. Current spectroscopic studies struggle to get useful spectra for more than a handful of individual stars. Increasing the number of spectra of  individual stars will allow more accurate dark matter models of these smallest known dark matter halos, and more detailed studies of the chemical abundances of their stellar populations.

Finally, the \textit{\gls{WST}} IFS will provide more detailed chemo-dynamical maps for galaxies in the local Universe than are possible with existing surveys thanks to its large field of view and small pixel scale and covering entire systems, often extending deep into the halo regions. These will provide more detailed observational tests for signatures associated with a variety of galactic chemo-dynamical processes. The MOS can be used simultaneously to sample  individual stars in the outer disc and halo of more nearby spiral galaxies.

\bigskip

The most fundamental feature of structure formation is the complex distribution of matter referred to as the {\bf cosmic web}. Galaxies form and assemble within this evolving network of dark matter haloes, filaments and voids. \textit{\gls{WST}}’s unique combination of a panoramic MOS and an on-axis IFS is ideally suited for understanding the baryonic process that occur in the cosmic web, both on small scales where gaseous flows regulate star formation and chemical enrichment, and in larger cosmic volumes where environmental trends can be charted. \textit{\gls{WST}}’s large aperture ensures that such synergies between galaxy-scale processes and larger cosmic structures can be studied at high redshifts (z$>2$) where star formation and AGN activities are at their peak, as well as at low redshifts (z$<1.5$) where finer details and improved S/N is possible. By the late 2020s, deep multi-band imaging over very large fields from ground and space-based facilities will ensure reliable target selection with minimal contamination from interlopers. In addition to undertaking a traditional {\bf photometrically-selected spectroscopic survey} defining the 3-D galaxy distribution over $0<$z$<7$, within a restricted redshift range $2<$z$<3$ a more powerful technique has emerged based on studies of the {\bf 3D topology of the Lyman alpha forest} seen in absorption along the line of sight to background sources. These clouds of intergalactic hydrogen trace the linear regime of density fluctuations and hence act as a valuable proxy for the dark matter distribution. Although current 6.5-10m telescopes have demonstrated the practicality of Lyman alpha tomography, only \textit{\gls{WST}}’s MOS has both the field-of-view and huge multiplex gain to chart the cosmic web systematically in much larger representative volumes. High quality spectra associated with both the traditional and tomographic redshift surveys can additionally be used to study the {\bf interface between galaxies and their circumgalactic medium}. Star-forming galaxies drive galactic-scale outflows and also accrete gas from the intergalactic medium. Currently, diagnostics of these gaseous flows via the kinematics and chemistry of absorption lines can only be achieved at high redshift through stacking spectra of selected sources. \textit{\gls{WST}} has the capability to accomplish such measurements for individual galaxies studied as a function of their location in the cosmic web.  

Simultaneously with both redshift surveys mentioned above, the IFS will trace {\bf the cosmic web in emission, charting gaseous flows in connecting filaments}. This will also be productive at redshifts z$\sim1$ where spatial sampling on scales $<1$ Mpc using familiar rest-frame optical lines is practical. Extending studies of how the cosmic web influences galaxy evolution over a large baseline in cosmic time will be a major advance, also enabling {\bf studies of how passive galaxies were quenched}. The IFS will also be instrumental in understanding the role that AGN play in governing galaxy assembly. Pixel-by-pixel continuum and emission line spectra can provide insight into how AGN are fuelled and how AGN feedback in the form of radiation pressure driven outflows or radio jets governs star formation in galaxies. \textit{\gls{WST}} will be operational alongside \textit{\gls{SKA}} and there is thus an obvious opportunity for a {\bf SKA-WST survey} aimed at establishing optical counterparts for \textit{\gls{SKA}} detections, not just to provide MOS redshifts but also for detailed IFS studies of the host galaxies. Current studies linking radio jets to star formation rates are restricted to low redshift galaxies whereas \textit{\gls{WST}} and \textit{\gls{SKA}} have the collectively capability to extend this to z$\sim1-3$ where such radio-mode feedback is thought to be dominant.

By the use of judicious pointings, the above high redshift MOS surveys can also be undertaken simultaneously with a {\bf comprehensive local volume (D$<25$ Mpc) IFS survey}. This will deliver spatially-resolved stellar and nebular spectroscopic data at exquisite resolution. This will complement similar imaging campaigns undertaken with ALMA, shortly with \lsst\, and ultimately with ultraviolet and radio facilities (e.g. UVEX, \textit{\gls{SKA}}). Studying the gas-phase metallicities of thousands of individual HII regions across different galaxies in combination with their local stellar populations will {\bf address key features in the baryonic cycle on $<100$pc scales for the first time}. By modelling spatially-resolved star formation histories, the local survey would also provide {\bf a detailed assembly history of galaxy sub-components}, a technique currently only possible for a handful of the closest large galaxies. Although \textit{\gls{WST}} has the capability to study high redshift galaxies in unprecedented detail, such a Nearby Galaxy Reference Survey will represent an equally innovative programme revealing the finer details of the physical processes governing how star formation and chemical enrichment occurs in galaxies.

\bigskip

Although it is often claimed we have entered an era of “precision cosmology,” the astrophysical community faces several fundamental puzzles. {\bf What causes the cosmic acceleration?} Must we resurrect the “cosmological constant” with an energy density that cannot be physically understood? Does its value evolve with cosmic time? Or is dark energy simply an illusion caused by an incorrect theory of gravity on large scales? Moreover, although the {\bf standard cold dark matter model} can successfully reproduce the observed growth of large scale structure from the epoch of recombination to the present day, anomalies on small scales remain raising further questions.  

These and other puzzles have led to a {\bf significant investment in new facilities} whose aim is to chart the 3D distribution of tens of millions of galaxies over large cosmic volumes and significant look-back times to constrain the evolving power spectrum of density fluctuations. Currently, there are dedicated facilities such as \desi\ and those that share time with other astrophysical applications such as \fourmost\ and \pfs. Data from these large redshift surveys will complement information derived from probes of the cosmic microwave background and imaging surveys which use weak gravitational lensing to trace the evolving dark matter distribution. Although the next few years will see improved weak lensing constraints from \lsst\ and \euclid\, associated spectroscopic redshifts of representative background galaxies will still be key in their interpretation. 

As the most powerful of the proposed next generation spectroscopic instruments, \textit{\gls{WST}} will lead to a dramatic increase in the number of galaxy redshifts, both in terms of sampling significantly increased cosmological volumes where signatures of non-Gaussianity on very large scales can test models of inflation, and over large look-back times, thereby directly probing the growth of structure and cosmic expansion history through measures of redshift space distortions and the baryonic acoustic oscillation feature.  Dark energy is usually considered to be a late-time phenomenon, so {\bf the primary \textit{\gls{WST}} legacy survey will focus on multiple galaxy tracers to redshift z$\sim1.6$}. However, exploiting \textit{\gls{WST}}’s large aperture, a higher redshift survey will provide the first measures of the expansion history and rate of structure growth in the redshift range $2<$z$<5$. This, in turn, can be extended to redshift z$\sim7$ using {\bf the distribution of Lyman-alpha emitting galaxies seen with the IFS}. These carefully structured surveys will benefit from close coordination with the extragalactic surveys discussed above, particularly at redshifts z$>3$ where higher quality spectra for astrophysical applications are more challenging to obtain.

The evolving power spectrum has many further applications beyond tracing the growth of structure and the expansion history. As neutrinos have mass, they suppress small scale power through free streaming and this can be used to constrain the {\bf total neutrino mass and its hierarchy}, independent of weaker limits from laboratory experiments. Likewise, by surveying the largest cosmic volumes, \textit{\gls{WST}}’s constraints on {\bf non-Gaussianity} will be $\sim$10 times better than those of current stage-IV experiments, making it possible to take a major step forward by achieving sub-unit precision on non-Gaussianity parameter ($f_{NL}$) and possibly detecting/confirming evidence of multi-field inflation.

As earlier probes such as \desi, \pfs, and \euclid\ may reveal cosmological surprises, some flexibility in survey design is essential. Fortunately, there are many additional cosmological applications possible with such a powerful MOS/IFS, each of which can be executed in conjunction with the extragalactic surveys mentioned earlier. These include studying {\bf correlations in the Lyman alpha forest} seen in absorption line spectra of quasars to provide stronger constraints on self-interacting and warm dark mater, {\bf direct measures of the intrinsic alignments} of close pairs of galaxies essential for full exploitation of \euclid\ weak lensing data, and providing {\bf redshifts for sources of gravitational waves and fast radio bursts}, both of which can be used in different ways to provide absolute cosmic distances and hence complement other measures of the cosmological parameters.

\clearpage

\section{WST in the scientific landscape of 2030s}\label{sec:introduction}
\paragraph{Authors} Vincenzo Mainieri$^1$, Richard I. Anderson$^2$, Jarle Brinchmann$^3$, Andrea Cimatti$^4$, Richard Ellis$^5$, Vanessa Hill$^6$, Jean-Paul Kneib$^2$, Anna McLeod$^{7,8}$, Cyrielle Opitom$^9$, Martin M. Roth$^{10}$, Paula S\'anchez S\'aez$^1$, Rodolfo Smiljanic$^{11}$, Eline Tolstoy$^{12}$, Roland Bacon$^{13}$, Sofia Randich$^{14}$, Luca Pasquini$^1$, Stefania Barsanti$^{19,20}$, Julia Bryant$^{28}$, Ian Bryson$^{34}$, Mark Casali$^{39}$, Matthew Colless$^{19,20}$, Warrick Couch$^{48}$, Camilla Danielski$^{14}$, Philippe Dierickx$^{13}$, Simon Driver$^{55}$, Adriano Fontana$^{25}$, Bianca Garilli$^{27}$, Roelof S. de Jong$^{10}$, David Lee$^{34}$, Matt Lehnert$^{13}$, Laura Magrini$^{14}$, Ben Montet$^{90,91}$, Ruben Sanchez-Janssen$^{34}$, Mamta Pandey-Pommier$^{98}$, Mark Sargent$^{106}$, Pietro Schipani$^{64}$, Matthias Steinmetz$^{10}$, Tony Travouillon$^{110}$, Laurence Tresse$^{29}$, Jo\"el Vernet$^1$, Christophe Yeche$^{118}$, Bodo Ziegler$^{71}$

\subsection{The demand for wide field spectroscopy}

Astronomy has definitely entered in the era of big data volumes: major current and upcoming facilities will provide astronomical data across the entire electromagnetic spectrum at a rate never seen before.  In particular over the next decade we expect a deluge of high-quality imaging data from upcoming ground-based (e.g., \lsst, \textit{\gls{SKA}}, CTA) and space (\gls{JWST}, \euclid, \nancy, \athena) telescopes. On the same timescale, the ESA space mission \gaia, which is revolutionizing our understanding of the Milky Way and its local environment, will have its final release. These rich imaging and astrometric data shall be complemented with spectroscopic information, with adequate spectral resolution and cadence, to fully understand the physical processes at play. Given the expected number of sources (e.g. 20 billion galaxies and 17 billion stars down to R$\sim27.5$ from the \lsst; 10 billion galaxies from \euclid; 1.5 billion stars down to the \gaia\  magnitude G$\sim21$ from \gaia) only a dedicated wide-field spectroscopy facility on a 10-meter class telescope can meet the challenge. The demand for a facility dedicated to spectroscopic surveys is an interest share worldwide and figures explicitly in many strategic science plans. In Europe, the recently released ASTRONET Roadmap 2022-2035 for Astronomy has clearly identified a "general-purpose, wide-field, high multiplex spectroscopic facility, for a telescope of 8-10m class" as one of the three high priority new ground-based infrastructures needed for the next decade. The same report highlights the key synergies between such new facility and other major astronomical capabilities (e.g. \jwst, \lsst, \euclid). An ESO appointed Working Group to study the scientific case for such new facility concluded that "it could enable transformational progress in several broad areas of astrophysics, and may constitute an unmatched ESO capability for decades" \citep{Ellis+2017}. The latest ESO pool among its users showed that $75\%$ of that community identified a "dedicated optical/near-infrared spectroscopic facility at a 10-meter class telescope" as the most crucial one for the future \citep{Merand+2021}. In the United States, the Decadal 2020 report recognizes how "massively multiplexed spectroscopy is required to fully realize the primary science goals of the VRO, the Roman Space Telescope, Gaia" and recommend to consider making a large investment of resources in a 10-meter class dedicated facility for the next decade. In the latest decadal plan for Australian astronomy, the "development of an 8-metre class optical/infrared wide-field spectroscopic survey telescope" was identified as a key area for future investments. Finally, the Long Range Plan 2010 in Canada states that a 10-meter class telescope, equipped with an extremely multiplexed spectrograph, "would be a unique resource for follow-up spectroscopy, both for the European Gaia satellite mission, and also for LSST and Euclid/Roman".

\wst\ is conceived to be the realization of the world-wide request for such new facility. This first version of the white paper aims to highlight the transformational science that \wst\ will achieve on its own, but at the same time show how \wst\ naturally complements many other major upcoming facilities and ultimately will contribute to a major progress in our understanding of the Universe. 

\begin{figure}
  \centering
  \includegraphics[width=\textwidth]{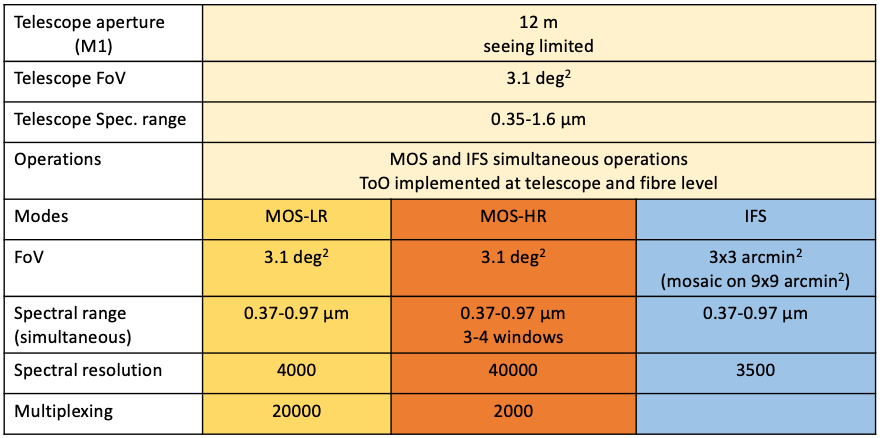}
  \caption*{\small Table 1: The baseline top-level requirements for \wst.}
  \label{fig:DMS}
\end{figure}

\subsection{Key capabilities of WST}

\begin{figure}
  \centering
  \includegraphics[width=0.8\textwidth]{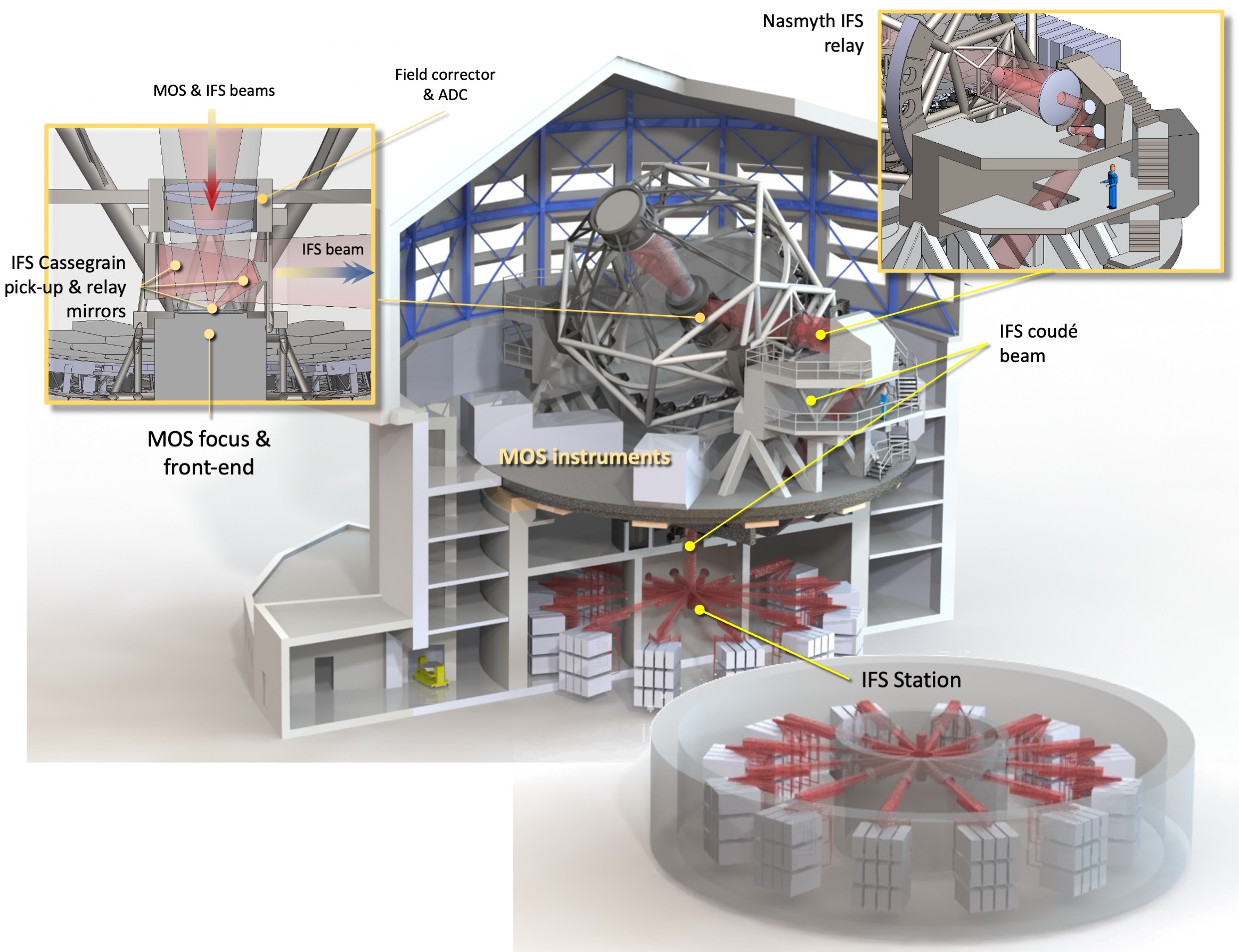}
  \caption{\small : Current Facility design. The MOS spectrographs are located on the azimuth floor, for minimal fiber length. The gravity-invariant IFS station is located in the pier, below the optical de-rotator. IFS sub-field extraction ($3\times3$ arcmin$^2$ out of the 13 arcmin diameter patrol field) is performed at the optical exit of the Nasmyth relay. Image credit to Gaston Gausachs.}
  \label{fig:Intro:DMS}
\end{figure}

\begin{figure}
  \centering
  \includegraphics[width=0.75\textwidth]{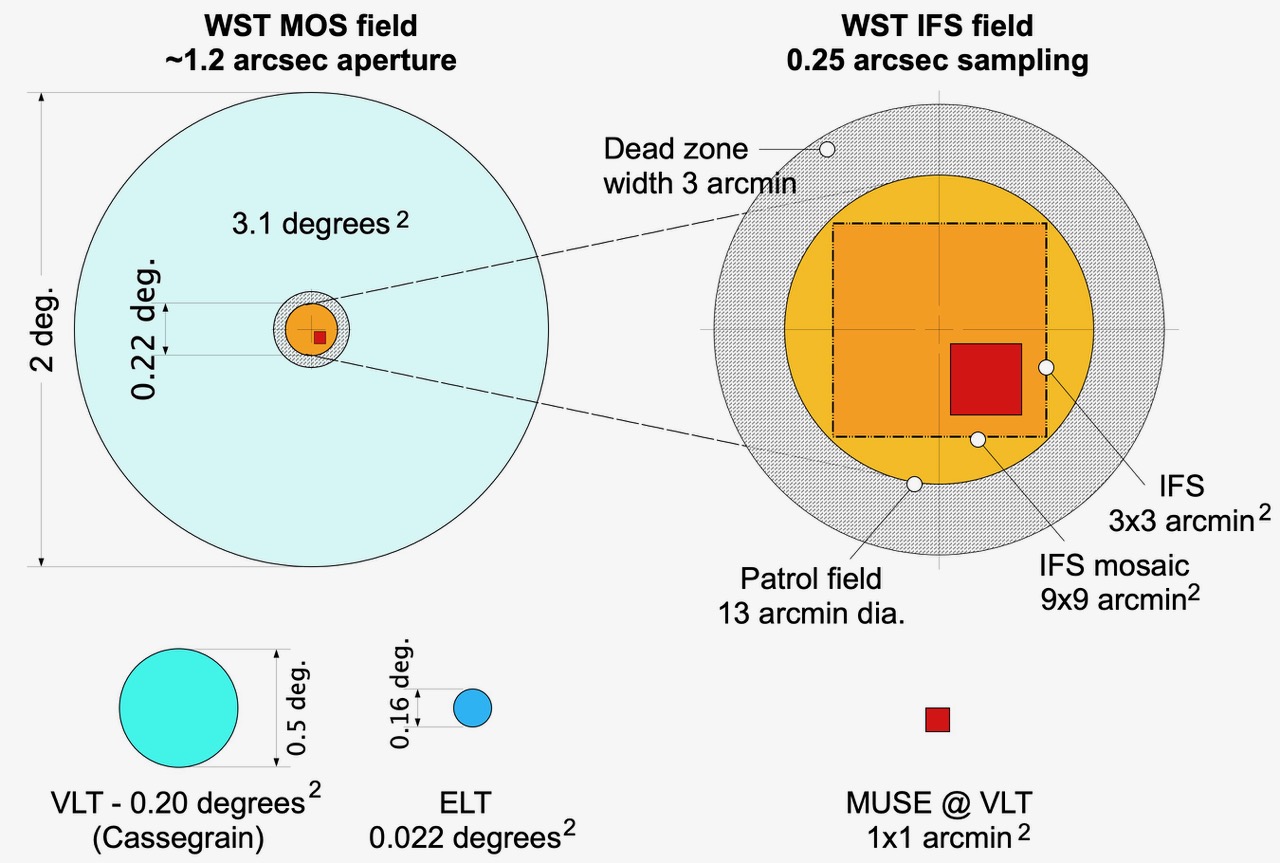}
  \caption{\small \textbf{Left:} \wst\ \gls{FoV}. The left panel shows the MOS \gls{FoV} and the central circular area available for IFS observations. The right panel offers a closer view of the latter. The IFS 3x3 arcmin$^2$ \gls{FoV} (in red) can be moved within the available area, providing the 9x9 arcmin$^2$ mosaic capability. The VLT, ELT and MUSE \gls{FoV} are represented for comparison.
}
  \label{fig:Intro:WST_FOV}
\end{figure}

The \wst\ observatory will be a facility dedicated to the efficient delivery of spectroscopic surveys (Figure~\ref{fig:Intro:DMS}). It will provide an end-to-end service from the preparation of surveys to the delivery of high quality calibrated science data product.  Uniquely, it will have the ability to simultaneously deliver fibre-fed multi-object spectroscopy over a large field of view (MOS) and integral field  imaging spectroscopy of a central field of $3\times 3$ arcminutes (IFS), see Figure~\ref{fig:Intro:WST_FOV}. While survey efficiency and quality will be a key performance and design driver, in the current global situation, considerable effort will be devoted to make such new facility sustainable, both during construction as well as across its operating lifetime. 

We highlight here some key and unique feature of \wst:
\begin{itemize}
    \item {\it Dedicated 12-meter facility with simultaneous operation of MOS and IFS.} \wst\ will be a facility fully dedicated to spectroscopic surveys. The telescope will be a 12-meter class segmented primary and will provide two concentric, simultaneously available fields: a baseline 3.1 deg$^2$ MOS field at a corrected Cassegrain focus located above the primary mirror, and a central 13 arcmin diameter IFS field that is re-imaged initially to a Nasmyth location. With a Cassegrain focus above the fast f/1 primary mirror, the optical design allows for a compact and stiff alt-azimuthal structure, with evident cost and performance benefits. The primary mirror will be made of 78 segments, which will be identical to the ELT ones, with identical supports and controls.  A Cassegrain corrector will provide also for atmospheric dispersion compensation. The overall optical design has been optimised to cover the 0.35-1.6 $\mu m$ wavelength range.
    \item {\it Survey speed and spectral performance.}  \wst\ will offer spectral capabilities covering a wide range of spectral resolutions with a high multiplexing. The MOS will have 20,000 independent fibres for low-resolution (R$\sim 4000$) and 2,000 high resolution (R$\sim 40,000$) fibres, the latter with larger patrol areas to cover the whole field. The monolitic IFS will provide R$\sim 3000$ spectroscopy for an entire  3$\times$3 arcmin$^2$ patch of the sky in one exposure. The high multiplexing, combined with the high sensitivity, large field-of-view and operational efficiency will ensure a high survey speed for \wst. At low spectral resolution, \wst\ will provide a S/N per Angstrom of three down to magnitude 24.5 (MOS) and 25.2 (IFS) in one hour exposure (point sources, AB magnitudes, 0.7 arcsec seeing, airmass 1.0) over $\approx90\%$ of the wavelength range. At high spectral resolution \wst\ will obtain a S/N per Angstrom of 30 (100) down to magnitude 19.4  (17.8) in one hour exposure (point sources, AB magnitudes, 0.7 arcsec seeing, airmass 1.0) in the selected wavelength windows.
    \item{\it Built-in capability for targets of opportunity observations.} Thanks to its large aperture, field of view and spectroscopic capabilities, \wst\ is uniquely located  to be the ideal follow-up facility for the dilution of triggers from upcoming imaging facilities (e.g. \lsst), the future generation of gravitation wave observatories (the Einstein Telescope, Cosmic Explorer, LISA), as well as astroparticle detectors, such as IceCube or Antares, which will issue real-time multi-messenger alerts across the full sky. Given the need for rapid follow-up of transient events, \wst\ will implement an operational model that would allow to issue \gls{ToO} observations at the telescope or the fiber level. Thanks to rapid data processing and analysis powered by machine learning, \wst\ will both \emph{issue} alerts (e.g. spectroscopic variability) and provide  spectroscopic observations of high-value transients reported by other facilities. 
\end{itemize}

\begin{figure}
  \centering
  \includegraphics[width=\textwidth]{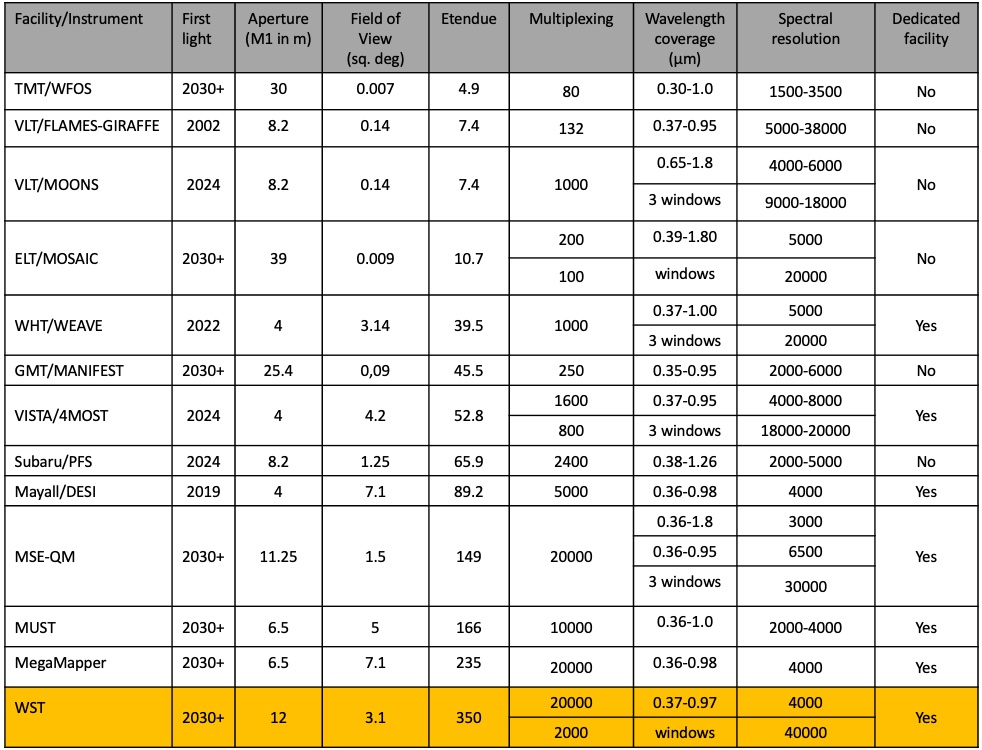}
  \caption*{\small Table 2: Summary of the main properties of current and upcoming MOS instruments. The list is sorted by etendue (i.e., aperture times field of view area). \wst\ MOS has the largest etendue, and combines a wide field of view, high multiplexing and a wide range of spectral resolution.}
  \label{fig:MOS}
\end{figure}

\begin{figure}
  \centering
  \includegraphics[width=\textwidth]{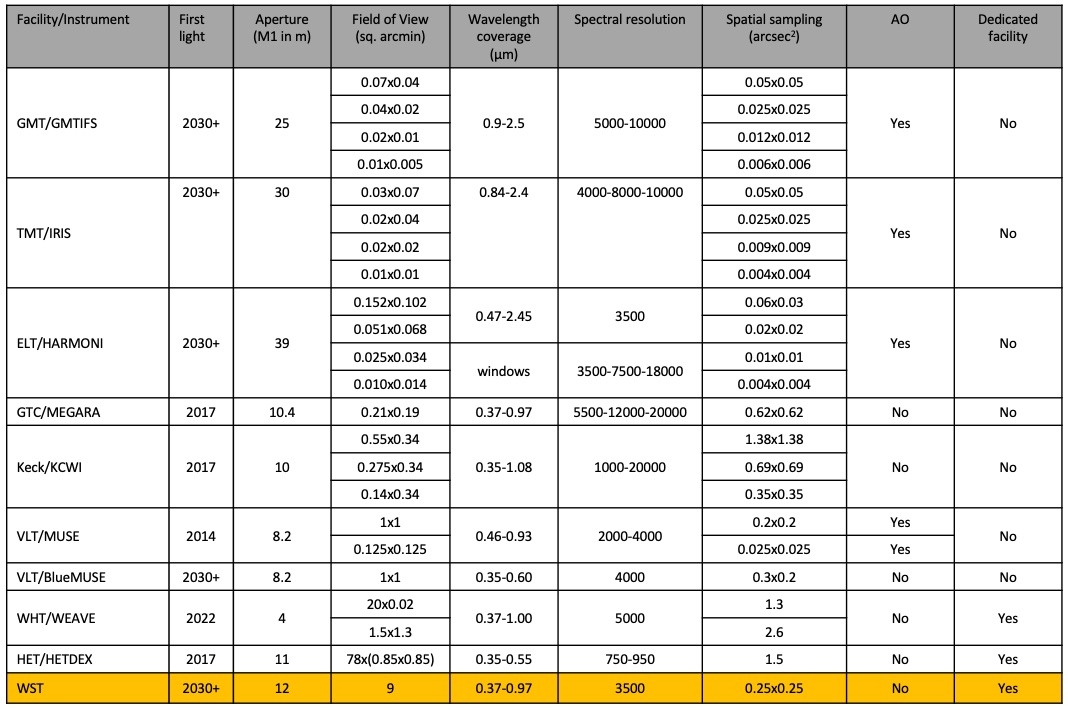}
  \caption*{\small Table 3: Summary of the main properties of current and upcoming IFS. The list is limited to panoramic (monolithic) IFS and sorted by the field of view.}
  \label{fig:IFS}
\end{figure}

\begin{figure}
  \centering
  \includegraphics[width=\textwidth]{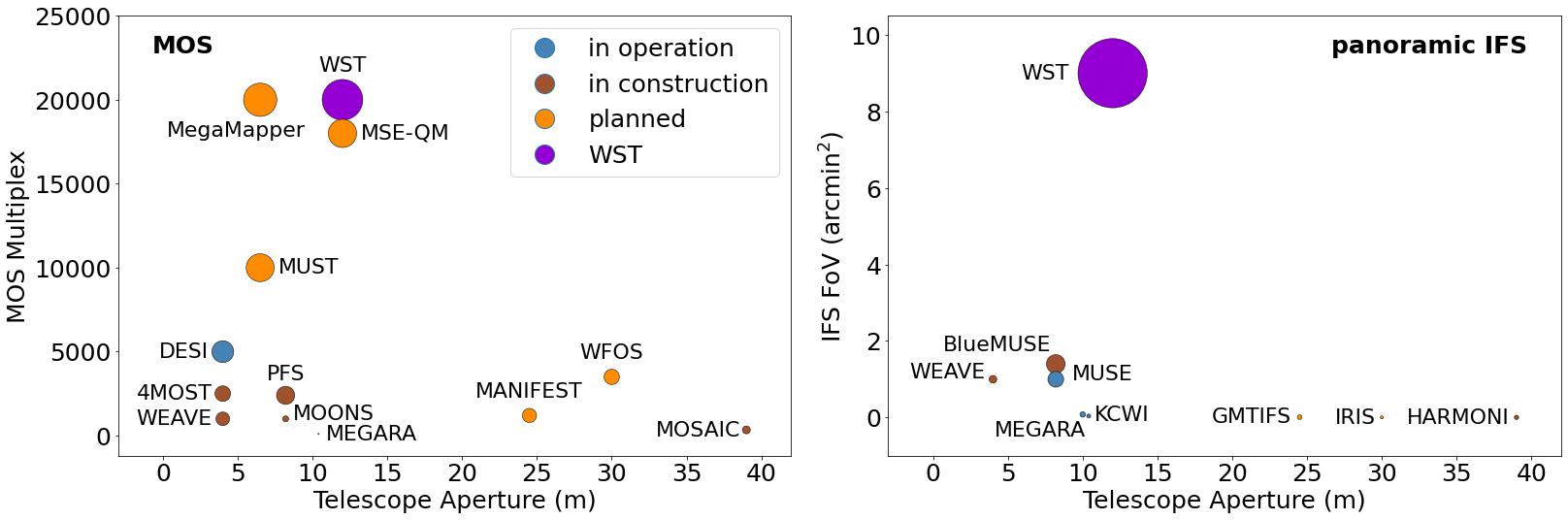}
  \caption{\small Comparison of \wst\ MOS (\textbf{left panel}) and IFS (\textbf{right panel}) capabilities with existing and proposed ground-based spectroscopic facilities. Circle areas are proportional to the etendue (i.e., aperture times field of view area). Note that for clarity, MOS or IFS with small multiplex or field of view are not shown in these figures. Multi-IFUs (e.g. KMOS, MOSAIC, Hector or MaNGA) or sparse-field panoramic IFS (e.g., HETDEX) are not shown in this comparison, which only considers panoramic (monolithic) IFS with $100\%$ fill-factor. However, multi-IFUs are considered as a possible upgrade to the MOS components (see Section \ref{intro:upgrade}).}
  \label{fig:Intro:MOS_IFS}
\end{figure}

\subsection{WST and other MOS and IFS facilities}

As described in the previous sections, \wst\ combines two very complementary approach to spectroscopic surveys: multi-object spectroscopy (MOS) based on fibre technology and panoramic integral-field spectroscopy (IFS) using glass slicers. Table 2 and Table 3 summarize the main characteristics of the current and upcoming MOS and IFS facilities, respectively. 

For the MOS, it is common practice to quantify their potential in terms of the so-called "survey speed" which is proportional to the telescope aperture, the field-of-view of the instrument, its multiplexing and the available observing time. In the left panel of Figure~\ref{fig:Intro:MOS_IFS} we compare three of these parameters (telescope aperture, field-of-view, multiplexing) of \wst\ and other proposed large-area MOS (MegaMapper and \mse) with current and upcoming MOS. It is clear from this figure the order of magnitude jump in survey speed represented by \wst. For example, current and upcoming dedicated 4-m facilities like \fourmost\ and \weave\ will certainly represent a powerful complement to \gaia\, but it will be only thanks to the large collecting area of \wst\ that we will be able to do the ultimate spectroscopic follow-up of this astrometric space mission. Upcoming optical/near-infrared MOS at 8-10m class telescopes (\moons\ and \pfs) will be powerful tool to study the physical properties of galaxies at z$>1$ and their connection with the environment, but their relatively small field-of-view (especially for \moons), small multiplexing and the fact that they have to share the telescope with other instruments will prevent them to be able to sample large fractions of the Cosmic Web with enough sensitivity and sampling to fully characterize the physical properties of galaxies in a wide range of environments at Cosmic Noon. \desi\ has been able to achieve a tremendous efficiency in collecting redshifts for large samples of galaxies already in the first years of operations to constrain Dark Energy models using BAO measurements, but \wst\ will be able to crucially extend these studies at higher redshifts, z$>2$, to probe neutrinos, inflationary models and modified gravity. \megamapper\ is a proposed new facility with a combination of large field-of-view and high multiplexing on an 8m class telescope, to be devoted to cosmological studies. \wst\ includes all the capabilities of \megamapper\ and in addition span a much wider range of science topics thanks to its multiple spectral resolution modes and simultaneous IFS operations. The Mauna Kea Spectroscopic explorer \citep[MSE]{MSE_Science,MSE_Instrument} has several science cases similar to \wst\ and it may be seen as a complementary facility located in the North hemisphere. A part from the geographical location, the complementary is also reflected in their instrumentation: the scientific potential of \wst\ is uniquely increased by the IFS at first light; \mse\ foresees an extension to the near-infrared (NIR) since first light while for \wst\ the NIR is considered as a possible upgrade (see Section \ref{intro:upgrade}).

The right-panel of Figure~\ref{fig:Intro:MOS_IFS} compares the field-of-view and telescope aperture of the \wst-IFS and other IFS currently in operation or planned. Note that multi-IFUs (e.g. KMOS, MOSAIC, Hector or MaNGA) are not shown in this comparison chart, which only considers panoramic (monolithic) IFS. However, multi-IFUs are considered as a possible upgrade to the MOS components of \wst\ (see Section \ref{intro:upgrade}). The limitation of most current or planned IFSs is their relatively small field-of-view. Using the etendue metric, the WST IFS will be 19 and 162 times faster than \muse\ and \kcwi\, respectively, which will allow to extend the plethora of science driven by IFS blind spectroscopic surveys to much larger patches of the sky.  

Finally, each of the future giant class telescopes (ELT, TMT, GMT) will have MOS and monolithic IFS in their suite of instruments. For technical feasibility they will be limited to very small field of view and consequently multiplexing (e.g. a multiplex of about 300 over a 0.01 square degree for MOSAIC/ELT, and a monolithic IFS of at most 0.02 square arcmin for HARMONI/ELT). They should therefore be regarded as highly complementary to \wst, which would be able to provide them follow-up targets.

\subsection{Synergies between \wst\ and other astronomical facilities}

\begin{figure}
  \centering
  \includegraphics[width=\textwidth]{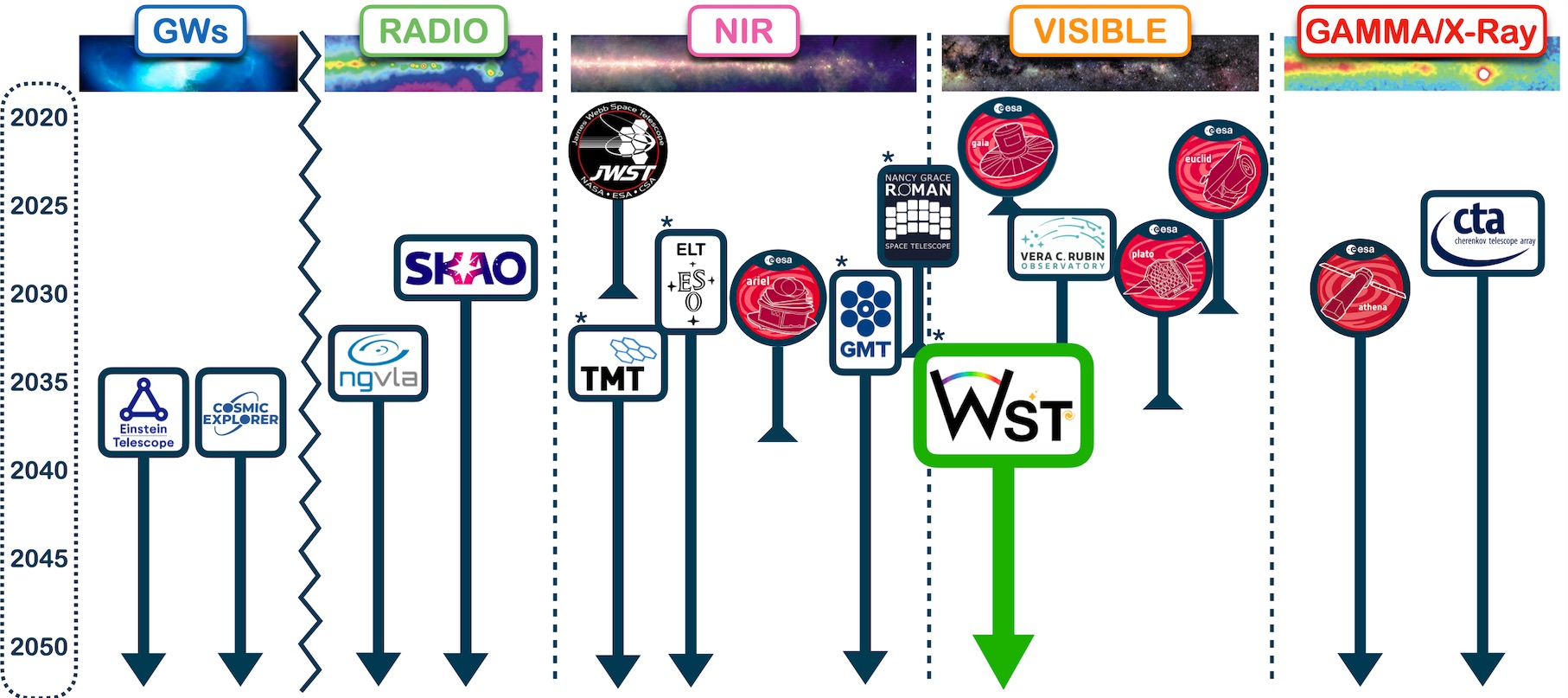}
  \caption{\small A graphical representation of the current and upcoming major astronomical facilities. The facility marked with an asterisk (e.g., ELT, TMT, GMT) have been listed in the spectral range they cover the most for visual purposes, but they also partly cover other bands. The duration of space missions reflect the publicly available nominal values. \wst\ will have strong synergies with many of those, filling in a gap in the current landscape.}
  \label{fig:Intro:Synergies}
\end{figure}

As detailed in this white paper, \wst\ has the potential to make a significant scientific impact on its own. In addition, \wst\ will fill a gap in the astronomical landscape and will have strong synergies with existing or forthcoming large ground-based and space-based facilities greatly
enhancing their scientific capabilities (see Figure~\ref{fig:Intro:Synergies}). In the following, we present a few examples of such synergies.

\subsubsection{The Gaia revolution}

\emph{Gaia} is an ESA astrometric space mission launched at the end of 2013 with the main aim to understand the formation and consequent evolution of the Milky Way galaxy, mapping the motions, luminosity, temperature and composition of more than a billion stars. The final data release of \emph{Gaia} is expected around 2030. By then, \emph{Gaia} will have provided photometry ($BP$, $RP$, and $G$), positions, parallaxes, and proper motions for more than 10$^{9}$ Galactic stars down to a magnitude $G$ = 20.7 mag \citep[see][for the contents of the data release 3 of \emph{Gaia}]{GaiaDR3}. \emph{Gaia} will also provide radial velocities (RVs) with precision between 1-15 km s$^{-1}$ for more than 150 million stars brighter than $G$ = 17. Chemical abundances of up to 13 species will be delivered for more than 2.5 million stars, mostly brighter than $G$ = 14 mag. \emph{Gaia} is already revolutionizing our understanding of the Milky Way evolution \citep[e.g.,][]{Belokurov2018,Helmi2018,Ibata2020}.

By mid of the next decade, on-going and upcoming spectroscopic surveys will have accumulated spectra for $\sim 50\times 10^{6}$ \emph{Gaia} stars but the vast majority at medium (R$<10,000$) spectral resolution. This will further enlarge the legacy of \gaia\ providing detailed kinematics for hundreds of millions of stars. But, even an accurate knowledge of their kinematics is not sufficient to trace these stars back to their initial position, due to the many unknowns, including the time variation of the Galactic potential, and transient structures such as bar and spiral arms. So, it is not possible to reassemble stellar associations now dispersed through the Galaxy with kinematics only. On the other hand, it is also well known that the chemical makeup of a star provides the fossil information of the environment where it formed. Under this premise, it should be possible to use chemical abundances to tag stars that formed within the same stellar association, if that association shows a recognisable pattern. Therefore, chemistry and kinematics together can maximize our capability to search for missing structures in the Galaxy, aiming at reconstructing the star formation history of the Galaxy, using the so-called {\it “chemical tagging”} \citep{FreemanBlandHawthorn2002}. It is crucial that the chemical information covers all nucleosynthesis chains, from $\alpha$ to neutron-capture elements when trying to illuminate how our Galactic components (bulge, halo, thick, and thin disk) formed and subsequently evolved. In addition, since the chemical differences between different structures/populations of stars are small, a high precision ($\sigma<0.05$ dex) will be required. The combination of R$=40,000$ spectral resolution ,  high-multiplex and large collecting area make \wst\ uniquely positioned to combine the chemistry and kinematics of the stellar populations in the Milky Way and finally understand the full formation history of our Galaxy.

\begin{figure}
  \centering
\includegraphics[width=\textwidth]{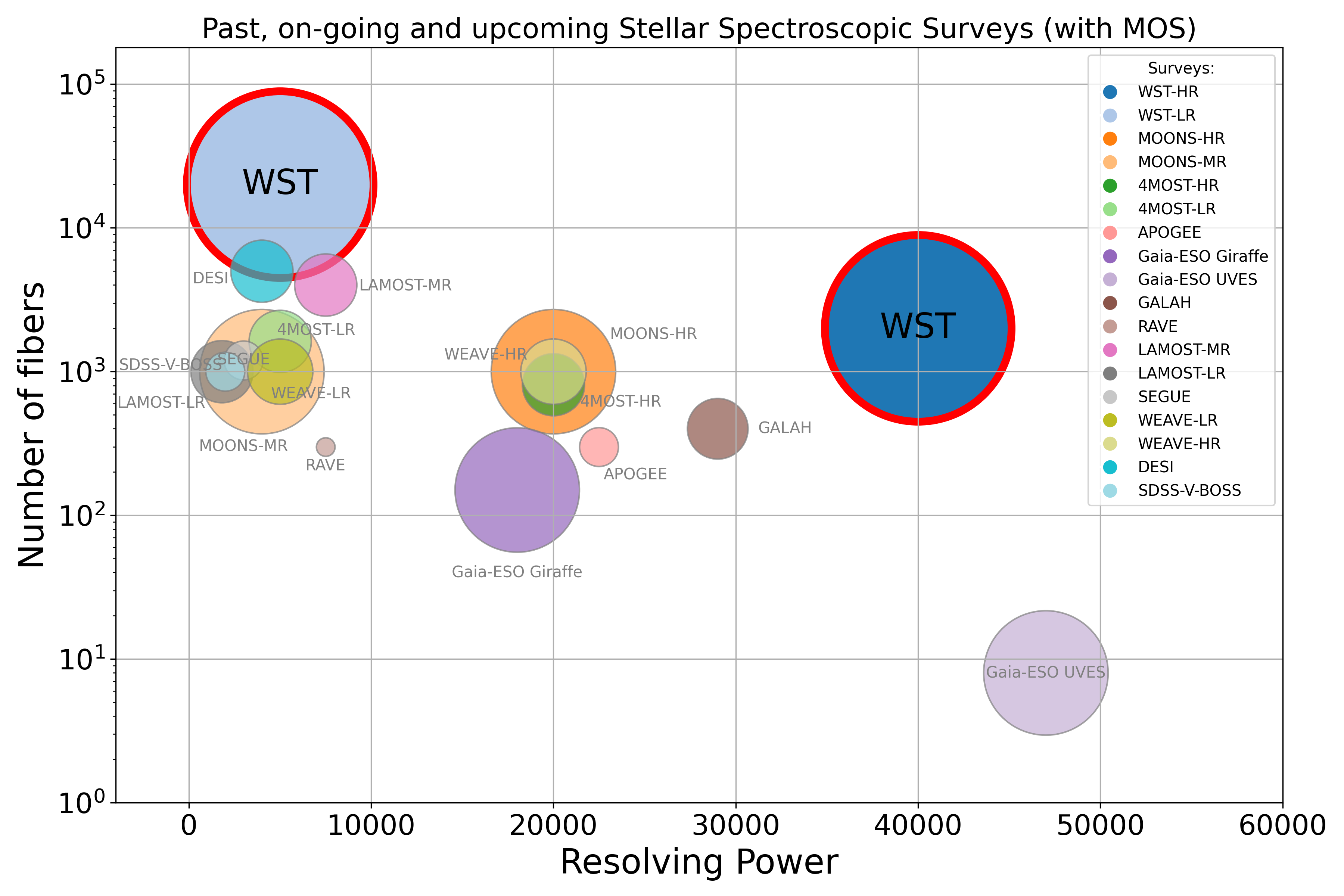}
  \caption{\small : \wst\ MOS (HR and LR) in the panorama of current  and upcoming stellar spectroscopic surveys. We have limited the comparison to surveys on already existing facilities. The size of the symbol is proportional to the collecting area of the telescope. \wst\ has a unique combination of multiplexing, spectral resolution and collecting area both at low and high spectral resolution.}
  \label{fig:Gaia}
\end{figure}

\subsubsection{Wide field imaging from space}

In the coming years, two new space missions will provide 0.2 arcsec imaging data over very large portion of the sky. The ESA \euclid\ mission features a 1.2m diameter telescope and two instruments covering the 0.55-1.2 $\mu$m wavelength range. Euclid was launched in July 2023 and during its nominal lifetime of six years will survey 15,000 deg$^2$ providing morphologies and photometric redshifts for billions of galaxies. The \nancy\ (formerly known as WFIRST) is a NASA-led space mission scheduled for launch in 2027. It will feature a 2.4m diameter telescope equipped with a wide-field instrument covering the 0.48-2.3 $\mu$m wavelength range. This rich imaging data-set will have numerous synergies with a dedicated spectroscopic facility like \wst. One key area for \wst\ will be the study of the physical properties and evolution of galaxies in the context of their environment at the peak epoch of star-formation. The large multiplexing, spectral resolution and sensitivity of \wst\ will allow to reconstruct the network of matter in the so-called {\it cosmic web}. It will be possible to trace these large scale structures both in emission as well as tracing the gas in absorption (i.e., IGM tomography). At the same time, high S/N ($\sim 30)$ spectra with appropriate resolution (R$>3000$) for $2<z<3$ galaxies down to i$_{\rm AB}\sim 25$ mag will allow to physically characterize the properties of the galaxies living in these structures (e.g., SFR, inflows and outflows of baryons). A key element to implement such a study with \wst\ will be a carefully selected galaxy sample covering the redshift of interest using the photometry provided by Euclid and Nancy Roman.

\subsubsection{Time-domain astrophysics\label{landscape:TD}}
Astronomers are passive observers who cannot control the behavior of the objects they are seeking to understand. However, the astronomical objects are not static, and the dimension of time creates the crucial ability to investigate how astronomical objects \emph{change}. Observed changes, or variability, provide invaluable information on astrophysical processes and enable a plethora of further applications of relevance for all other science directions in this white paper. Thanks to very significant investments in large imaging surveys from the Ground and from Space, astronomy is entering a golden era for studying variability-related phenomena. The Vera C. Rubin Observatory's Legacy Survey of Space and Time (henceforth: \lsst) at Cerro Pach\'on in Chile, will acquire low-cadence multi-band \emph{video} of the sky at an approximate frame rate of 0.33/day. While \lsst\ will already have completed its first project phase, its next phases would likely consider and build on synergies with \wst\ (Z. Ivezic January 2024, priv. comm.). Moreover, lessons learned as part of \lsst's transients and variable stars science collaboration (LSST-TVSSC\footnote{\url{https://lsst-tvssc.github.io/}}) and from \lsst's first phase would inform scheduling and planning of variable phenomena with \wst. A useful current project that underlines the synergies of time-domain photometry and large-scale spectroscopic survey is the array of 40cm telescopes TMTS \citep{TMTS,TMTS_variability_I} that targets variability of sources observed by LAMOST \citep{LAMOST2022}. At present, large spectroscopic surveys, such as \fourmost, \weave, and SDSS-V provide the best opportunities for collecting spectra of many variable sources. However, these facilities remain limited in terms of scheduling, number of targets, and the depth that can be observed, which becomes a problem for alerts issued by \lsst\ due to limited overlap in magnitude. An analogous situation arose in the late 2000s, when the {\it Kepler} spacecraft reported a huge number of exoplanet candidates that could not be characterized due to a lack of spectroscopic follow-up capacity. While even \wst\ cannot remedy this situation fully in the \lsst\ era, it is the currently best suited facility capable of providing the crucial spectroscopic follow-up of a sizeable fraction of faint variable sources over large regions of the sky.

Multi-messenger astronomy will be mainstream in the \wst\ era, and entire facilities dedicated to rapidly evolving phenomena, spanning energy ranges from  TeV (CTE) through $\mu$eV (\textit{\gls{SKA}}). The \textit{\gls{SKA}} will issue time-sensitive alerts at sub-mm wavelengths also from the Southern Hemisphere, while the successors of current gravitational wave observatories (LIGO, Virgo, KAGRA, etc.), such as the Einstein Telescope, Cosmic Explorer, and ESA's LISA mission, as well as astroparticle detectors, such as IceCube or Antares, will issue real-time multi-messenger alerts across the full sky. \wst\ will be located at a strategic location, allowing it to play a crucial role in this era of multi-messenger astronomy, which has been identified as a key priority for the 2030s and 2040s by all major planning exercises (including the US 2020 Decadal Survey, the UK STFC Roadmap, and the Astronet Roadmap). \wst's unmatched ability of providing time-resolved optical spectroscopy for a large number of faint objects will render this facility truly irreplaceable. MSE would create wonderful synergies by sampling the Northern hemisphere complement. Given the need for rapid follow-up of transient events, \wst\ will regularly issue \gls{ToO} observations at the telescope or the fiber level. Thanks to rapid data processing and analysis powered by machine learning, \wst\ will both \emph{issue} alerts and provide  spectroscopic observations of high-value transients reported by other facilities. Considering the complexity of the data and their volume, this will further present exciting new challenges for big data analysis.

\subsubsection{Synergies with future radio facilities\label{Intro:radio}}

To complete this overview on synergies between \wst\ and other major facilities we cannot forget the upcoming changes at radio frequencies. The \gls{SKA} is a new-generation radio telescope which will operate on a wide range of frequencies (50 MHz-15 GHz) and when completed it will have approximately a total collecting area of one square kilometer. The \textit{\gls{SKA}} will be implemented in two phases: \textit{\gls{SKA}} Phase-1 (SKA1) will have $\approx 10\%$ of the total collecting area and will cover both LOW (50-350 MHz) and MID (350 MHz - 15 GHz) frequencies and it should be operational by 2030. \textit{\gls{SKA}} Phase-2 with the full collecting area and frequencies coverage will probably be in operation in the second half of next decade. With the current proposed array design (subject to change), a 1-hour integration time, the SKA1-LOW and SKA1-MID arrays are projected to operate at rms sensitivities of 82.48 $\mu Jy/beam$ at 200 MHz up to 11 $\mu Jy/beam$ at 1.4 GHz (from the official \textit{\gls{SKA}} sensitivity calculator\footnote{https://www.skao.int/en/ska-sensitivity-calculators}). Further, with a 2-year integration time for an all-sky survey at 1 GHz, SKA1 is anticipated to achieve a 3 $\mu Jy$ rms sensitivity, detecting approximately 4 galaxies per arcmin$^{2}$ (at an object detection threshold of S/N $>$10) and over 0.5 billion radio sources. In addition, in Band 1 SKA1$-$MID can in principle detect neutral hydrogen (HI) and OH masers line in emission out to a redshift of z $\sim$2 \citep[][]{Verdes2015} and absorption up to z$\sim$6 \citep[][]{Morganti2015,Wagh2024}, as well as CS [1- 0], CO[1- 0], HCN[1- 0], HCO+[1- 0] extragalactic lines in Band 5 up to z$\sim$8 \citep[][]{Wagg2015} and trace the cold molecular gas reservoirs in star-forming galaxies \citep[][]{Carilli2013}. The imaging surveys will cover an effective \gls{FoV} per pointing of 5.45 deg$^{2}$ at 11 arcsec resolution at 110 MHz, 1 deg$^{2}$ at 0.4 arcsec resolution at 1.4 GHz, and 6.7 arcmin$^{2}$ at 0.04 arcsec resolution at 15 GHz. An SKA1-MID band 2 survey will allow the detection of SFR $>$ 10${M}_\odot$/yr galaxies out to $z\,{\sim}$\,4  in ultra-deep survey tiers (area of $\sim$1\,deg$^2$) and galaxies with ULIRG-like SFRs ($\geq$100\,${M}_\odot$/yr) out to similar redshifts even in a wide survey tier covering of order 1000\,deg$^2$ \citep[][]{Prandoni2015,Coogan2023}. 

Towards the end of the 2030s, the planned northern hemisphere \gls{ngVLA} observatory will additionally provide sensitive continuum and spectral imaging capabilities at $\sim1-116$ GHz down to equatorial and somewhat sub-equatorial fields also covered by \wst. Science with the \textit{\gls{ngVLA}}  will focus on thermal gas/dust emission on milli-arcsecond scales and redshifted molecular line emission.

\noindent
There are several synergies between WST and these future radio facilities, as examples:
\begin{itemize}
    \item \wst-MOS will be uniquely placed to understand the spectral properties and redshift distribution of the new population of sources that will be discovered by the \textit{\gls{SKA}}, therefore providing valuable information about large-scale structure formation, galaxy evolution, and the origin of the large-scale coherent magnetic fields in galaxies over cosmic times.The MOS capability of \wst\ will also allow to compare the distribution of dark matter against that of baryonic matter in the intra-cluster medium, as mapped by the \textit{\gls{SKA}}. This will enable to investigate the interaction between dark and baryonic matter and its influence on the dynamics and distribution of matter within the cosmic-web \citep[]{Pommier2016}.
    \item \wst-IFS will provide the ionized gas phase kinematics of the inter-stellar medium of galaxies, complementing the neutral hydrogen component traced by \textit{\gls{SKA}}. We will be able to perform spatially resolved, multi-phase gas kinematics analysis on a wide range of galaxies and redshifts. This will be crucial to understand how gas flows into galaxies,  is consumed by star formation and regulated by nuclear activity, via feedback mechanisms, and recycled into new stars, and how all these processes depend on galaxy properties and their environments.

\end{itemize}

\subsection{Future upgrades\label{intro:upgrade}}

The anticipated lifespan of \wst\ will cover several decades, as for many others ground-based facilities. It is therefore crucial to anticipate long-term upgrades in the telescope, instruments and data processing capabilities. These upgrades could enhance the scientific capabilities of planned surveys, or address potential new “questions”. Possible upgrades include the extension to the NIR, the incorporation of deployable multi-IFUs, and the addition of a medium resolution MOS mode. 

The telescope's spectral coverage extends from the UV to the H-band (350 - 1600 nm) while the current baseline top-level requirement for the \wst\ instrumentation is constrained to the UV-visible range (370 - 970 nm, see Table 1). This decision is driven by two main factors: a) the economic aspect, as the current cost of the few hundred of IR astronomical detectors required would be prohibitively expensive; b) the feasibility of achieving the required high accuracy of sky subtraction with MOS fibre-fed spectrographs on a large telescope like \wst, particularly in the NIR range, has yet to be demonstrated. Nevertheless, current technological constraints may become obsolete in the future with ongoing advancements. Similarly, we will learn more on the accuracy achievable at those wavelengths using two upcoming instruments (\moons\ at the \vlt\ and \pfs\ at Subaru).

Another possible upgrade could concern the use of deployable multi-IFUs, which could open new science capabilities, e.g. investigating the alignment of the galaxy spin axis with respect to the orientation of the filament within which it resides. The most probable path to this new mode will be to re-use the existing spectrographs, it would be therefore important in the core design decisions on the \wst\ Facility not to preclude such possibility, e.g. not having fibre connectors.

The possibility to have in the future a medium resolution (R=10-20\,000) MOS mode will be explored. This new mode within \wst\ would allow to reach 0.1--0.2km/s accuracy radial velocities to much greater depth than the current generation of surveys will provide (down to G$\sim$20-21, while \fourmost\ or \weave\ are planning to reach G$\sim$15-16 at R=20\,000). This would enable {\it for the first time} extreme precision Galactic dynamics in large volumes, including tests of gravity and dark matter content/clumpiness of the Milky Way halo. Such a mode would also allow good quality chemical information in much larger volumes than any survey so far, including for example dwarf and ultra-faint galaxies, and stellar streams down to their main-sequence members. In such a mode, a pointed survey of thousand(s) of deg$^2$ probing areas of specific interest (dwarf galaxies, Magellanic Clouds, streams) at intermediate resolutions (R=10-20\,000) would allow to fulfil these important science cases.

\clearpage

\clearpage

\section{Time Domain\label{sec:timedomain}}
\paragraph{Authors} Richard I. Anderson$^2$, Paula S\'anchez S\'aez$^1$, Cyrielle Opitom$^9$, 
Marc Audard$^{18}$, 
Patricia Ar\'evalo$^{17}$,
Amelia M. Bayo Aran$^1$, 
Sofia Bisero$^{26}$, 
Susanna Bisogni$^{27}$, 
St\'ephane Blondin$^{29,1}$, 
Rosaria Bonito$^{30}$, 
Marica Branchesi$^{33}$, 
M\'arcio Catelan$^{42,43,44}$, 
Filippo D'Ammando$^{40}$,
Maria G. Dainotti$^{52}$, 
Annalisa De Cia$^1$, 
Suhail Dhawan$^{54}$, 
Valentina D'Orazi$^{31},50$, 
Ulyana Dupletsa$^{33}$, 
Nandini Hazra$^{33,95}$,
Dragana Ili\'c$^{60,60}$,
Valentin D. Ivanov$^1$, 
Luca Izzo$^{64}$,
Darshan Kakkad$^{65}$,
Andjelka B. Kova\v cevi\'c$^{60}$, 
Hanindyo Kuncarayakti$^{67,68}$, 
Yuna Kwon$^{69}$, 
Fiorangela La Forgia$^{45}$, 
Monica Lazzarin$^{45}$, 
Paulina Lira$^{74}$,
Eleonora Loffredo$^{33,95}$, 
Kate Maguire$^{75}$,
Fatemeh Z. Majidi$^{77}$, 
Marcella Marconi$^{64}$, 
Sean McGee$^{80}$, 
Alessandra Migliorini$^{85}$, 
Dante Minniti$^{86,87,88}$,
Matt Nicholl$^{94}$, 
Francesca Onori$^{95}$, 
Anna F. Pala$^{96}$, 
Swayamtrupta Panda$^{97}$, 
Micha\l\ Pawlak$^{99}$, 
Priscila J. Pessi$^{15}$, 
Luka \v C. Popovi\'c$^{102}$,
Roberto Raddi$^{103}$, 
Alberto Rebassa-Mansergas$^{103}$, 
Martin M. Roth$^{10}$,
Graham P. Smith$^{80}$,
Susanna D. Vergani$^{26}$, 
Giustina Vietri$^{27}$,
Quanzhi Ye$^{117,118}$,
Tayyaba Zafar$^{39}$

\subsection{Introduction}

\wst\ will provide unmatched spectroscopic capabilities in the era of panchromatic all-sky video (cf. Sect. \ref{landscape:TD}) thanks to its wide \fov, large collecting area, and high multiplexing, all of which mitigate necessary trade-offs between survey area, depth, and cadence. This era will come as a natural progression from the first all-sky surveys, the {\it Durchmusterungen} \citep{BonnerDurchmusterung,CapeDurchmusterung,CordobaDurchmusterung}, initiated in the mid 19th century. The next logical step in this 200-year progression is the addition of spectroscopic resolution, which could be done within the next 20 years to yield crucial information concerning the physics of astronomical objects. 

The defining feature of the time-domain section is that the timing of the observation is crucial to the ability to interpret the phenomenon of interest, since time-variable phenomena can occur on any physical scale, spanning Solar System Science to Cosmology, and to fundamental physics, and are thus relevant for all other science cases presented in this document. As we are interested in maximizing the information offered by time-resolved observations, we are both considering the scientific interests as well as any operational requirements and opportunities. 

\begin{figure}
    \centering
    \includegraphics[width=1\textwidth]{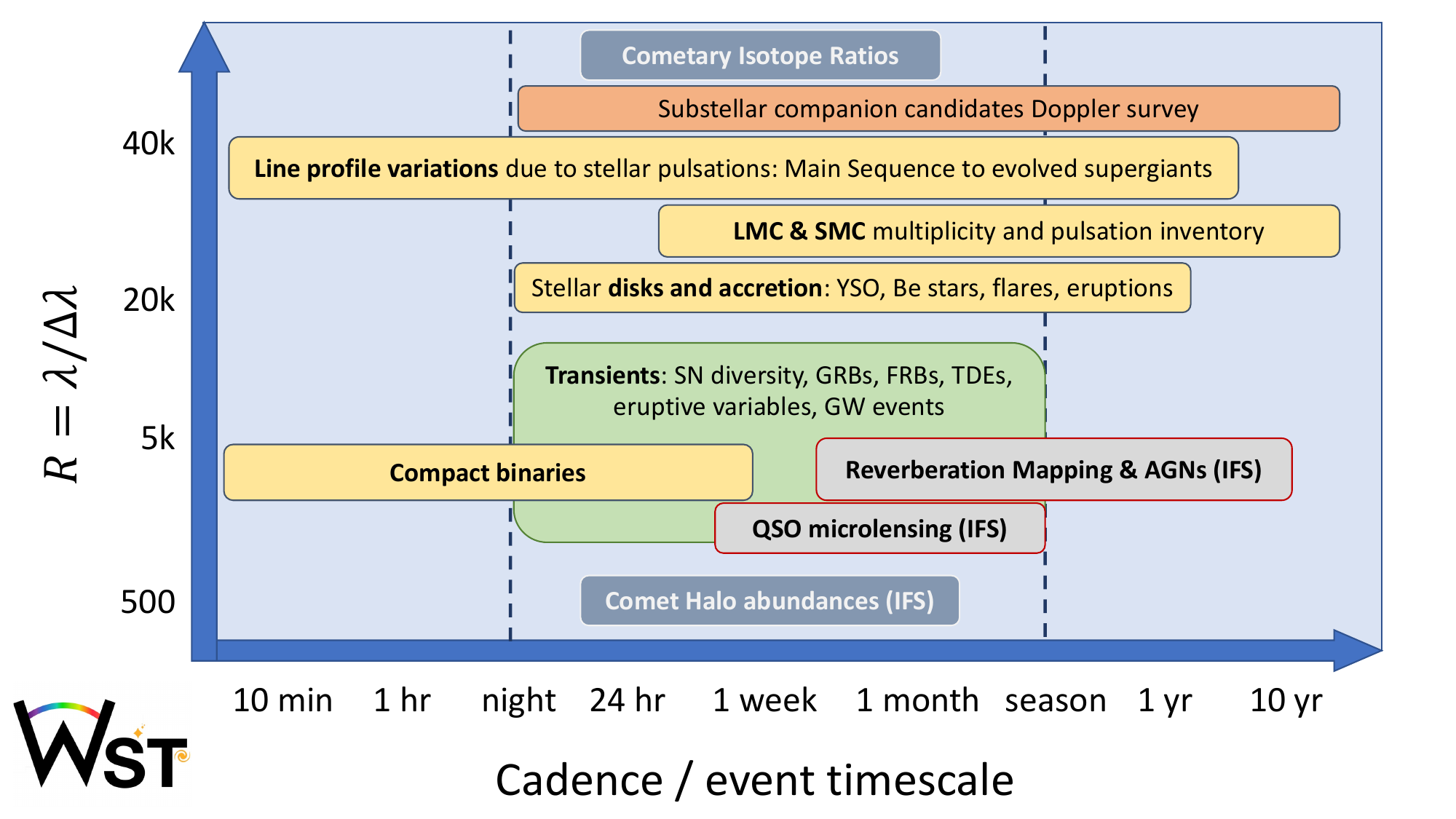}
    \caption{Sketch of time-variable phenomena that \wst\ will observe alongside approximate indications of suitable spectral resolution and variability timescales.}
    \label{fig:td:res_timescale}
\end{figure}

The variations considered here may occur on very different  timescales, mainly between hours and years, and can appear suddenly, sporadically, or periodically, and may be repeating or be observable only at very specific times (e.g., supernovae or comets). Given a plethora of phenomena, different spectroscopic resolving power is required to investigate these phenomena. Figure\,\ref{fig:td:res_timescale} provides a schematic overview of this. Moreover, the time domain opens new windows for serendipitous discoveries that we seek to optimize by considering from the start a survey strategy that maximizes the ability to collect time-resolved information. In this way, \wst\ seeks to enable new and exciting avenues for investigating time-variable phenomena on all scales, from the Solar system to Active Galactic Nuclei (AGN).
\wst's simultaneous IFS and MOS capabilities will provide unprecedented opportunities for studying time-resolved phenomena in sky regions marked by both high and low stellar densities, such  globular clusters, nearby galaxies, and in the search for electromagnetic (EM) counterparts to gravitational wave (GW) events.

\wst's specific strength is its ability to collect high-quality and homogeneous (repeat) spectroscopic observations of a very large number of variable objects, which can be further followed-up using either \wst\ or other facilities. Where its high multiplexing or simultaneous MOS+IFS operations are required, \wst\ will also be used for telescope-level target of opportunity (T-ToO) observations that would be disruptive to other surveys. However, in the majority of cases, fiber-level ToO (F-ToO) will suffice to collect precious data on variable sources within each pointing of \wst, which shall be kept flexible enough to maximize the ability to observe high-value transient phenomena. \wst\ will therefore play a unique role in classifying objects, building up statistics, creating training sets, and sorting or filtering specific objects for further study. With this approach, \wst\ will retain the crucial ability to detect exceedingly rare events of particular importance. 

This section presents a necessarily incomplete overview of possible time-domain areas, which will be expanded upon in future work. It is organized as follows: Section\,\ref{td:GW} presents multi-messenger observations of GW events; Sect.\,\ref{td:AGNs} the variability of active galactic nuclei; Sect.\,\ref{td:transients} the field of astronomical ``transients''; Sect.\,\ref{td:stars} stellar variability; Sect.\,\ref{td:solarsystem} Solar system objects. The final Sect.\,\ref{td:serendipitous} briefly discusses serendipitous opportunities and the need for \wst's own alerts system.

\subsection{Electromagnetic counterparts of gravitational wave events\label{td:GW}}

Multimessenger astronomy has been identified as a priority in most major astrophysics planning exercises and will be mainstream in the \wst\ era.  
In particular, the next generation GW observatories Einstein Telescope \citep[ET]{Maggiore2020} and Cosmic Explorer \citep[CE]{CE} are expected to revolutionize  the future of multimessenger astrophysics \citep{ETscience2023}, e.g., by detecting $\sim 10^5$ binary neutron star (BNS) mergers per year, reaching redshifts well beyond the star formation peak.
Importantly, ET's access to low frequencies will enable the detection of BNSs well before their merger and, in turn, to follow their inspiral  for up to several hours in the case of nearby events. This leads to the exciting possibility of catching BNS mergers during the prompt emission that will be particularly informative of the extreme physics during this phase.

\wst's contributions to the follow-up of GW sources fall into two main categories. On the one hand, \wst's extragalactic and cosmology surveys will create an unprecedented catalog of galaxy redshifts out to $z < 2$, cf. Sect.\,\ref{sec:legacy_survey}, allowing the correlation with a majority of GW events, even without optical counterparts. On the other hand, GW events with EM counterparts require disruptive T-ToO observations with rapid reaction times. The following focuses on events where time-critical observations with \wst\ are required: BNS coalescence events leading to kilonovae (KNe) (Sect.\,\ref{td:kilonovae}), the application of KNe as tracers of cosmic expansion (Sect.\,\ref{td:H0}), and gravitationally lensed EM counterparts to gravitationally lensed BNS (Sect.\,\ref{td:lensedBNS}). Dark sirens (without EM counterparts) do not require time-critical observations and are discussed in Sect.\,\ref{cosmo:GW}.

\subsubsection{GW events with electromagnetic counterparts, kilonovae\label{td:kilonovae}}
ET is expected to detect approximately $10^5$ BNS systems per year, reaching redshifts well above the star formation peak ($z \sim 3$). 
Beyond the local Universe, EM counterparts are expected to be quite faint and have rather large sky localization regions. Target of Opportunity observations by \lsst\ will provide useful photometric constraints for the exact localization of the EM counterparts of GW events  \citep{Andreoni2022,Smith2023}, as well as a plethora of non-GW EM alerts within the same \fov. The correct identification and characterisation of the GW EM counterparts requires spectroscopic observations and galaxy redshift catalogs, as shown spectacularly by the discovery of GW170817 \citep[][and references therein]{Margutti2021}. ET, \lsst, and \wst\ together will  unravel the population of BNS systems, the physics of neutron stars \citep{GW170807-EOSRadii}, and create an incredibly rich astrophysical data set for understanding other astrophysical processes, such as the chemical enrichment of the interstellar medium by r-process elements \citep{Kasen2017,Levan2023}. 

\begin{figure}
    \centering
    \includegraphics[width=1\textwidth]{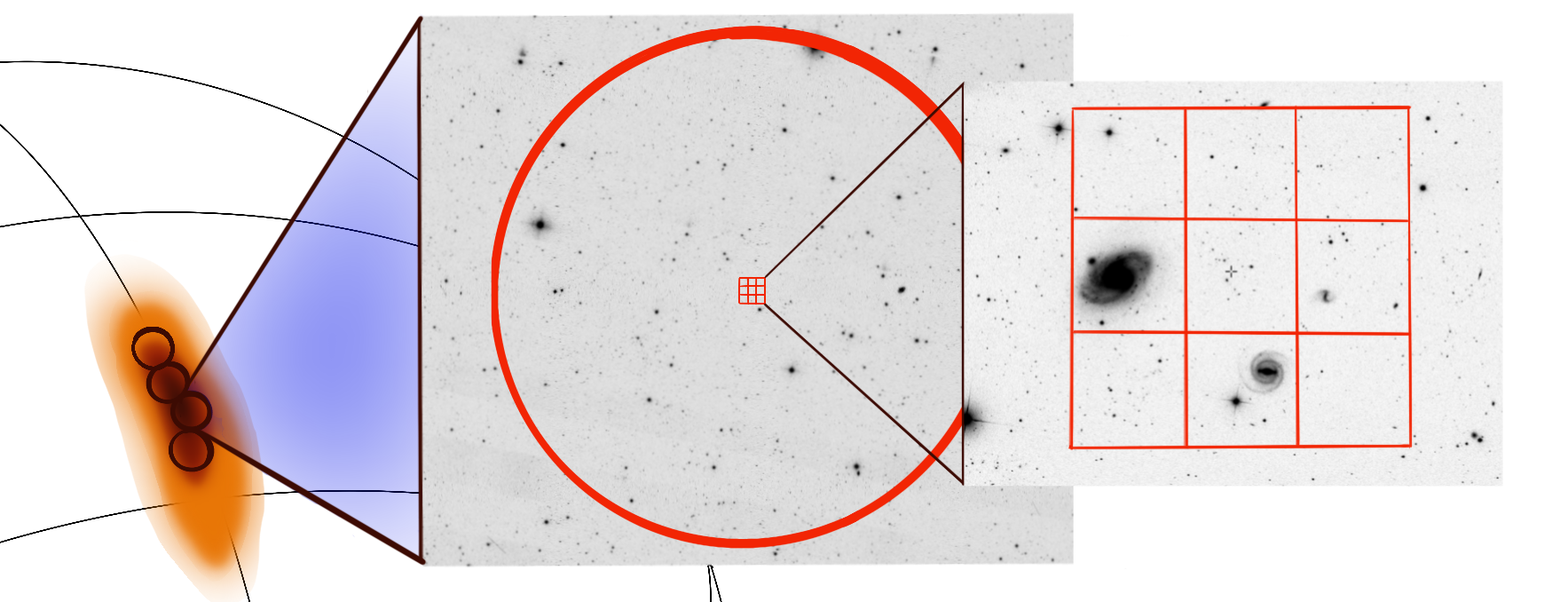}
    \includegraphics[width=0.49\textwidth]{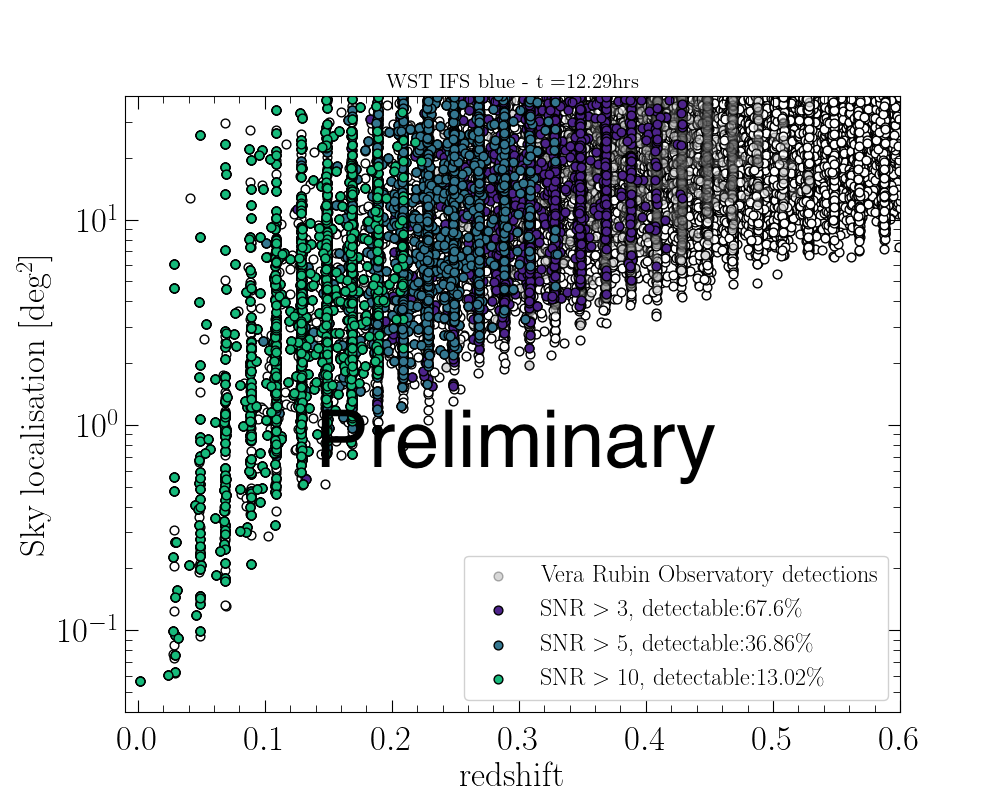}
    \includegraphics[width=0.49\textwidth]{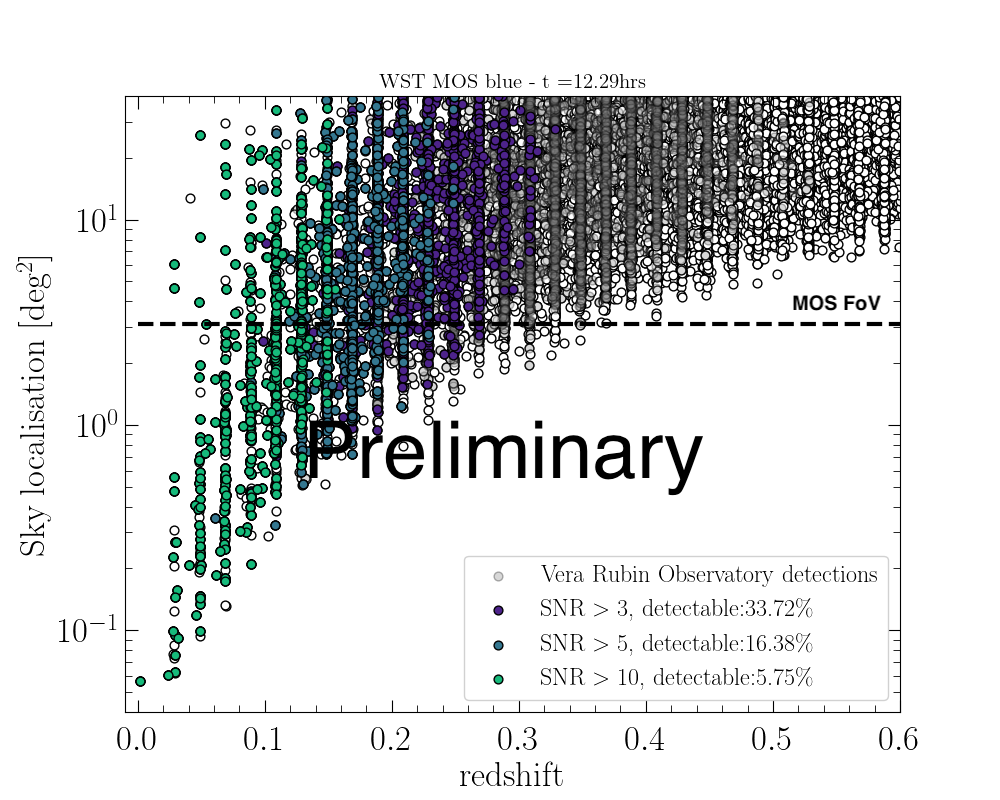}
    \caption{{\it Top:} Conceptual sketch of how \wst's MOS and IFS targeting high probability regions of a BNS event reported by ET. \wst's instrumentation will be superbly suited for the detection and characterization of EM counterparts to GW events, notably kilonovae such as the one associated with GW170817. To enable early-time characterization, the IFS will be placed on high-probability and high-density regions, while the wide-field low-res mode collects spectra of galaxies at distances consistent with that of the BNS event and live EM alerts, e.g., based on targeted \lsst\ observations of the sky localization region. Credits: Bisero/Anderson/Vergani {\it Bottom:} Simulated performance of \wst's capabilities to follow-up KNe. The sensitivity of \wst's IFS and the large \fov\ and high multiplexing of the MOS provide crucial and complementary capabilities to ensure that the GW events will be observed (Bisero et al., in prep.; Loffredo et al., in prep.; developed within the ET OSB.}
    \label{fig:td:ETsketch}
\end{figure}

The simultaneous use of \wst's sensitive IFS and MOS provide transformational capabilities in this game as illustrated in Figure\,\ref{fig:td:ETsketch}, which shows \wst's \fov\ over a sketched sky localization region (``banana'') of a BNS merger observed by ET (top of Fig.\ref{fig:td:ETsketch}), and provides performance predictions (bottom) based on simulations of BNS events provided by the ET Observational Science Board (OSB) Multimessenger division. 
According to these simulations, the sensitivity of the IFS will allow to detect GW EM counterparts at a signal-to-noise ratio (S/N) $> 10$ at $z< 0.2$ provided observations are collected within $\sim 12$ hours of the merger. Over a ten-year survey,  \wst\ would thus characterize on the order of $1000$ events. At lower S/N ( $3 - 5$), the IFS will enable detections out to $z \lesssim 0.4$ for a correspondingly much larger number of events. At the same time, the MOS  will provide the crucial ability to map large parts of the sky localization region, ensuring rapid spectroscopic characterization of EM alerts within that footprint.
Additionally, \wst's IFS will be particularly powerful for nearby KNe, such as the one assiociated with GW170817, that can appear significantly off-center from their host galaxies. 
Together, the IFS and MOS will be unbeatable at ensuring a maximum number of GW EM counterparts are spectroscopically characterized.

Since KNe are brighter and fade slower in the NIR than in the optical, extending \wst's capabilities into the NIR would further increase the volume in which \wst\ could characterize KNe, resulting in a fast increase in number of targets. Additionally, the NIR contains specific emission lines for diagnosing heavy element ejecta \citep{Levan2023}. Nevertheless, a rapid response mode with an optical-only \wst\ would already be transformational for characterizing KNe.

Of course, GW EM counterparts require disruptive T-ToO. However, despite introducing operational challenges, a limited number of T-ToO observations would neither challenge, nor significantly delay, the completion of the other surveys, also because other surveys can be completed during T-ToO observations, e.g., by conducting galaxy redshift surveys with the majority of the fibers. 

\subsubsection{Direct measurements of the Hubble constant $H_0$ and  parameter $H(z)$\label{td:H0}}
The KN AT2017gfo associated with GW170817 has been a uniquely important GW event thanks to the detection of its EM counterpart. An important application that has received particular attention is the direct late-Universe measurement of the Hubble constant, $H_0$, from the GW luminosity distance combined with the spectroscopic redshift of its host galaxy, NGC\,4993 \citep{GWH0}. This ``standard siren'' measurement has fuelled high expectations for an accurate direct $\sim 1\%$ late-Universe measurement of $H_0$ that is independent of stellar standard candles, such as classical Cepheids and type-Ia supernovae \citep{Riess2022}, although the single recorded event currently limited the accuracy of the measurement to $\sim 7\%$ \citep{Palmese2023}. Spectroscopically characterized KNe are likely to yield the most accurate $H_0$ measurements, especially as the diversity of the underlying sources may contribute to the systematic uncertainties \citep[potentially analogous to SNeIa diversity][]{Benetti2005}. 

Accurately mapping the Hubble-Lema\^itre law using kilonovae will have significant impact on the interpretation of the Hubble constant Tension and its proposed solutions \citep{DiValentino2021,Verde2023}.  Using gravity as a messenger for mapping cosmic expansion could expose further limitations in the current cosmological paradigm, perhaps even ones not previously considered. However, the spectroscopic KN characterization and unambiguous redshift measurements will be crucial to accurately interpreting the results. 

A ``dark siren'' measurement of $H_0$ \citep{SoaresSantos2019} is possible when host galaxies of GW events can be (probabilistically) assigned to GW events without EM counterparts, such as coalescing binary black holes (BBHs), which have been the dominant type of GW event since LIGO's first detections. \wst\ will very significantly improve galaxy redshifts out to a significant redshift. Next-generation GW facilities, such as ET, CE, and LISA (adopted\footnote{\url{https://www.esa.int/Science\_Exploration/Space\_Science/Capturing\_the\_ripples\_of\_spacetime\_LISA\_gets\_go-ahead}} for launch in 2037), will measure GW events in the first galaxies. \wst's ability to map the cosmic web in unprecedented precision and number will thus enable dark siren measurements of the Hubble parameter $H(z)$, and of cosmological parameters, notably the deceleration parameter, $q_0$, and the dark energy equation of state $w$ independently of SNeIa. 

\subsubsection{Multi-messenger gravitational lensing\label{td:lensedBNS}}
The \wst/ET/\lsst\ era will also be transformational for discovering gravitationally lensed BNS and their KN counterparts. Such multi-messenger gravitational lensing discoveries will make epoch-defining contributions to tests of General Relativity (simultaneous tests of GR pillars of lensing and GWs) and probe a broad range of other physics including $r$-process nucleosynthesis in the distant Universe \citep[e.g.][]{Baker2017,Goyal2021,Smith2023}. 

The expected rate of lensed BNS detections by ET is $\simeq 10-1000$ per year, of which $\simeq10\%$ will be localised to a solid angle comparable with the \wst\ field of view \citep{Magare2023}. This opens up the exciting opportunity for \wst\ to be the key facility via which multi-messenger gravitational lensing grows from  single object discovery in to statistical samples and impactful science. 

The required mode of operation will be T-ToO with lead times from between immediately after an alert is received up to $\lesssim 36$\,h after the alert, with two complementary scenarios. In the first, relatively conventional mode, the \wst\ ToO would follow-up candidate gravitationally lensed EM counterparts that have been identified photometrically (for example by Rubin ToO's, as discussed by \citealt{Smith2023}). The second, more ambitious (not mutually exclusive), mode would place a \wst\ fiber on every plausible lensed host galaxy within the GW sky localisation, independent of the existence of a photometric candidate lensed EM counterpart. This new ``direct to spectroscopy'' mode will leverage the synergy between \emph{Euclid}, \lsst\ and the \fourmost\ Strong Lensing Spectroscopic Legacy Survey \citep[4SLSLS]{Collett2023}. In the upcoming decade 4SLSLS will deliver a highly complete sample of gravitationally lensed galaxies that are plausible hosts of gravitationally lensed explosive transients of all kinds in the \wst\ era, including lensed BNS/KNe. We are collaborating with 4SLSLS, \emph{Euclid} Strong Lensing Science Working Group, and \lsst\ Strong Lensing Science Collaboration colleagues on optimising the synergy of these highly complementary facilities.

Simultaneous IFS and MOS observations will be very powerful within the T-ToO observations outlined above. This is because every \wst\ pointing will contain at least one moderately rich cluster of galaxies, which will be a plausible gravitational lens responsible for lensing a BNS/KN, regardless of whether it has previously been identified as a lens in previous photometric surveys \citep{Ryczanowski2020}. The T-ToO strategy would therefore aim to place the wide-field IFS on the most massive gravitational lens in each \wst\ field, thus benefiting from not requiring to choose where to deploy fibres for the most powerful and largest Einstein radius lens within the \wst\ footprint.

\subsection{AGN variability and Black Hole masses\label{td:AGNs}}

The nuclei of some galaxies produce large amounts of electromagnetic radiation that cannot be explained by stellar activity. These are known as Active Galactic Nuclei (AGNs). Soon after the discovery of the first AGN it was hypothesized that this radiation must be powered by the release of gravitational energy during the accretion of material onto a Supermassive Black Hole (SMBH), with a typical mass M $\sim  10^{6-10} M_{\odot}$, located at the center of its host galaxy \citep{Lynden-Bell69}. 

AGNs are among the most energetic phenomena in the Universe and are characterized by time-variable emission in every waveband in which they have been studied, from radio to very high energy (VHE) \citep{Peterson001}. Variability studies are fundamental to understanding the extreme physical conditions of accretion disks near supermassive black holes (SMBHs). Recent studies indicate that AGN variability can be well described as a stochastic process, with characteristic timescales ranging from days to years \citep{Kelly09,Burke21}. For a single object, the shortest timescales are associated with shorter emission wavelengths (e.g., \citealt{Lira15}). 

Even though variability is one of the defining characteristics of AGN, we do not completely understand the mechanisms that drive such variations. Several physical models have been invoked to explain them, including accretion disk instabilities and obscuration by orbiting or infalling dusty clouds \citep{Stern18,Ross18}. Previous studies on the origin of AGN variability have been limited by the lack of large and diverse AGN samples \citep{MacLeod10,Simm16,Arevalo23}. This problem might be partially solved with recent photometric time-domain surveys, like  ZTF, as well as the upcoming \lsst. However, multi-epoch spectroscopy is vital to fully understand the physical properties and phenomena behind the variable nature of AGNs, and their impact on galaxy evolution (e.g., AGN feedback).

\subsubsection{Measuring Black Hole Masses using Reverberation Mapping\label{sec:td:RM}}
Previous studies have demonstrated that AGN variations in the X-ray and the UV/optical continuum are observed at a later time in the infrared range and the UV/optical broad emission lines (BEL). These delayed variations can be understood as the reprocessing of the X-ray and UV/optical emission from the corona and the accretion disk, respectively \citep{Arevalo08,McHardy18,Cackett21}.

\begin{figure}[t]
  \centering
  \includegraphics[width=0.8\textwidth]{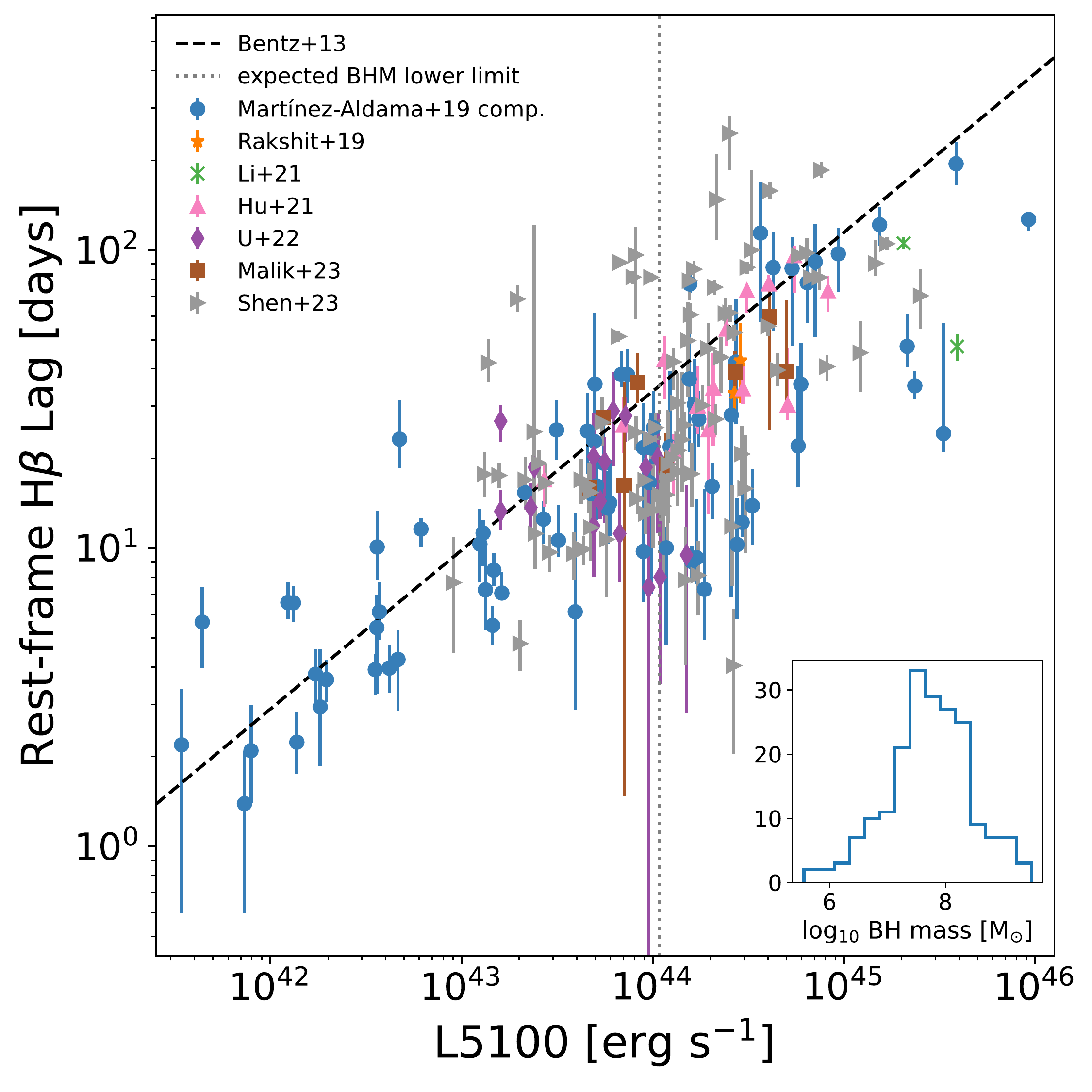}
  \caption{\small Radius–luminosity (R-L) relation for the H$\beta$ broad emission line. Literature compilation, including compilation by \cite{Martinez-Aldama19}; \cite{Rakshit19,Li21,Hu21,U22,Malik23,Shen23}. The black dashed line shows the R-L relation derived by \cite{Bentz13}. The grey dotted vertical line shows the expected luminosity lower limit for the SDSS Black Hole Mapper (BHM; \citealt{Kollmeier17}). The bottom-right panel shows the BH mass distribution of the complete sample.  }
  \label{fig:RM}
\end{figure}

Reverberation mapping (RM) is a unique technique to study the spatially-unresolved structures of AGNs \citep{Peterson93,Kaspi00,Lira18,Edelson19,Cackett21}. RM measures the observed time lags between contemporaneous light curves of AGN observed in different wavelength ranges, by cross-correlating the light curves and associating these lags to the light travel time between the different structures of the AGN. In particular, we can use RM to estimate the AGN broad line region (BLR) size by measuring the time lag between UV/optical continuum emitted by the accretion disk and the BEL variations (e.g., \citealt{Lira18}), and from this, we can derive a radius-luminosity relation (R-L; e.g., \citealt{Bentz13}) for each BEL. An example of this is shown in Figure \ref{fig:RM} for the H$\beta$ BEL. We can use these BEL lags to measure direct SMBH masses \citep{Peterson99,Peterson00}. RM is one of the few methods that allow us to determine SMBH masses directly (besides the dynamically resolved methods used in the local Universe), and any other determinations rely on RM as the primary calibrator of the R-L relationship \citep{Shen11,Rakshit20}. However, as we show in Figure \ref{fig:RM}, current RM samples are limited in size (with only a few hundred objects studied), and in range of BH masses and AGN luminosities covered. 

So far, from the spectroscopic side, large size RM campaigns have yielded (or will yield) up to thousands of AGNs, such as \fourmost, which will perform an RM study of up to 1000 AGNs with $0.1<z<2.5$, and $r<22.5$ \citep{Swann19}, and the Black Hole Mapper (BHM), which is conducting a RM campaign of over 1,000 AGNs, spanning a very wide range of redshifts from 0.1 to 4.5, with $i<20$, and a range of bolometric luminosities L$_{\text{bol}} \sim 10^{45}$ to $10^{47}$ erg s$^{-1}$ \citep{Kollmeier17}. These studies will considerably improve the samples available for RM studies, but will still miss the low luminosity and BH mass end of the currently known AGN samples.

Leveraging the 12-meter diameter and wide field of view of the \wst, we can undertake an extensive spectroscopic variability and RM study involving more than 100,000 AGNs over five to 15 years. These AGNs will be identified through \lsst\ \citep{LSST,Panda19,Czerny2023}, which is anticipated to detect $10^7$ AGNs with a limiting magnitude of approximately $r\sim24$, and will span various ranges of redshift, black hole mass, luminosity, and accretion rate. With this campaign we will fill the gaps (in terms of AGN physical properties) of previous RM campaigns, we will study if the R-L relation can be extended to lower (higher) masses and luminosities, and we will identify possible dependencies of this relation with other factors (e.g., host properties, metallicity, redshift, among others). With the proposed optical coverage (up to 970 nm), we can conduct RM studies with \civ up to $z\sim5$, with \mgii up to $z\sim2.3$, and with Balmer lines up to $z\sim1$. Although \civ can also yield mass estimates, this line is usually contaminated by non-virialized components, leading to much greater uncertainties. A future NIR upgrade (up to H band) coverage would allow us to perform studies of Balmer lines up to $z\sim2.7$, \mgii line up to $z\sim5.4$, and with \civ even up to $z\sim10$ (here we are assuming that by the start of the operations of \wst, there will be several AGNs detected at this redshift range, from, in particular, \jwst\ campaigns).

To conduct RM studies efficiently, we need to detect structure in the light curves of AGNs (e.g., peaks and valleys). Considering the typical timescale of the optical stochastic variations of AGNs, ranging from days to a few years depending on the BH mass \citep{Burke21}, and considering the effects of time dilation ($t_{\text{rest}}=t_{\text{obs}}/(1+z)$), the ideal program length would be 15 years, especially for very high redshift sources ($z>4$), however, 5 or 10 years is acceptable for the more local sources. Moreover, to be able to trace the shortest possible time delays (order of few days, see Figure \ref{fig:RM}), the ideal temporal cadence per target would be one observation per week for low redshift sources ($z<1$), and one observation every two weeks for the high redshift ones. We also require a S/N larger than 20 in the continuum so that we can detect low amplitude variations (e.g., \citealt{Cackett21}). Considering the synergy with other \wst\ campaigns, we propose to conduct most of the observations in regions like the LSST Deep Drilling Fields (LSST DDFs), since these areas will be frequently visited in the context of other programmes, in particular the \wst\ extragalactic surveys.

Recent non-targeted RM campaigns have shown that possibly there is a need to add other physical parameters to the R-L relationship to find more accurate masses (e.g., \citealp{Martinez-Aldama19,Du19,Panda22,Maithil22}). This most likely seems to be the accretion rate. Since bright distant quasars are known to be accreting at high rates, it seems of great importance to test the R-L relationship at cosmic dawn, by directly conducting RM experiments in these objects. These tests will be possible in the context of this proposed campaign. Our RM campaign will also allow us to use AGNs as standard candles for cosmological studies at high redshifts (e.g., \citealt{Watson11,Wang14,Panda19,Martinez-Aldama19,Czerny2023Ap&SS}), and also to conduct spectral variability studies for a diverse (in terms of their physical properties) population of AGNs, which will help us to better understand the origin of the AGN variations (e.g., \citealt{Peterson88,Barth15,Shen19}).

\subsubsection{Changing-look AGNs and AGN flares\label{sec:td:clagn}}
AGNs are commonly classified in the optical range by the presence or absence of broad permitted emission lines ($\text{FWHM} \gtrsim 2000$ km s$^{-1}$), into Broad Line AGN (BL or type 1) and Narrow Line AGN (NL or type 2), respectively. Changing-look AGNs (CLAGNs; also known as changing-state AGNs) correspond to sources that change their classification as type 1 or type 2 AGN within a time scale of months or years \citep{LaMassa15,Ricci22,Temple2023}. During this transition phase, broad lines in the optical spectrum of CLAGN can appear or disappear, accompanied by a drastic change in the AGN continuum flux, by orders of magnitudes which is not expected considering the typical variability amplitudes observed in large AGN samples ($\sim 0.1 - 0.2$ magnitudes per year). Only a few dozens of CLAGNs have been identified so far \citep{Wang24}, with most discovered quite recently as time-domain surveys have grown. This is mainly related to the lack of spectroscopic observations at different phases of the transitional phenomena.

Drastic variations in AGNs have also been related to AGNs displaying anomalous flaring activity (e.g., \citealt{Trakhtenbrot19NatAs,Frederick21}), and to recently discovered ambiguous nuclear transients (ANTs; e.g., \citealt{Hinkle22} and references therein). The nature of these transient events remains open. Still, it has been proposed that they could be related to rejuvenated SMBHs that experienced a sudden increase in their accretion rate \citep{Trakhtenbrot19NatAs}. 

All these phenomena challenge the standard physics of thin accretion discs and the unification model for AGN. Even if they are rare phenomena, their study is fundamental in investigating the physics of accretion taking place in the unresolved central engine. However, most of the known CLAGNs have been detected using archival data, and thus, great effort is needed to catch them during the transition phase. 

\wst\ will facilitate the study of CLAGN and flaring AGNs in two ways. First, serendipitous discoveries will be allowed by splitting visits to AGNs into two or more sub-exposures. Second, spectral monitoring campaigns of CLAGN and flaring AGN candidates selected from photometric surveys like \lsst\ will allow a more detailed understanding of the physical phenomena behind these events, by observing their spectral evolution in real time.

\subsubsection{Binary AGNs\label{sec:td:dualagn}}
Most, if not all, galaxies hold a SMBH in their center. Since galaxies evolve through galaxy major mergers, the formation of  binary SMBH systems is postulated. Dual AGNs have been identified through direct imaging at separations of several parsecs to kiloparsecs. The evolution of SMBH binaries with separations a few parsecs is debated, leaving the presence and distribution of close binaries (CB) of SMBHs  uncertain. Several electromagnetic characteristics are employed to indirectly search for CB-SMBHs, with periodic variations of the flux being  the most promising \citep{Gutierrez22} and peculiar broad emission line profiles being one of the most widely used methods \citep{Komossa06,Popovic12,DeRosa19}.  The CB-SMBHs have not yet been confirmed, though many candidates have been found (e.g., \citealt{Jiang22}). Moreover, they are particularly interesting as they are prime targets for low-frequency GW signatures by Pulsar Timing Array and space-based GW observatories, such as LISA \citep{DeRosa19}.

We propose an extensive multi-epoch follow-up with \wst\ of \lsst\  identified periodic AGNs, expected to range between 10,000 to 100,000. This estimate is derived from the anticipated $10^7$  AGNs up to 24 mag expected to be detected by \lsst\ and the CB-SMBH detection probability of $10^{-3}$ \citep{Volonteri09} to $10^{-2}$ \citep{DOrazio18}. With a dedicated spectroscopic monitoring using \wst\, we aim to provide radial velocity curves with enough cadence (minimum 4 epochs) to confirm the binary orbit, exclude the red-noise variability, and test more complex physical models on observed peculiar broad line profiles and line shape variability \citep{Popovic21}.

\subsubsection{AGN outflows\label{sec:td:agnout}} 

Variability is essential for studying outflows in AGN observed through absorption, up to very high velocity of $v\sim0.1-0.2 \: c$ \citep{Rodriguez-Hidalgo20}. AGN outflows are believed to play a significant role in injecting energy into the surrounding interstellar medium, influencing the evolution of the AGN host galaxy (e.g., \citealt{Fabian12,Fiore17,Laha21}). The temporal variability of absorption line profiles serves as a powerful tool for investigating the origin and propagation of such outflows (e.g., \citealt{Misawa14,Vietri22}). It imposes constraints on their density and distance, thereby enabling the measurement of outflow kinetic power. Variability becomes detectable when the recombination time (t$_{\rm\,rec}$) of an absorption line is shorter than the timescale of the ionizing continuum variation and the time between consecutive observations \citep{Barlow92}.

Considering a mean t$_{\rm\,rec}$ of approximately 2 days in the rest-frame, as observed in the majority of SDSS quasar populations \citep{He19}, a temporal cadence of two weeks per target at high redshift (e.g., $z\sim 3$), as outlined in the RM science case, seems ideal for initially monitoring the variability of outflows. Subsequently, increasing the time between visits will enable a diverse sampling of variability, focusing on monitoring changes in absorption strength, which are found to be more prominent in long-term variability \citep{Capellupo13}.

\subsection{Transients\label{td:transients}}
Thanks to modern all-sky surveys, including for instance the the Panoramic Survey Telescope and Rapid Response System \citep[Pan-STARRS]{Chambers2016}, the All-Sky Automated Survey for Supernovae \citep[ASAS-SN,][]{2014AAS...22323603S},  the Asteroid Terrestrial-impact Last Alert System \citep[ATLAS,][]{2018PASP..130f4505T,2020PASP..132h5002S}, Gaia \citep{GaiaDR3}, the Zwicky Transient Facility (ZTF; \citealt{Bellm19}), and others, new transients are detected at an unprecedented rate of currently a few tens per night. With the advent of \lsst\ \citep{LSST}, this number will rise by $\sim$five orders of magnitude. Already now, the available spectroscopic facilities are not adequate to keep up even with the current discovery rate \citep[current classification rate is $\sim 10\%$]{Kulkarni2020}. This situation is bound to worsen with the deluge of transients from \lsst; moreover, the majority of those will occupy the faint end of the magnitude distribution, making spectroscopic observations increasingly challenging and resulting in a low single-digit percentage of transients that will be followed-up spectroscopically \citep{Bellm2016volumetric}.

\wst\ is poised to be a game-changer in this challenge thanks to its high multiplexing and 12m aperture. \wst\ will observe faint transients down to 24 mag using the low-res mode (fainter if resolution is reduced to R $\sim 500-1000$, e.g., by binning). For example, this will allow to study classical novae in nearby ($D_L < 10$\,Mpc) galaxies and permit to quantify their role as factories  of lithium, CNO isotopes, and dust in thee galaxies, as well as possible progenitor of type Ia SNe \citep[e.g.,][and references therein]{DellaValle2020}. A spectroscopic survey with this depth of observation is unattainable with existing or planned facilities. 

While the transient nature of the target objects typically requires ToO observations, the large number of alerts expected from \lsst\ renders it unfeasible to trigger T-ToO observations for a significant fraction of events. However, F-ToO observations will be collected for a fraction of all fibers (possibly $\sim 10\%$, tbd) in each pointing, ensuring that a large fraction of the transients falling within the \fov\ will be captured. Where faster reaction times are needed, T-ToO observations with \wst\ may be considered against the option of triggering ToO observations on other facilities with specialized instrumentation. Additionally, many science topics would very significantly benefit from slow-response or irregular observations with \wst. These include, for example, general transient classifications, host galaxy redshift/stellar population studies, spectroscopy of SN impostors and intermediate-luminosity transients, spectroscopy of SN progenitor analogs in the local Universe, search for SN signals in galaxy spectra, SN remnants study, late-time nebular phase followup, and lensed SNe studies. Such observations could be conducted without disrupting other surveys due to time-insensitive scheduling. Yet, the need for repeat observations of faint objects means that spectroscopy is usually the bottleneck and hence \wst\ would very significantly improve over the current situation.

\subsubsection{Supernovae\label{sec:td:sne}}
\begin{figure}
   \centering
   \includegraphics[height=0.8\textheight]{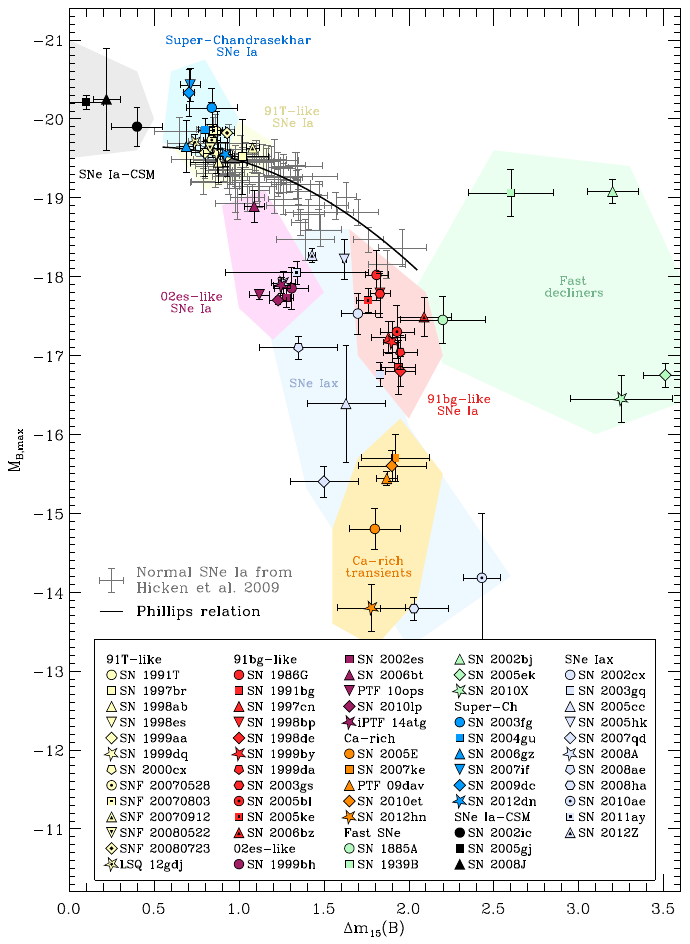}
   \caption{Absolute magnitude vs light curve decline rate $\Delta m_{15}(B)$ for potential thermonuclear transients. Part of this large diversity is identifiable by light curve evolution. Spectroscopy is required for other cases and will very substantially aid to unravel supernova diversity. Taken from \citet[Fig.~1]{Taubenberger2017}. For an overview of the diversity of all SNe, see Fig.~7 in \citet{Perley2020}}
   \label{fig:td:SNdiversity}
\end{figure}

Supernovae (SNe) are among the most common transient events to be discovered by state-of-the-art all-sky photometric surveys. SNe are understood to be either terminal explosions of massive stars at the end of their evolution or explosions of white dwarf stars in binary systems \citep[see e.g.][and references therein]{Langer2012,maoz14}. They are visible across cosmic scales and are crucial in many branches of astrophysics, such as stellar evolution, galaxy evolution, heavy element nucleosynthesis, and cosmology. Traditionally, SNe have been empirically classified using lines observed near the epoch of the light curve peak \citep{filippenko97}. In recent years, however, a huge unexpected diversity has been discovered \citep[their Fig.~7]{Perley2020}. Figure\,\ref{fig:td:SNdiversity} illustrates the diversity of thermonuclear explosions by showing the main classes of known extreme objects relative to the ``vanilla'' Type Ia supernovae (SNe Ia). While \lsst\ and \wst\ will therefore discover many SNe of familiar types, there is enormous potential for discovering novel SNe types that will provide unique and crucial insights into the extreme physics of supernovae and their explosion mechanisms. 

This could, for the first time, include luminous and long-lived pair-instability supernovae (PISNe) that have been predicted to originate from massive ($140 - 260 \, M_\odot$ zero age main-sequence) metal-poor stars \citep[e.g.,][]{Fowler1964,Kasen2011}.  The characteristics of the PISNe progenitor stars are consistent with those of population III stars thus, a PISNe detection would push the limits of our understanding on stellar evolution and the early Universe. Additionally, superluminous supernovae \citep[SLSNe, e.g.][]{2019ARA&A..57..305G} are $10 - 100$ times more luminous than normal SNe and evolve on timescales of months \citep[e.g.][]{2018ApJ...860..100D} and generally occur in host  host galaxies with extreme properties \citep[e.g.][]{2011ApJ...727...15N,2016ApJ...830...13P}. Understanding their explosion mechanisms and relations to PISNe and PopIII stars would present an important breakthrough that will require \wst's capability of providing early time spectra of a very large number of distant and faint transients. The same goes for many very faint and fast evolving events discovered at the faint end of SN luminosity distributions.
Rapid spectroscopic follow-up in particular is crucial to characterizing SN classes and to unraveling explosion physics and the diversity of SN progenitors (see e.g., \citealt{Nugent2011} for the case of SN2011fe). \wst's F-ToO would be perfectly suited to these endeavors, allowing classification within hours to days after explosion.

\paragraph{Fast-response observations of SNe} are useful to catch the objects at the early stage, within days or even hours after the explosion \citep[e.g.][]{bruch23,jacobsongalan23}. At this very early phase of SNe, the interaction between the SN ejecta and the confined circumstellar matter (CSM) formed by pre-SN progenitor mass loss may occur, signaled by narrow emission lines and higher-ionisation transitions. These spectral features disappear quickly with time, and higher spectral resolution will be beneficial to the detection and characterisation of the spectral lines \citep[e.g.][]{smith23} \hbox{---} \wst\ will be well suited for this purpose. Fast response will also be useful to catch rapidly-evolving objects related to SNe, such as the so-called fast blue optical transients \citep[{FBOT}, e.g.][]{perley19} and the recently identified class of luminous fast-cooling transients \citep{Nicholl2023}. Type Ia SNe \citep{maoz14,leibundgut01} dominate in a magnitude limited survey, whereas core-collapse supernovae dominate by volume. By pushing the surveyable volume, \wst\ will open the window to investigating diversity of core-collapse supernovae.  Spectroscopy is required to distinguish SNe from contaminating objects, such as intermediate-luminosity optical transients \citep[e.g.][]{cai21}, including luminous blue variable eruptions and luminous red novae, as well as cataclysmic variables, stellar flares, and even asteroids.
Normally, the early phase following detection is when the spectral classification is performed, and it is also the phase when the most important information about stellar progenitors can be inferred from the analysis of emission and/or absorption lines that originate in the surrounding circumstellar medium (CSM). 

\paragraph{Slow-response observations of SNe} will cover the later evolution of the objects within a timescale of months to years. SNe reach peak brightness after an early rising phase of typically $2-3$ weeks. While this so-called main photospheric phase has been often captured and studied, \wst\ will be useful for studying the physics of the objects and training photometric classifiers. It is also possible that the initial classification during the early rise is inconclusive, and a re-classification during this bright phase will be required to ascertain the type and nature of the explosion.
During the late-time nebular phase, SNe spectra are dominated by emission lines arising from the inner ejecta, which are critical for investigating the properties of the exploding stellar core \citep[e.g.][]{kuncarayakti20,fang24}. As the ejecta plow outward and encounter the detached CSM resulting from the previous progenitor mass loss episodes, CSM interaction spectral signatures will appear in the spectrum and this can be used to trace back the mass loss history of the progenitor star \citep[e.g.][]{kuncarayakti22,kuncarayakti23}.
Eventually, the SN will fade away and transition into SN remnants. Studies of SN remnants as IFS-resolved sources \citep{larsson21} or as a population of MOS-suitable targets in host galaxies \citep{Long+2022} are both within \wst\ capabilities, as well as studying SN light echoes in the local Universe which is very useful for assessing the geometry of the explosion and viewing angle effects \citep[e.g.][]{rest08}.

\paragraph{Host galaxies of SNe} offer more opportunities to study the SNe and progenitors. The SNe must have been born from a stellar population, and thus the parent stellar population parameters such as age and metallicity can be used to constrain those of the SN progenitor star including initial mass and metallicity \citep{anderson15,kuncarayakti18,lyman20}. Making use of the IFS of \wst\, a study of nearby SN host galaxies in the local Universe \citep{pessi23,galbany18,Kruehler2017} can be done in conjunction with the nearby galaxy survey (Sect.\,\ref{sec:legacy_survey}). This may cover potential progenitor stars in the resolved stellar populations \citep[e.g. SN 1987A, see review by][]{mccray16}, providing an unprecedented legacy spectroscopic data of the progenitor for future SN explosions in these backyard galaxies.
At higher redshifts, host galaxy studies using MOS will amass a large statistical sample useful for cosmology and galaxy evolution studies. This will also provide the opportunity to search for SN signals diluted in the integrated host galaxy spectrum \citep[e.g. in SDSS,][]{graur13}.

\begin{figure}
    \centering
    \includegraphics{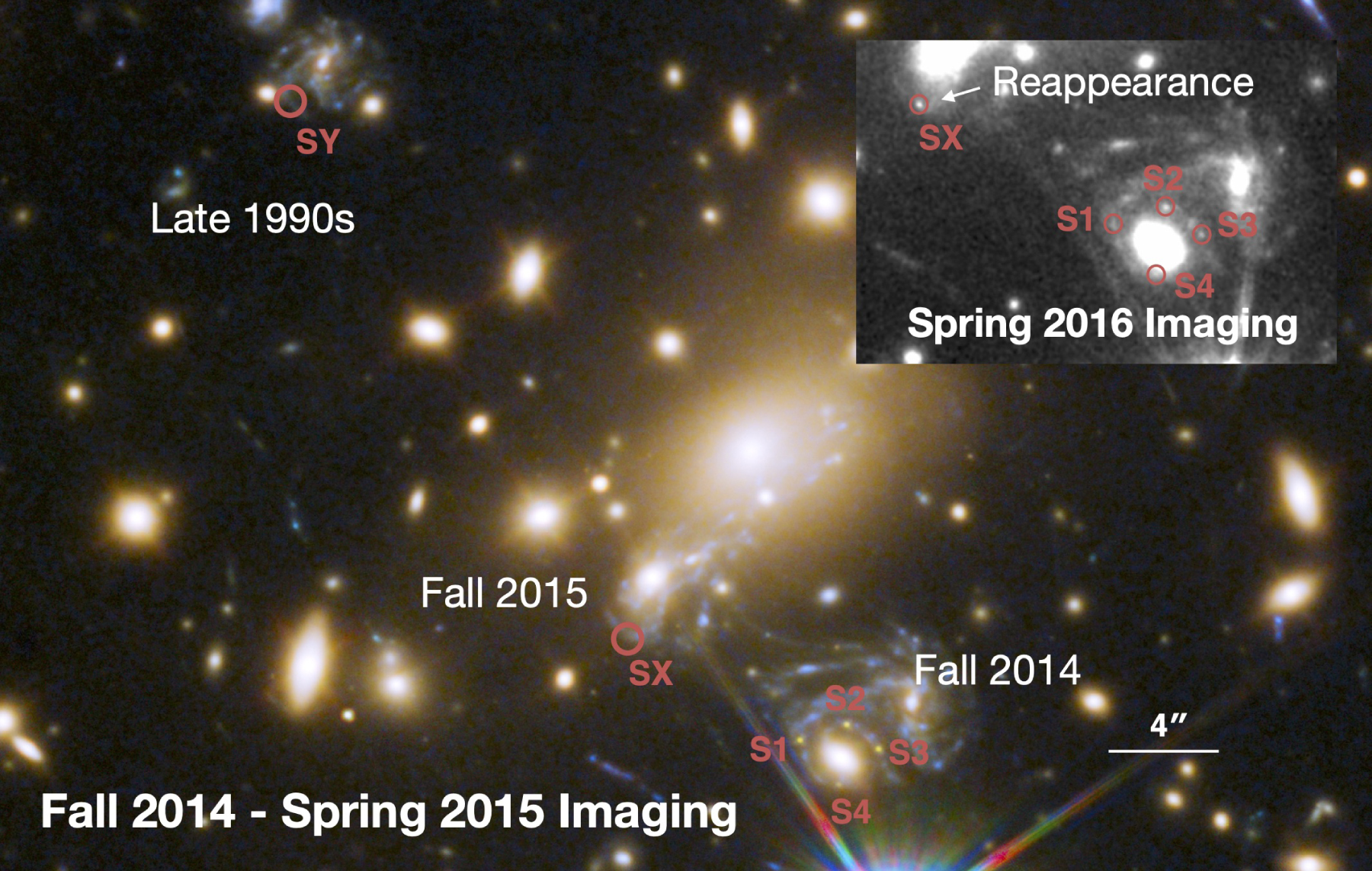}
    \caption{The lensed supernova `Refsdal' and its appearance at different times. Figure taken from \citet[their Fig.~1]{Kelly2023}.}
    \label{fig:td:refsdal}
\end{figure}

\paragraph{Lensed supernovae} are of particular interest for supernova cosmology and for studying thermonuclear explosions at high redshift thanks to the magnification by the gravitational lens that can boost the signal by several magnitudes \citep{Goobar2023}. While many lensed galaxies and QSOs have been detected, the first strongly lensed Type Ia supernovae  with resolved multiple images have been discovered only very recently \citep[e.g., SN Refsdal][]{Kelly2023}, cf. Fig.\,\ref{fig:td:refsdal}. \lsst\ will discover tens of such events per year, and a large fraction of which will have long time-delays suitable for precision cosmology.
A key ingredient in understanding lensed SNe is spectroscopically typing their prompt emission before maximum light in the leading image, such that high-resolution photometric follow-up can be obtained promptly. Spectroscopy can even allow to robustly determine time-delays from a single observation of a mulitply imaged SN \citep{Johansson2021}. \wst\ will be an excellent facility to get high quality spectra of lensed supernovae both for classification and in cases where the multiple images can be resolved, an independent estimate of time-delays.

\subsubsection{Gamma-ray bursts (GRBs)\label{sec:td:grb}}
Gamma-ray bursts (GRBs) emit intense flashes of gamma-rays followed by radiation across the electromagnetic spectrum. They are classified into two main types based on their gamma-ray duration: short-duration (with T90$<2$s), and long-duration (with T90$>2$s), where T90 denotes the time over which a burst emits from 5\% of its total measured counts to 95\%. Short-duration GRBs are typically associated with the merger of compact objects such as neutron stars or black holes, potentially accompanied by the emission of gravitational waves \citep{GW170817GRB} and the formation of kilonovae, cf. Sect.\,\ref{td:kilonovae}. On the other hand, long-duration are signposts of star formation owing to their association with the deaths of massive stars \citep{Woosley1993}.

GRBs serve as cosmic beacons, offering a unique glimpse into the early Universe \citep{Tanvir2009} and the properties of intergalactic and interstellar media. The simple intrinsic power-law spectrum \citep{Sari1998} of GRB afterglows provides a wealth of information about the environments through which their light traverses, and \wst\ will allow dissecting these signatures imprinted on GRB spectra.

In F-ToO mode, or serendipitously, \wst\ could observe the first few GRBs using optical spectroscopy within minutes after their trigger. Such observations will provide breakthrough insights into GRB progenitors and central engines and allow to study the rapid spectral and temporal evolution of these energetic events \citep{Gao2015GRB,Zhu2023GRB}. Additionally, \wst\ will comprehensively sample GRBs across various timescales, creating strong synergies with facilities providing alerts and follow-up at high-energies, notably the ESA {\it Athena} mission \citep{Barcons2017}.

\paragraph{Intergalactic Medium (IGM) Properties:} absorption features in the spectra of GRB afterglows (or quasars) probe the physical conditions and chemical composition of the IGM \citep{McQuinn2008,Peroux20}. Variations in absorption lines with redshifts trace the line-of-sight distribution of neutral gas in the Universe, enabling us to trace the cosmic web and study its evolution over cosmic time. F-ToO observations within minutes to hours after the initial trigger will provide crucial initial spectra, with subsequent observations conducted days after the explosions to study afterglow evolution, including decay rate, spectral evolution, and late-time flares.

\paragraph{Host Galaxy Environment:} The absorption and emission features in the spectra of GRB afterglows encode information about the properties of the interstellar medium (ISM) within the host galaxy \citep{Zafar2011,Selsing2019}. Analysis of metallicity, dust content, and kinematics of the ISM can offer insights into the star formation history, chemical enrichment processes, and dynamics of galaxies hosting GRBs \citep[e.g.][]{2019A&A...623A..43B}, and out to the epoch of reionization \citep[e.g.][]{2023A&A...671A..84S}.
\wst's observations within hours to days after explosion will provide insights into ISM properties (metallicities, dust content, \& kinematics), star-formation histories, and dynamics of host galaxies. Additionally, non-time-critical IFS observations of spatially-resolved GRB hosts out to $z \sim 0.2$ will help to understand the role of environments properties, such as merger and star formation histories and metallicity gradients, in shaping the conditions from which the GRB originated \citep{Kruehler2017,Izzo2017}.

\paragraph{Progenitor Properties:} Spectroscopic observations of GRB afterglows  and their associated SNe constrain progenitor properties, such as mass, metallicity, and evolutionary stage \citep{Hjorth2003}. This information is crucial for understanding the mechanisms leading to GRB formation and the role of massive stellar populations in shaping the Universe.
F-ToO observations within hours to several days post explosion will be suitable to this end.

\paragraph{GRBs as cosmological probes}
GRBs can serve as cosmological tools alone \citep{DainottiLenart2023} or combined with other high-z probes \citep{Bargiacchi2023GRBQSO,Dainotti2023comb}. They provide SN-independent information on the Hubble parameter \citep{Dainotti2021SNeIa} and for investigating the evolution of the Dark Energy equation of state, $w(z)$ \citep{Muccino2021}.
However, the number of GRBs with measured redshifts is currently too small, their features too broad, and there is an unfortunate (Malmquist-type) bias toward more luminous GRBs at a given $z$. Much more spectroscopic data is thus needed to render GRBs competitive (with SNe Ia) cosmological tools \cite{Dainotti2016,Dainotti2017,Dainotti2020,Dainotti2022ApJS}.

\wst\ will provide spectra of $10^7$ galaxies (among those the GRB host galaxies) and the spectrum of optical GRB afterglows with the IFS in the range $z=2-7$, down to a magnitude of $24(AB)$. 
\wst\ will thus fill in the observations of faint GRBs with plateaus from up to $z=7$ and balance the GRB sample in redshifts and luminosities to enable accurate cosmological inferences.

\subsubsection{Tidal disruption events (TDEs)\label{sec:td:tde}}

The new time-domain photometric surveys are aiding in the detection of nuclear transients within galaxies that host SMBHs. Tidal Disruption Events (TDEs) involve the disruption of a star as it approaches a BH with a mass of $\lesssim 10^8 M_{\odot}$ \citep{Rees88,vanVelzen20,Gezari21}. With so many physical processes at play, occurring on a human-friendly timescale of a few months and at virtually all wavelengths, TDEs have been heralded as a unique laboratory for studying black holes. They can reveal dormant SMBHs and probe the occupation fraction in galaxies of all types. As disruption occurs inside the event horizon for SMBHs $\gtrsim 10^8 M_\odot$, TDEs may be the best way to find 
intermediate mass black holes (IMBHs; \citealt{Greene20}). They directly constrain the mass and even the spin of SMBHs, as well as the physics of launching jets and outflows. Understanding the TDE rate and its evolution over cosmic time may solve the mystery of how SMBHs can exist as early as $z>7$.

It is still a puzzle how TDE optical emission is produced, but the most important physics is encoded in the TDE spectrum. In particular, the Bowen fluorescence lines (N III $\lambda$4640 and O III) are a direct probe of hidden UV continuum from the accretion process. In a few TDEs, disk-like line profiles have been detected directly in H I and/or He II 4686 \citep{Short2020,Wevers2022}, indicating not only that accretion began early but that it was mediated by a disk. The ratio of He II to H I constrains the radius of the stellar debris \citep{Roth2016}. Outflows have been detected from blueshifts in both H and He II. To use these clues in the spectrum, we must de-blend the He-Bowen line
complex, with separations of only $\gtrsim50$ km/s. To do this for a large statistical TDE sample requires a classification survey with at least R$\sim3000$.

A \wst\ survey of TDEs would be transformational for BH physics. At a depth of $\sim24$ magnitude, we expect $\sim0.1$ ‘live’ TDEs per square degree, assuming $~3\times10^{-5}$ TDEs per galaxy per year \citep{Yao23}, and a 50 day flare peaking at an absolute magnitude -19 to -22 \citep{Hammerstein23}. A few $\times$ 1000 TDEs could be observed spectroscopically in 5 years, assuming few sq deg \fov, $\sim1$ hour exposures and 80\% observing efficiency. Candidates would be provided by \lsst\ and identified by blue colours and in galaxy nuclei. A typical TDE would be visible to $z\sim1$, and the brightest to $z\sim2$. This is sufficient to measure the rate evolution over cosmic time \citep{Kochanek16}. The slope of this evolution reveals the mechanism governing TDE formation: e.g., nuclear star formation or SMBH binaries \citep{Stone18}, and is therefore a unique probe of galaxy/SMBH mergers. At these redshifts, optical spectra from \wst\ would probe rest-frame UV emission. Only a handful of TDEs have UV spectra to date, and TDEs appear to be even more diverse in the UV than in the optical. The ratio of UV absorption to emission lines constrain the inclination of the accretion disk, while the strength of lines reveals the structure of the disk wind, including the angular width and internal clumping \citep{Parkinson20}. Comparing the covering fraction of the wind to the relative numbers of optical to X-ray TDEs would be the most direct test of the reprocessing paradigm for TDE optical emission, but requires a robust sample of UV spectra only possible from a wide-field survey targeting TDEs at $z\sim1$.

\subsection{The spectroscopic variability of stars\label{td:stars}}

The study of stellar variability benefits from centuries of observations and investigations and yet stands to be revolutionized in the next decade. Variability searches were systematically introduced across the full sky in the late 1900s using photographic plates. Charge Coupled Devices (CCDs) revolutionized the ability of large imaging surveys, initially targeting gravitational microlensing \citep{Paczynski1986,EROS,MACHO,OGLE},  to observe large sky areas and significantly improved precision and survey speed. This combination resulted in large and homogeneous variable star data sets that have shaped our current understanding of stellar variability \citep[cf.][for a recent review]{Catelan2023}. However, the ESA mission \gaia\ \citep{GaiaMission} took the lead in the number of variable stars detected as part of its third data release \citep[DR3][]{GaiaDR3,Eyer2022}. The next data releases, DR4 and DR5, will benefit from a doubled temporal baseline, better calibration, additional data types, and improved variability detection and characterization. This fortuitous combination will increase the number of known variable objects with $G \lesssim 20.7$\,mag by an order of magnitude to $\gtrsim 100$\,million. Starting in 2024, the \lsst\ with its significantly fainter magnitude limits will further amplify the deluge of variable objects, even well beyond the local group \citep{LSST,LSST-TVS}. Overlap in \gaia's fainter magnitude range ($16 < G < 20.7$) will be extremely useful for several reasons, notably including extended temporal baselines (light curves, proper motion), significantly different window functions due to sampling differences, additional photometric bands, and the precision gain provided by \lsst\ at \gaia's faint end. Specifically, \lsst's deep-wide-fast survey is projected to collect 56, 80, 184, 184, 160, and 160 observations in ugrizy filters, respectively, over its $10$\,yr duration \citep{Bianco2022} and with a per-epoch precision of $\sim 5$\,mmag \citep{LSST-SB-V2}. Interestingly, this sampling compares to \gaia's average number of observations in three bands ($G$, $G_{BP}$, $G_{RP}$) across the sky ($\sim 140$) assuming a total baseline of $\sim 66$ months, while \lsst\ improves on the per-epoch photometric uncertainties for objects $G \gtrsim 17$\,mag or $G_{BP} \gtrsim 15.8$\,mag (or \lsst's bright limit), cf. Fig.\,21 in \citet{Ivezic2012}. Moreover, \gaia's $Bp$ and $Rp$ low-resolution spectra and the RVS instrument are currently collecting the largest spectroscopic datasets to date\footnote{\url{https://www.esa.int/Science\_Exploration/Space\_Science/Gaia/Gaia\_factsheet}}, whereas DR3 and the Focused Products Release (FPR) already demonstrated the usefulness of \gaia\ RV time series data for pulsating variable stars \citep{Katz2023,GDR3-SOS-CEP,GDR3-SOS-RRL,GDR3-FPR-LPVs}. 
As a result, combining data from \gaia\ and \lsst\ to study variable stars will usher in unprecedented opportunities for using variable stars as astrophysical laboratories, and optical spectroscopy is crucial to this effect. 

\begin{figure}[t]
    \centering
    \includegraphics[width=\textwidth]{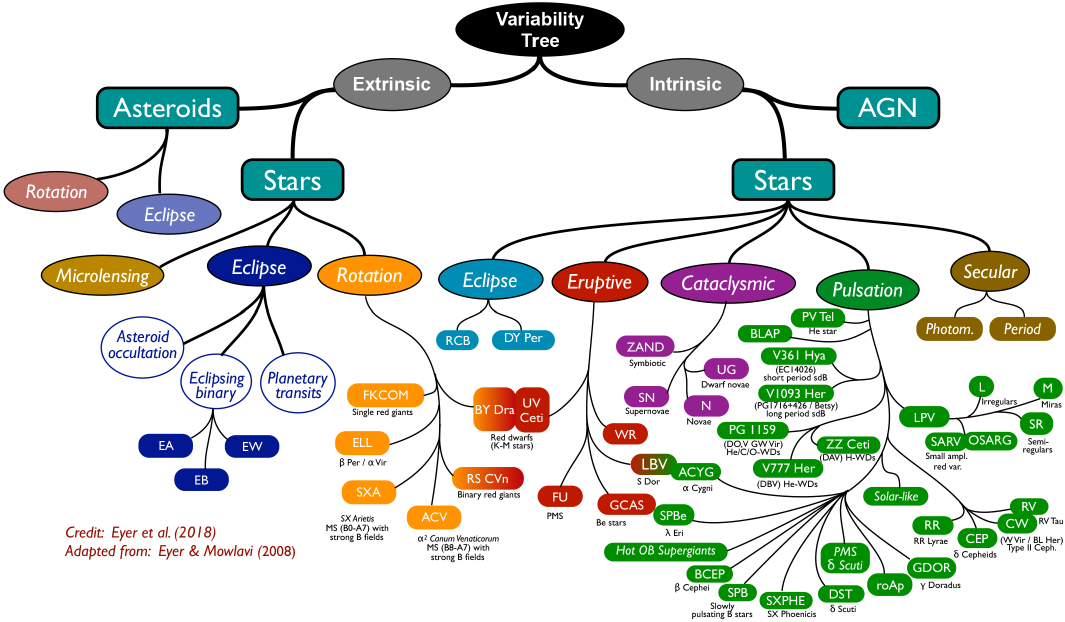}
    \caption{The Variability Tree illustrates a plethora of variable astronomical objects, grouped into intrinsic and extrinsic types \citep{GaiaDR2variability}. \wst\ will make fundamental contributions to virtually all variability types by providing time-resolved low and high-resolution spectroscopy within the Milky Way, its satellites, and (beyond) the Local Group.}
    \label{fig:td:variabilitytree}
\end{figure}

Only \wst, located on the Southern hemisphere with its 12m collecting area, large field of view, and high multiplexing, will be able to collect high-quality (and higher resolution than \gaia) spectroscopy for a significant fraction of the variables to be discovered by \gaia\ and \lsst. Thus, \wst\ will be crucial to characterize variable objects of diverse types (cf. Fig.\,\ref{fig:td:variabilitytree}), especially when photometry alone leads to ambiguous results, and to filtering the most interesting objects for in-depth analysis and observational follow-up.
\wst's spectra will enable stellar astrophysical investigations using a broad range of spectroscopic quantities, listed in order of increasing complexity in the following Sect.\,\ref{sec:td:stars:data}. A brief overview of specific types of variable stars is presented in Sect.\,\ref{sec:td:stars:types}.

\subsubsection{Spectroscopic quantities and variability types\label{sec:td:stars:data}}

\paragraph{Radial Velocity (RV)} variations are the most common changes measured using stellar spectra. Doppler shifts induced by orbital motion can range from amplitudes in the hundreds of \kms\ in the case of close-in compact binaries to well below the detection limit in case of low-mass exoplanets. The former are readily detected using even low-resolution spectra, whereas the latter require high resolution in conjunction with long-term line profile stability. In this spectrum of extremes, \wst's low-res mode will cater to faint sources as well as those where a precision of a few to 10 \kms\ is sufficient. Conversely, the ability of the high-res mode to detect small RV signals will ultimately be limited by the stability of the wavelength solution and the recorded line shapes.  Specific instrumental design choices could allow improvements beyond the performance of FLAMES/UVES, which reaches a statistical precision of $\sim 0.25$\,\kms \citep{Jackson2015}. For example, improved light scrambling properties of the optical fibers could improve line shape stability, whereas drift monitoring (e.g., using laser frequency combs) would allow to monitor the RV stability of the instrument over both long and short timescales. Improving RV precision beyond the capabilities of FLAMES/UVES would open new scientific lines of inquiry. For example, there are currently $223$ ($137$) confirmed exoplanets\footnote{\url{www.exoplanet.eu} accessed on 2023-12-08} with semi-amplitudes $K > 0.050$\,\kms\ ($K > 0.100$\,\kms), ranging from $0.4\ (0.71) - 60\,M_{\mathrm{jup}}$. Given the likely significant observer bias towards low-mass exoplanets, there may be a large population of high-mass planetary companions waiting to be found. Additionally, a long-term RV precision of $0.1$\kms\ or better would provide sensitivity to brown dwarfs. Striving towards a long-term RV precision of $0.05-0.1$\kms\ for the high-res mode would thus open up a potentially transformative avenue for \wst\ to achieve an unbiased survey of heavy substellar companions (jupiters and brown dwarfs) to go well beyond the groundbreaking study of solar-like multiplicity by \cite{Duquennoy1991}, which was conducted using the CORAVEL instrument's RV precision of $\sim 0.3$\,\kms. In between these two extremes, \wst\ will record a plethora of RV signals due to orbital motion as well as large scale pulsations that will allow to map stellar multiplicity across different stellar populations and galaxy environments, notably in focus regions, such as the Magellanic System and the \lsst\ deep drilling fields, as well as in resolved populations (Sect.\,\ref{sec:respop}) as far as a couple of Mpc in the case of RGB stars with observations separated by $\sim 10$\,yr. 

\paragraph{Line shape variability} can arise from multi-lined spectroscopic binaries (SB2), pulsations, shocks, accretion, and stellar activity, among other phenomena. \citet{LSST-EBs} estimates that \lsst\ will identify $\sim 1.7$\,million eclipsing SB2 systems, whose spectroscopic orbits will provide information crucial to understanding the stellar mass-luminosity relation across the Hertzsprung-Russell Diagram (HRD). In particular, detached late-type eclipsing binary systems of red giants are of crucial importance to measuring accurate geometric distances, e.g., to the Magellanic Clouds \citep{Pietrzynski2019,Graczyk2020}. SB2 systems involving pulsating components are of exceptional value to measuring mass and thus to understanding the nature of the variables \citep[e.g.,][]{Debosscher2013,Pilecki2021}. Line shapes can be evaluated for individual lines \citep{Perdelwitz2023}, using cross-correlation profiles that combine information from a specified set of lines \citep{Baranne1996}, or on a pixel-to-pixel level across a wide wavelength range \citep{Binnenfeld2022}. \wst's high resolution mode with $R = 40\,000$ (factor 2 better than \fourmost) will allow to straightforwardly identify multi-lined binaries for velocity differences exceeding $\sim 15$\,\kms, while SB2 systems may be identified even at smaller velocity differences using machine learning techniques \citep{ElBadry2018,Traven2020}. Compared to planned or ongoing large spectroscopic surveys operating on smaller telescopes (e.g., \fourmost), \wst\ will benefit both from higher maximum spectral resolution and from higher resolution at each magnitude, avoiding the loss of information induced by binning as typically applied to increase S/N. The combination of \wst's larger collecting area and higher resolution will therefore allow to study phenomena far outside the range of \fourmost. Even just a few or handful of epochs of spectral observations provide indispensable information for variable star classification because, for example, pulsating stars exhibit variable line depth, width, and asymmetry due to temperature changes, turbulence, rotation, and other atmospheric effects. Variability classifiers can easily incorporate markers sensitive to this type of variability to distinguish between intrinsic and extrinsic variability based on just a few observations. Last, but not least, significant spectral changes, e.g., due to variable emission lines, will be detectable in a variety of objects using the low- and high-res mode.

\paragraph{Spectroscopic indices} provide important information of stellar variability and activity. For example, the Ca H\&K $S-$index, and variants thereof, plays a vital role in mitigating the impact of stellar activity in the detection of extrasolar planets \citep{Wilson1968,Baliunas1995}. The $\Delta S$ index, derived from the same lines, can serve as metallicity indicator for RR~Lyrae stars \citep{Preston1959}, and the relative strength of spectral lines can also be used to detect the presence of hot companions in classical Cepheids \citep{Kovtyukh2015CaHKbinaries}. Measuring the temporal variations of spectroscopic indices for statistically significant and unbiased stellar samples will thus provide invaluable information for population studies targeting exoplanet habitability, stellar multiplicity, and the distance scale. 

\paragraph{Pixel-by-pixel comparisons} among spectra recorded at different times enable variability studies analogous to difference imaging analysis, albeit across thousands of pixels simultaneously. Thanks to advanced machine learning techniques, and building on the body of spectra previously collected, it will thus be possible to detect variability without resorting to derived quantities, such as RV or indices. \citet{Lemasle2020} demonstrated the use of pixel-to-pixel variations to determine temperature variations. In particular variability occurring within narrow wavelength intervals, such as emission line strengths, are more readily detected by spectroscopy than by wide-band photometry. 

\begin{figure}[t]
\centering
\includegraphics[height=0.5\textheight]{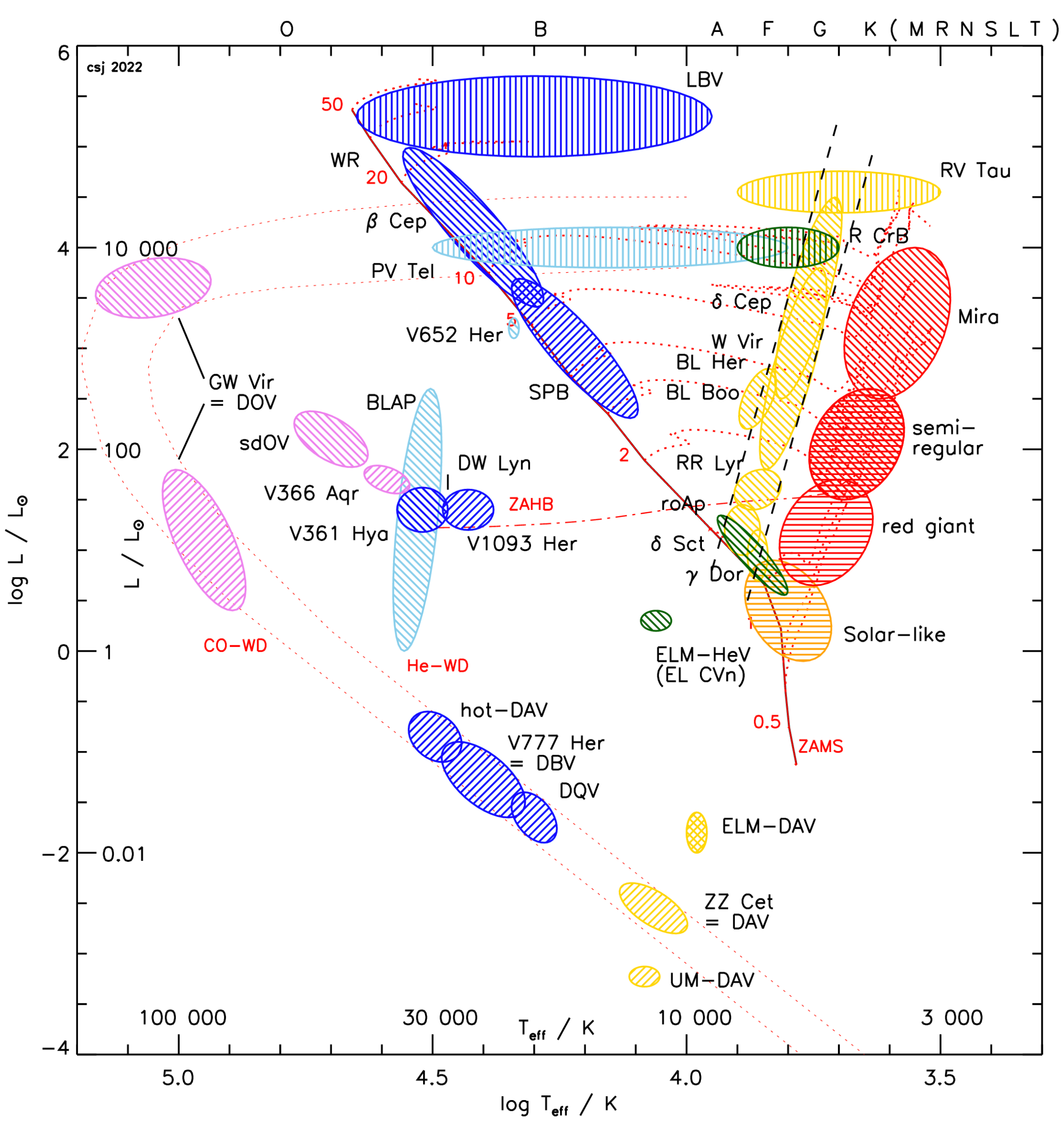}
\caption{Stellar pulsations occur across all regions of the Hertzsprung-Russell Diagram, and \wst\ will provide time-resolved spectroscopy for all of these types, including (nearby) faint sources as well as very distant (even extragalactic) luminous sources. Taken from Fig\,3.4 in the LSST-TVS Handbook \citep{LSST-TVS} with original credit to J.~Christensen-Dalsgaard and C.S.~Jeffery}
\label{fig:td:HRD}
\end{figure}

\paragraph{Atmospheric parameters and elemental abundances} are crucial to understanding the astrophysics of variable stars. In particular, effective temperature ($T_{\rm{eff}}$), surface gravity  ($\log g$), and chemical abundances can be determined from optical spectra, cf. Sect.\,\ref{sec:galactic} for details. Determining these parameters significantly aids classification because pulsations occur in specific regions of the HRD, the so-called instability strips (or regions) illustrated in Fig.\,\ref{fig:td:HRD}. In turn, atmospheric parameters are crucial to test predictions from stellar models, notably with respect to the instability strip boundaries \citep[e.g.,][]{Anderson2016rot,Groenewegen2020,Espinoza-Arancibia2023} and the purity of instability strip regions \citep[e.g.,][for ZZ~Ceti, $\delta$~Sct, and RR~Lyrae stars, respectively]{Castanheira2007,Murphy2019,CruzReyes2024}. Importantly, high-amplitude pulsating variables, such as Cepheids and RR Lyrae stars, exhibit time-variable atmospheric parameters and broadening effects driven by variations in temperature and turbulence \citep[e.g.][]{Luck2004,Kovtyukh2005}.

\paragraph{Simultaneous IFS \& MOS observations} will provide unique and transformational capabilities for detecting stellar variability phenomena in dense resolved stellar populations (cf. Sect.\,\ref{sec:respop}), such as Galactic (globular) clusters and nearby galaxies. In the crowded centers, point-spread function spectroscopy applied to the giant IFS system would be capable of collecting RV time series data for on the order of 10,000 stars within a single 3x3 arcmin$^2$ IFS \fov\ with a precision as good as $0.5$\,\kms, depending on stellar types and brightness \citep[assuming \muse-like performance]{Kamann+2013,Giesers+2018}. These dense regions would also be where the majority of variable objects would be found for a given pointing. In parallel, observations by the high-res mode would probe variability, kinematics, and other spectroscopic information (abundances, etc.) in the less crowded outer regions, while the low-res mode can collect useful kinematics for cluster membership and rotation, for example \citep[e.g.,][]{Meylan1986,Sollima2019}.

\subsubsection{Some highlights from the Variability Tree\label{sec:td:stars:types}}
Stellar variability (SV) occurs on all timescales observable with \wst\ and can denote anything from a single one-off variation, e.g., due to microlensing or eruptive behavior, to well-behaved periodic variations, such as orbital motion or pulsations, cf. Fig.\,\ref{fig:td:variabilitytree}. Whether or not sources are observed to be variable is largely a question of observational power, since all stars are variable at some minor level. 
For many low-amplitude phenomena, single-epoch spectroscopy will be sufficient to characterize and understand the source of variability. However, spectra provide hundreds to thousands of independent, simultaneous flux measurements that can be compared among epochs to discover new forms of variability or to remove ambiguity of variability classification, notably in the case of weaker signals. In the following, we present a short and incomplete selection of stellar variability science where \wst\ will change the game.

\begin{figure}
    \centering
    \includegraphics[width=0.55\textwidth]{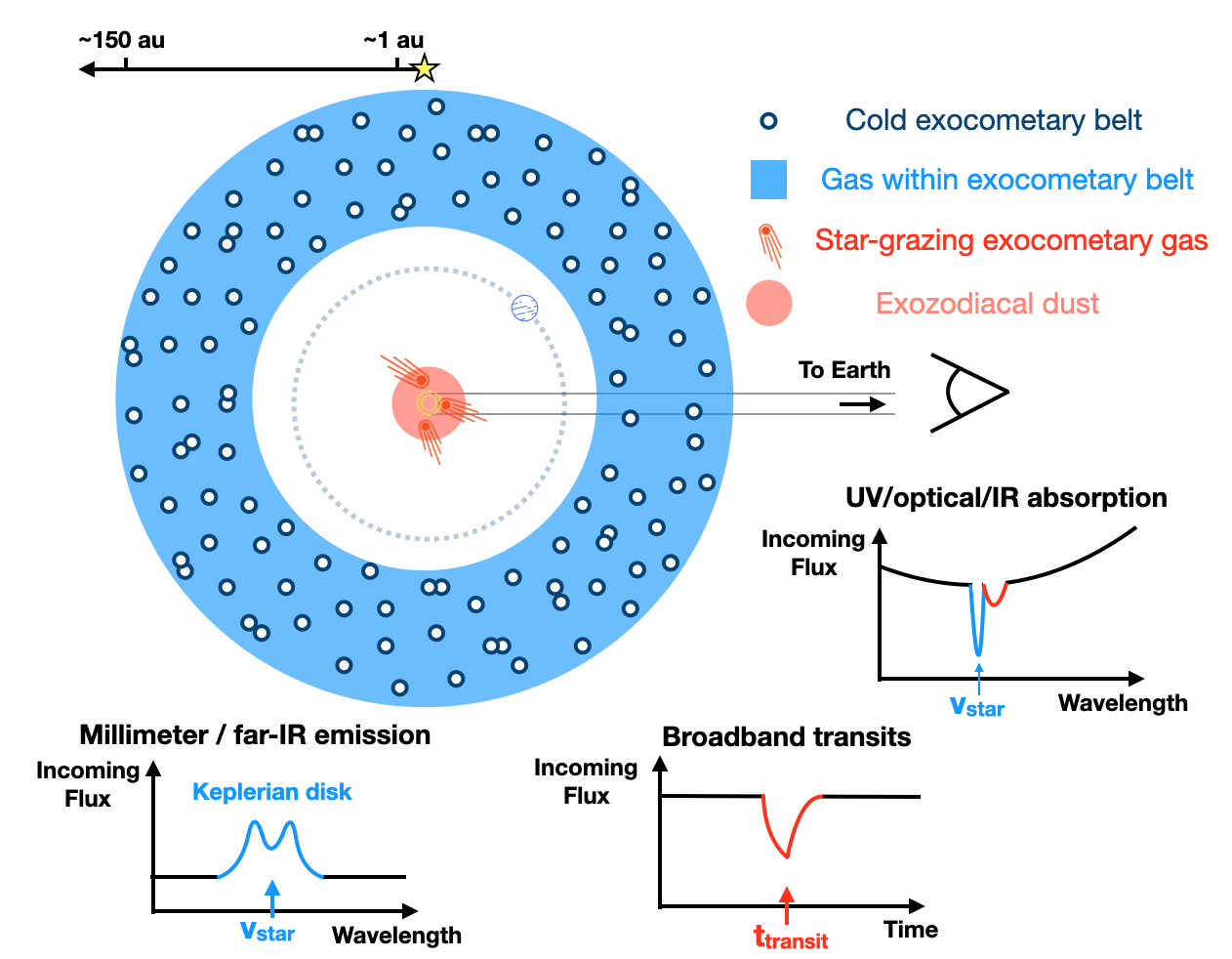}
    \includegraphics[width=0.44\textwidth]{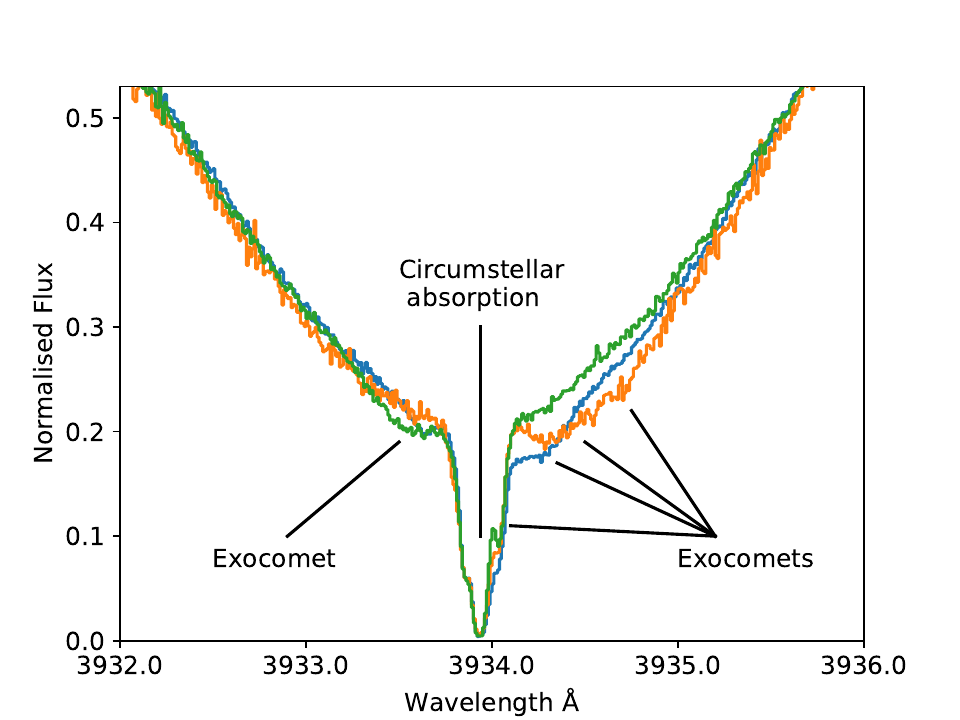}
    \caption{{\it Left:} Cartoon illustration of exocomet signatures in stellar spectra. {\it Right:} exocomet signatures observed in the Ca H\&K lines of $\beta$ Pictoris. Both figures from \citet{Fitzsimmons2023}}
    \label{fig:td:exocomets}
\end{figure}

\paragraph{Exocomets\label{sec:td:exocom}} are the extension of ``ice sublimating bodies'' happening in their approach of stars other than the Sun. Even though the first exo-comets (a.k.a "falling evaporating bodies", FEBs) were observationally discovered before the first exoplanets orbiting sun-like stars \citep{Ferlet87}, their discovery rates have remained low. The presence of FEBs was first inferred from transient components in metallic lines in the Beta Pictoris system. Temporal sequences of many very high quality spectra of the star Beta Pictoris showed spectral lines associated to calcium, displaced by Doppler shifts, and moving in time: one could see the "the star" as one stable contribution, but also some random contributions that appeared and disappeared in time scales of hours and days, cf. Fig.\,\ref{fig:td:exocomets}. These "capricious" additional pockets of gas between us and the star are interpreted as the gaseous tails of comets.

Since this first exciting discovery, targeted individual star monitoring campaigns have reported three other bona fide FEB-hosting stars \citep{exocomets2022}. More recently, larger survey-like initiatives have been conducted taking advantage of spectroscopic archival data originally obtained in the context of exoplanet searchers. Despite these larger efforts, the census remains in the few tens and is severely affected by biases in the samples and data treatment \citep{Fitzsimmons2023}. \wst\ will be able to revolutionize this field mostly because of three technological aspects and its envisioned observation modes: i) extremely large multiplexing capability, ii) current higher spectral resolving power, iii) inclusion in wavelength coverage of the Ca II doublet. These aspects, together with a large number of potentially FEB-hosting targets distributed homogeneously across the whole sky (early type pre or main sequence stars) will allow to investigate the properties of exocomets through simple repeat observations of targets without specific requirements on cadence, etc. 

\paragraph{Young Stellar Objects (YSOs)\label{sec:td:yso}} exhibit significant and diverse variability behaviors from their early phases until they reach the main sequence. The time scales and amplitudes of those variations cover the 10$^{-2}$-10$^{5}$ days and 0-6 magnitude ranges, respectively.
Accretion processes populate the upper area of this parameter space (see Fig. 3 from \citealt{Fischer23}), with episodic accretion causing extreme outbursts that can last longer than decades in the case of FU Orionis objects, and a scaled down and shorter version in EX Lupi-type objects, although the origin of such outbursts and the different between the objects still remain a mystery (e.g., \citealt{Fischer23, Audard2014}). In addition, pre-main sequence variability can encompass a wide range of further cause, such as disk and/or dust occultation, magnetic activity, shocks in jets, etc, which can be investigated thanks to photometric variability
(see, among others, \citealt{Cody14,Venuti2021,Bonito2023}, also focusing on \lsst\ survey strategy optimisation for short-term to long term variability in YSOs).
However, spectroscopic monitoring of accretion-sensitive lines and continuum at various timescales is required to distinguish between the proposed theoretical scenarios that can explain the pile up of material in the disk and the triggering of its accretion onto the star in the case of episodic accretion. These main theoretical scenarios include: combinations of gravitational instability (GI) and magnetohydrodynamic turbulence at the inner accretion disc \citep{Armitage01,Kratter16,Bourdarot23}, self-regulated thermal instability \citep{Bell94}, thermal instability introduced by a massive young planet inside the accretion disc \citep{Lodato04, Clarke05}, the imbalance between gravitational and magnetorotational instability (MRI; \citealt{Zhu09, Elbakyan21}), fragmentation of massive protostellar discs \citep{Vorobyov05, Vorobyov10}, infalling of piled-up disc materials outside the star–disc corotation radius \citep{DAngelo10, DAngelo12}, and flybys or gravitationally bound stellar perturber \citep{Cuello19, Borchert22}. 

Long baseline observations of accreting YSOs utilising intermediate resolution spectroscopic facilities are highly in demand but have never been carried out in a homogenous, systematic way. Continuous monitoring of accretion variability in these stellar systems would allow to quantify various phases of accretion processes using a good variety of magnetic diagnostics in \wst's wide wavelength range. Only recently it has been revealed the accretion variability in Classical T Tauri Stars (CTTS) is relevant in a space of 48 hours and continues varying over a year. The dramatic dimming events monitored in some CTTS, due to occulting materials or differentiated planetesimals in the accretion disks of both CTTS and Herbig Ae/Be stars have also been monitored using photometric facilities but never accompanied by simultaneous spectroscopy. \wst's unprecedented support would lift the major obstacles affecting the measurement of CTTS' accretion properties as well as line variability of major magnetic activity diagnostics - such as Hydrogen Balmer series and Ca II H\&K re-emission lines. H$\alpha$ emission line as well as forbidden emission lines (like e.g. [NII] doublet near H$\alpha$ and [SII] doublet near Lithium line) can be crucial to both discriminate the presence of accretion/ejection processes at work in YSOs (cf. also Sect.\,\ref{gal:starformation}), also in those cases where a strong and variable nebular contribution dominates the emission (\citealt{Bonito2020}). This would benefit from a future IR upgrade to enable complementary studies of low accretors representing the latest stages of accretion. Such low accretors are identified mainly using the He I 10,830\AA line, and previously were misclassified as weak-lined T Tauri stars (WTTS) showing little to no signs of accretion, while in fact, they play a crucial role in our understanding of when the accretion process in YSOs would end which is still an open question \citep{Thanathibodee_2023}.

\paragraph{Classical pulsating and long-period variable stars} are invaluable calibrators of astronomical distance scales and laboratories of stellar evolution. Spectroscopic observations resolve Doppler shifts due to orbital and pulsational motion, line shape variations due to the chromatic pulsations, time-variable atmospheric parameters, and are required to standardize luminosity distances.

Classical Cepheids trace ages on the order of a few tens to several hundreds of Myr \citep[e.g.,][]{Anderson2016rot}, and reliably trace  Galactic abundance gradients, notably of [Fe/H] and $\alpha-$elements, such as oxygen and sulfur \citep[e.g.,][]{Luck2011,Genovali2014,Ripepi2022}, rendering them crucial for piecing together the Milky Way's recent evolutionary history \citep[e.g.,][]{Luck2018,Lemasle2022,daSilva2023}. Additionally, the chemical composition of classical Cepheids is relevant for their use as standard candles \citep[e.g.,][]{Breuval2022,Anderson2024bookchap}, for understanding internal mixing processes, e.g., due to rotation \citep{Anderson2014rot}, and for testing pulsation models \citep{DeSomma2022,Marconi2024}. There are $\sim 9700$ classical Cepheids \citep{OGLE-MCs-CEP,GDR3-SOS-CEP} in the Magellanic system, nearly all of which will be accessible to detailed and time-resolved spectroscopic characterization using \wst's MOS-HR since $96\%$ of Cepheids in the Magellanic system have average $V-$band magnitudes $\lesssim 18$\,mag. A similar number of classical Cepheids is expected to reside in the Milky Way, although extinction within the Galactic disk will limit the spectroscopically accessible fraction. By contrast, there are on the order of $\sim 500$ classical Cepheids whose abundances have been measured using high-resolution spectroscopy. High-res mode observations with \wst\ will thus radically change this situation, enabling detailed abundance analyses for more than an order of magnitude increase in the sample of classical Cepheids. Additionally, they will greatly improve the classification of pulsating stars in the MW when parallax information is insufficiently precise by detecting variable line shapes. 

RR Lyrae stars are population-II stars and among the most numerous high-amplitude pulsating stars \citep[$\sim 45000$ in the Magellanic Clouds, $> 270,000$ in the Milky Way]{OGLE_RRL,GDR3-SOS-RRL}, and yet, their origin remains poorly understood. Typically considered to be part of the Galactic halo, where they trace stellar streams \citep[e.g.,][]{Belokurov2018unmixing,Prudil2021}, there has recently been intense debate concerning RR Lyrae stars whose metallicity and kinematics suggest membership of the Galactic disk \citep{Iorio2021}. Furthermore, the discovery of several counter-rotating RR Lyrae stars in the Galactic halo hints at origins in Galactic merger events \citep{Feuillet2022,Medina2023}. Determining accurate elemental abundances spectroscopically is thus critical for understanding the origin of RR Lyrae stars, for calibrating relations between chemical abundances and light curve parameters \citep[e.g.,][]{Jurcsik1996,Li2023RRL}, and for using RR Lyrae stars as standard candles \citep{Garofalo2022}, notably in the infrared \citep{Bhardwaj2022,Marconi2015}. \wst\ IFS observations of RR Lyrae stars in globular clusters \citep{CruzReyes2024} would be very useful to this end and could further provide an interesting complementary view of the problem of multiple populations in globular clusters. 

More than $100,000$ Southern (DEC $ < 20$ deg) RR Lyrae stars brighter than $G \lesssim 18$\,mag \citep{GDR3-SOS-RRL} will be fully characterizeable using \wst's high-resolution MOS ($R = 40,000$), and more than twice this number of stars will be chemically tagged across all parts of the Galaxy. This will result in a gain of three orders of magnitude compared to the currently largest homogeneous catalog of high-resolution spectroscopy of RR Lyrae stars \citep{Crestani2021}, which contains merely 208 RR Lyrae stars. Thanks to such transformative capabilities, \wst\ is bound to identify exciting new peculiarities that will drive new discoveries for stellar physics and near-field cosmology. An extensive spectral coverage is essential for determining elements from various nucleosynthesis channels, ranging from light elements like carbon to $\alpha$-elements (e.g., Mg, Ca), iron-peak elements (e.g., V, Sc, Cr, Co), and both slow- and rapid-neutron capture elements (e.g., Sr, Y, Ba, La, Eu, Nd).

\begin{figure}[t]
    \centering
    \includegraphics[width=\textwidth]{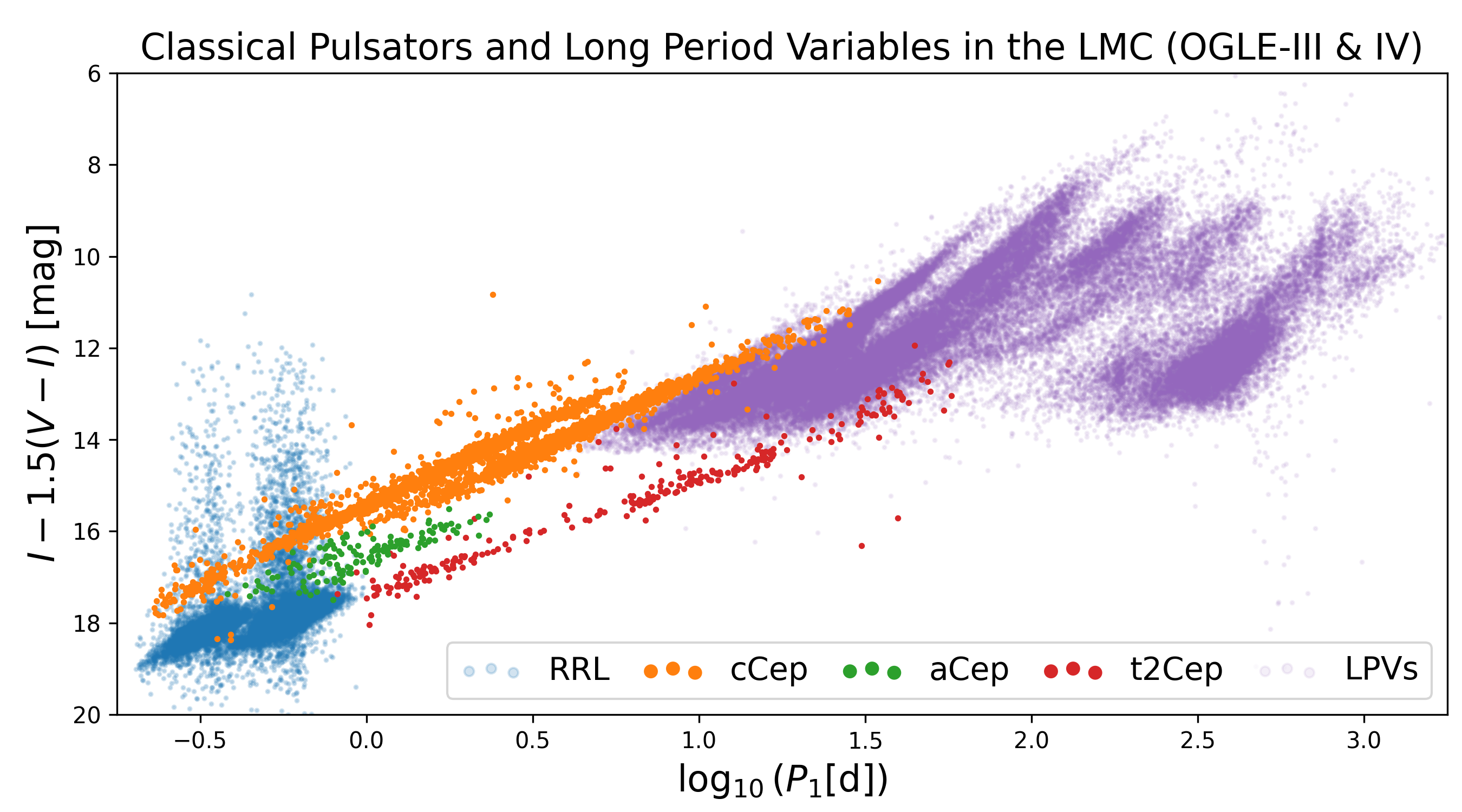}
    \caption{Leavitt laws of classical pulsators and long-period variables in the LMC from the OGLE collection of variable stars. Reddening-free Wesenheit magnitudes \citep{Madore1982} are shown versus the logarithm of the dominant pulsation period for 140 anomalous Cepheids \citep{OGLE_Cepheids}, 4435 classical Cepheids \citep{OGLE_Cepheids,OGLE-MCs-CEP}, 38853 RR Lyrae stars \citep{OGLE_RRL}, 275 type-II Cepheids \citep{OGLE_T2Cep}, and 91374 LPVs \citep{OGLE_LPV}. Different pulsation modes are not distinguished but show as parallel sequences, and single value of the reddening coefficient, $R^W_{VI}=1.5$, is adopted for the sake of simplicity and for illustration only. \wst\ will feature unprecedented capabilities for collecting time-resolved high-quality spectroscopic observations for all of these sequences (and more) in the entire Magellanic system and other high-cadence focus regions. Indeed, \wst\ will be the only facility capable of  collecting high-resolution spectroscopy of $10^5-10^6$ pulsating stars across the Southern sky to understand the physics of these stars and calibrate their use as standard candles.}
    \label{fig:td:pulsatorsLMC}
\end{figure}

\wst\ will further unravel the chemical composition of hundreds of thousands of long-period variable (LPV) stars that exhibit several different period-luminosity sequences \citep{Wood1999,OGLE-SMC-LPV}. For example, spectroscopy is required to understand population differences among small amplitude red giant stars used to measure distances using the Tip of the Red Giant Branch (TRGB) method \citep{Lee1993TRGB,Anderson2023TRGB}, which is the most common stellar standard candle in the local Universe and particularly suited to infrared observations with the \jwst\ \citep{Anand2021EDD,Anand2024jwst}. \wst\ will be unmatched in its ability to provide high-quality spectra and abundances required to unravel the multi-periodicity and the multiple period-luminosity sequences exhibited by LPVs. \wst\ will unravel the effects of chemical composition on the application of semi-regular variables and Miras as standard candles \citep{Whitelock2008,Huang2018,Huang2020,Trabucchi2021} as well as to understanding their evolution and pulsation driving mechanisms \citep{Wood2015,Trabucchi2019,Trabucchi2022}.

The Magellanic Clouds play a crucial role for understanding pulsating star populations thanks to stars residing at nearly a common distance \citep{Leavitt1908,OGLE_Cepheids,Pietrzynski2019} and allow the most precise determination of metallicity effects on distance measurements \citep{Breuval2022}. Figure\,\ref{fig:td:pulsatorsLMC} illustrates Leavitt laws \citep{Leavitt1912} of several pulsating star classes in the LMC from the OGLE collection of variable stars\footnote{\url{https://ogledb.astrouw.edu.pl/~ogle/OCVS/}}. Pulsating stars also unravel the line of sight structure and geometry of the Magellanic Clouds  \citep[e.g.,][]{Bhuyan2023lmc}. Understanding the internal kinematics of the Magellanic clouds using pulsating stars will require dealing with metallicity effects, pulsations, and multiplicity all at once.  For the first time, \wst\ can perform a complete and homogeneous  inventory of time-resolved high-resolution spectra for $> 10,000$ classical Cepheids, $> 100,000$ RR Lyrae stars, and several $10^5$ long-period variable stars in the Galaxy and the Magellanic Clouds ($V \lesssim 18$\,mag). Only \wst\ is capable of thus completing the spectroscopic complement to Henrietta Leavitt's seminal work begun more than a century ago by fully characterizing classical pulsators in the Magellanic System based on optical spectra.

\paragraph{White Dwarfs} (WDs) are the end stage of stellar evolution for the vast majority of stars. 
Around 20 per cent of the currently observed WDs are members of (detached or accreting) binaries and multiple systems \citep{Torres+2022}, and are excellent probes for testing (i) the current models of star formation and evolutionary mechanisms \citep{Badenes+2018}, (ii) the correlation among age-related stellar properties such as their activity, rotation, and metallicity \citep{Rebassa-Mansergas+2021} and (iii) the key ingredients (angular momentum losses, mass transfer process and response of the donor star to the mass loss) of the models describing the evolution of compact binaries \citep{Belloni+2023}.

While $\simeq 150,000$ WDs in binaries will be observed by \fourmost, the majority of the targets will be observed with the low-resolution instrument, since they are intrinsically faint. Only the large aperture of \wst\ will allow to study the faint ($G > 20$ in LR and $G > 16$ in HR) targets, i.e. those that are not accessible by \fourmost, in detail and to obtain phase-resolved observations with sufficiently short exposure times to adequately sample the orbital phase for short-period compact binaries.		

\begin{figure}[t]
  \centering
  \includegraphics[width=0.9\textwidth]{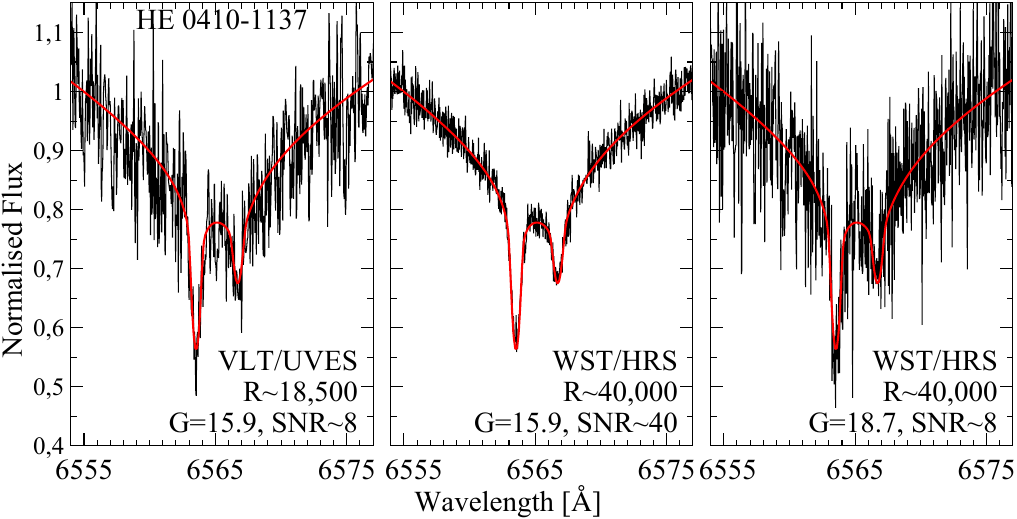}
  \caption{\small The VLT/UVES spectrum of the double WD HE 0410-1137 observed by the SPY program is shown on the left. Short exposure times (10 minutes) are required to avoid orbital smearing, and high spectral resolution is required to separate the two narrow non-LTE H$\alpha$ cores of the binary WD. Current efficiency estimates predict significant gains over VLT/UVES, so that \wst\ will achieve higher S/N at the current limiting magnitudes (central panel, simulated spectrum), while also enabling high-resolution observations of targets fainter than those accessible to current 8-meter class telescopes (right panel, simulated spectrum). 
  The data are shown in black while over-plotted in red are the best-fitting model spectra.}
  \label{fig:NLTE_Ha}
\end{figure}

Detached double WDs have been proposed as Type-Ia Supernova (SN\,Ia) progenitors in the double-degenerate channel but, so far, their characterization through observations has been limited to $\simeq 100$ bright ($G < 16$) objects that can be observed with the current high-resolution spectrographs at 8m-class telescopes, see e.g. the SPY (SN\,Ia Progenitor surveY) survey with the VLT/UVES spectrograph \citep{Napiwotzki+2020}. Only \wst's high-res mode RVs will allow for the identification of $\simeq 8000$ detached double WDs down to $G \simeq 19$ via the detection of the narrow non-LTE core of the Ha line (Figure\,\ref{fig:NLTE_Ha}) and will be invaluable for understanding the progenitors of SNe\,Ia.

\wst’s low-res mode will achieve at least an 1-kpc volume-limited characterization of $\simeq 4000$ accreting WDs ($G \lesssim 23$). Conducting a detailed census of these binaries based on low-res mode RVs will be crucial for understanding the pathway towards SNe\,Ia via the single-degenerate channel \citep{Whelan+1973}, constraining accretion physics by Doppler tomography \citep{Marsh2005}, and for testing the models of binary evolution \citep{Pala+2020}. In particular, by unveiling the population of compact WD binaries ($P_\mathrm{orb} < 2\,$hours), \wst\ will also allow to characterise the foreground noise from unresolved Galactic binaries that will eventually limit the sensitivity of the ESA \textit{Laser Interferometric Space Antenna} (\textit{LISA}) mission \citep{Nelemans2009,Kupfer+2019,Scaringi+2023}.

\paragraph{Black Holes} (BHs) are the fascinating end products of massive star evolution. Prior to the first detections of gravitational waves by LIGO/VIRGO, the majority of stellar-mass BHs were detected as X-ray binaries \citep[which exhibit a plethora of variability features over timescales of seconds to decades, e.g.,][]{Gandhi2016}, and only approximately 20 of them have been dynamically confirmed \citep{Corral-Santana2016}. However, a much larger population of dormant, non-active, BHs is waiting to be discovered \citep{Mazeh2008}. Two first candidates have been identified from \gaia\ data recently  \citep{ElBadry2023}, and much more will be detectable by \wst. Additionally, tidal deformation in close-in systems can lead to ellipsoidal variability, which will be known from ground-based surveys, \gaia, and \lsst\ \citep{Gomel2023}, and whose orbital periods are measured photometrically. Using 5 to 10 epochs per star, \wst\ will unambiguously identify the dormant BH companions in these systems via their extreme orbital velocities. A pilot study will be carried out using \fourmost\ \citep{Pawlak2023}. However, \fourmost\ will only cover the systems with OB-type primaries in the Magellanic Clouds. Conversely, \wst\ we will be able to go both whole-sky and deeper, increasing the size of the sample by two orders of magnitude (from ~700 to possibly tens of thousands) and also covering most possible companion spectral types. \wst's large collecting area and time-domain focus are required to collect a statistically significant sample of stellar mass BHs in the MW and LG that will unravel the dormant BH population, its mass distribution, and formation history. However, the simultaneous use of the IFS and MOS in nearby galaxies and star clusters will be particularly powerful for detecting BH companions to ordinary stars, see Sects.\,\ref{subsec:respop-clusters} and \ref{subsec:respop-LMCSMC}.
Detecting dormant BHs is largely complementary to studying BH populations using GW events. BH binaries are detected at earlier evolutionary stages, and some of these systems will eventually become BH-BH binaries and potential GW sources. Studying dormant BHs with \wst\ will therefore provide a much broader perspective on the evolution of the massive stars and their end products.

\subsection{Solar system\label{td:solarsystem}}

Our own solar system is the only known example of a planetary system harbouring life. It is also the only planetary system for which we can study a range of bodies resulting from the planetary formation process (from asteroids to giant planets) in detail and sometimes even in-situ. In that context, understanding how our own solar system formed and evolved is crucial to understand the formation and evolution of planetary systems. Small bodies of the solar system, like asteroids and comets, are remnants from the planetary formation process. They are thus essential tools to understand the history of our solar system. 

Comets are among the most pristine relics of the protoplanetary disc, where planets formed and evolved. When a comet approaches the Sun, the ices contained in its nucleus sublimate to form an atmosphere of gas and dust around the nucleus called the coma. Interstellar comets, formed around another star and crossing the solar system are also of great interest, as they have the potential to inform us about the planetary formation process in other planetary systems. By observing the composition and morphology of the coma of (interstellar) comets, we can probe cometary ices retaining precious clues about the conditions prevailing in the early stages of our solar system. 

A key indicator of the conditions prevailing in the early solar system is isotopic ratios. Several isotopic ratios can be measured in the coma of comets at optical wavelengths ($^{14}$N/$^{15}$N and $^{12}$C/$^{13}$C in particular), providing precious clues about nitrogen reservoirs in the early solar system for example \citep{Rousselot2014}. A multi-fibres instrument with a high spectral resolution will  enable measurements of these isotopic ratios simultaneously at different spatial positions in the coma to search for variations, which has never been done before and is impossible with current facilities.

One of the outstanding issues in the study of comets is to link small molecules (radicals) produced in the coma by the photo-dissociation of more complex molecules and observed at optical wavelengths to their precursor present in the nucleus ices. It has been shown in recent years that key steps toward answering that question can be made using IFS to produce simultaneous maps of  the spatial distribution of different radicals in the coma \citep{Opitom2019}. IFSs mounted on large telescopes have also proven very efficient to detect faint levels of activity around solar system  and interstellar objects \citep{Opitom2020}. Even if cometary activity has been studied for decades, we still lack a consistent picture of the activity of small bodies across the solar system, especially far from the Sun.. A key issue with current IFSs is the field of view covered. The coma of comets can spread over arcminutes, and the limited \fov\ on \muse\ and likely the future BlueMUSE means we are unable to sample the whole coma. A large \fov\ IFS on a 8m-class telescope will provide an incredible opportunity to detect and map the emission of CN, N$_2^+$, CO$^+$, C$_2$, NH$_2$, and [OI] in a range of comets and interstellar objects permitting to significantly increase our understanding of cometary activity and the release of species in the coma. 

While comets contain ice and are thus mostly observed with a atmosphere, asteroids are not. Observing them can reveal a wealth of information about the mixture of materials making up their surface. These measurements are key to understanding the evolution of the inner solar system \citep{DeMeo2014}. However, asteroids are faint and to this day we are missing a large sample of spectroscopic observations of different type of asteroids. With its large filed of view IFS, \wst\ offers the opportunity to observe multiple asteroids at the same time and quickly build a representative sample of asteroid surface spectra for different types of objects. 
\begin{figure}[t]
    \centering
    \includegraphics{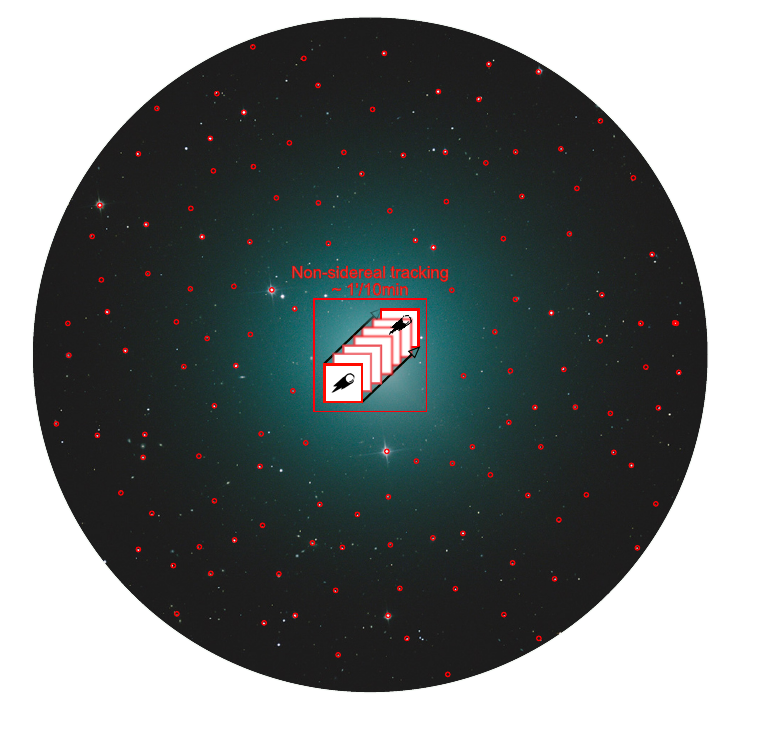}
    \caption{Illustration of observation of a solar system object with \wst (not to scale). Motion of the IFS inside a box could be used to track the motion of solar system objects. This illustrates the advantage of the large \fov\ of \wst\ for the observation of extended objects. Photo credit: Alex Cherney.}
    \label{fig:td:solarsystem}
\end{figure}

Further away in our solar system, Kuiper Belt Objects (KNOs)  are another class of objects that are thought to have remained relatively unaltered and spectroscopy of their surface has been used to study their composition. However, these objects are faint and far away from the Sun so these measurements remain very difficult to perform with currently available facilities. The surface of Triton, the biggest satellite of Neptune, that is also known to be a captured KBO \citep{Agnor2006}, is composed of volatile-rich ices, including N$_2$, H$_2$O, CO$_2$, CO, and CH$_4$ \citep{DeMeo2010}, with N$_2$ as the dominant species. Ongoing ground-based campaigns determined short \citep{Hicks2004} and long-term variability \citep{Grundy2004} of the major components. \cite{Grundy2004} and \cite{Grundy2010} show that the depth of the  N$_2$ spectral absorption varies with Triton’s rotation. CO has a similar variability with respect to Triton’s rotation, suggesting that CO and N$_2$ co-occur, while CH$_4$ behaves differently. One CH$_4$ band, centred at 0.89 nm, has been used to study the variability of this species and is observable with \wst. Triton being a captured KBO \citep{Agnor2006,mccord1966,mckinnon1984}, we expect at least some KBOs to have a similar surface composition and variability. Hence, a systematic survey of a large number of KBOs in the visible range would help understand the composition of these objects, providing a statistically significant and homogeneous sample. The expected large \fov\ of \wst\ would allow us to observe several KBOs at the same time rapidly building a significant sample.

Finally, because IFSs are so efficient for the study of faint extended emissions, a large IFS on \wst\ could also be used to study the exosphere of the Galilean satellites. Galilean satellites have been shown to have an exosphere \citep{McGrath2004} resulting from sputtering and sublimation of its surface. The study of their exosphere is thus key to constraining the composition of the galilean satellites and the processes at play on their surface. Na and K have been also observed at Europa, using ground-based facilities \citep{Brown1996,Brown2001}, that could have an endogenic origin and hence be used as a proxy of the composition of the supposed subsurface ocean of the satellite \citep{Ozgurel2018} if their endogenic nature were confirmed.  Optical auroral emissions of neutral oxygen at 630.0 nm, 636.4 nm, 557.7 nm, 777.74 nm and 844.46 nm were detected on Ganymede’s sub-Jovian hemisphere in eclipse \citep{deKleer2023}. These exospheric emissions and their temporal variations can be studied using \wst. The search for exospheric compounds is a key scientific goal in the near future for the outer planets satellites, with a special attention to the Galilean moons. Considering the ESA Juice mission arrival to the Jovian system in 2031 \citep{grasset2013} and the NASA Europa Clipper to be launched this year with an arrival scheduled for 2030, \wst\ will be perfectly in time to complement space-based observations with JUICE's vis-nir spectrometer MAJIS (Moon And Juice Imaging Spectrometer, \cite{piccioni2013}) and the EIS (Europa Imaging System, \cite{turtle2023}) camera on board Europa Clipper.

\subsection{Serendipity and Alerts\label{td:serendipitous}}

Certain science cases require dedicated time-resolved programs. However, the need to split the observations of  almost all surveys anticipated with \wst\ into multiple sub-exposures naturally gives rise to repeat observations that create unprecedented opportunities for  serendipitous and unprejudiced time-domain science. A significant number of astronomical sources in the sky are variable on time-scales varying from a few seconds to several years and optimising the breakdown and temporal sampling of observations for all surveys with \wst\ enables additional studies of expected (and unexpected) variable sources. 

All of the proposed instrument modes have a high potential for serendipitous discoveries by opening new discovery spaces when exposures are separated in time by a useful amount\hbox{---}even without adopting a specific cadence. Hence, there exist enormous synergies between the time-domain science case and all others. For example, extra-galactic IFS surveys would enable us to detect variability in resolved galaxies (cf. Sects.\,\ref{subsec:respop-clusters} and {subsec:respop-massivestars}). High spectral resolution observations of stellar fields with the MOS would strongly help classify and characterize (e.g., radially pulsating) variable stars and detect multi-lined binaries. Low resolution observations of white dwarfs at random epochs would enable the discovery of short period binaries, such as white dwarf binaries, by detecting radial velocity variations. Furthermore, the variability of emission features or flares \citep{Kowalski2024} in all sorts of objects could be quantified statistically and as a function of wavelength. To enhance the likelihood of serendipitous discovery, we recommend that all instrument modes are used at all time, even if no target has been identified for the IFS or some of the MOS fibers at a given time. Incidentally, this will include untargeted IFS observations that could lead to particularly interesting surprises.

To maximize serendipity, the total exposure times for all surveys should be split into at least three sub-exposures each. While the optical temporal separations  will depend on specific science cases, non-identically (and possibly logarithmically) spaced exposures spread throughout at least a season would allow us to sample a variety of temporal scales from weeks to months.

To take advantage of this discovery space, it is essential to not only optimise the scheduling of the various surveys, but also to process and analyse observations in real time. Thus, \wst\ will for the first time \emph{issue} spectroscopic alerts via a dedicated broker system. Developments in the study of astronomical transients from alert streams generated by all-sky photometric surveys, particularly by the \lsst\ alert brokers \citep[e.g.,][among others]{alerce,antares,lasair,fink}, will be the guide for \wst. Variability signals detected by \wst\ will occur on timescales of hours to days, which requires real-time data reduction and calibration to detect the event while it is happening in order to coordinate further follow-up. These requirements place constraints on the data transfer and reduction pipeline for the facility. To facilitate the detection of all types of variable sources, the \wst\ archive should be conceived as a repository for external (non-\wst) reference spectra that is VO compliant. Moreover, alerts processing should be done on-site, requiring significant additional computational and energy infrastructure.

Of course, the number of alerts issued by \wst\ will be much smaller than the number issued by \lsst. However, \wst's data complexity will be much higher, and spectroscopic training sets available for classification will be significantly less comprehensive. In addition to data reduction, basic data analysis should be performed soon after the observations, to compare spectra of objects to existing templates and detect changes (e.g., \citealt{Green22}). Coupled with a system to generate alerts when changes or anomalies are detected and pass the information to the community for potential follow-up, this type of infrastructure would transform \wst\ into a true transient-detection and -characterization facility. 

\clearpage

\clearpage


\section{Exoplanet, Stellar and Galactic Science Case}\label{sec:galactic}

\paragraph{Authors} Rodolfo Smiljanic,$^{11}$ Eline Tolstoy$^{12}$, Vanessa Hill$^6$, Tadafumi Matsuno$^{12}$, Georges Kordopatis$^6$, Laura Magrini$^{14}$, Richard I.~Anderson$^2$, Francesca Annibali$^{16}$, Amelia Bayo$^1$, Michele Bellazzini$^{16}$, Maria Teresa Beltran$^{14}$, Leda Berni$^{24}$, Simone Bianchi$^{14}$, Katia Biazzo$^{25}$, Joss Bland-Hawthorn$^{28}$, Henri M.~J.~Boffin$^1$, Rosaria Bonito$^{30}$, Giuseppe Bono$^{31}$, Dominic Bowman$^{32}$, Vittorio F.~Braga$^{25}$, Angela Bragaglia$^{16}$, Anna Brucalassi$^{14}$, Innocenza Bus\`a$^{35}$, Giada Casali$^{19}$, Viviana Casasola$^{40}$, Norberto Castro$^{41,10}$, Lorenzo Cavallo$^{45}$, Cristina Chiappini$^{10}$, Laura Colzi$^{46}$, Francesco Damiani$^{30}$, Camilla Danielski$^{14}$, Ronaldo da Silva$^{25,51}$, Roelof S.~de Jong$^{10}$, Valentina D’Orazi$^{31,50}$, Ana Escorza$^{21,22}$, Michele Fabrizio$^{25}$, Giuliana Fiorentino$^{25}$, Francesco Fontani$^{25}$, Patrick Fran\c{c}ois$^{26}$, Francisco J.~Galindo-Guil$^{57}$, Daniele Galli$^{14}$, Jorge Garcia-Rojas$^{21,22}$, Mario Giuseppe Guarcello$^{30}$, Amina Helmi$^{12}$, Daniela Iglesias$^{59}$, Valentin Ivanov$^1$, Pascale Jablonka$^2$, Sergei Koposov$^9$, Sara Lucatello$^{50}$, Nicolas Martin$^{66}$, Davide Massari$^{16}$, Jaroslav Merc$^{81}$, Thibault Merle$^{82,83}$, Andrea Miglio$^{84}$, Ivan Minchev$^{10}$, Dante Minniti$^{86,87,88}$, N\'uria Miret Roig$^{71}$, Ana Monreal Ibero$^{89}$, Ben Montet$^{90,91}$, Andres Moya$^{93}$, Thomas Nordlander$^{19}$, Marco Padovani$^{14}$, Anna F.~Pala$^{96}$, Loredana Prisinzano$^{30}$, Roberto Raddi$^{103}$, Monica Rainer$^{104}$, Sofia Randich$^{14}$, Alberto Rebassa-Mansergas$^{103}$, Donatella Romano$^{16}$, Germano Sacco$^{14}$, Jason Sanders$^5$, Lorenzo Spina$^{14}$, Matthias Steinmetz$^{10}$, Grazina Tautvai\u{s}ien\.{e}$^{108}$, Yuan-Sen Ting$^{19}$, Maria Tsantaki$^{14}$, Elena Valenti$^1$, Mathieu van der Swaelmen$^{14}$, Chistopher Theissen$^{109}$, Guillaume Thomas$^{21}$, Sophie Van~Eck$^{82}$, Carlos Viscasillas V\'azquez$^{108}$, Haifeng Wang$^{45}$, Martin Wendt$^{121}$, Nicholas J.~Wright$^{116}$\\

\subsection{Introduction}\label{sec:gal:introduction}

Our knowledge of exoplanet, stellar, and Galactic astrophysics has recently experienced an impressive transformation, thanks to the overwhelming amount of data produced by several large stellar surveys that reached maturity in the last decade or so. The \textit{Gaia} mission of the European Space Agency (ESA) is perhaps the most extraordinary and complete example. \textit{Gaia} has provided so far precision astrometry (positions, parallaxes and proper motions), photometry, and spectrophotometry for almost 2x10$^9$ stars down to a magnitude of $G$ = 20.7 \citep{GaiaMission, GaiaDR3}. Facilities such as the \lsst\ \citep{LSST} and the \euclid\ space mission \citep{Euclid} will soon detect ten times more stars down to fainter magnitudes ($r$ $\sim$ 27 mag) and push the availability of astrometry to objects as faint as $r$ $\sim$ 24 mag. The \nancy\ (to be launched in late 2027) will carry out a deep IR survey of the entire Milky Way plane and the Bulge \citep{Paladini23} to unprecedented depth and resolution, covering a significant fraction of the Milky Way stellar population. These projects and missions will provide targets to address a rich variety of science cases with \wst\ follow-up spectroscopy.

Spectroscopy is needed to realise the full scientific potential of photometric and astrometric surveys. It is the main tool with which we can study the chemical composition and 3D motions in the Universe. Spectra are needed for chemical abundances, radial velocities (RVs), and deeper insights into the physical processes that take place in stars and their environment, including their evolutionary stage, masses and ages. \textit{Gaia} itself has so far (DR3) provided constraints on ages, abundances (of up to 13 chemical species) and RVs (with precision between 0.1-15 km s$^{-1}$) from its spectra (wavelength range 846$-$870 nm and R$\sim 11~500$), with detection and accuracy depending on the spectral type, for 33 million stars with $G \leq$ 14 mag \citep[][]{Katz2023, RecioBlanco2023, Fouesneau2023} of these $\sim$5.6 million stars have atmospheric parameters and abundances. In the next releases, this limit might go down to $G \sim$ 16.2 mag, for a sample of up to 100 million stars with radial velocities. To perform high-quality spectroscopy of fainter stars, and so increase the volume of the Milky Way sampled in detail and with higher precision, new facilities are required. Only a small fraction of the stars in existing and future catalogues could ever be followed up with the current range of facilities. 

To complement \textit{Gaia} with precision RVs and/or a more varied (and precise) inventory of chemical abundances for a large number of stars, wide-field multi-object ground-based spectroscopic surveys are needed. For example, the \textit{Gaia}-ESO Survey \citep{GaiaESO2022a,GaiaESO2022b}, the GALactic Archaeology with HERMES \citep[GALAH,][]{GALAH}, and the Apache Point Observatory Galactic Evolution Experiment \citep[APOGEE,][]{APOGEE}, have observed several million stars at moderate spectral resolution (R $\geq$ 20\,000), providing precise measurements of chemical abundance over a large number of elements. Others, such as the Large sky Area Multi-Object fiber Spectroscopic Telescope \citep[LAMOST,][]{LAMOST2022} and the Sloan Extension for Galactic Understanding and Exploration \citep[SEGUE][]{SEGUE}, have provided RVs for stars almost as faint as the limiting magnitude of \emph{Gaia} through low-to-medium resolution spectroscopy (R $\gtrsim$1500). Although these surveys (in terms of resolution and wavelength coverage) and the samples they targeted (in terms of magnitude limits and stellar populations) are extremely heterogeneous, they have contributed significantly to the progress in exoplanet, stellar, and Galactic astrophysics.

Surveys and instruments that have recently started or will start soon include the WHT (William Herschel Telescope) Enhanced Area Velocity Explorer \citep[WEAVE,][]{WEAVE}, the 4-meter Multi-Object Spectroscopic Telescope \citep[4MOST,][]{4MOST}, the Milky Way Mapper (MWM), part of the Sloan Digital Sky Survey V \citep{SDSS_V}, the Dark Energy Spectroscopic Instrument (\desi) Milky Way Survey \citep[MWS,][]{Cooper2023}, the Prime Focus Spectrograph (PFS) at the Subaru 8.2-meter telescope \citep{Sugai2015}, and the Multi-Object Optical and near-Infrared Spectrograph \citep[MOONS,][]{MOONS2022} at the VLT. Together, these new efforts will more than double the number of stars with available ground-based spectroscopy by 2035, compared to today, for a total of $>$2--3 $\times$ 10$^7$ objects. However, these facilities will have limitations. This is particularly relevant when it comes to the ability to follow up the fainter targets from future major photometric surveys. It is also important for the intrinsic accuracy of the spectroscopic measurements and the number of chemical elements that can be detected at fainter magnitudes and in a range of stellar types, especially for low-metallicity stars.

\begin{figure}[!t]
  \centering
  \includegraphics[width=0.8\textwidth]{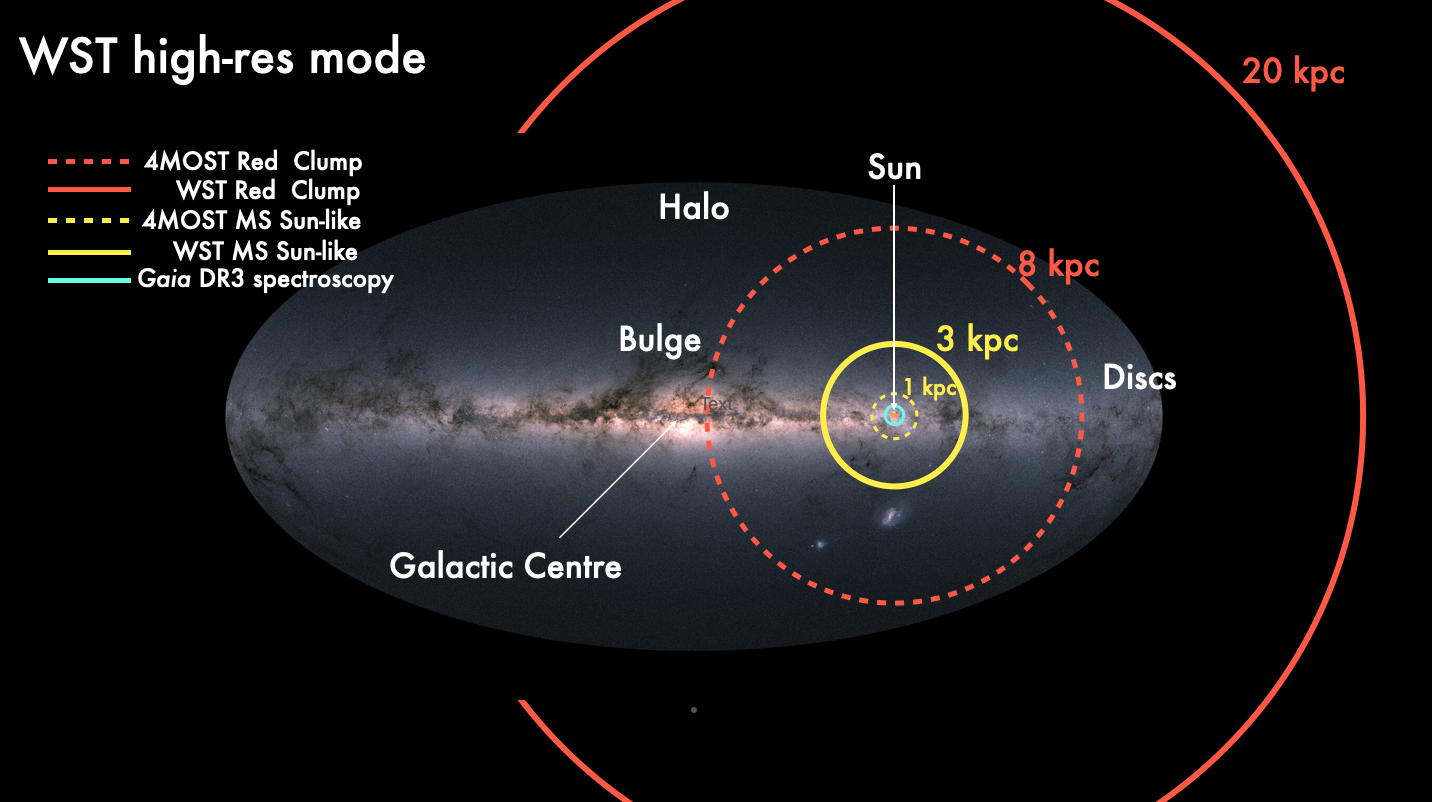}
  \caption{Sky map with circles comparing the regions up to where a certain stellar type can be observed in high-resolution mode with \wst (red and yellow solid lines) and \fourmost\ (red and yellow dashed lines). The region where \emph{Gaia} DR3 spectroscopy is available is shown in cyan
}
  \label{fig:volume}
\end{figure}

Apart from the relatively limited sky coverage \textit{Gaia}-ESO Survey and \moons\ at the VLT and the new \pfs\ instrument at the Subaru telescope, all previous and currently planned surveys make use of telescopes with a 4 metre aperture or smaller. This significantly restricts the spatial volume of the Milky Way accessible to precision spectroscopy (Fig. \ref{fig:volume}). In terms of spectral resolution, most projects take spectra at or below R~$\sim$~20\,000, except for GALAH, with R~$\sim$~30\,000, and the $\sim$7\,000 stars observed by the \textit{Gaia}-ESO Survey with R $\sim$ 40\,000 \citep{Smiljanic14, Worley24}. 

The MSE project \citep[][]{MSE_Instrument}, which is still in the planning stages, includes a similarly large multi-object spectroscopy facility. This is the only planned facility that is similar to the \wst\ and its position in the northern hemisphere makes it highly complementary. The 39-metre Extremely Large Telescope (ELT) currently being built by ESO, and other large telescope projects, will eventually provide access to various spectroscopic facilities, all with limited fields of view focussing on relatively small numbers of very faint targets, and thus very different science cases.

Therefore, in current and planned surveys, the combination of depth, completeness, sky coverage, and spectral resolution that \wst\ will offer is lacking. \wst\ will be able to overcome the limitations mentioned above, providing a unique opportunity of opening new parameter spaces to undertake comprehensive, new ground-breaking exoplanet, stellar, and Galactic science that is beyond the reach of current existing and planned facilities. The capabilities of the \wst\ MOS will be critical to the advancement of astronomical research within the facility landscape in the next decade (see Fig.\ \ref{fig:Intro:Synergies} in Section \ref{sec:introduction}). \wst\ will be fundamental for the systematic observation of new stellar targets discovered by \lsst\ and \euclid\ in the Milky Way and beyond, using the low-resolution mode. \textbf{In this mode, \wst\ has the potential to obtain almost 10$^8$ stellar spectra in 10 years}. The high-resolution spectroscopy mode is unique and essential for enabling precision chemical abundance analysis of weak lines in targets that cannot be explored at this level of detail for large samples by other facilities. In this mode, \wst\ has the potential to obtain almost 3.3$\times$10$^7$ stellar spectra in 10 years. That is one order of magnitude more than will be available from previous MOS surveys at the same resolution. Furthermore, the careful choice of wavelength coverage will be crucial to open access to key spectral features that are absent from other surveys and will significantly improve the accuracy and range of stellar types with measurements of numerous additional chemical elements. The IFS will also enable detailed investigations of resolved stellar populations in densely populated regions such as the Bulge and the central regions of dense stellar systems such as star clusters and dwarf galaxies (see more details of such cases in Section \ref{sec:respop}).

In the following sections, we discuss four broad cases that have driven the current technical requirements of the \wst\ instruments in terms of spectral resolution, efficiency, target numbers, radial velocity stability, and field of view. They form the core of the \wst\ requirements for exoplanet, stellar, and Galactic science cases. 
 
It is also important to note that the cases presented in this document are not exhaustive and that numerous other exciting scientific opportunities in exoplanet, stellar, and Galactic astrophysics will also be possible with a \wst\ built to these specifications.

\subsection{Origins of the elements}\label{sec:gal:elements}

The seminal works of \citet{Burbidge57} and \citet{Cameron1957} laid the basis for our modern understanding of the nucleosynthesis of the chemical elements. All chemical elements heavier than H, He \& Li (and some fraction of He and Li themselves as well as Be and B) have their origins connected to the life and death of stars and are returned to the interstellar medium either by stellar winds or in different types of stellar explosions and mergers. Since the late 1950s, our understanding of nucleosynthesis has progressed tremendously, proving many of the early ideas correct, refining others, and also adding new processes to those that explain the origin of the elements \citep[see e.g.][for recent reviews]{Diehl2022, ArconesThielemann2023}. However, major uncertainties still remain regarding the exact sites and stellar evolutionary stages at which several of the key physical processes that chemically enrich the Universe take place. Pinning them down will have a significant impact across numerous areas of astrophysics, as stellar and chemical evolution impacts many diverse fields of physics, astrophysics, and chemistry, and this is the main goal of this science case.

Among the fundamental open questions motivating this case, one may mention: \textit{Do the known nucleosynthesis processes explain the origin of all of the elements in the Universe? Are there nucleosynthetic processes that we have not yet identified? How do stellar yields change as a function of stellar metallicity and mass?} Achieving a more comprehensive understanding of the origin of the elements and the properties of all nucleosynthetic processes is essential for a variety of other astrophysical problems. Indeed, this science case permeates the other exoplanet, stellar, and Galactic cases discussed here, supporting the search for answers to questions such as: \textit{How does star and planet formation change as the Universe becomes enriched with metals? How do stars lose mass? How do (massive) stars explode? How does the evolution of stars in binary or multiple systems differ from the evolution of a single star? Can any type of explosive event be used as a standard candle?} These complex questions can be addressed by observational tests that explore trends and scatter in the measurements of a range of different chemical elements. The large samples allow for the accurate tracking and differentiation of various Galactic components. Large samples also offer the best chances to find rare objects, such as extremely metal-poor stars, which are tracers of the earliest times and to capture short-lived objects, such as novae or born-again giants like Sakurai's object, enabling us to probe key rapid phases of stellar evolution.

\textbf{The key requirement to make a fundamental breakthrough in answering these questions is to obtain precise and accurate abundances with errors better than 0.05 dex.} Chemical information must cover all nucleosynthesis processes for a large number of long-lived stars formed in a range of environments at distinct epochs. This is imperative for uncovering the integrated effect of stellar nucleosynthesis on the chemical enrichment history of different Galactic stellar populations as a function of time, from the formation of the Milky Way in the early Universe until the Galaxy we see today. The required level of abundance precision can only be achieved with spectra with high signal-to-noise ratio (S/N) and high spectral resolution (R $\sim$ 40\,000). A high spectral quality is indispensable since key elements are represented only by weak lines, which are often blended with other features. Careful selection of wavelength windows will ensure that the absorption lines of elements originating from different nucleosynthesis channels are covered. We stress the need to also cover spectral regions with key elements that are missed by other surveys (such as Zn and Th) and to push toward blue wavelengths (to 380-390 nm), to include several important tracers of neutron capture processes. The density of spectral lines is well known to increase dramatically towards bluer wavelengths, informing the requirement for sensitivity and spectral resolution.

\subsubsection{The sources of neutron-capture elements} 

The heavy elements produced by various neutron capture rates (rapid, slow, and intermediate) provide fundamental insights into the energetic physical processes that create neutron fluxes and into their importance throughout the history of star formation in the Universe. Most elements can be formed by more than one of these processes, but some key elements trace only one formation channel. In the following, we list some open questions and mention examples of elements that can be used to trace the neutron capture processes in stars of different ages. Some of these elements are outside the wavelength range covered by current or planned surveys or cannot be measured because of the low resolution. Other listed elements are being or will be measured in these surveys, but \wst\ in MOS-HR mode will be able to measure them more precisely and/or for more stellar types and metallicity regimes. This will allow for a more accurate understanding of these physical processes.

\paragraph{r-process elements}
Quantifying the relative importance of the various sources of r-process elements is still missing. We know that neutron-star mergers (NSMs) produce r-process elements\citep{Watson2019}, but are they the dominant source \citep{Cowan2021}? Can NSMs alone explain the abundance of $r$-process elements in the first, second, and third peaks \citep{Cote2019NSM}? Are there other sites for the r-process, such as magneto-rotational supernovae (MRSNe) \citep{Nishimura2006} and collapsars? Are there stars that do not have any r-process elements \citep{Cescutti2015}, which one may expect if all r-process elements are produced by rare (early?) events? \\
    \textbf{Key elements:} Sr, Y, Zr, Eu, Gd, Dy, Sm, Os, Th

\paragraph{s-process elements} 
Low- and intermediate-mass AGB stars ($\sim 1-8\,\mathrm{M_\odot}$) are the main production sites of $s$-process elements. Do we know enough about the evolution of such AGB stars, particularly about the mass loss and mixing mechanisms \citep{Karakas2014, Kamath23}? High-mass stars ($\gtrsim 8\,\mathrm{M_\odot}$) also produce neutron-capture elements through the so-called weak $s$-process. The efficiency of this nucleosynthesis channel depends on stellar rotation, especially at low metallicity \citep[e.g.,][]{Limongi2018}. Can we constrain the rotation speeds of metal-poor massive stars through the abundances of $s$-process elements? How does the efficiency of the s-process vary with mass, mixing mechanisms, and metallicity? \\
    \textbf{Key elements:} C, N, Sr, Y, Ba, La, Ce, Pb

\paragraph{$i$-process elements}
The origin of the peculiar abundance patterns in stars (mostly metal poor) showing enhancements in both s- and r-process elements (r/s-stars) is still an open question \citep[e.g.][]{BeersChrislieb05, Masseron10, Gull18}. Several scenarios have been explored to explain the hybrid abundance properties, and one is the so-called intermediate neutron capture process \citep[e.g.][]{Choplin21}, where the neutron densities are intermediate between those of the s- and r-processes. This process may take place during the early AGB phase of low-metallicity low-mass stars and explain the elemental distribution of most of the r/s-stars. \\
\textbf{Key elements:} C, Mg, Sr, Y, Ba, La, Ce, Nd, Sm, Eu, Gd, Dy, Pb

\begin{figure}
    \centering
    \includegraphics[width=0.90\linewidth]{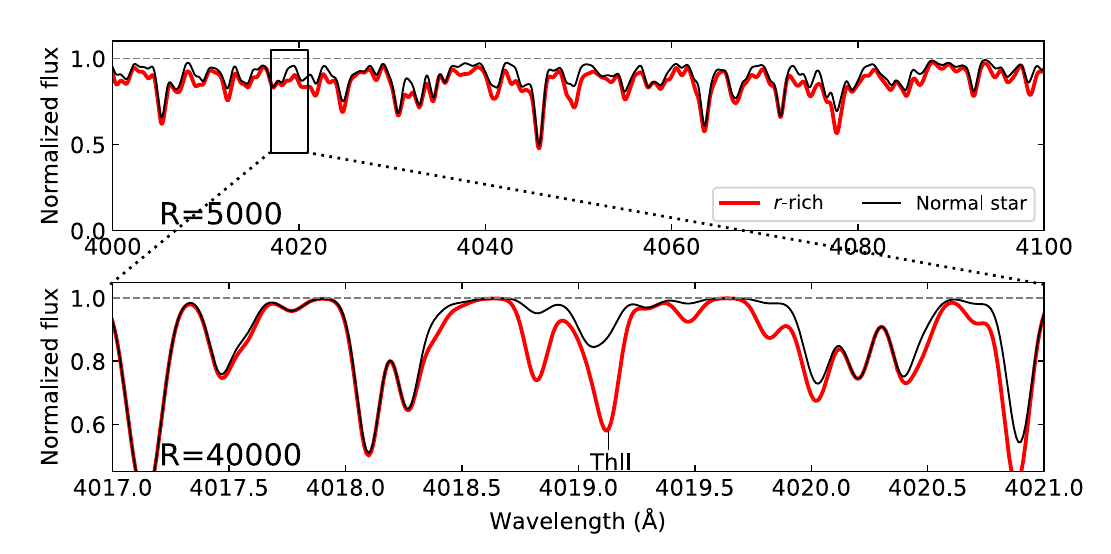}
    \caption{Synthetic spectra for a typical metal-poor star and an $r$-process enhanced star at $R=5000$ (top) and $40000$ (bottom). The bottom panel shows the region shown as the box in the top panel. 
    \label{fig:Th-detectability}}
\end{figure}
\vskip0.3cm
\noindent{These high-quality abundances will allow us to address the following scientific cases.}

\subsubsection{Insights on supernovae nucleosynthesis} 
Type Ia supernovae (SNeIa) are the final explosions of binary star systems with one or two electron-degenerate carbon-oxygen white dwarfs (see also Section~\ref{sec:td:sne}). However, it is not yet known exactly what explosion mechanisms are dominant and whether most SNeIa are the result of a system with one or two compact objects \citep{LivioMazzali2018}. On the other hand, the explosion mechanisms of massive stars ($\gtrsim 8\,\mathrm{M_\odot}$), including the so far hypothetical zero-metallicity population III (Pop III) stars, are also not well understood. The questions include the following: How many subtypes of SNeIa exist? What SNeIa subtypes are important as nucleosynthesis sources? Are there iron-rich metal-poor stars that can help constrain thermonuclear nucleosynthesis \citep{Reggiani2023}? How frequently do pair-instability supernovae occur \citep{Salvadori2019,Xing2023}? (see also section \ref{sec:td:sne}). How aspherical are the explosions? What is the typical mass of the first stars and their explosion energies \citep{Ishigaki2018}? How did the metals from the first stars start polluting the Universe? Can we identify kilonovae (neutron star mergers) as the key source of neutron-capture elements? \\
    \textbf{Key elements:} C, N, O, Na, Mg, Si, Ca, Sc, Mn, Co, Cu, Zn

\subsubsection{The role of star formation history on the chemical enrichment}
To uncover the origin of the elements, it is not enough to study the abundance patterns of individual stars. These also need to be placed in the context of the entire history of star formation and chemical enrichment of the entire Galaxy \citep{Kobayashi2020}. The Galactic disc is thought to be formed mainly by stars dispersed from clusters and associations, where each one may carry its own chemical signature. The halo, on the other hand, is known to be formed by a combination of stars formed in situ and those accreted from satellite galaxies. By disentangling stars of different origins, we will be able to study how chemical enrichment processes change in different environments with different star formation histories, which, in turn, will tell us about the sites of nucleosynthesis processes. The most effective way to disentangle stars of different origins is to combine kinematic and dynamical information with high-precision chemistry (see also Section \ref{sec:originsMW}). The key here is to increase the number of mapped stars and decrease the uncertainties on the derived chemical abundances. This will maximise our ability to connect abundance variations with the local details of the star formation history and efficiency in different substructures of the halo and disc of the Galaxy. \\
    \textbf{Key elements:} All those listed above

\begin{figure}[t]
    \centering
    \includegraphics[width=0.50\linewidth]{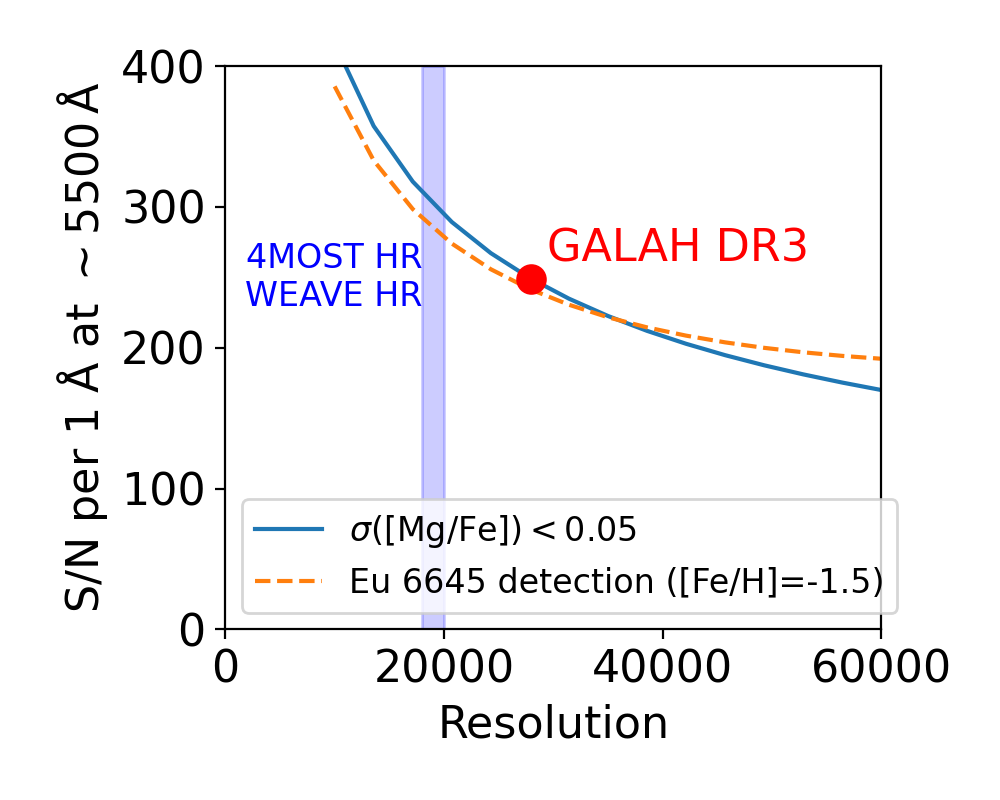}
    \includegraphics[width=0.40\linewidth]{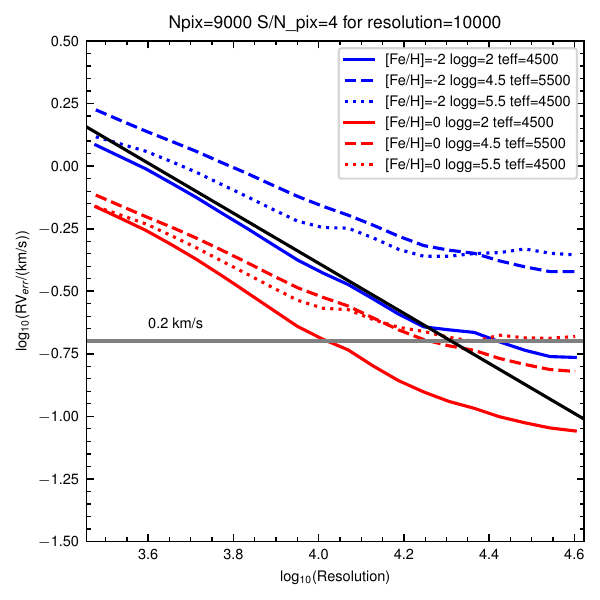}
  \caption{\small \textbf{Left:} The required resolution and S/N to measure Mg abundance with a precision of 0.05 dex and to detect Eu from the line at 6645 \AA~for a red giant with [Fe/H] = $-$1.5. For the measurement of Mg abundance, we scaled the result of the GALAH survey \citep{Buder2021}, assuming photon-noise-limited measurements. We note that although the normalisation might change depending on the survey's wavelength coverage and pipelines, the overall trend should not change. For the detection of Eu, we assume 3$\sigma$ detection. \textbf{Right:} For the current properties of the \wst\ spectrograph, for a range of stellar properties, we show how the velocity precision increases as the spectral resolution increases for a star of the same brightness and the same exposure time. It is assumed that the signal-to-noise ratio per pixel scales as 1/resolution which is appropriate for read-out noise-limited observations, while the wavelength coverage of the spectra is assumed to decrease with increasing resolution by keeping a constant total number of pixels (N=9000) with a constant spectral sampling of 2.2 pixels per resolution element. The solid black curve shows a velocity floor corresponding to 0.03*pixels. The horizontal black line shows a constant velocity accuracy of $\pm 0.2$km/s, suggesting a spectral resolution, R$\gtrsim 22 000$ to achieve this precision. Note that in the case of Poisson noise, the trend of decreasing velocity error with increasing resolution continues all the way to the highest resolutions.} 
  \label{fig:Matsuno1}   \label{fig:koposov}
\end{figure}

\subsubsection{The \wst\ requirements for precise chemical abundances}

\begin{figure}[t]
  \centering
  \includegraphics[width=0.8\textwidth]{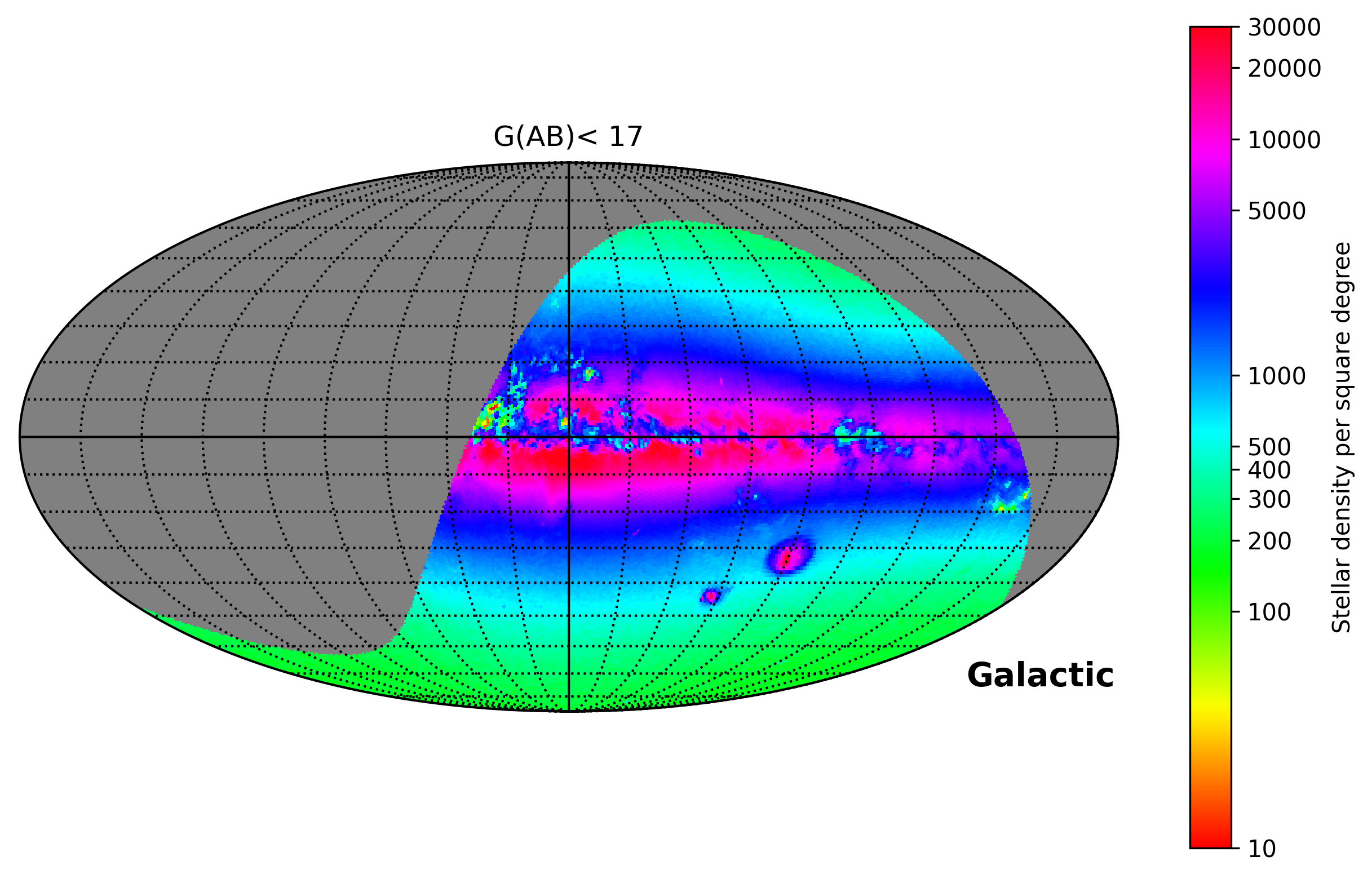}
  \caption{Mollweide map of stellar density from {\it Gaia} DR3 in Galactic coordinates selected with Dec $<$ 20$^{\circ}$ on the sky, showing the stars accessible from the southern hemisphere. The map is colour-coded with stellar density per square degree and it shows stars brighter than G$=17$  with extinction A0$<2$. }\label{fig:densitymap}
\end{figure}

The high-resolution mode (R $\sim$ 40\,000) of \wst\ will be the key to advance our understanding of the origin of chemical elements. High resolution will allow us to observe elements that have not been detected in current survey data, such as Th (Fig.~\ref{fig:Th-detectability}). The main advantage of \wst\ lies in the powerful combination of high multiplex (1000-2000 objects) over a $\sim$3.1 deg$^2$ field of view with the large telescope collecting area. In a survey lasting 5$-$10 years, the \wst\ would be able to collect $\sim$3.3$\times$10$^7$ high-quality stellar spectra that can provide better than 0.05~dex precision (Fig.~\ref{fig:Matsuno1}, left). This will be an unprecedented number of stars with high-precision abundances, one order of magnitude more than the number of stars with high-resolution spectra from previous surveys, and increasing the accuracy and range of chemical elements. 
The power of larger samples and the increased accuracy are two-fold. First, it provides statistically significant information to precisely trace the evolution of the elements with time in different stellar populations. Second, it increases the number of rare and short-lived objects in the sample. These rare objects are exactly those that can provide critical insight to validate or discern between models and theories. Figure \ref{fig:densitymap} shows a density map of \textit{Gaia} DR3 sources brighter than $G$ = 17 mag with low extinction, showing that there are numerous targets throughout the southern sky.

\begin{figure}
    \centering
    \includegraphics[width=0.75\linewidth]{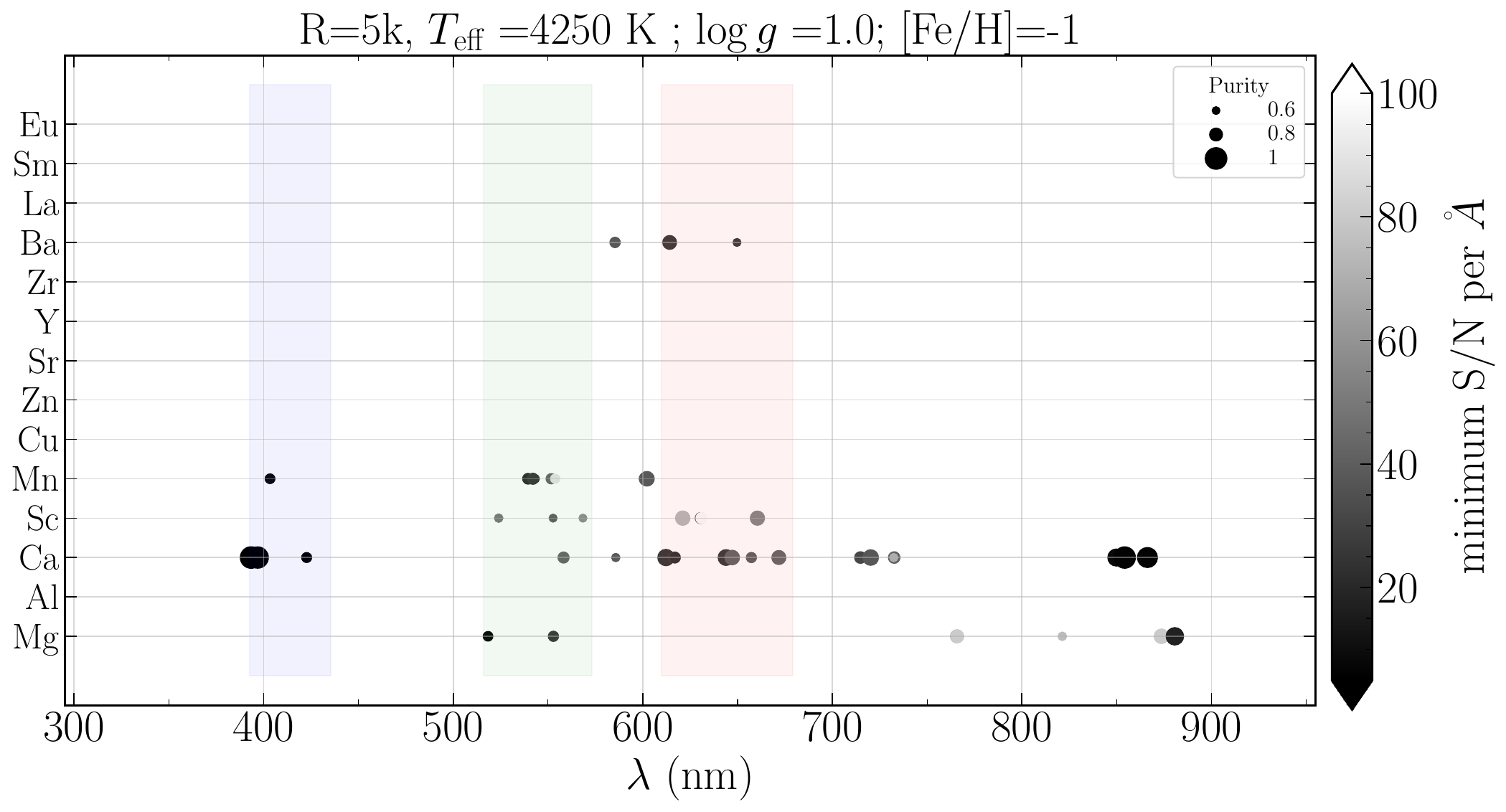}
    \includegraphics[width=0.75\linewidth]{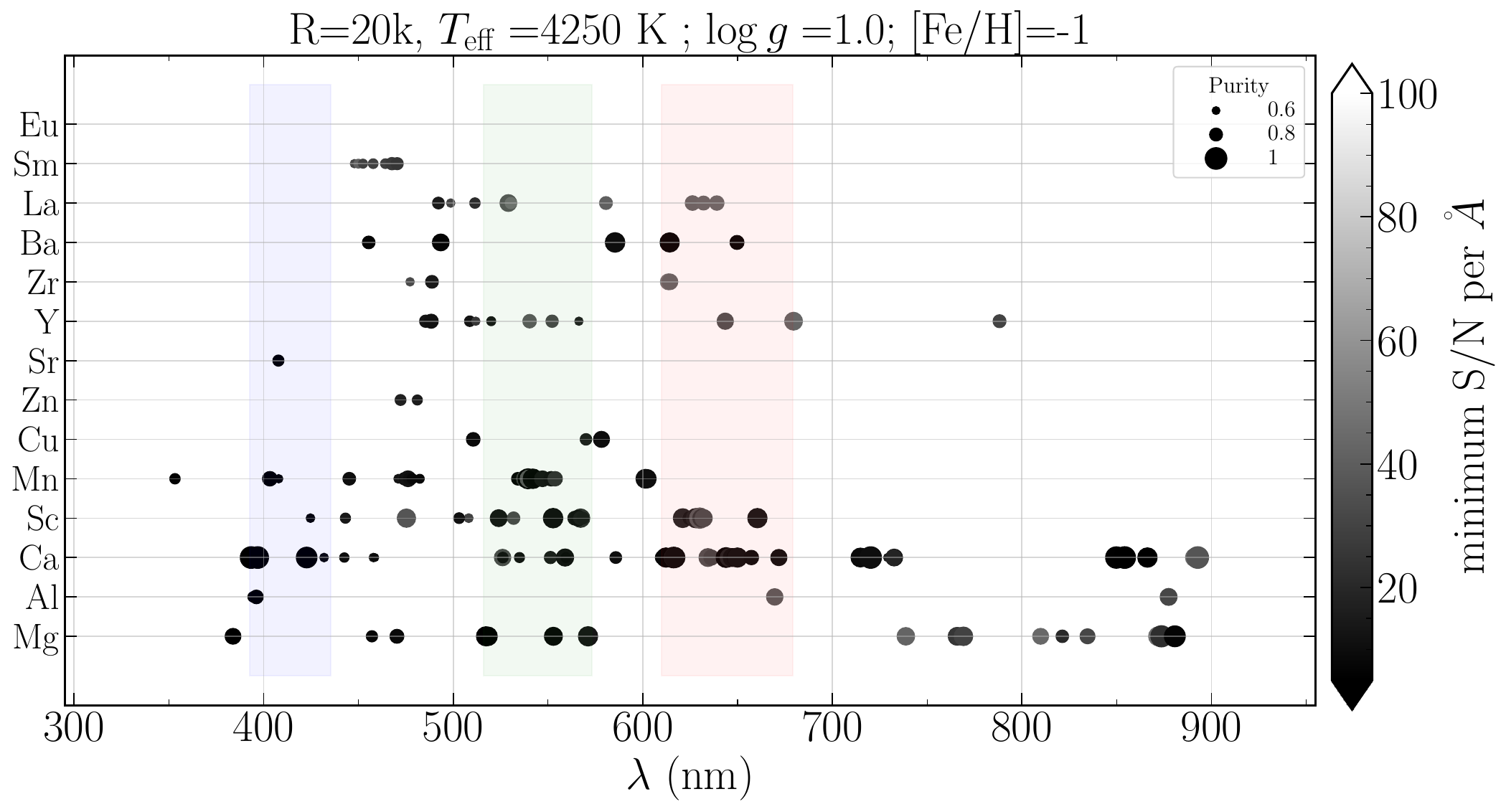}
    \includegraphics[width=0.75\linewidth]{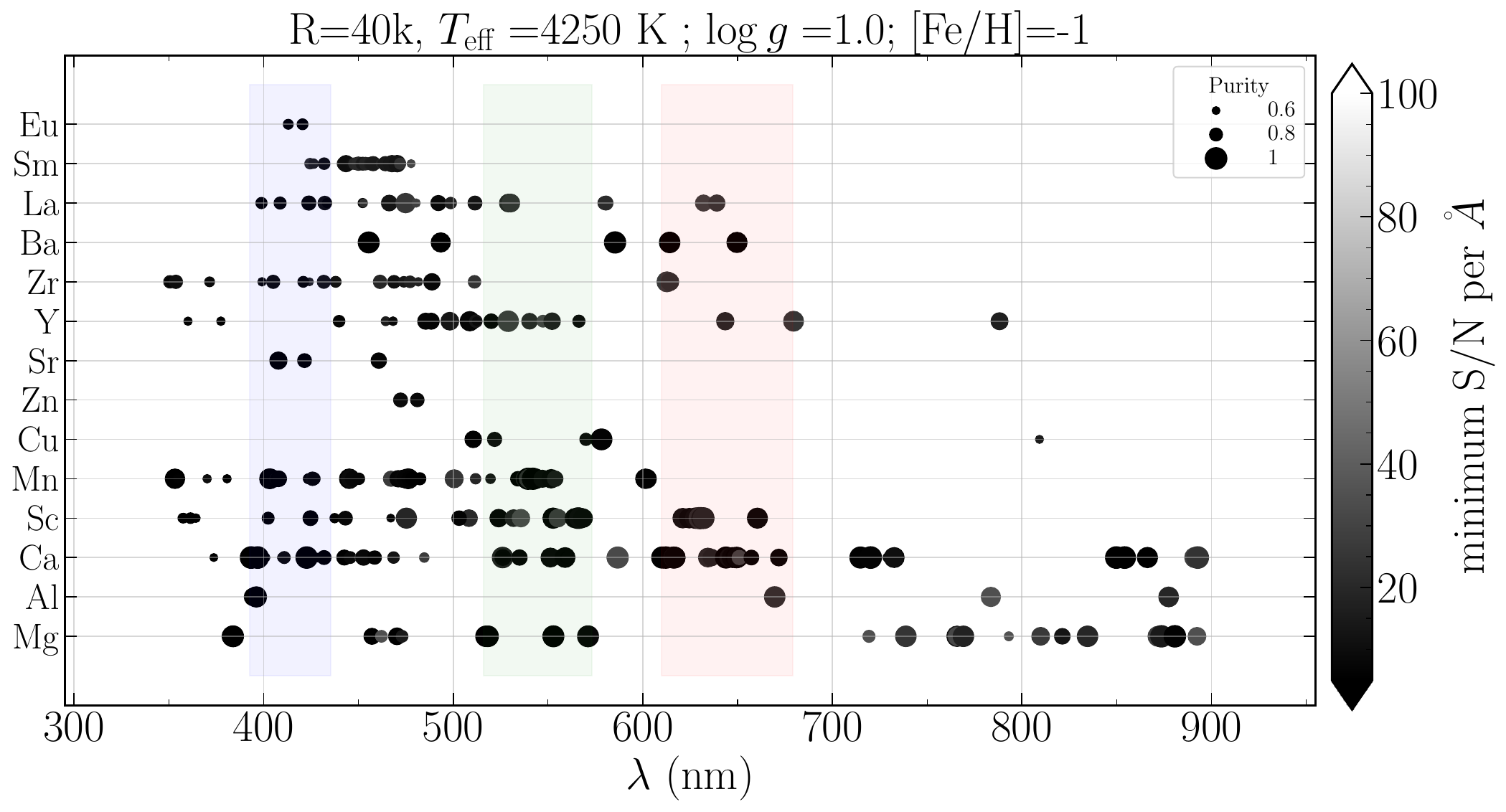}
    \caption{Line-detectability for a few cherry-picked  elements, for a $\feh=-1$ giant with solar-scaled abundances (except for alpha elements, where [$\alpha$/Fe]=+0.4 has been adopted). Top plot is derived for spectra with $R=5\,000$, middle one with $R=20\,000$ and bottom one with $R=40\,000$. Blue, green and red colour boxes represent \fourmost's HR windows. A purity threshold of 0.6 has been chosen \citep[adapted from][]{Kordopatis23b}. }
    \label{fig:line-detectability}
\end{figure}

It is essential to carefully select the \wst\ high-resolution mode wavelength regions to ensure the inclusion of elements that lower resolution next-coming spectrographs (such as \fourmost) either cannot access or which would benefit from improved precision. This will enable measurements of chemical abundances for a wide range of different stars with \wst\ (e.g. Eu, Sm, La, Zr, Y, Mn, Zn for low-metallicity stars; see Fig.~\ref{fig:line-detectability}). The wavelength coverage will be divided into different arms with a spread of $\sim$50\,nm each. High resolution is also key here to decrease the problem with blending of lines (especially toward high metallicities and blue wavelengths) and to give access to a plethora of weak atomic lines (especially toward low metallicities). 

Since \wst\ will cover the southern sky, it is expected that most of the targets for a \wst\ high-resolution survey will already have a spectrum observed by \fourmost\, in either of its lower-resolution modes (R = 20\,000 or R = 6\,000). This removes the need for a wide wavelength range, including features for the determination of atmospheric parameters, and leaves \wst\ free to focus on the determination of specific abundances and to prioritise the detection of elements and perhaps some molecular bands (e.g. CH, NH) that are absent (or poorly measured) from (by) other surveys.

\subsection{Origins of the Milky Way system}\label{sec:originsMW}

The detailed study of Milky Way stellar populations requires all-sky surveys with accurate distances to many millions of stars to pick out the patterns that occur on large scales in the volume around and beyond our Sun (Fig. \ref{fig:volume}). This is undergoing a revolution with {\it Gaia}, which has enabled a paradigm shift in our understanding of stars, their three-dimensional (3D) positions (including distance), 3D velocities, and intrinsic properties (stellar parameters, chemical abundances, ages). This has enabled a wealth of advances in our understanding of how the Milky Way has been assembled \citep[e.g.][]{Belokurov2018, 2018A&A...616A..10G, Helmi2018, Ibata2020}. 

As we discussed earlier, {\it Gaia} is delivering parallaxes and proper motions for $\sim$2$\times$10$^{9}$ stars down to $G$ $\sim$ 20-21mag. Abundances (e.g. N, Mg, Si, S, Ca, Ti, Cr, Fe, Ni, Zr, Ce, and Nd) and RVs (with a final precision better than 5-10 km s$^{-1}$) are or will be available from \emph{Gaia} for certain types of stars as faint as $G$ $\sim$ 14-16, but only for the wavelength range 846$-$870nm and at R$\sim 11~500$. Importantly, however, this is a restricted parameter space and also leaves a significant gap of $\sim$ 6 magnitudes where \textit{Gaia} will provide constraints on parallaxes and proper motions for stars without RVs or detailed chemical abundances. This means that we cannot carry out detailed studies of the dynamical and chemical structure beyond the immediate vicinity of the Sun. 

Furthermore, in the coming years \lsst\ will provide a new revolution with homogeneous $ugrizy$ photometry down to $r$ = 24.5 (in a single visit, with co-added maps down to $r \sim$ 27.5 over a 10-year period) together with proper motions (allowed by the multiple visits). This will provide samples $\sim$4 magnitude fainter than \textit{Gaia} and so extend our ability to study 3D kinematics of the Milky Way further into the Bulge and the outer disk regions, as well as probing resolved galaxies beyond the Milky Way halo. Crucially, large-scale, targeted ground-based spectroscopic surveys are required to close the gap in stellar RVs and chemical abundances at the faint end of both the \textit{Gaia} and \lsst\ catalogues (15 $<$ $G$ $<$ 24).

A high-resolution MOS-HR spectrograph (R $\sim$ 40\,000) capable of observing with more than 1000 fibres on a FoV $\sim$3.1 deg$^2$ offers an exciting platform for advancing our understanding of the Milky Way, building on the results of previous large surveys in the southern hemisphere, such as \fourmost and \moons. Moreover, a low-resolution MOS-LR mode in a telescope like \wst, reaching at least R $\sim$ 5000-7000 at the calcium triplet (CaT) region ($\sim$ 850-870 nm), gives the unique opportunity to observe faint stars in the \textit{Gaia} and \lsst\ catalogues. With this, it is possible to obtain RVs, CaT metallicities, and, in some cases, reliable abundances of several interesting elements \citep{Bellaz08,Kordopatis2011, muccia12,muccia17}.

The combination of the quality of the high-resolution sample and the depth of the low-resolution sample will significantly enhance our understanding of the Milky Way's stellar populations and its system of neighbouring galaxies:

\subsubsection{Dissecting the Milky Way disc with chemical tagging}
Stars migrate across the Milky Way, eventually reaching regions that are very different from those in which they formed \citep[e.g.][]{Sellwood+2002,  MinchevFamaey10}. 
The time variation of the Galactic potential due to disk asymmetries, such as spiral arms and the central bar, as well as infalling satellites, precludes the use of kinematics alone as means of tracing back stars to their birth position. In contrast, chemical abundances, combined with age estimates, provide unique fingerprints that identify the environment in which stars formed. This means that chemical abundances combined with kinematics can be used to tag stars that form within the same stellar association, when that association shows a unique and recognisable pattern \citep[the so-called ``chemical tagging'',][]{FreemanBlandHawthorn2002}, which can then be placed to their birth radii using a novel approach \citep[e.g.][]{Lu22, Ratcliffe23}.
The large number of stars with precise abundance and age measurements are the key in chemical tagging \citep{Price-Jones2019}, which \wst\ can achieve through its high multiplex, high resolution capability, and large aperture. The combination of chemistry and kinematics is not always straight forward to interpret \citep[e.g.][]{Casamiquela21} and HR spectroscopy will maximise our ability to accurately identify related groups of stars and reconstruct the star-formation history of the Milky Way disc in extremely fine detail. \wst\ in addition to more accurate radial velocities, will also make the link to a larger range of chemical elements, and especially heavy elements not available in current or planned surveys.

\weave\ and \fourmost\ are expected to study extensively the disc in both R$\sim20~000$ and R$\sim5000$ and potentially discover new chemo-kinematic groups that would explore the disc’s non-equilibrium state and non-axisymmetries. \wst, on the other hand,  will allow to have access to more elements to perform a more thorough chemical tagging, orbital determination and overall date better the stars (e.g. via isochrone fitting). \wst\ will also improve the study the very metal-rich stars ([M/H]$>0.2$) expected throughout the disc, coming from the innermost parts of the Galaxy \citep{Kordopatis15}  and for which the spectra exhibit heavily blended lines making the elemental abundance determination challenging  at lower resolutions. We can also expect new unknowns to enter the more detailed picture in the future, and although we can't be sure what form they may take, enlarging the parameter space (to higher spectral resolution) will undoubtedly help to take the next steps in a deeper understanding.

\subsubsection{Open clusters, their tidal tails, and stellar streams}
Star clusters are embedded in the Galactic gravitational field and, over time, lose their members because of the influence of tidal forces. The lost stars form elongated structures pointing in opposite directions, the tidal tails, which are expected to be equally populated. Detailed analysis of \textit{Gaia} data has revealed asymmetries in the tidal tails of several open clusters \citep{Roser19, MeingastAlves19, Jerabkova2021A&A...647A.137J, Boffin2022MNRAS.514.3579B}, challenging the common assumption of symmetric tidal tails. The asymmetry might be due to an external perturbation such as, for example, the encounter with a molecular cloud \citep{Jerabkova2021A&A...647A.137J} and not solely a function of the age of the cluster at work in the decay that acts on shorter timescales than the mass segregation \citep{Tarricq22}. It can also be explained as asymmetries in the star cluster potential larger than Newtonian dynamics can provide \citep{Kroupa2022MNRAS.517.3613K} or the possibility that populations from different clusters may be intermingled due to Galactic tidal forces \citep{Nikiforova20}. To investigate the nature of tidal stars, and in general of stellar streams, the membership of the stars should be confirmed using their chemistry. Observations require a MOS with a large FoV that can cover tidal tails and streams that often extend tens of degrees across the sky down to faint magnitude limits. This is required to obtain sufficient statistics on the low-surface-brightness structures and provide the precise abundances of many elements needed. \vskip0.2cm 
It has also been shown that the morphology and dynamics of tidal streams are good probes of the properties of the Galactic bar, such as its strength or its pattern speed \citep{Hattori2016MNRAS.460..497H, Pearson2017NatAs...1..633P, Thomas2018}. 
\begin{figure}
    \centering
    \includegraphics[width=0.99\linewidth]{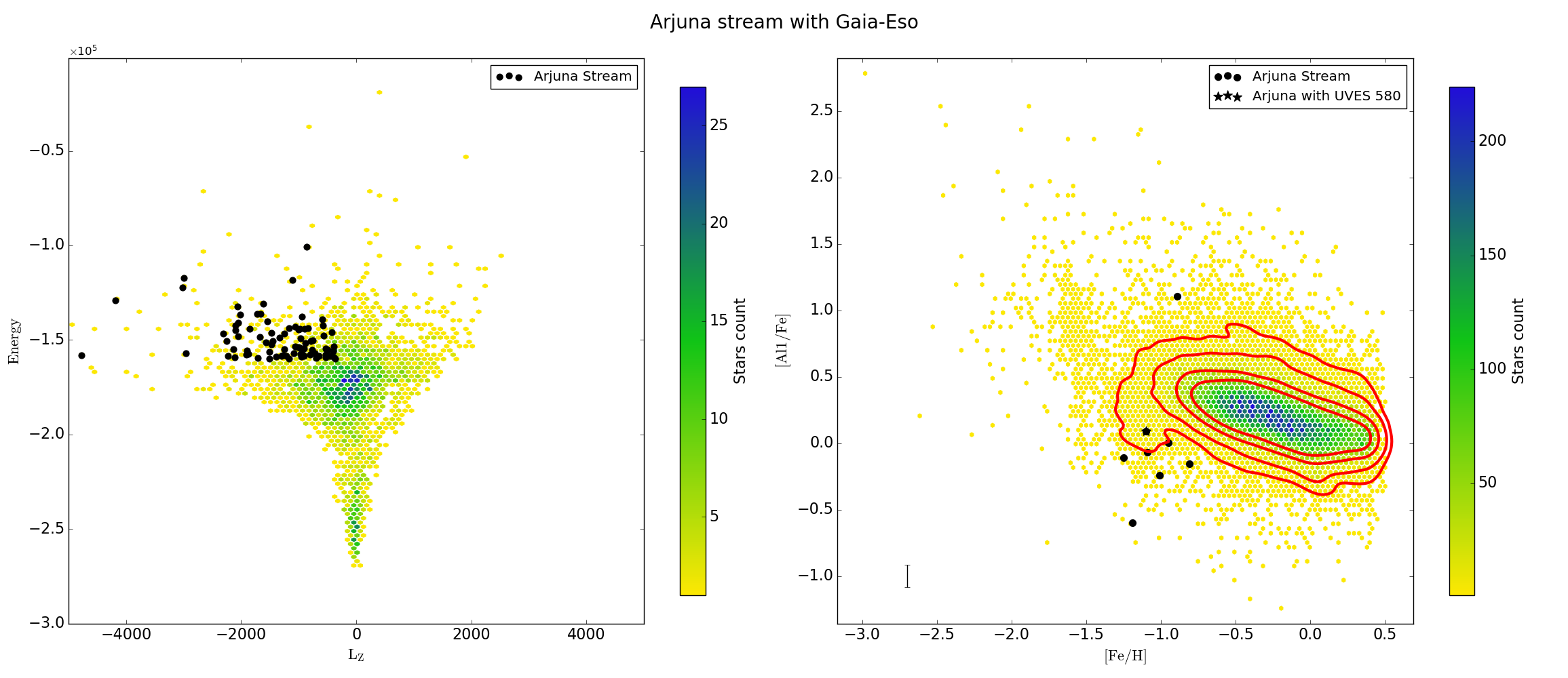}
    \caption{Orbital properties (left panel) and abundances ([Al/Fe] vs [Fe/H], right panel) of the Gaia-ESO full sample. The accreted stars belonging to the Arjuna stream are in black. 
    \label{fig:arjuna}}\end{figure}

\begin{figure}[t]
  \centering
    \includegraphics[width=0.45\textwidth]{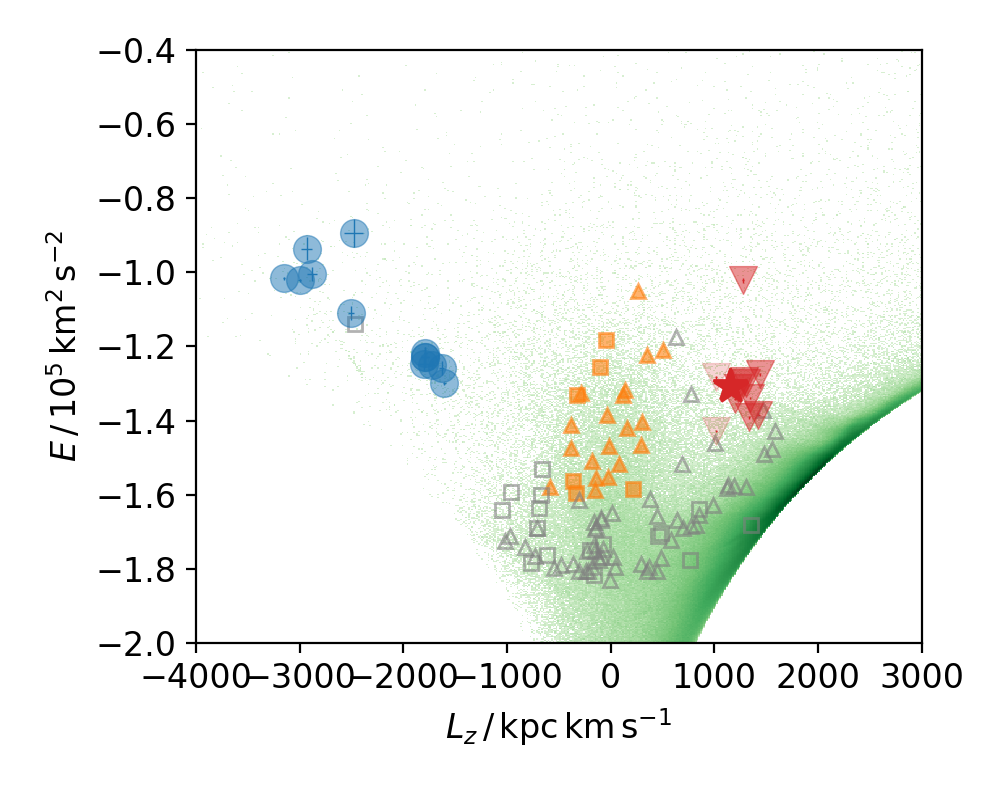}
  \includegraphics[width=0.45\textwidth]{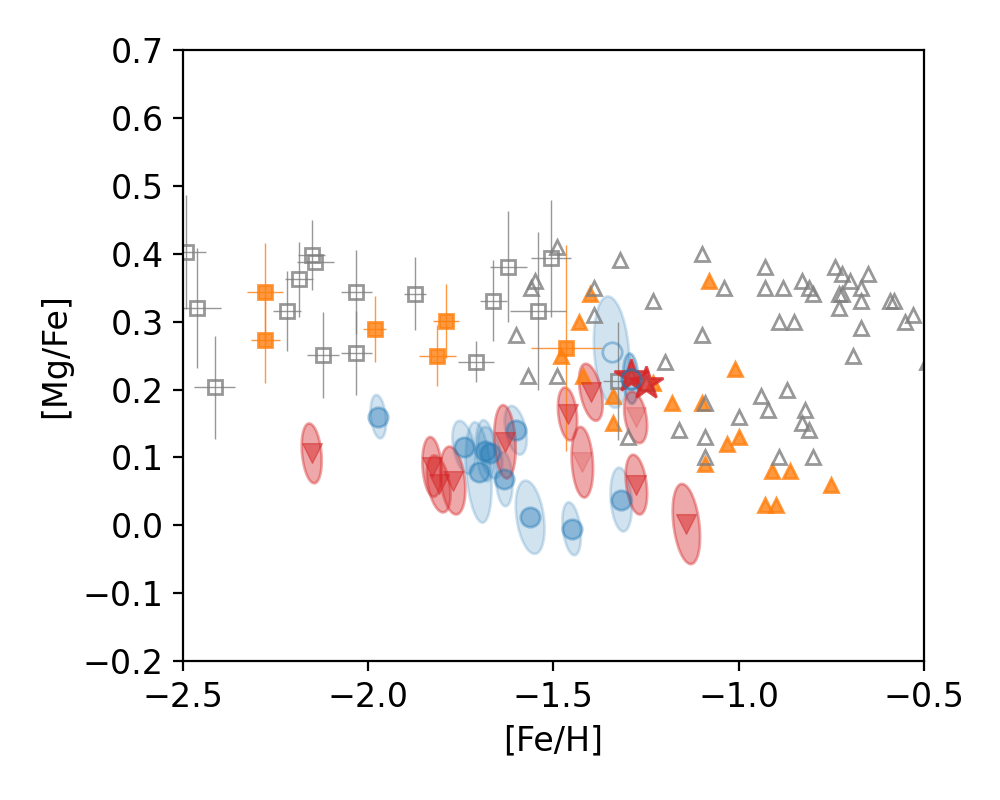}
  \caption{\small \textbf{Left:} Substructures of the Galactic halo identified through their kinematics, tracing distinct accreted galaxies (red: Helmi streams, blue: Sequoia, orange: Gaia-Enceladus-Sausage). \textbf{Right:} Chemical abundance of stars of different origins, with the same colour code. Data are from \citet{NissenSchuster2010}, \citet{Reggiani2017}, and \citet{Matsuno2022a, Matsuno2022b} }
  \label{fig:HelmiStream}
\end{figure}

\subsubsection{Measuring stellar ages with chemistry: chemical clocks} 
Ages remain a huge challenge in determining the properties of the Milky Way going back in time and the order events in the early Milky Way. \wst\ will greatly enhance our ability to measure abundance ratios that are sensitive to age (the so-called ``chemical clocks'', which involve abundances of neutron capture elements and/or of light elements such as C and N). The exploitation of these clocks requires careful calibration, based on samples of stars with reliable age determinations, such as open clusters or asteroseismic targets \citep{Casali2019, Casali2020, Casali2023, Magrini2022b, ViscasillasVazquez2022, Kordopatis23a}. Well-calibrated chemical clocks can be used to produce extensive age maps of the disc, crucially expanding our understanding of the various different stellar populations and how they have moved with time. The higher resolution and larger telescope collecting area and thus greater flux sensitivity of \wst, compared to \fourmost\ or \weave, will be fundamental for deriving neutron-capture abundances with high precision for a statistically significant number of stars in all the different components of the Milky Way, enabling a ground-breaking leap in this field. 

\subsubsection{Going deeper into the chemodynamical view of the Bulge} 
The inner Galaxy is composed of a superposition of several populations that differ in chemical composition, age, and kinematics. Disentangling these and determining how they are connected to the rest of the Galaxy is an ongoing quest. \fourmost\ and \moons\ will begin collecting critical spectroscopic data sets that are needed, but they will be limited by one or all of field of view, sensitivity, and spectral resolution. With \emph{Gaia} and complementary photometry, a major achievement has been to reveal the Galactic bar directly in stellar density maps \citep[e.g.][] {Anders19, Anders22, BJ21}, hence giving a hint of the tremendous impact that the full \emph{Gaia} data set will have on our knowledge of the innermost regions of the Galaxy \citep[see][]{Barbuy18}. Adding spectroscopic information provides critical extra dimensions to these photometric maps. Distance uncertainties as low as 10\% are achievable from the Bulge to the outer disk \citep{Queiroz20, Queiroz23}, making it possible to perform orbital analyses of the stellar populations in the innermost kpcs of the Galaxy, for the first time. Current studies have begun to reveal a complex mix of stellar populations, providing new observational constraints to chemodynamical models of the Bulge and inner disk \citep{Queiroz21, Arentsen22}.
   
Current data for the Bulge still lack accurate ages. This determination requires spectroscopic measurements of stellar properties to identify main sequence turnoff and subgiant stars, for which isochrones can be used to determine precise ages \citep[e.g.][]{Kordopatis23a}. \wst\ in MOS-LR mode, with its large FoV, multiplex, and sensitivity, is uniquely able to obtain the statistically significant spectroscopic samples of faint stars in crowded regions\footnote{In the innermost Bulge regions, in which crowding for faint main sequence turnoff and subgiant may be too high given the size of the \wst\ MOS fibres, the IFS of \wst\ will become instrumental to allow disentangling spectra.} needed to probe the Bulge and obtain large samples of accurate ages. \wst\ in MOS-HR mode will also obtain detailed and precise chemical abundances for many elements otherwise unavailable, to allow the disentangling of the co-spatial stellar populations in the Bulge region. The combination of \wst\ MOS-LR and MOS-HR data for the MW Bulge promises to be transformative. It will open a window into this critical component of our Milky Way that is seen in the formation stages in other galaxies at high redshift with JWST. The complementarity will be extremely powerful.

\subsubsection{Characterising the assembly and accretion history of the Milky Way} 
The observation of a diverse sample of stars, near and far, contributes to a detailed understanding of all phases of Galaxy formation. \wst\ will give a substantial contribution to identify and characterise the assembly and accretion history of the Milky Way, in particular: {\em i)} High-quality abundances to be used as chemical fingerprint identification of past accretions through their abundances (see, e.g., Fig.~\ref{fig:arjuna} in which the orbital properties and abundances of the Gaia-ESO sample are shown, highlighting the unique chemical characteristics of the accreted stars in the Arjuna stream). The distinctions in chemical abundance ratios in the Galactic populations (see Fig. \ref{fig:HelmiStream}) open an opportunity to identify and characterise both large-scale stellar populations, such as thin/thick discs, halo, Bulge, and small-scale stellar populations, including remnants of past encounters or disrupted star clusters; {\em ii)} Chemodynamical identification of past accretions, through large samples that extend to fainter magnitudes to probe further out in the halo ($>$10\,kpc). This will enable reconstruction of the full assembly/accretion history of the Milky Way, discriminating between stars formed in situ and those formed in progenitor galaxies of different masses and star formation efficiencies \citep{NissenSchuster2010,FernandezAlvar2018, Mackereth2019, Horta2020, Minelli2021,Matsuno2022b,Matsuno2022a,daSilvaSmiljanic2023, Cecc24}. Since chemical abundance differences between populations can be small, a precise chemical abundance measurement is necessary. In a photon noise limited situation, the $S/N$ per angstrom required to achieve certain precision scales with $1/\sqrt{R}$ (left panel of Figure~\ref{fig:Matsuno1}), and hence the high-resolution capability of \wst\ will be essential to conduct these studies.

\subsubsection{Kinematic mapping of the MW components}

Dark Matter is one of the most fundamental unknowns in astronomy and it is only detected indirectly in a galaxy through the analysis of the movement of gas or stars.  Spectra allow for precise measurements of stellar velocities and, in combination with proper motions and mass models, stellar motions can be converted into orbital parameters within the Milky Way \citep[e.g.][]{Antoja18}. The ability to simultaneously observe numerous stars over the full sky provides a unique opportunity to create a detailed kinematic map of the Milky Way, revealing the dynamics of its various components and thereby the underlying mass distribution in the Galaxy \citep[e.g.][]{Johnston02, McMillan17, Posti19} and its sub-structures \citep[e.g.][]{Belokurov2018, Helmi2018, Battaglia22}. A key result of these measurements is our ability to determine the spatial distribution of dark matter in the Milky Way \citep[e.g][]{Reid14}; in no other galaxy will we be able to derive its 3D mass distribution and its small scale mass structure in comparable detail. By mapping the stellar kinematics with WST we will have larger samples than is possible with any other facility \citep[e.g.][]{Bensby19}. These will range from the very central regions of the Galaxy to its distant halo outskirts, and a large number of (dark) mass measures can be made to constrain the formation history of the Milky Way and the nature of Dark Matter \citep[e.g.][]{Cautun20}. The global 3D shape of the Milky Way dark matter halo can be determined (e.g., radial profile, flattening, triaxiality) as well as perturbations due to the in-fall of the Magellanic Clouds \citep[e.g.][]{Gomez15, Cautun19, Garavito21, Vasiliev23, Foote23}, and other past events. These studies can also be extended to dwarf galaxies as sub-structure in the outer halo and also as objects of interest in the their own right \citep[e.g.][]{Simon2019, Errani23}. Studies of the mass of the Galactic bar and spiral arms, their pattern speed and temporal evolution can be performed, and whether the dark matter responds to them. Another important element is to improve the measurements of kinematic perturbations in cold stellar streams due small dark matter halos passing by, thereby constraining the power spectrum of dark matter sub-halo masses in comparison to the predictions of various dark matter models \citep[e.g.][]{Malhan22}. 

Building on the results of \fourmost\ and \desi\  among other surveys, we will deepen our understanding of the Milky Way's role in the broader cosmological context and how our Galaxy can help us to understand the properties of other galaxies across cosmic time. For example, using stellar streams to probe the nature and minimum mass of dark matter clumps \citep[e.g.][]{Banik21, Li22} is presently extremely challenging. There are very few stars at the magnitude limits of current surveys and the expected velocity variation created by density perturbations caused by dark-matter halos passing through these  streams is predicted to be extremely small. Using the formalism defined by \cite{Erkal16} it can be calculated that a stream orbiting the Milky Way at a distance of 14~kpc can expect a maximum velocity kick over its lifetime of 5~Gyr to result in a velocity change of $\sim$(0.6, 0.3, 0.1)km~s$^{-1}$ for sub-halos in the range (10$^{7.5}$, 10$^{6.5}$, 10$^{5.5}$M$_\odot$), respectively. For the lowest mass dark halos very few if any stars are expected, and thus this approach is the only way to probe this scale. This important study requires a combination of faint magnitude limits to increase the number of stream stars and sufficient velocity precision to detect the small changes predicted.

At every spectral resolution, there is a floor to the achievable precision for the velocity measurements (see Fig.~\ref{fig:koposov}). The advantage of a large telescope is that we can push for higher precision measurements for a large sample of targets. We can only improve the velocity accuracy however if we also increase the spectral resolution compared to previous surveys. For example, DESI has a precision floor of $\pm 1$km/s, \citet{Cooper2023}. A spectral resolution in the range of R$\sim 20-25~000$, will result in a velocity error of the order of $\sim0.2$km/s, see Fig.~\ref{fig:koposov}. At R$\sim 40~000$ the velocity errors will be very similar, but with a brighter magnitude limit at which the required S/N can be achieved. These measurements require a trade-off between S/N and velocity precision that ideally requires an intermediate spectral resolution compared to that needed for precise abundances (R$\sim40~000$ MOS-HR) and basic metallicities (MOS-LR).

\subsection{Origins of stars and planets}\label{sec:gal:origin}

Understanding how stars form is fundamental to address several open issues in modern astrophysics. On a large scale, the evolution of galaxies is regulated by the rate at which stars form and by the initial mass function \citep{Krumholz2014PhR...539...49K}. Looking at a more local scale, star formation is the way to trace the origin of the solar system \citep{Manara2023ASPC..534..539M}.  

Decades of multi-wavelength studies of young stellar systems in our Galaxy led to the conclusion that stars form from giant molecular clouds, not in isolation but in groups composed of a number of stars ranging from a few tens to several thousands \citep{lada03, mcke07}. Such stellar groups in most cases disperse in the field within a few tens of million years after the parent clouds are dispersed \citep{Krumholz2019}. During these early stages of their evolution, stars are characterised by the presence of a circumstellar disk that dissipates within a few million years, forming planetary systems.

Within this broad scenario, many issues are still open. In particular, it is not clear what the roles of various physical mechanisms (turbulence, magnetic fields, stellar feedback) in regulating the formation of stellar systems and their dispersion into the field are, and what the role of environment (e.g. density and metallicity of molecular clouds) in defining the star formation process at both small and large scales is.
Furthermore, as proved by spectroscopic surveys of stars hosting planets, the property of planets emerging from the star formation process may depend on the metallicity of the parental cloud, while the strong magnetic activity that characterises young stars can influence the properties of the planetary atmosphere.

Optical and near-infrared spectroscopy play an important role in addressing these issues. Spectroscopic observations are in many cases the only available tool to confirm the youth of photometrically selected young stars and also provide radial velocities that combined with astrometry are used to study 3D stellar kinematics \citep[e.g., Wright et al. 2024, submitted; and ][]{Miret2024NatAs...8..216M}. Multiple spectroscopic indicators can be used to determine ages and masses and to study star-disk systems by tracing accretion and outflow.

\begin{figure}
\centering
\resizebox{0.7\hsize}{!}{
\includegraphics{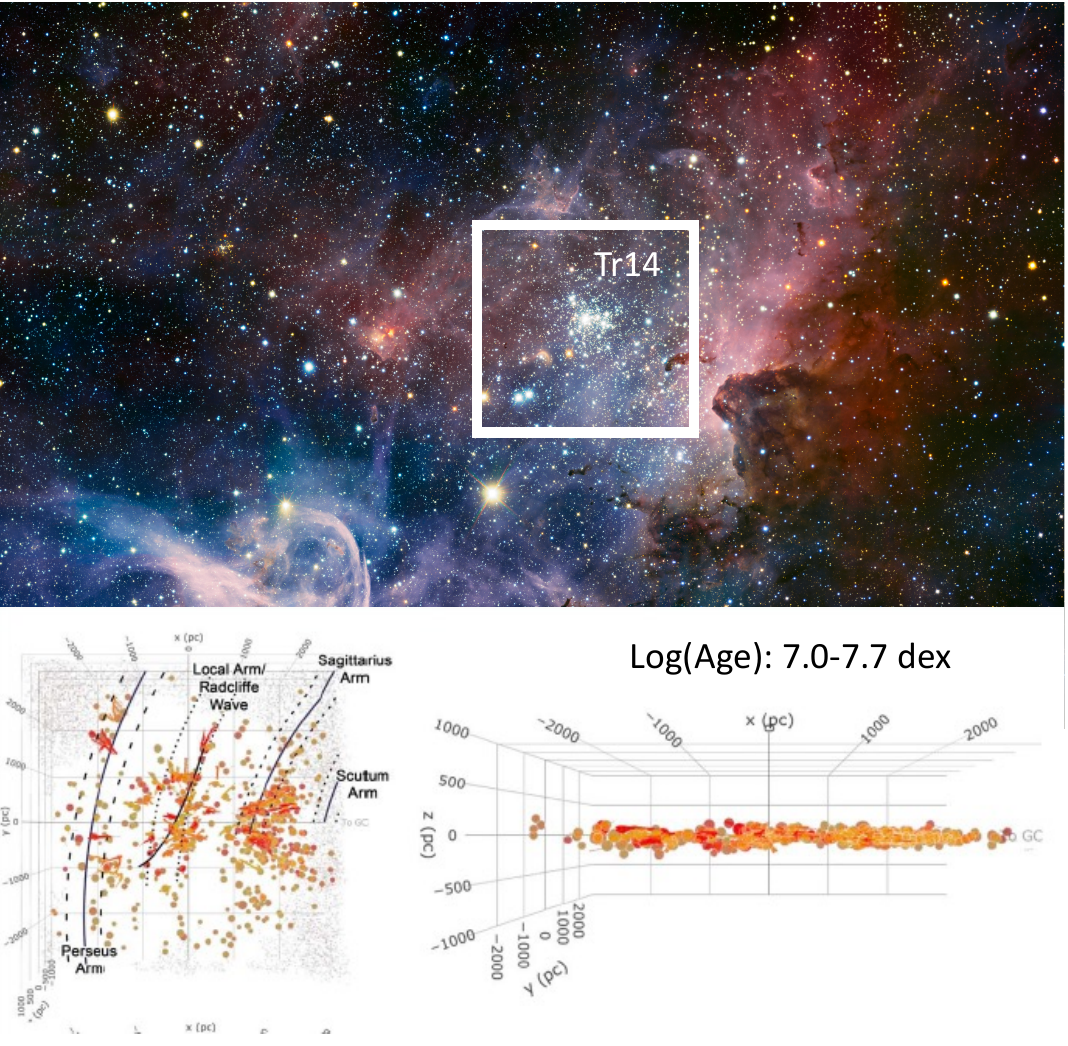}}
\caption{The top panel shows an infrared image of the Carina nebula (Credits: ESO/T. Preibsich), over-plotted with a  $9\times9~arcmin^2$ square, corresponding to a mosaic of 3$\times$3  \wst\ IFS centred on the massive cluster Tr~14. The bottom panel shows the distribution of the young stellar groups identified by \cite{Kounkel2020AJ}.
\label{fig:Sfr_fig}
}
\end{figure}

Thanks to its large collecting area and the combination of an IFS with a large-field fibre-fed spectrograph, the \wst\ can play a transformational role in the field. Specifically, the IFS will allow us to study with unprecedented detail the star formation processes in massive and dense environments. The low-resolution MOS will be used to investigate the properties of dispersed populations \citep{wrig23}, see Fig. \ref{fig:Sfr_fig}. Finally, the high-resolution MOS is an ideal instrument for measuring abundances and studying activity in large samples of planet-hosting stars. In the following, we outline some specific science cases.

\subsubsection{Investigating the nature of Galactic strings}
Current photometric surveys offer an opportunity to investigate star formation (SF) in environments of various densities and have challenged the traditional view that high-density clusters are the primary sites of SF. Theoretical models, such as those of \cite{krui12}, propose that the efficiency of cluster formation increases with the gas surface density, suggesting that star clusters are not a fundamental unit of SF, but one of many potential outcomes. Observations of high-density star clusters have long shaped our understanding of SF, but recent studies suggest a paradigm shift. Low-density associations, once considered remnants of dense clusters, are now seen as coherent populations formed in their observed spatial configuration. The ``strings'', identified by \cite{koun19} and also recognised by \citet{jera19,becc20,wang22}, are weakly bound filamentary structures, lying parallel to the Galactic plane (see bottom panel of Fig. \ref{fig:Sfr_fig}). They challenge previous notions of SF, displaying coherent and coeval structures. However, in recent years, several authors \citep[e.g.][]{mane22,zuck22,wrig23} offered contrasting views on the physical nature of these structures. Some reported chemical homogeneity, but others questioned their existence as coeval structures. A large-scale spectroscopic survey exploiting the \wst\ capabilities (thousands of fibre allocations for targets with G$<$20) would allow one to explore a large statistical sample in the range of low-mass stars down to the spectral types M0-M1, up to $\sim$2\,kpc \citep{pris22}. This could be crucial in resolving these controversies by validating these structures using spectroscopic signatures of youth (e.g., Li equivalent width, activity) and testing their chemical homogeneity and dynamics. 

The latter two requiring high spectral resolution.

\subsubsection{Effects of the star formation environment on the properties of stars and planetary system}
\label{gal:starformation} 

Stellar clusters younger than $\sim$10 million years are ideal targets for studying feedback from young stars on the star and planet formation process and early stellar evolution. The first few million years of stellar evolution are, in fact, crucial for the formation of planets around young stars, since this time scale corresponds to the dispersal of proto-planetary disks around low-mass stars \citep{Pollack1996Icar..124...62P, Manara2023ASPC..534..539M}. During this period, dynamical interactions can end with the ejection of some planets, increasing the population of brown dwarfs and free-floating planets \citep{Miret2023Ap&SS.368...17M}. 
Although this issue has been dealt with in detail in the Solar neighborhood \citep[see, e.g.,][for a recent review]{Manara2023ASPC..534..539M}, the question of the effect of the environment is still open. Indeed, all the most studied star-forming regions are located in the proximity of the Sun, easily accessible by current instrumentation, and they host low- to intermediate masses (i.e. are less massive than a few $10^4\,$M$_\odot$). Since the most massive star-forming regions are found in the inner part of the Galactic disk at southern latitudes, \wst\ will be able to obtain high-quality spectra for thousands of young stars in distant and massive star clusters, encompassing a wide range of star-forming environments. \par

Accretion and outflow play a fundamental role in shaping the evolution of the star+disk systems and the planet formation process. They regulate the evolution of the angular momentum of the system and the inward migration of material from the outer regions of disks. In both cases where disk evolution is driven by viscosity \citep{Lynden-BellPringle1974MNRAS.168..603L} or by magneto-hydrodynamical winds \citep[e.g.,][]{Armitage2013ApJ...778L..14A}, the incidence of energetic photons and particles onto proto-planetary disks is expected to play an important role in disk dispersal. This is because they can influence the level of ionisation, the disk temperature profile, the effectiveness of the coupling with the local magnetic field, and disk instabilities. In young stellar clusters, energetic radiation and relativistic particles are provided by the local environment, for instance, by nearby massive stars, in particular if in massive binary systems with colliding wind zones. \wst\ will allow, for the first time, detailed studies of the effects of feedback on disk accretion and outflows (by measuring accretion and mass-loss rates in proto-planetary disks using diagnostics such as the H$\alpha$ line and the forbidden emission lines typically observed in T~Tauri stars, \citealp{Hartigan1995ApJ...452..736H}). It will be possible to determine the typical time scale at which the intensity of disk accretion \citep[e.g.][]{Alcal2017A&A...600A..20A, Gangi2022A&A...667A.124G} and outflows \citep[e.g.][]{Nisini2018A&A...609A..87N, Giannini2019A&A...631A..44G}  decreases as a function of both stellar and environmental properties. Measurement of the intensity and velocity fields of the nebular lines themselves, for clusters embedded in H~{\sc ii} regions is therefore very important to determine the ionising flux irradiating circumstellar disks around low-mass stars \citep[e.g.][]{Damiani2017A&A...604A.135D}.\par

Numerous studies suggest that energetic radiation feedback can influence both the shape of the Initial Mass Function (IMF) and the expected characteristic stellar mass of the IMF. This has the important consequence that the final products of the star-formation process would differ in energetic environments with respect to less massive star-forming regions, such as those typical of the solar neighbourhood. Energetic radiation, for instance, can raise the temperature of the surrounding gas, and thus the Jeans mass in the collapsing cloud, which strongly suppresses fragmentation, leading to a decrease in the formation of low-mass objects \citep[e.g.:][]{Krumholz2006ApJ...641L..45K}. Strong radiation from massive stars can evaporate the outer layers of an accreting core, halting accretion and producing brown dwarfs and free-floating planets \citep{Whitworth2004A&A...427..299W, Bouy2009A&A...493..931B}. From the spectroscopic determination of the properties of the hundreds of thousands of young stars that will be accessible by the \wst, it will be possible to test the universality of the IMF over a set of different star-forming environments that to date have never been explored with high-quality spectra down to low-mass stars. 
In investigating  accretion/ejection processes  in young stellar objects (YSOs), it is crucial to take into account the contribution due to the nebular emission, dominant in embedded environments \citep[e.g.,][]{Bonito2013, Bonito2020} and possibly affecting the estimate of the mass accretion rate \citep[e.g.][]{Natta2004A&A...424..603N}. 
This could be accounted by monitoring the emission lines characteristic of accretion and ejection processes in YSOs and of nebular emission \citep{Bonito2020}. 

The possibility of observing young clusters with a large IFS will allow us to address this issue, since it allows the simultaneous mapping of emission from the stars and the surrounding interstellar medium. In addition, we could measure the intrinsic intensity of emission lines from young stars, i.e., formed in the immediate circumstellar environment and not in the nebula. These lines are important diagnostics of residual accretion from the circumstellar disk, and a correct separation between nebular and stellar components is fundamental to understand the link with the environmental conditions, as mentioned above.

\begin{figure}
\resizebox{0.7\hsize}{!}{
\includegraphics{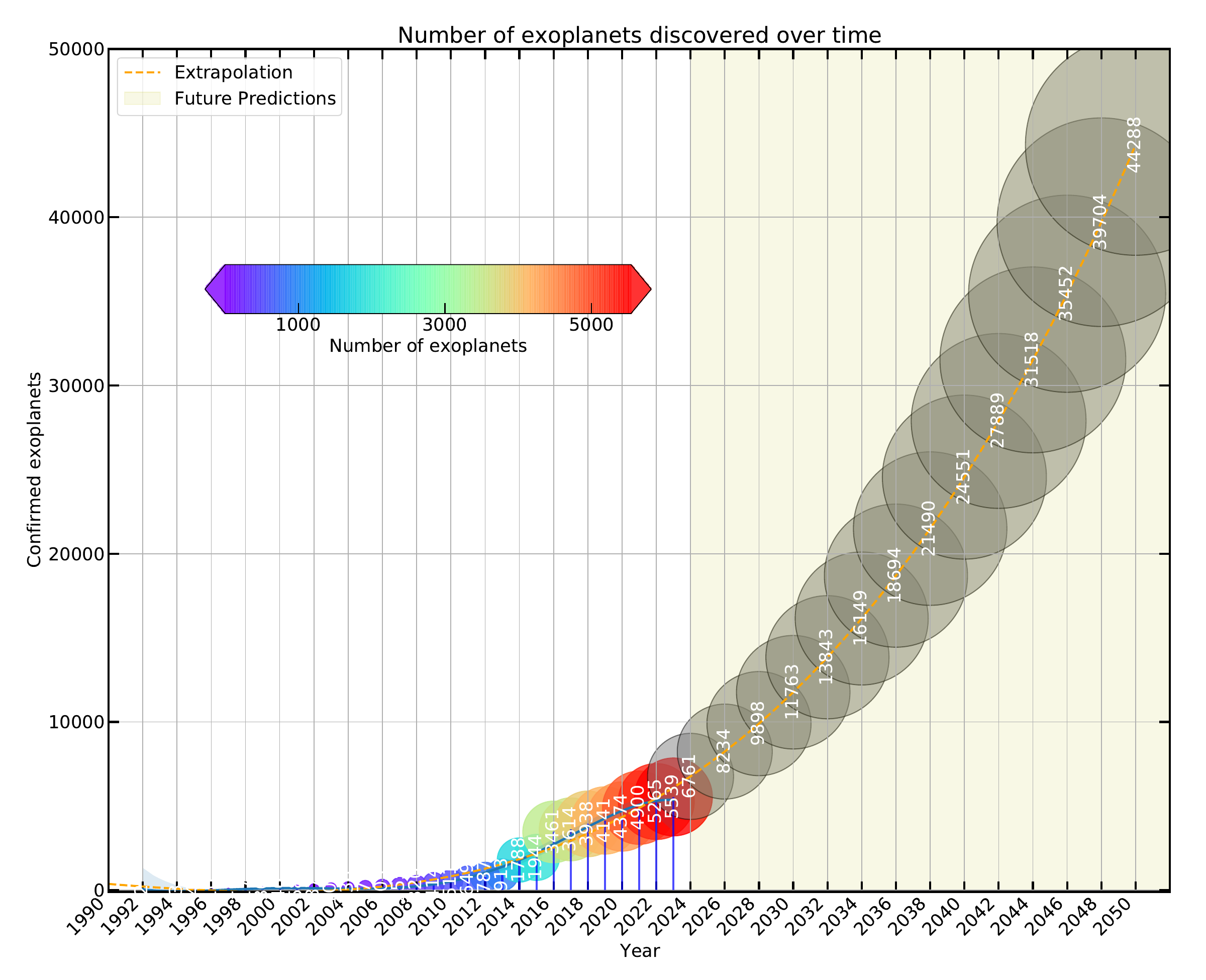}}
\caption{Number of discovered planets as a function of time from the NASA exoplanet catalogue (November 2023). Numbers after 2023 are extrapolated with exponential fit derived with the data collected so far.
\label{fig:n_planets}
}
\end{figure}

\subsubsection{The role of host-star chemical composition in shaping  planetary systems}

Ongoing and future space missions ({\it Gaia}, {\it TESS}, {\it CHEOPS}, {\it PLATO}, \euclid, and \nancy;  \citealt{GaiaMission}, \citealt{TESS} \citealt{CHEOPS}, \citealt{PLATO}, \citealt{Bachelet2022A&A...664A.136B}, \citealt{RomanTelescope}) will increase our knowledge of the number, type, and characteristics of planets in our Galaxy. Most of the data mentioned here will be collected by the end of the first half of the 2030s. Assuming that \wst\ will be ready in about 10 years, we might expect to have discovered $\sim$15\,000 planets at the beginning of the survey and 50\,000 at its end (see Fig.~\ref{fig:n_planets}). These are conservative estimates because new methods and instruments may allow many more planets to be found. For instance, the Galactic Bulge Planet Survey with the Roman telescope is expected to reveal $\sim$100\,000 transiting planets \citep{Montet2017}. 

The systems discovered will be well matched for follow-up with \wst\ due to its unique capabilities of collecting thousands of spectra in the Southern sky. 
Those large number of exoplanet hosts discovered by these missions, with 16 $<$ r $<$ 20 mag, which will make up the bulk of then-known exoplanets, cannot be followed up in high resolution by any other facility. 

Stars and their planets are formed roughly at the same time from the same molecular cloud. Therefore, a strong link between the composition of the star and that of the planet is expected. The present-time statistics of planet-host stars indicate the relevant role played, for instance, by global metallicity and other mineralogical diagnostics in the shaping of planets, especially giant ones \citep[e.g.][]{fischer05, Adibekyan2019,  Taylor2021, Magrini2022, Biazzo2022, gratton23}. With its high-resolution MOS mode, \wst\ will allow precise stellar abundances that can be used to seek correlations with the properties of the planetary system (multiplicity, mass, density, temperature, atmospheric composition, orbit), inferring, e.g.,  the dominant planet formation mechanisms in early disks. 

In addition, the study of planet-host stars belonging to different parts of the Galaxy, or observed in the solar neighbourhood but coming from other parts of the disc through the process of stellar migration \citep[e.g.][and references therein]{Dantas2023}, may extend our knowledge of the Galactic Habitable Zone (GHZ). New observations suggest that terrestrial planets can be found around stars of all metallicities \citep{MaliukBudaj2020}. This, together with the detection of (complex) organic molecules in the (far) outer Galaxy with abundances similar to those measured in the vicinity of the Sun \citep{Fontani2022a,Fontani2022b}, clearly extends the boundaries of the GHZ. The high-resolution observations of \wst\ will shed light on the driving mechanisms forming stars that host Earth-like planets and, thus, can harbour life.

\subsubsection{The role of magnetic field in formation and evolution of  planetary systems}

In its low-resolution MOS mode, \wst\ will characterise the activity, structure, and dynamics of tens of thousands of planet host stars \citep[e.g.][]{KrejcovaBudaj2012} by covering the calcium H\&K lines, the sodium D lines, H$\alpha$, and the calcium infrared triplet. These data will enable investigations of the effects of star-planet interactions and magnetic enhancement, which can have significant implications for photo-evaporation of young planetary atmospheres. By comparing these measurements with non-planet hosts and across clusters of known ages, it will also be possible to understand the evolution of stellar activity and its implications for planets across stellar evolutionary timescales. In its high-resolution MOS mode it will enable inference of the geometric and energetic configuration of the outer atmosphere (from the chromo-sphere to the corona), which is decisive in triggering star-planet interactions such as Auroral-Radio-Emission.

\subsubsection{Interstellar medium} 
The MOS-HR mode of \wst\ will be able to probe the interstellar gas by means of absorption lines. The combination of the collecting area and the target density will allow us to map the spatial variations of the interstellar medium (ISM) better than currently, piggybacking on the HR Galactic surveys. The strongest lines in the optical range are from Na, K, and Ca. These features make it possible to probe the detailed gas kinematics of the ISM \citep[e.g.][]{Ritchey24}, allowing the study of the clumpiness of the ISM on a range of different scales. These studies may also provide a connection to the high-redshift quasar absorption line work and the local chemical abundance estimates of specific elements in the Milky Way and other nearby galaxies \citep[e.g.][]{Stawinski23}. 
\wst\ spectra will also allow for a detailed study of Diffuse Interstellar Bands (DIBs). It will be possible to resolve the velocity structures for a large statistically significant sample and thus contribute to the general understanding of their environmental conditions, especially with regard to dust content and gas densities \citep[e.g.][]{Bailey15, Wendt17}.

Furthermore, the \wst\ IFS covering a contiguous area of 9 arcmin$^2$ will be the best instrument to study molecular clouds and perform extended source spectroscopy. These regions are pervaded by cosmic rays (CRs) that regulate the chemical and dynamical evolution of star-forming regions. CRs can penetrate into the densest regions of molecular clouds, where UV radiation is blocked by the absorption of dust grains and molecular species, determining the ionisation fraction, and hence the degree of coupling between the gas and the magnetic field. In addition, CRs control the heating of the gas, the excitation and dissociation of H$_2$, and the formation of complex chemical species observed at different scales. 

Secondary electrons produced by the ionisation of ambient atomic and molecular gas from the ISM as well as primary cosmic-ray protons can produce excited hydrogen atoms
by direct H$_2$ dissociation and electron capture, resulting in Lyman and Balmer emission. 
Recently, \citet{Padovani2024A&A...682A.131P} calculated the emission spectrum of atomic and molecular hydrogen, including the Lyman and Balmer series, as a function of incident cosmic-ray flux, the H$_2$ column density passed through, the isomeric H$_2$ composition, and the dust properties. The observations carried out with the IFS, combined with the model predictions, will make it possible to map the distribution of low-energy cosmic rays in the Galaxy, identifying the sources of cosmic rays, and constraining the propagation models. Furthermore, the FWHM of the H$\alpha$ broad component will provide information on the proton velocity and the post-shock temperature to interpret possible new acceleration mechanisms.

\subsection{Evolution of stars}

Stellar evolution is an important pillar of modern astronomy. Stellar and chemical evolution models rely on assumptions of our understanding of all stages of the life of stars. Although several aspects of stellar evolution are well understood, many unresolved questions persist that impact all areas of astronomy focussing on stellar populations. Stellar evolution models are widely used to interpret the properties of individual stars and the integrated properties of galaxies anywhere in the Universe. To improve our models of stellar and chemical evolution, precise measurements of the detailed properties of large samples of stars at all stages of evolution are needed. Many of these stages are short-lived and/or difficult to observe. Moreover, physical processes such as mass loss remain poorly understood, even though that is a critical parameter in the life of stars. The way in which stars reach the final stages of their life cycle influences our understanding of the processes through which stars enrich the universe with chemical elements over the course of cosmic time. Binary stars play a critical role in affecting the stellar evolution outcomes, but also remain poorly studied beyond a relatively small number of special cases. Large surveys are essential to change this scenario. Current surveys are contributing to these goals, but important details will remain hidden until HR spectra (R$\simeq 40$k) of large samples of all types of stars are available.

\subsubsection{White Dwarfs} 
White dwarfs are the end fate of low- to intermediate-mass stars ($< $10 M$_{\odot}$), a major fraction of stars in the Universe. White dwarfs can be found either in isolation or as members of binaries (detached or accreting) and multiple systems. These objects are also studied in the transient science case, see Section~\ref{sec:td:stars:types}. 

The \emph{Gaia} mission identified hundreds of thousands of WDs, allowing for the first time both magnitude- and volume-limited population studies from statistically significant samples. With \fourmost, the White Dwarf Binary Survey \citep{Toloza2023}, in close collaboration with the \fourmost\ Milky Way Disc and Bulge Low-Resolution Survey \citep[4MIDABLE-LR,][]{Chiappini2019}, will observe $\sim$150\,000 close and wide \emph{Gaia} WD binaries and also a large number of single WDs. These observations will be a major breakthrough in the field, but these studies will mainly be performed at low spectral resolution, partly because these samples tend to be faint, but also because this is what is currently available. HR spectra of WDs and their progenitors with \wst\ will provide more detailed insight into their evolution and are the natural next step.

A total of $\sim$115,000 \emph{Gaia}-detected WDs \citep{GentileFusillo2021} can be targeted for observations in the Southern hemisphere. They are within 500~pc of the Sun and have cool enough effective temperatures to ensure the presence of lines in their spectra. It may even be possible to resolve the Zeeman splitting caused by weak magnetic fields. From previous studies, the spectral types and stellar parameters will be known to 100~pc. This will be extended to 500~pc with \wst. The WD sample observed at HR with \wst\ will make it possible to:

\begin{enumerate}

\item Establish firm statistics on the occurrence of exoplanets around WDs, by means of the detection of minute quantities of planetary material in the optical spectra of their otherwise pristine hydrogen or helium atmospheres. The fraction of WDs that show metal lines is directly related to the resolution of the spectra, with an increasing fraction for increasing resolution \citep{Koester2014}. To date, no large-scale high-resolution programme has been performed or is planned. 
\item Derive precise RVs by resolving the core of Stark broadened lines, where the core is formed under non-LTE conditions. This is only possible from HR spectra \citep{Raddi2022}. Such an analysis will enable an accurate age-velocity-dispersion relation for these old stellar remnants. Multi-epoch spectroscopy would be useful for identifying hidden companions such as another WD, a low-mass star, or substellar objects (see Section~\ref{td:stars}).

\item Accreting white dwarfs (AWDs) are compact interacting binaries in which a WD is accreting matter from a companion star via Roche lobe overflow. Their study will constrain the key ingredients (i.e., angular momentum loss, mass transfer process, and response of the donor star to mass loss) of models describing the evolution of all types of binaries \citep{KolbStehle1996, SchreiberGansicke2003, Knigge2011}. The final fate of AWDs is intimately linked to Type Ia supernovae, but it is not yet clear which configurations can lead to a successful explosion. With the HR spectra, it will be possible to resolve sharp metal lines from the photosphere of the WDs in this sample, from which it will be possible to measure their rotation rate. This is a critical parameter, as spinning up could allow the WD to exceed the Chandrasekhar mass limit without triggering a thermonuclear explosion \citep{King1991}.

\end{enumerate}

\subsubsection{Stellar Multiplicity}
Stellar multiplicity is ubiquitous across the HR diagram and across the Galaxy. Multiple systems have been seen for a long time as key objects for constraining theories of stellar evolution and stellar nucleosynthesis. They also now appear as key objects to constrain planet formation and survival: there is extensive evidence that binary systems frequently host exoplanets \citep{Schwarz16}, making them ideal laboratories for studying complex dynamics \citep{FabryckyTremaine2007}. Moreover, multiple systems contribute to shaping the chemical enrichment of the Galaxy. They include the progenitors of type~Ia supernovae that are efficient producers of intermediate-mass and iron-peak elements and the progenitors of compact binaries (NS-NS, NS-BH, BH-BH) that are efficient producers of neutron-rich r-elements. Binaries are needed to understand the chemical peculiarities of some classes of stars, such as the carbon-enriched metal-poor (CEMP), S, or Ba stars. \emph{Gaia} DR3 gives a teaser with the identification of astrometric, spectroscopic, and photometric binaries \citep{2022arXiv220605595G}. 

Future \emph{Gaia} releases will consolidate these samples and upcoming surveys such as the LSST will identify even more systems due to their deeper footprint \citep[of the order of 100s per sub-spectral type,][]{Best2021}. Wide binaries are excellent probes for testing models of star formation and evolution \citep{Badenes+2018} and the correlation between age-related stellar properties such as their activity, rotation, and metallicity \citep{Rebassa-Mansergas+2021}. Most known systems are in the Southern hemisphere \citep{ElBadry2021} and consist of main sequence pairs ($\sim$500~000). However, a significant fraction ($\sim$10~000 objects) also include WD companions. In addition to that, more exotic evolved systems, such as WDs in wide triples \citep{PerpinyaValles2019} and hierarchical triples containing WDs and brown dwarfs \citep{RebassaMansergas2022} are starting to emerge. 

High resolution \wst\ spectra of all these different types of binary systems will allow the determination of ages, also for main sequence companions, and the derivation of radial and rotational velocities, activity indices, and metallicities. Close binaries are typically detected as spectroscopic binaries (SBs) and are a natural result of spectroscopic surveys. Their identification is described in more detail in Section~\ref{sec:td:stars:data}.

The MOS-HR mode of \wst\ will be ideal for performing spectral fitting of composite spectra of SBs to constrain the physics and the chemistry of the photospheres of both components. This will provide valuable information for studying the statistics of SBs as a function of the spectral type, the luminosity class, or the metallicity. Properly recognising and flagging spectroscopic binaries will always be essential in the quality assessment of the stellar parameter and chemical abundance determinations.

\subsubsection{Ionised nebulae: the initial and final stages of stellar evolution}

H~{\sc ii} regions and planetary nebulae represent different moments in stellar evolution, as well as in Galactic chemical evolution. The former are clouds of hydrogen ionised by young massive stars \citep[e.g.][]{2015A&A...582A.114W}. The latter represent one of the final stages in the evolution of low- and intermediate-mass stars and are key tracers of Galactic chemical evolution \citep[e.g.][]{Magrini2016A&A...588A..91M}. Despite these differences, they share very similar spectra, characterised by intense recombination lines of H and He, collisional lines of O, N, Ar, S and other elements, and weak recombination lines of some metals \citep[see, e.g.][]{rojas2007ApJ...670..457G, rojas2013A&A...558A.122G, rojas2022MNRAS.510.5444G}. 

The field of view of the IFS of the \wst\ is perfectly suited to sample even the largest nebulae with a single pointing. Furthermore, the spectral resolution is perfectly suited not only for the determination of the physical and chemical properties of the nebulae, allowing the measurement of the weak metal recombination lines, but also for resolving the expansion velocities of PNe and distinct kinematical components \citep{2020A&A...634A..47M,Monreal2022Galax..10...18M}, since under proper sampling of the emission lines the line centroid can typically be determined with an accuracy of $\lesssim$1/5 the spectral resolution \citep{2008A&A...479..687A,2017A&A...603A.130M}.
In addition, the extension of the spectral range to blue wavelengths, from about 3700 \AA\, will allow the observation of both the [O~{\sc II}] lines that are fundamental for the determination of total oxygen abundance, and the [O~{\sc III}] 4363\AA\, line, which is a key electronic temperature diagnostic. Simultaneous determination of collisional and recombination lines, together with electron temperature sensitive auroral lines, in a large number of H~{\sc II} and PNe regions, will allow a breakthrough in understanding the discrepancy between the abundances derived with the two sets of lines \citep{mendez2023arXiv231110280M}. 

Finally, (many) PNe do possess a faint halo (i.e. $\sim$3 orders of magnitude fainter than the main nebula) carrying the footprint of the last stages of the star evolution during the AGB phase. Given their low surface brightness, their spectroscopic characterisation is extremely challenging today, with only a few examples of attempts to do so \citep{2005ApJ...628L.139M,2008A&A...486..545S}.
The high \wst\ sensitivity together with its large field of view would allow one to characterise (and in many cases detect for the first time) these structures simultaneously to the main PN.

\subsection{Galactic survey strategy}\label{sec:gal:survey}

\subsubsection{A three tier Galactic survey}

A preliminary strategy for a Galactic survey with \wst\ would entail designing a survey with several tiers, using different spectral quality regimes (resolution and signal to noise) of the MOS mode, that would serve a collection of science cases outlined in the previous subsections. The following tiers would capture the most novel aspects of \wst\ for our Galactic, stellar, and exoplanet science: 

\begin{enumerate}

\item {\bf a survey of a large fraction of the Southern sky at R=40\,000 with very high S/N down to G$\sim$17}, allowing us to build an ultimate precision elemental abundance sample of relatively nearby stars ($\sim$2-3\,kpc for a main-sequence turnoff star and up to $\sim$20-70\,kpc for a red giant star). 

Typical densities of such targets are given in figure \ref{fig:densitymap}, and cover essentially the Galactic discs and inner halo.

\item {\bf within the same footprint, a survey at lower S/N down to G$=18$}, preserving the superb resolution to enable the best radial velocity precision while still enabling good quality elemental abundance measurements (similar or better than \fourmost\ or \weave). This will cover a much larger Galactic volume ($\sim$30-100\,kpc for a red giant star), thus reaching all stellar populations of the Milky Way and its nearest neighbours, including the diffuse stellar halo and its substructures, and the brighter stars in the Galactic Bulge.

\item {\bf a survey of a large fraction of the Southern sky in low-resolution (R$\sim$5\,000) down to G$\sim$22-23}, enabling one to enlarge the volume in which chemo-dynamics of stellar populations will be accessible (the Bulge down to its main sequence, the deep Milky Way halo, substructures to large distances, and local group galaxies). Typical stellar densities within the magnitude range within reach of this survey are of 4,000 to 8,000 per deg$^2$, at intermediate Galactic latitudes (based on deep imaging from the UNIONS/WHISHES survey, number counts between g$=$18 and 23). This survey would be unique compared to any ongoing or planned spectroscopic survey because of its completeness and Galactic volume covered, matching the depth expected for where LSST and Roman will provide proper motions.

\end{enumerate}

\noindent{Such} a strategy would allow us to serve the specific surveys/science goals outlined in the following sub-sections.

\subsubsection{Mapping the Galaxy discs at high-resolution}

    Considering the \wst\ multiplex for the MOS-HR (2000 fibers in a $\sim$3.1 deg$^2$ field) and a typical observation time of 1 hr (which allows to have a S/N=100/A for a star of mag=18), for 100 nights per year for 5 years, about 10$^7$ stars could be observed, allowing the detection of about 10$^4$ fossil groups/clusters, now disrupted. The detection rate can also be increased by pre-selecting samples belonging to macro populations such as the thick disc, or the outer thin disc \citep[e.g.][]{Ting15}. The pre-selection can benefit from current and forthcoming surveys, in particular of the \fourmost\ Galactic surveys. In addition to searching for and characterising destroyed clusters, \wst\ will also map the homogeneity within known open clusters and develop successful strategies for chemical tagging. The large number of fibres makes it possible to study a larger number of fainter members in relatively nearby clusters, especially in their tidal tails (where the large field of view is critical). This will allow us to follow the effects of stellar evolution on photospheric abundances and to study anomalies possibly related to the presence of planets or their engulfment. Furthermore, because of the large FoV it will be possible to study the presence of chemical members which are kinematically dispersed in the peripheries and streams of star clusters. The numbers of stars per square degree over the sky is shown in Fig.~\ref{fig:densitymap}, for stars with G$<$17 with low extinction.
    
\subsubsection{Searching for the stellar halo at high-resolution 
}

    The stellar halo of the Milky Way is a very sparse stellar population, requiring all-sky wide-field surveys to hope to collect significant samples. Previous surveys (e.g. \fourmost) will provide the needed target list in advance, and so it will be possible to follow up these samples at high spectral resolution down to fainter magnitudes. This will dramatically increase the sample observed at this resolution and thus also the overview of the properties of this critical component of the Milky Way, which keeps traces of past interactions with external stellar systems. Enlarging the sample of HR measurements is critical to trace the rare examples of the most metal-poor stars, which allow us to probe further back into the past of the Milky Way and also the star formation history in the Universe.
    
    The low density of the targets requires deep observations with an instrument with a sensitivity higher than that of current generations of surveys, reaching down to $G>16$.    

    In addition, a large field of view is also needed because the estimates predict about 100 targets per square degree at G$\sim 18$ and 50 targets per square degree at G$\sim 16$. Low-metallicity stars require HR to get reliable abundances of all critical elements, as the lines will always be weak. It is also especially interesting to make a full analysis of these ancient populations to provide insight into early events (e.g. neutron capture processes). There will always be relatively few targets in any \wst\ FOV (a few hundred), so this survey will be one part of the overall Galactic surveys, adding these rare objects as high-priority targets to the general survey.
    
\subsubsection{Surveying nearby dwarf galaxies at high-resolution 
}
    There are a number of dwarf galaxies (ultra-faint and also dwarf spheroidal and dwarf irregular) within $\sim250$~kpc of the Sun, which put them at reach of long pointed observations with \wst. These systems are quite concentrated at equatorial and southern latitudes. They are distant, and so making complete surveys of large samples requires the sensitivity gains from a large telescope and efficient MOS instruments. The HR allows for a survey with a range of chemical elements similar to those observed in the Milky Way, probing the same formation channels for chemical elements in a different environment from the Milky Way. The comparison to the \wst\ Galactic survey will be critical to finally resolve the differences and similarities between the properties of stars in dwarf galaxies and the stellar populations of the Milky Way. The \wst\ field of view allows most of the dwarf galaxies of interest to be covered in a single pointing, although multiple pointings may be needed to compensate for the crowding of the targets, depending on the fibre system spacing. This will be a relatively small number of pointed observations, compared to the Galactic survey, but with considerably longer exposure times to get as deep as possible. Providing spectroscopy of fainter stars and with a higher resolution compared to previous surveys, this effort will provide significantly better abundance measurements and much better sampled kinematics. See also Section~\ref{subsec:respop-LGdwarfs}.
    
\subsubsection{Mapping the Bulge at high-resolution and low-resolution 
}
    The complexity of the Bulge is the extremely dense stellar population making crowding the main issue. In addition a large fraction of the stars have very high metallicity, requiring high spectral resolution to overcome severe line blending and determine accurate abundances. The large number of stars that can be mapped at lower spectral resolution will contribute to the kinematic survey of this extremely complex region going down to fainter magnitudes over wider areas than will be possible with current or planned facilities. As with the other Milky Way samples, the \wst\ high-resolution follow-up will assume previous spectroscopy with ongoing and planned surveys (primarily \fourmost\ and \moons).

\clearpage


\section{Resolved stellar populations from the Milky Way to nearby Local Volume galaxies}\label{sec:respop}

\paragraph{Authors} Giuseppina Battaglia$^{21,22}$,
Julia Bodensteiner$^1$,
Jarle Brinchmann$^3$,
Norberto Castro$^{41,10}$, 
Maria-Rosa Cioni$^{10}$,
Miriam Garcia$^{46}$, 
Robert Grand$^{58}$,
Artemio Herrero$^{21,22}$,
Pascale Jablonka$^2$,
Sebastian Kamann$^{58}$,
Anna McLeod$^{7,8}$,
Ryan Leaman$^{71}$,
Martin M. Roth$^{10}$,
Andreas A.C. Sander$^{122}$,
Jorick S. Vink$^{115}$, 
Peter M. Weilbacher$^{10}$

\subsection{Introduction}
\label{subsec:respop-intro}

\begin{figure}[th!]
  \centering
  \includegraphics[width=1.0\textwidth]{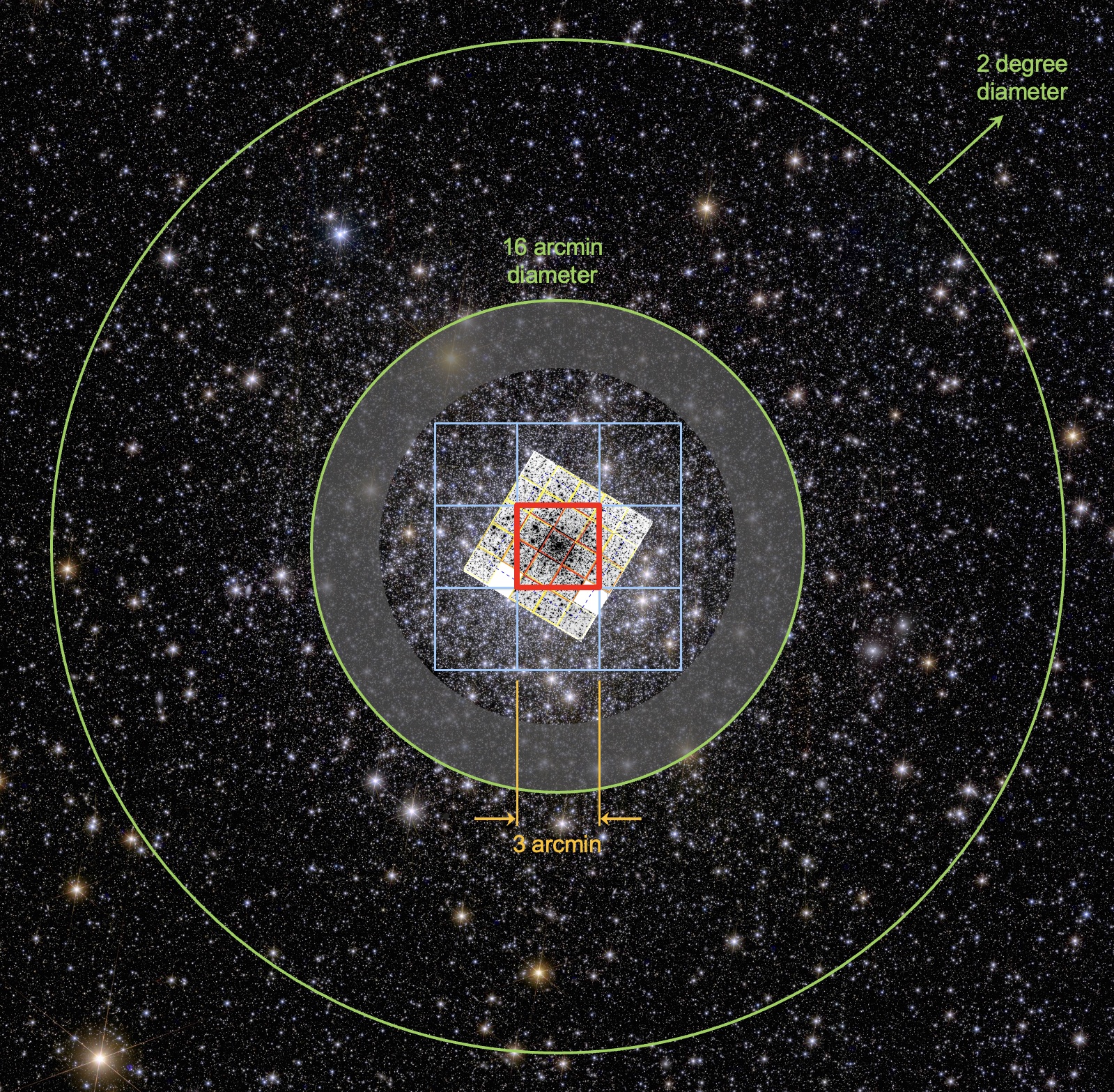}\\
  \caption{\small NGC\,6397 with footprint of \wst\ MOS and IFS (see Fig. \ref{fig:Intro:WST_FOV}), showcase for \wst\ transformative science with resolved stellar populations (background image: Euclid). Center: reconstructed image with 12.000 deblended stars from mosaic of $5\times5$ MUSE exposures, adapted from \citet{Husser+2016}. Note: the full \wst\ MOS field is four times larger than the outer circle. Credit: M. Roth.} 
 \label{fig:NGC6397}
\end{figure}

Light collecting power and high angular resolution of modern telescopes have enabled 
star-by-star studies of the formation and evolution of stellar systems such as Milky Way globular clusters, the Magellanic Clouds, and other Local Group galaxies. Specific examples include HST surveys, e.g., ANGST \citep{Dalcanton+2009}, GHOSTS \citep{Radburn2011}, or LEGUS \citep{Calzetti+2015}.\\

Technically, such studies have relied on software tools like DAOPHOT \citep{Stetson1987} that employ PSF-fitting techniques to de-blend overlapping stellar images in heavily crowded fields. However, the information content for astrophysical interpretation has been limited to photometry.\\ 

After the advent of integral field spectroscopy at 8m-class telescopes, this situation has changed dramatically.  An analogous tool to DAOPHOT for data cubes, PampelMuse, was presented by \citet{Kamann+2013} and subsequently applied to numerous MUSE observations of crowded fields in star clusters and nearby galaxies, see review by \citet{Roth+2019}. MUSE at the VLT has become a game-changer for high multiplex crowded-field spectroscopy, with no competitor worldwide.
Nevertheless, the 1~arcmin$^2$ field-of-view (FoV) remains a limiting factor for resolved stellar systems 
in very nearby galaxies which, due to their large sizes on the sky, require a FoV that is at least one order of magnitude larger than the one of MUSE.\\

As a major step forward, \wst\ will become a unique, unrivaled facility for the spectroscopic study of resolved stellar populations by providing a wide-field IFS for crowded fields, and simultaneously a 
2-degree diameter MOS for surrounding areas where crowding is not an issue.

\subsection{The binary content of star clusters across a Hubble time}
\label{subsec:respop-clusters}

\wst\ will be used to study the binary populations of massive ($>10^5$ M$_{\odot}$) clusters. By targeting both young massive clusters, which are abundant in the Magellanic Clouds and other nearby star-forming galaxies, and old globular clusters in the Milky Way, we will be able to understand how binary stars impact the evolution of massive clusters and vice versa. Extending the scope from the Milky Way to other nearby galaxies is important to encompass different metallicity environments.\\

Each binary in a star cluster will evolve through a multitude of interactions with other cluster members. These interactions strongly alter the primordial populations via binary disruption, flyby and exchange interactions, or the Kozai-Lidov mechanism, and result in systems that are endemic to clusters, such as dynamically formed binary black holes or low-mass stars orbiting degenerate companions \citep{Giesers+2018}. The detection and characterization of the latter hold crucial information about the kick velocities resulting from supernova explosions, a major unknown limiting our capabilities to understand the growing number of gravitational wave detections.\\

\begin{figure}[h!]
  \centering
  \includegraphics[width=0.90\textwidth]{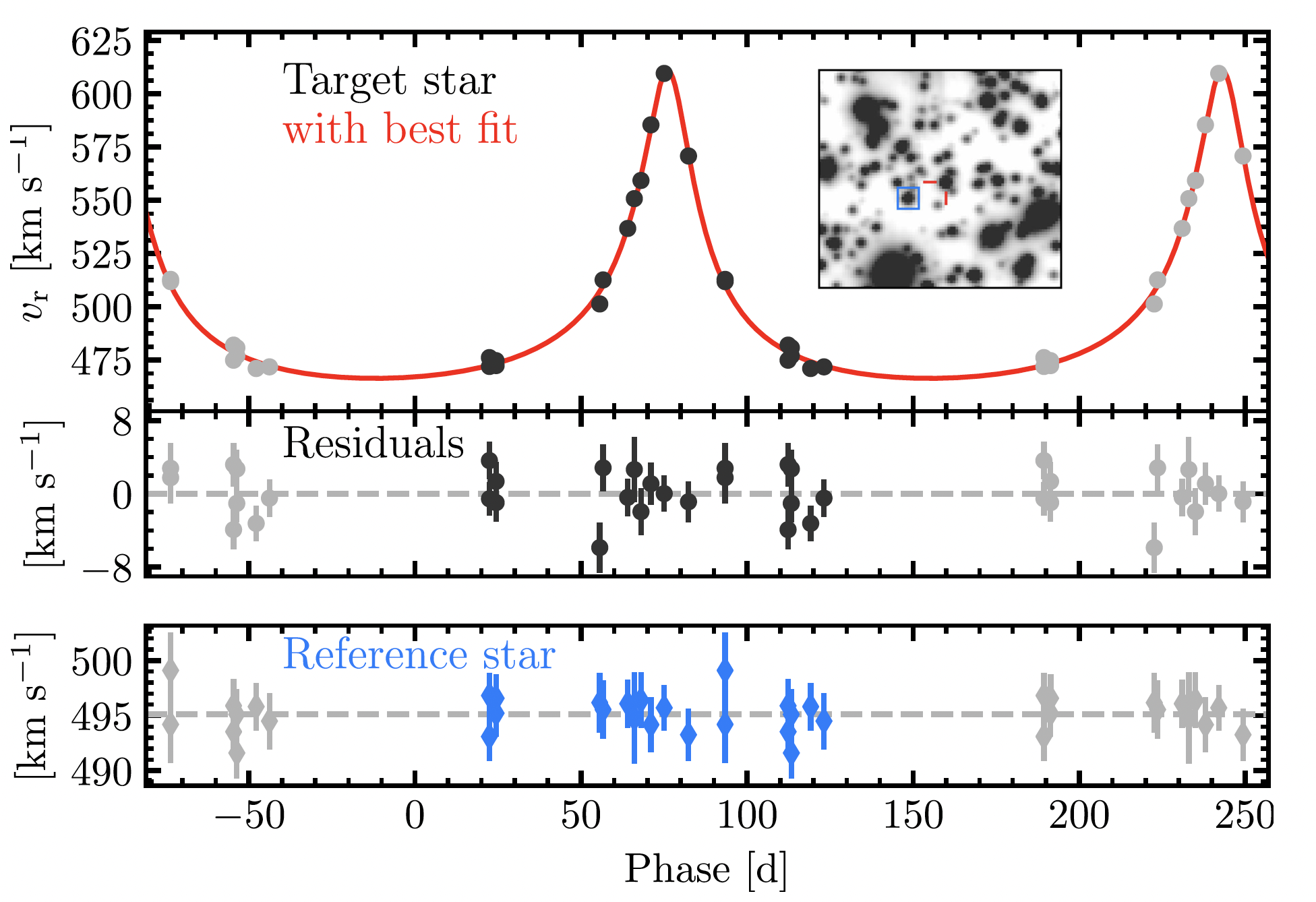}\\
  \caption{\small Showcase for \wst\ transformative science in the search for spectroscopic binaries:
  MUSE discovery of a stellar-mass black hole in NGC\,3101. Top: radial velocity measurements 
 v$_r$ of the target star (red marker in the insert, showing reconstructed continuum image from 
 MUSE data cube), phase folded for the 167 d period. Error bars are smaller
than the data points. The red curve shows the best-fitting Keplerian orbit. 
The middle panel contains the residuals after subtracting this
best-fitting model from the data. The bottom panel shows the radial velocity
measurements of the reference star (blue marker). Adapted from \citet{Giesers+2018}.} 
 \label{fig:NGC3201}
\end{figure}

Thanks to their well-known ages and metallicities, star clusters play a key role in improving our understanding of post-interaction products, like blue stragglers, stripped stars, or classical Be stars, e.g., \citet{Bodensteiner+2020}. Different formation mechanisms for these objects make predictions that can best be tested in clusters, for example, the formation of blue stragglers via stellar collisions or mass transfer \citep{Ferraro+2023}.

Not only is the evolution of binaries strongly altered by the surrounding cluster members, but binary stars also have a crucial impact on the evolution of clusters, as they are reservoirs that balance the energy budgets and hence the lifetimes of their hosts. Binaries reverse and moderate the core collapse of long-lived clusters, while few-body interactions involving binaries efficiently eject stars from clusters. Binaries are further considered as a solution to the open questions in our understanding of massive clusters, like the origin of the bimodal spin distribution observed in young open clusters, or the multiple populations phenomenon observed in old globulars.\\

While high-precision photometry can provide us with statistics on the binary population inside a cluster via the so-called-binary main sequence, e.g., \citet{Milone+2012}, a detailed characterization of the overall energy stored in binary systems and the nature of interacting systems requires spectroscopy. However, owing to the crowded fields, relative faintness of the stars, and the low binary fractions of evolved clusters, this is a challenging task. While MUSE has been a game-changer in this respect (see \citealt{Giesers+2019}, \citealt{Bodensteiner+2021}, \citealt{Saracino+2023}), its low spectral resolution and limited blue wavelength coverage are fundamental limitations, especially in the study of young clusters. BlueMUSE will largely erase these limitations, but a fundamental limitation will remain, namely the operation mode: to detect and characterize binaries across a wide period range, the same cluster fields must be observed with varying cadence over a period of years.\\

Owing to the enhanced sensitivity of \wst\ compared to current and upcoming facilities, we expect that we can probe deeper into the cluster stellar populations by at least 1~mag. In many cases, such as young or intermediate-age massive clusters in the Magellanic Clouds, this will enable the crucial step across the main-sequence turn-off, thereby exposing populations of tightly bound binaries before they are altered by stellar evolution (e.g., mass transfer, common-envelope evolution).\\

Most clusters have half-light radii of a few arcminutes, while their tidal radii are tens of arcminutes. Therefore, we will be able to target the dense cluster centres of dozens of clusters with different ages in different metallicity environments with the IFS of \wst, while its MOS capabilities will enable us to probe the cluster outskirts, where the densities are low enough such that crowding is not a problem. Arguably, the zone of avoidance between IFS and MOS will result in a gap in our coverage. To mitigate this problem, we will explore IFS mosaic-ing options.\\

Finally, it should be noted that many young clusters (such as R136 in the LMC) are embedded in star-forming regions that fill the entire \wst\ field of view. Hence, we will be able to study the binary properties of these regions using the same observations dedicated to the star clusters.

\subsection{The Magellanic Clouds}
\label{subsec:respop-LMCSMC}

\noindent

\noindent
Despite being the best-studied satellites of the Milky Way, the Magellanic Clouds still hide many secrets. These include, among others, their history of interactions, both mutually and with the Milky Way, the existence of central massive black holes, or the chemical enrichment with time and across the faces of the galaxies. The Magellanic Clouds are ideal cosmic laboratories for multiple reasons. Both systems are actively star-forming, and the Large Magellanic Cloud hosts the most massive star-forming region known in the Local Group, 30~Doradus. Furthermore, they cover a substantial range in both stellar mass \citep[LMC: $1.5\times10^9\,{\rm M_\odot}$, SMC: $4.6\times10^8\,{\rm M_\odot}$,][]{McConnachie2012} and mean metallicity (LMC: -0.5, SMC: -1.0). Hence, they enable us to perform detailed studies of star formation in low-mass galaxies and as a function of metallicity (see also Sec.~\ref{subsec:respop-massivestars}).

Owing to their vicinity and fortunate orientation, we can perform such studies on a wide range of scales, from detailed analyses of individual star-forming regions or star clusters to panoramic views of the entire galaxies. Therefore, it is not surprising that throughout the last years, the Magellanic Clouds have been targeted by a number of photometric surveys, such as VMC \citep{Cioni+2011}, or SMASH \citep{Nidever+2017}. While spectroscopic censuses of similar breadths are currently missing, this picture will soon drastically change. New multi-object spectroscopic facilities such as \fourmost\ or \moons\ will perform dedicated surveys of the Magellanic Clouds \citep[e.g.,][]{Cioni+2019,Gonzalez+2020}. In the course of SDSS-V \citep{Kollmeier2017}, the Local Volume Mapper will create a panoramic spectroscopic view of the Magellanic Clouds at a low spatial resolution of $\sim36^{\prime\prime}$ per spaxel. At the same time, \textit{Gaia} provides invaluable data to decipher the Clouds both chemically and dynamically. \\

Owing to the large apparent dimensions of the Magellanic Clouds of several degrees on the night sky, it is currently barely possible to cover their entire footprints using 8m-class telescopes. For example, the coverage of the LMC foreseen for the \moons\ survey will include only a few percent of the total extent of the galaxy \citep[see Figure 2 in][]{Gonzalez+2020}. On the other hand, this limitation is largely erased with smaller facilities, and indeed, the \fourmost\ survey designed by \citet{Cioni+2019} will cover the Magellanic Clouds in their entirety. The survey's limiting magnitude of G=20, however, is several magnitudes brighter than what \wst\ will deliver in single-hour exposures. In terms of stellar populations, this implies that with \wst\ we will be able to probe age-sensitive regions of the colour-magnitude diagram, such as the sub-giant branch, for even the oldest stellar populations in the Magellanic Clouds. At an age of $\sim10$~Gyr, the sub-giant branch corresponds to optical magnitudes of $V\sim22$ at the distance of the LMC, more than 1~mag brighter than the limiting magnitude of the \wst\ for 1~h exposures at low resolution.\\

The unique combination of IFS and MOS capabilities opens the possibility to overcome crowding limitations impacting current and upcoming spectroscopic facilities. The restrictions on fiber placement (e.g., a minimum fiber-to-fiber distance of $15^{\prime\prime}$ in the case of \fourmost) severely limit their capabilities to study crowded fields. While repeated visits provide a possibility to mitigate fibre-placement limitations, contamination by nearby stars will remain a concern. Using archival HST imaging, we estimate that the projected density of stars brighter than $V=23$ is about $1/{\rm arcsec^2}$ towards the centre of the LMC. Compared to typical MOS fibre sizes (1.45~arcsec for \fourmost, 1~arcsec for \moons), this implies that the average fibre spectrum collected in this region will be contaminated by 1-2 sources brighter than the \wst\ limiting magnitude. The panoramic IFS of \wst\ will therefore be vital to probe dense regions inside the Clouds, such as star clusters (cf.~Sec.~\ref{subsec:respop-clusters}) or the central regions of the galaxies. We envision an optimized observing strategy where the IFS is placed on a massive star cluster while the MOS fibers are used to sample the surrounding field. We further highlight that the panoramic IFS of \wst\ will enable us to solve a fundamental problem for the spectroscopy of stars that are embedded in nebulosities, especially massive stars near H\,II regions. In star-forming galaxies containing substantial amounts of gas such as the Magellanic Clouds, critical diagnostic (stellar) lines are contaminated by strong nebular emission lines that are difficult to subtract accurately with slit spectrographs, and quite impossible to avoid with fiber-based MOS, see, e.g., \citet{Bestenlehner+2024}. As demonstrated by \citet{Becker+2004} for stars in M33 and \citet{Roth+2004} for PNe in M31, IFS is particularly capable to disentangle point source spectra from a nebular environment.\\ 

With regards to the centre of the LMC, \wst\ will enable a dedicated search for a central massive black hole. The occupation fraction of massive black holes in low-mass galaxies is still poorly constrained, yet it has strong implications on our understanding of galaxy formation \citep[see review by][]{Greene20}. Previous searches remained inconclusive \citep[e.g.,][]{Boyce+2017}. Due to the complex kinematics of the LMC, different tracers result in different estimates of the location of the galaxy centre, some differing by more than 1~deg \citep[see Fig.~9 in][]{Niederhofer+2022}. In addition, many dwarf galaxies host off-axis, or ``wandering'' black holes \cite[e.g.,][]{Reines+2020}, further complicating the search. The IFU of \wst\ will provide a unique possibility to sample the entire central region of the LMC and to uncover the kinematics of the LMC to unprecedented precision.

\clearpage

\subsection{The population of massive stars in metal-poor galaxies}
\label{subsec:respop-massivestars}

\noindent

Galaxies are chemically and dynamically shaped by massive stars (M$>8\,$M$_\odot$).  As principal sources of heavy elements and UV radiation, the most massive stars - despite their brief lives - play a fundamental role in the composition and ionization of the Universe \citep{Langer2012}. Their mighty ends, as supernova explosion (SN), make their impact in the interstellar medium (ISM) more acute \citep{Kennicutt1998}, and label massive stars as the most plausible progenitors of long $\gamma$-ray bursts (GRBS, \citet{Woosley-Bloom2006}) and gravitational waves (GW) \citep{Marchant+2016,Abbott+2017}.\\

However, the formation and evolution of massive stars are far from being understood. Stellar evolution is mainly controlled by the initial mass of the stars, but other factors shape their evolutionary paths. Metallicity, rotational velocity, strong stellar winds, binary interaction, magnetic fields, and mergers affect the evolutionary channels and the lifetime of a star \citep{Maeder+2000,Langer2012,Kudritzki+2000,deMink+2013,Schneider+2016}. 
The role of these parameters is both more important and uncertain for the most massive stars $>100\,$M$_\odot$ \citep{Vink+2015,Crowther2019}. In a Universe of ever-increasing chemical complexity, understanding how the physics of massive stars depends on metallicity (Z), particularly in metal-poor environments (low-Z) is crucial to assess their impact on their host galaxies along cosmic history. Massive stars in the SMC (1/5 Z$_\odot$) have so far set the standard for the low-Z regime. However, their metallicity is only representative of the relatively late Universe: the lessons learned from SMC massive stars do not apply to important epochs such as the peak of the SFR of the Universe (z$\sim$2, \citet{Madau+2014} and references therein) or earlier times. We need to reach further.\\ 

Models of massive stars at very low-Z fail to provide answers to outstanding challenges because of the paucity of observational constraints.
At very low-Z, e.g., chemically homogeneous evolution is possible, which could quadruple the production of ionizing photons within galaxies and enable the formation of massive double black holes and gravitational waves. Models cannot connect the evolution of massive stars with GRBs and SLSNe, preferentially found in low-Z galaxies, hampering the interpretation of future LSST data. A fundamental unknown is whether more  massive stars can form in metal-poor environments, a crucial hypothesis to confirm that Pop~3 stars were massive, and to quantify their contribution to re-ionization.\\

All these unknowns propagate to the simulations of unresolved populations and impact our interpretation of the high redshift Universe, including the frontier observations that JWST is delivering today.
We need large systematic surveys to unveil the evolution of the most massive
stars and the role of metallicity. Systematic studies will provide homogeneous results and unbiased empirical anchors for the theory of stellar evolution as they have done for the Milky Way and 30 Dor, e.g., \citet{Evans+2011}; \citet{Castro+2014};
\citet{Holgado+2022}.\\

State-of-the-art IFS have enabled the deblending of crowded stellar fields \citep{Kamann+2013,Roth+2018}, e.g. dense star clusters \citep{Castro+2021}. However, small fields-of-view (e.g., 1 arcmin$^{2}$ of MUSE) need to be superseded by the next generation of multi-object spectrographs.
\clearpage

The \wst\ allows the revolutionary concept to systematically study and resolve stellar populations in low-Z galaxies, breaking the metallicity and distance frontier of the Magellanic Clouds for the first time in a systematic way. The proposed high multiplex (20,000) multi-object spectrograph combined with a giant panoramic IFS presents a unique instrument for stellar physics. \wst’s proposed spectral resolution and wavelength coverage are better suited to unveil massive stars at $\ge$1 Mpc than current facilities. This is a revolutionary concept that will allow us to study the massive stellar content of galaxies at the outer reaches of the Local Group and beyond in low-Z environments.\\

\begin{figure}[th!]
  \centering
  \includegraphics[width=0.9\textwidth]{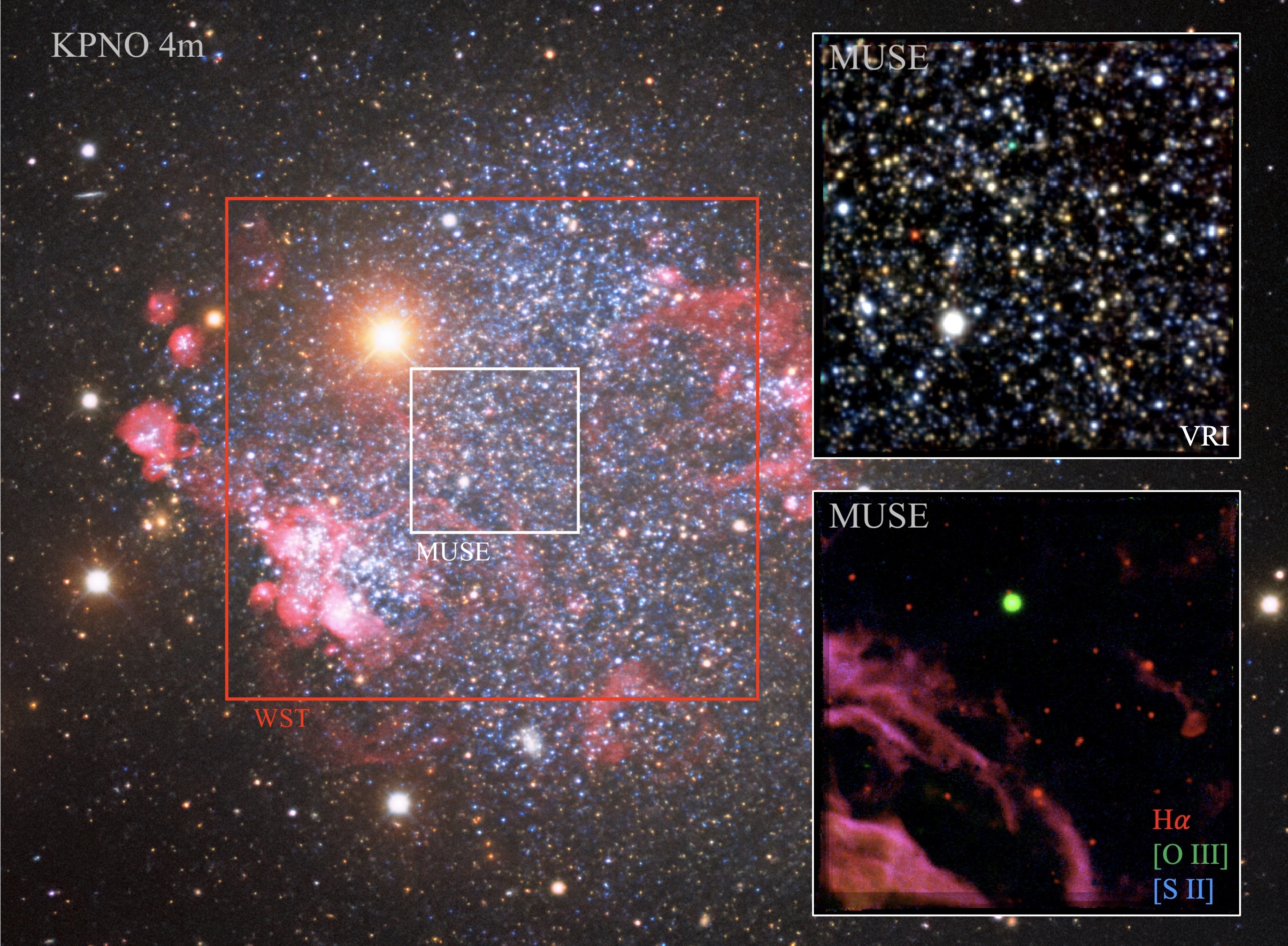}\\
  \vspace{3mm}
  \includegraphics[width=0.85\textwidth]{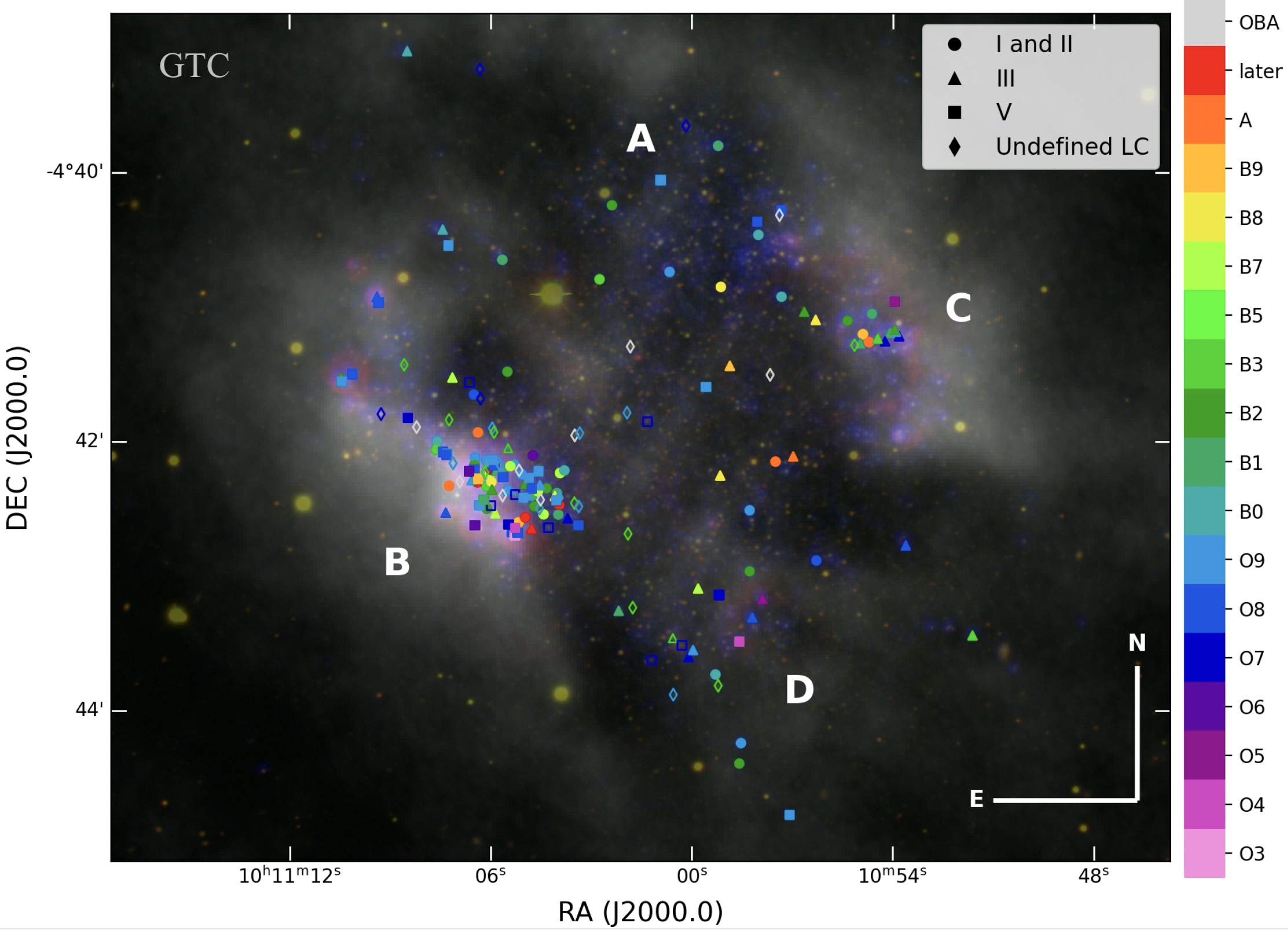}
  \caption{\small Nearby dwarf irregular galaxy Sextans A (d=1.32 Mpc), showcase for \wst\ transformative science in metal-poor dwarf galaxies. \textbf{Top:} Direct image from KPNO Mayall Telescope (NOIRlab).   Inserts: reconstructed images with $\sim$500 stars and gas from a single 4~h MUSE data cube. Credit: ESO archival data, M. Roth.  \textbf{Bottom:} 150 OB stars from 44.7~h MOS and longslit observations with the 10\,m GTC \citep{Lorenzo+2022}.} 
 \label{fig:Sextans-A}
\end{figure}

A prototypical galaxy to be targeted by \wst\ is Sextans A (Fig.~\ref{fig:Sextans-A}). Bringing the 10m telescope GTC to its limits, \citet{Lorenzo+2022} recently presented a catalog of 150 OB-stars in this galaxy with 1/10Z$_\odot$, the first extensive list of massive stars significantly more metal-poor than the SMC. Sextans A is an ideal repository to study the evolution of low-Z massive stars, however, i) the census of massive stars is still incomplete, and ii) extensive, detailed spectroscopic analyses are unfeasible with present-day instrumentation.\\

The inserts in the top panel of Fig.~\ref{fig:Sextans-A} illustrate the capability of integral field spectroscopy (MUSE datacube, Program ID: 108.221U, PI: Tramper, exposure time 4~h). The reconstructed VRI image reveals that all of the stars visible in the direct image are recovered.  About 500 stars yield spectra for spectral type classification, determination of stellar parameters and abundance determination for a subset of stars.
The emission line insert shows H\,II shells (pink), emission line stars (red), and a bright planetary nebula (green).  The \wst\ IFS will provide a field-of-view 9 times as large, allowing to sample the entire galaxy and the measurement of $\sim10.000$ stellar spectra as well as faint gaseous nebulae in merely 4 exposures (total exposure time: 
24~hours). The \wst\ MOS will allow to simultaneously record thousands of halo star spectra at medium or high spectral resolution. For comparison with the current state-of-the-art, a total of 44.7~h of MOS and long-slit spectroscopy exposure time at the 10\,m GTC between 2014 and 2021 has permitted the classification of 150 OB stars \citep{Lorenzo+2022}.\\
 
Massive O stars in Sextans A have apparent magnitudes from V=18 to 22. \wst\ can reach the ZAMS of O9.5 V stars with S/N$\sim60$ at R=5.000 in about 6 hours. The same exposure will also include bright supergiants of different types, returning S/N of 240 for the brightest members. Since the galaxy can be fully covered by 4 IFS pointings, only 24 hours of observation would give us information about stellar parameters and binary fractions (together with LSST light curves) of a complete population of 1/10Z$_\odot$ massive stars. This is a major step forward in our knowledge of massive stars in low-Z environments, breaking the barrier represented by the SMC. Providing proxies for the high redshift Universe requires surveying the massive star population of nearby galaxies with higher star-formation rates and/or low-Z. There are numerous other nearby, metal-poor galaxies with resolved populations of massive stars reaching down to 1/50 Z$_\odot$, e.g., NGC\,6822, IC\,1613, WLM, NGC\,3109, Sextans~B, Leo\,A, etc. Spectroscopy of the massive stars in these galaxies would offer an unprecedented legacy to study massive star evolution, feedback, and synthesis of populations.

\clearpage

\subsection{The baryonic and dark matter halo properties of Local Group dwarf galaxies}
\label{subsec:respop-LGdwarfs}

Dwarf galaxies, i.e. systems with stellar mass at least one order of magnitude smaller than that of the Milky Way (MW), are the most common type of galaxies found in the Universe today, therefore learning about their formation and evolution implies learning about the most common mode of galaxy formation. They are also widely considered some of the best systems from which we can gather constraints on the nature of dark matter (DM) and on the effect that baryonic processes can have on modifying the inner structure of DM haloes. 
The Local Group (LG) hosts a sizable population of these galaxies, about 35 of which with luminosities between $10^5$ and $10^9$ L${\odot}$ and distances $<2.5$~Mpc would be visible from Paranal.\\ 

The determination of radial velocities, metallicities, and chemical abundances of large samples of individual low-mass stars in these systems allow us to reconstruct the evolution of their stellar component, probing from the earliest phases of their formation, all the way to z=0, constrain the properties of their DM halo, and perform studies of their binary star population in an unprecedented, efficient way. 

\begin{figure}[h!]
  \centering
  \includegraphics[width=\textwidth]{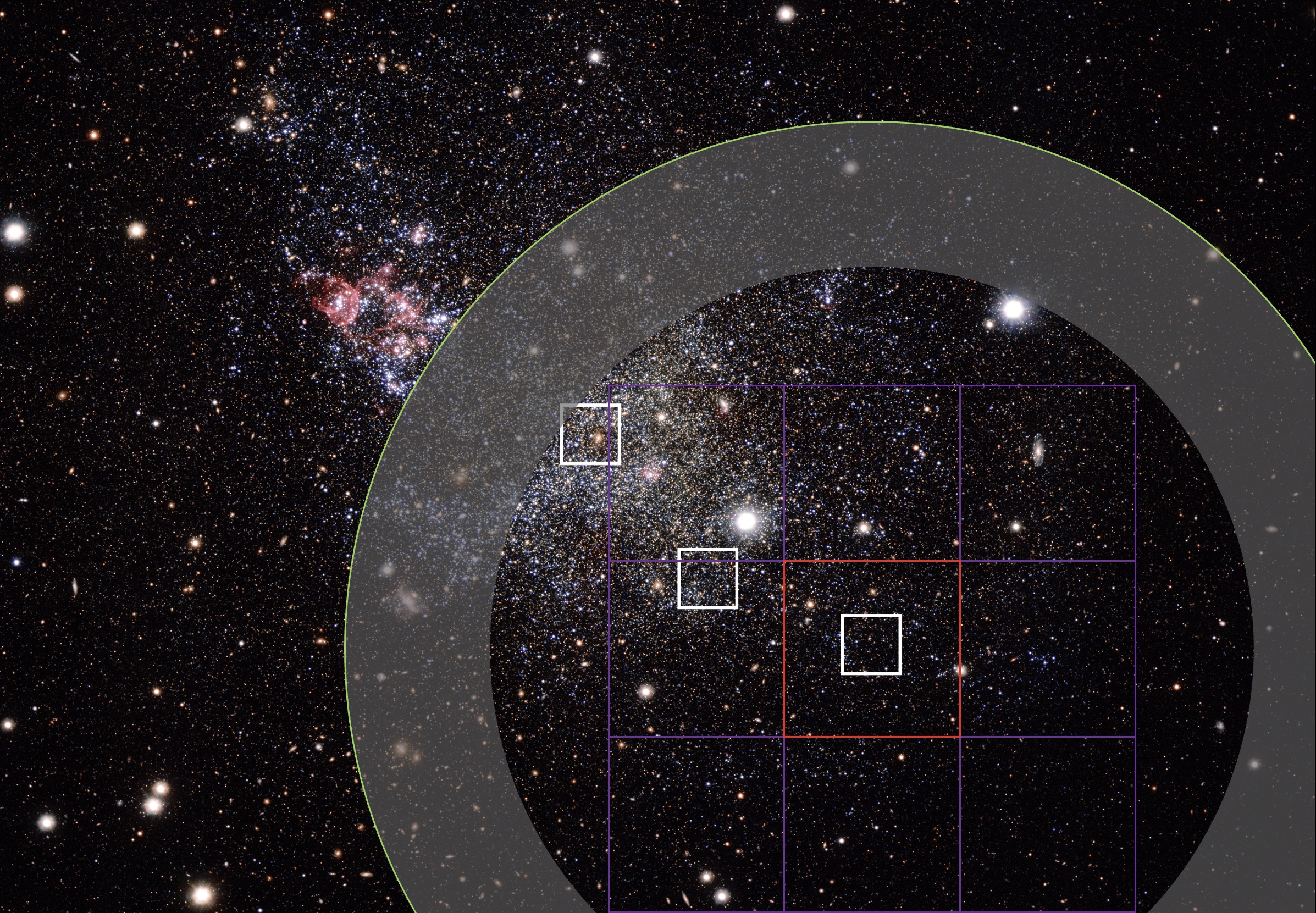}
  \caption{\small WFI color image of IC1613, roughly covering one effective radius. The red square is one \wst\ IFS pointing, to be compared to the white squares, indicating the current MUSE coverage available for this galaxy \citep{Taibi+2024}. The main body of the galaxy could be fully covered with 12-15 \wst\ IFU pointings. The grey ring is the \wst\ „zone of avoidance“. The external parts would be simultaneously covered by the MOS. Credit: M.~Roth.} 
 \label{fig:IC1613}
\end{figure}

\clearpage

\noindent {\bf The inner density profile of DM haloes:}
As discussed in detail in \citet{Bullock+2017}, it is well known that the rotation curves of DM-dominated galaxies prefer DM haloes that are less dense and with an inner density profile slope less steep than those predicted for pure DM-only $\Lambda$CDM halos (cusp/core problem). Also, the internal kinematics of local dwarf galaxies indicate a dearth of DM-haloes with virial masses of the order of $10^{10}$~M$_\odot$ which however should have been massive enough to form a stellar component ("too-big-to-fail" problem). Modern hydrodynamical simulations from several groups find that, in certain regimes of stellar mass over DM-halo mass, stellar feedback due to supernovae explosions can alter the inner density profile of DM halos, e.g., \citet{Governato+2012}, \citet{DiCintio+2014}, \citet{Fitts+2017}, providing an appealing solution to these problems.  The determination of the DM halo density profile of Local Group dwarf galaxies covering a range of masses ($10^5\ldots10^8$~M$_\odot$) offers the opportunity to test these models, as these stellar masses encompass both the regime where stellar feedback should be inefficient as well as that in which it should be at its maximum efficiency of transformation. The availability of LG dwarf galaxies of similar masses but exhibiting short vs very extended star formation histories also open a window to test models that claim that such alterations of the inner DM halo density profile can also occur at small stellar masses, as long as the star formation history is extended enough \citep{Read+2016}. Probing out to the stellar masses in which the cusp/core transformation is expected to be most efficient as well as faint systems with extended star formation histories, requires targeting large samples of individual stars in LG dwarf galaxies found outside of the MW virial radius, with a dense sampling of the inner regions and precise radial velocities.    

\vspace{3mm}
\noindent {\bf Hierarchical accretion down to the smallest masses:}  One of the predictions of the currently favored cosmological framework is that galaxies partially form through the accretion and merging of smaller galaxies. This mechanism should occur at all galactic scales, even at those of dwarf galaxies; for example, it is expected that about 15-20\% of dwarf galaxies found outside of the host virial radius (e.g. $> 300$~kpc from the MW) might have experienced a major merger since z=1 \citep{Read+2016}.  Unambiguous signs of this process are routinely found around large galaxies, like our own Milky Way and M31, where deep imaging surveys clearly show dwarf galaxies being tidally disrupted near massive hosts. The galaxy holding the record for being the smallest system in which direct signatures of this process have been detected is DDO68 \citep{Annibali+2016}, of mass similar to the LMC. 

At lower stellar masses, it has not been possible yet to detect such direct signs of this hierarchical formation process. However, a growing number of LG dwarf galaxies, including several MW satellites, shows peculiarities/anomalies in the spatial distribution and/or kinematic properties of their stellar component, e.g., Ursa Minor: \citet{Bellazini+2002}, Fornax: \citet{Coleman+2005}, \citet{Battaglia+2006}, AndII: \citet{Amorisco+2014}; Phoenix: \citet{Kacharov+2017}; Sextans: \citet{Cicuendez+2018}, due to accretion/mergers, as shown with cosmological simulations by \citet{Cardona-Barrero+2021} for Phoenix.

The census of these features in LG dwarf galaxies is far from complete, in particular for what concerns those systems found at $> 300$~kpc from the MW, for which spectroscopic samples of individual stars are of relatively low statistics (typically of the order of $10\ldots100$ stars).  As for the nearby MW satellites, which have angular extents of several degrees on the sky, one of the surprises that Gaia DR2/eDR3 has brought about is the finding of potential member stars out to tens of half-light radii, typically in elongated configurations (TucanaII: \citet{Chiti+2021}; Fornax: \citet{Yang+2022}; Hercules: \citet{Longeard+2023}; Sculptor: \citet{Sestito+2023}; various systems: \citet{Jensen+2024}). Interestingly, in several cases, the Gaia-based orbital histories, e.g., \citet{Battaglia+2022}, \citet{Pace+2022}, would not appear to support the hypothesis of tidal disturbance from the MW as the origin of these features; alternative intriguing possibilities are that these are the remains of past accretion events,  e.g., \citet{Genina+2022}, \citet{Goater+2024}, and there are indications that detailed chemical abundances could be revealing of their accreted origin, e.g., \citet{Waller+2023}. These extended features are 1-2 degrees away from the center of these galaxies. With its wide MOS field-of-view, as a simultaneous complement to the IFS, \wst\ could provide spectroscopic follow-up of potentially all the target members in just a couple of pointings and with modest exposure times down to the depth of the Gaia magnitude limit at high spectral resolution, or spectroscopic follow-up of potential members that will be discovered with \lsst.

\vspace{3mm}
\noindent {\bf Chemical libraries:} Chemical properties of the stellar populations of dwarf galaxies carry important information on their evolution and assembly. It is expected that the signature in neutron capture elements is a tracer of the formation and merger history of dwarf galaxies \citep{Mashonkina+2017}, but we are still lacking a clear view of these abundances in any of the LG dwarfs. Besides, it will be impossible to conclude on the processes driving their mass assembly, particularly in their early days \citep{Sanati+2023}. While many LG dwarf galaxies already benefit from detailed determinations of their star formation histories from deep photometry, e.g., \citep{Gallart+2015}, our knowledge of the chemical properties of the stellar populations of these systems is poor, due to their small number of stars. This can be efficiently tackled by \wst, with detailed chemical tagging at high resolution and by the derivation of metallicities at lower spectroscopic resolution for the fainter stars, by probing far out for the most massive of LG dwarf galaxies, as well as deep for the smaller ones and out to greater distances. 

Determination of detailed elemental abundances from high-resolution spectroscopy for hundreds of stars in individual LG dwarfs enables building a ”chemical abundances library” for systems that cover a range of stellar masses and of known star formation histories. Abundances of elements that have different nucleosynthetic sites (core-collapse supernova, AGB winds, neutron star mergers) can provide a window into how the star formation history and mass of a particular galaxy influenced its efficiency of chemical enrichment. These ”chemical libraries” would represent templates with which to interpret abundance ratios from integrated light observations of accreted stellar halos around thousands of galaxies at low redshift - providing a way to chemically dissect those mixed halos and recover the likely mass distribution of satellites that have merged. The \fourmost\ survey 4DWARFS \citep{Skuladottir+2023}
will collect high-resolution spectra for large samples of stars in all southern MW dwarfs and streams accessible to a 4m-telescope; but the chemical properties of LG dwarfs galaxies beyond the MW virial radius will remain unexplored.

\vspace{3mm}
\noindent {\bf Binary and variable stars:} The determination of the properties of binary and variable stars in LG dwarf galaxies offers the opportunity to understand their formation in metal-poor, low-density environments (see also Section~\ref{subsec:respop-clusters}). Assuming first light around 2040, \wst\ will provide future observations for monitoring of large samples of individual RGB stars in the nearby Milky Way satellites with reference to FLAMES/GIRAFFE observations in 2000-2010 and with \fourmost\ in 2025-2030, assembling a baseline of 40+ years. This will be an enormous leap forward for studies of the binary population of low-mass stars, which will be able to break degeneracies between the fraction of binary stars and the properties of the period distribution, and detect possible deviations from the one of the solar neighborhood.

\subsection{Extragalactic Archaeology: data mining resolved stellar populations in nearby galaxies to test cosmological simulations}
\label{subsec:respop-ExtraArch}

\noindent
Over the past decade, cosmological simulation within the $\Lambda$CDM paradigm have advanced from a description of the large-scale structure of the Universe and its evolution merely based on Dark Matter to now include baryons, the cycle of matter, and chemical evolution down the scale of individual galaxies.  However, because of practical limitations of computing power, there is a dichotomy between simulations covering large scales, e.g. Illustris \citep{Vogelsberger+2014}, or Eagle \citep{Schaye+2015} on the one hand, and the smallest scales, e.g. LYRA \citep{Gutcke+2021}, or EDGE \citep{Agertz+2020} on the other hand. Table~\ref{tab:Simulations} presents an overview of simulations that have become available over the past 10 years, covering volumes of up to 10$^6$ Mpc$^3$ with a resolution of M$_{gas} \approx 3\times10^6$M$_\odot$ on the larger scales, down to 10$^{-4}$ Mpc$^{3}$  at a resolution of M$_{gas} \approx 4$~M$_\odot$ on the smallest scales. Crucially, the small scales are necessary to adequately resolve the structure of the ISM and small-scale physics of feedback processes, such as supernova explosions, and a proper modeling of cooling. Future improvements of these descriptions will include the importance of stellar winds from massive stars, their ionizing H and He continua, and the interplay with the ISM \citep{Gutcke2023}. This type of simulation is, for the moment, limited to the smallest dwarf galaxies (up to $\sim 10^9$ $\rm M_{\odot}$ in halo mass) with future extension to larger objects anticipated. While cosmological simulations typically follow the stellar component at the same resolution as the gas, Auriga Superstars \citep{Grand+2023} employs a novel approach to boost the resolution of star particles in massive spiral discs by almost 2 orders of magnitude (to $\sim 800$ $\rm M_{\odot}$), which in the coming years will be increased further. This will enable predictions for stellar chemo-kinematics at the star-by-star level
for these objects.

\begin{table}[h!]
\centering
{\footnotesize
\begin{tabular}{ | l |  l | c | c | } 
\hline
Name        	& Reference 		            &    log~(V/Mpc$^3$)     & log~(M$_{gas}$/M$_\odot$)		\\
\hline\hline
Illustris~~~~~  & \citet{Vogelsberger+2014}~~~~~ &      6             &        6.2           \\   
Eagle          & \citet{Schaye+2015}          &         6             &        6.3           \\ 
IllustrisTNG50  & \citet{Pillepich+2019}~~~~~ &      5             &        5.0           \\ 
NIHAO          & \citet{Wang+2015}            &         1             &        5.5           \\ 
Romulus        & \citet{Tremmel+2017}         &         1              &        5.4           \\ 
Apostle        & \citet{Fattahi+2016}         &          2             &        4.0           \\                   
Auriga         & \citet{Grand+2017}            &        $\sim 1$       &        3.8           \\
FIRE           & \citet{Chan+2015}            &        1.2            &        3.8           \\    
VINTERGATAN    & \citet{Agertz+2021}          &        0.5            &        3.8           \\    
ERIS           & \citet{Guedes+2011}          &          0            &        4.2           \\   
MARVEL         & \citet{Munshi+2021}          &          1            &        3.2           \\ 
Halo6          & \citet{Jeon+2017}           &          -2.2         &        2.7           \\  
EDGE          & \citet{Agertz+2020}          &           -1.0         &        1.8           \\
LYRA          & \citet{Gutcke+2021}          &           -4.0         &        0.6           \\
\hline
\end{tabular}
  \caption{\small Cosmological Simulations} 
  \label{tab:Simulations}
}
\end{table}

Observational tests for models describing the chemical and kinematic evolution of
galaxies have initially relied on global spectroscopy e.g., from SDSS which, however, is limited by the selective spatial coverage of the aperture and the associated bias. Significant progress over this limitation has been achieved with integral field spectroscopy (IFS), and surveys like CALIFA \citep{Sanchez+2012}, SAMI \citep{Bryant+2015}, or MaNGA \citep{sdss-manga}, but still limited to physical areas on kpc scales, owing to the angular resolution of the IFS spaxels, and the distance of the galaxies.

\noindent
The development of crowded field integral field spectroscopy \citep{Kamann+2013} has provided the tools to resolve galaxies out to distances of ~2 Mpc and beyond at pc~scales into individual stars,  H\,II regions, diffuse ionized gas (DIG), 
supernova remnants, planetary nebulae, etc., at unprecedented sensitivity.
A first attempt with MUSE observations of the nearby galaxy NGC\,300 (d=1.88 Mpc) has demonstrated this capability, e.g., pilot study by \citet{Roth+2018}, study on BA-type supergiant stars by \citet{Gonzalez-Tora+2022}, faint H\,II regions by \citet{Micheva+2022}, and PNLF by \citet{Soemitro+2023}.

\begin{figure}[h!]
  \centering
  \includegraphics[width=0.56\textwidth]{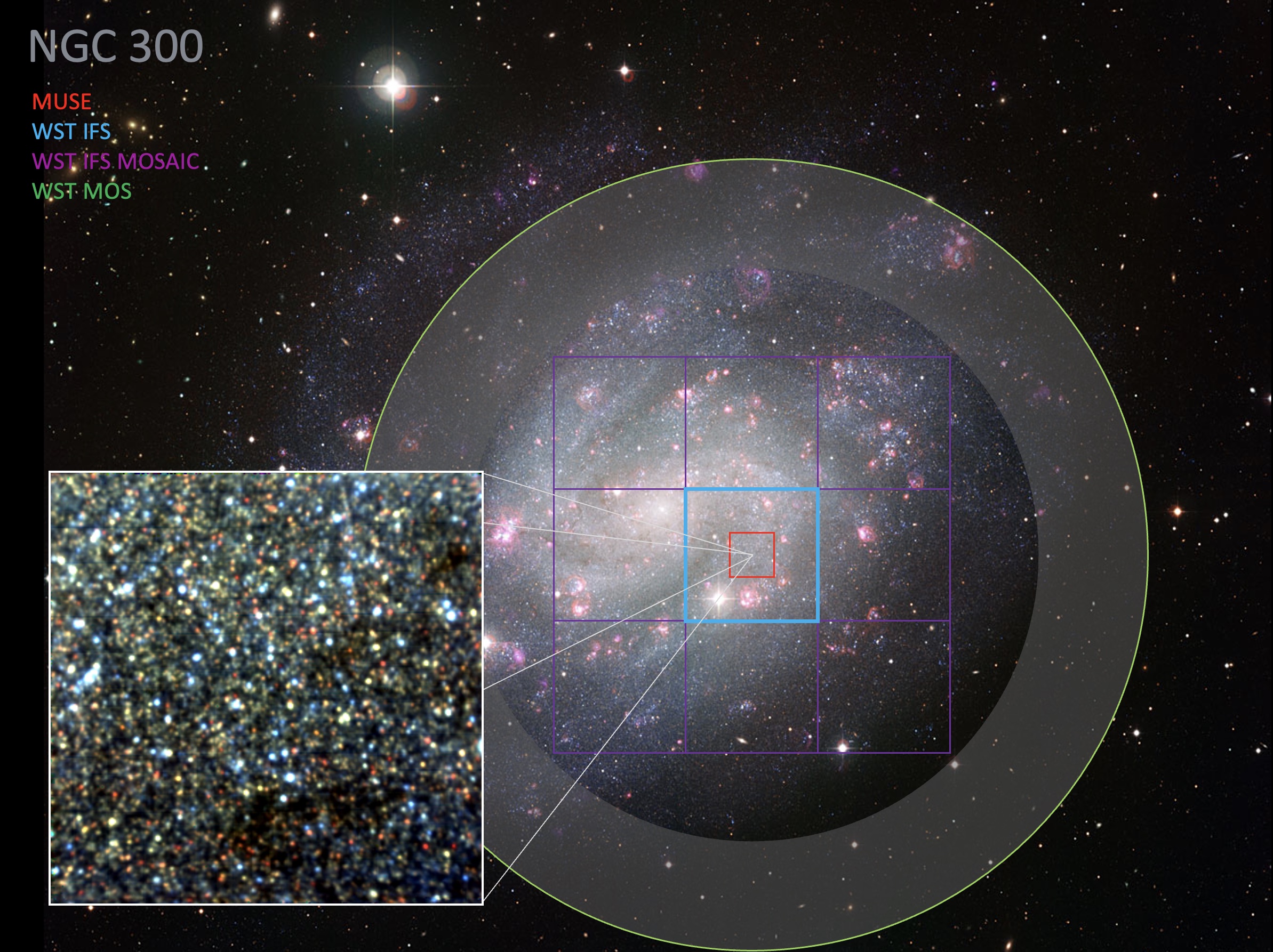}
  \includegraphics[width=0.43\textwidth]{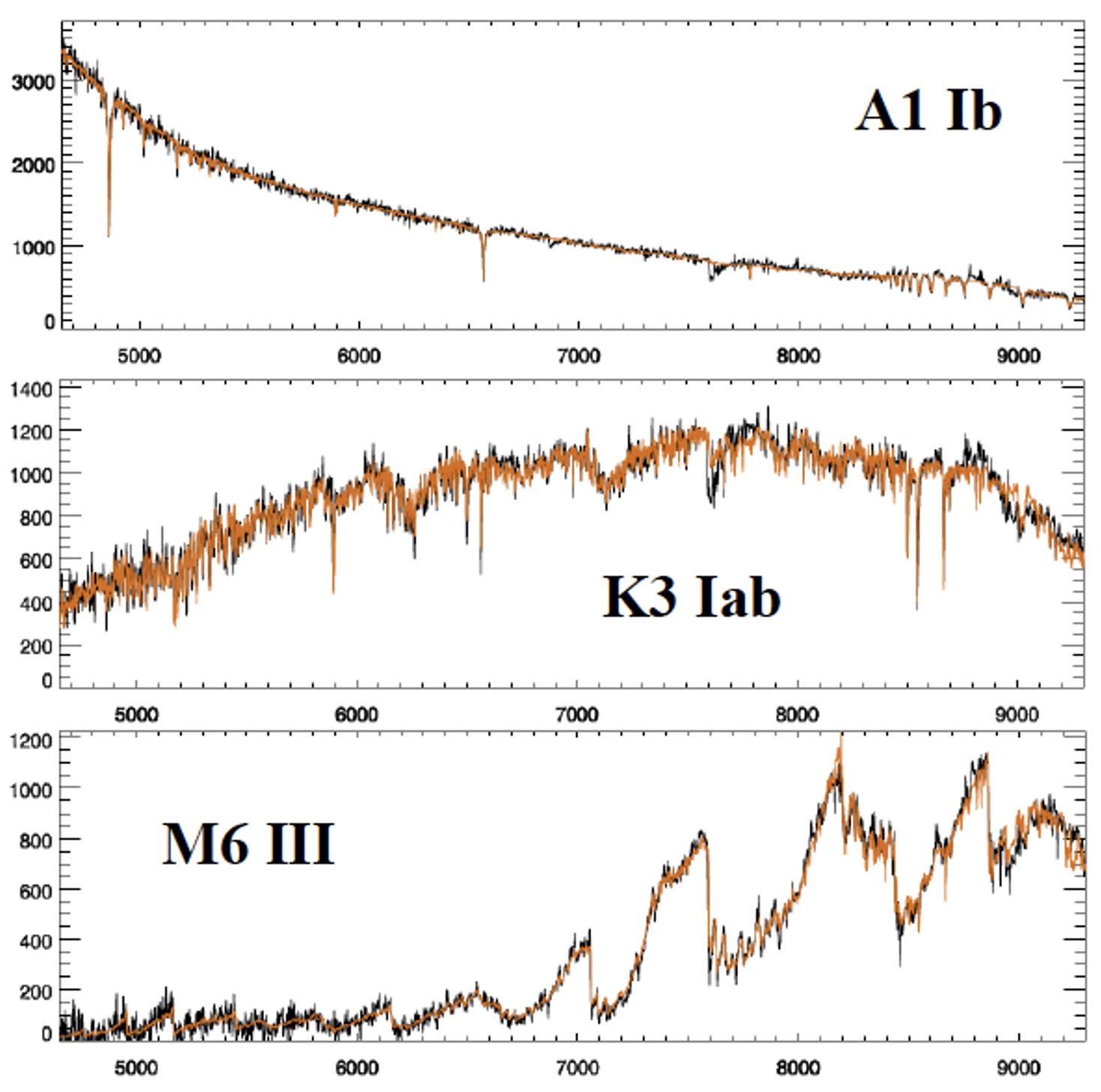}
  \caption{\small NGC\,300 (d=1.88 Mpc), showcase for \wst\ transformative science on resolved stellar populations in Local Volume galaxies. \textbf{Left:} Direct WFI image(ESO).   \textbf{Inserts:} reconstructed image with 517 high S/N stars from a 1.5~h MUSE exposure.  \textbf{Right:} Example for high-quality spectra from this exposure. Black: observed spectra, orange: library template spectra \citep{Roth+2018}.} 
 \label{fig:NGC300}
\end{figure}

\noindent
A review on crowded field IFS by \citet{Roth+2019} provides an overview of the technique and further studies, e.g., R136 in 30Dor \citep{Castro+2018}, Leo P \citep{Evans+2019}, NGC\,300 \citep{McLeod+2020}, and others. The study of massive stars in nearby galaxies in particular was identified as a highlight science case for BlueMUSE \citep{Richard+2019}. However, a major limitation of 
MUSE/BlueMUSE is the relatively small field of view in comparison to the angular size of nearby galaxies.  For NGC\,300, the proposed \wst\ project would enable IFS coverage of the entire galaxy, including the faint outskirts, at moderate spectral resolution in 30 pointings with the unique additional capability to obtain medium or high-resolution spectra for pre-selected objects 
with the MOS. This capability would enable unprecedented studies of the assembly history and secular evolution of nearby galaxies.\\

\noindent
{\bf Signatures of secular evolution and the nature of galactic spiral arms:} One of the most prominent processes in the secular evolution of spiral galaxies is the migration of stars away from their birth radii as a result of dynamical interaction with a bar and spiral arms \citep{FreemanBlandHawthorn2002,Sellwood+2002}. Not only may radial migration shape several galactic chemo-dynamic trends, such as the flat age-metallicity relation observed in the Milky Way, e.g., \citet{Casagrande+2011}; the chemically distinct Galactic discs, e.g., \citet{Schoenrich+2009}; and U-shaped radial age/metallicity/colour profiles of galaxies, e.g., \citet{Roskar+2008}, but its efficiency and observational signatures are linked to the physical nature of spiral arms – a still debated and unsolved problem.

The increase in detail and sophistication of numerical simulations (see Table~\ref{tab:Simulations}) have brought forth novel observable predictions for radial migration induced by transient spiral arms \citep[see][]{Grand+2016}. Some of the main signatures are peculiar streaming motions of approximately 20 km/s along spiral arms: the velocity field is locally tangentially slow and outward moving on the trailing edges of spiral arms, and vice versa for the leading edges. In addition, stars and gas on the trailing edges of spiral arms are predicted to be more metal-rich compared to stars on the leading edges at the same radius for galaxies that possess a negative radial metallicity gradient: thus, the amplitude of this variation informs the radial metallicity profile as well as the strength of migration. 

So far, these predictions have not been confronted with many observations: for example, the study of \citet{Sanchez-Menguino+2016} found evidence for peculiar gas motions in the spiral galaxy NGC 6754. As a major step forward,  a key advance of \wst\ will be the simultaneous use of IFS and MOS operation to provide detailed chemo-kinematic maps on the basis of individual stars with a large field of view covering entire discs of galaxies up to distances of $\sim8$~Mpc. This will provide much-needed observational tests for predicted signatures associated with galactic structural components and provide insight into the nature of spiral arms.\\ 

\noindent
{\bf Bulgeless galaxies:} Classical bulges are a common galactic component and a natural prediction of the hierarchical merging scenario given by $\Lambda$CDM. However, about half of nearby Milky Way-mass galaxies do not have a classical bulge \citep[e.g.][]{Gadotti2009}. Cosmological simulations based on $\Lambda$CDM have difficulty in forming bulgeless galaxies \citep{Peebles2020}; whether this discrepancy would require a fundamental change in our standard cosmological model or improved modeling of sub-grid physics is unclear. It is therefore critical to understand how bulgeless galaxies form compared to those that contain classical bulges. The ability of \wst\ to resolve stellar populations in the crowded regions of galactic centres and up to and beyond the galaxy outskirts simultaneously will enable detailed characterisation of the star formation histories of different components as well as the detection of accretion events (particularly of smaller mergers given their larger predicted offset from the age-metallicity relation of in-situ stars). This will provide constraints on the formation pathways of classical bulges and bulgeless galaxies that will be necessary to understand the existence of the latter.\\ 

\noindent
{\bf Haloes of galaxies:}
Globular clusters (GC) are powerful tools to study galaxy formation, but as yet not studied well in the haloes of nearby galaxies \citep{Usher+2024}. Simultaneous MOS observations in parallel to IFS coverage of the galaxies proper will provide invaluable information on GC metallicities and kinematics in addition to the main program. The same applies to halo planetary nebulae (PN) that were found to be excellent tracers of the metal-poor haloes of nearby  galaxies \citep{Hartke+2023}.\\ 

\noindent
{\bf Serendipitous discovery potential:} It is well known that novel instrumentation exploring new parameter space is almost guaranteed to lead to discoveries that were not anticipated at the time of planning. In particular, the \wst\ IFS as an unbiased, untargeted survey tool can be expected to yield such discoveries. Experience with MUSE as a precursor has shown that crowded field IFS in nearby galaxies is very sensitive to make such discoveries, e.g., faint supernova remnants \citep{Roth+2018,Evans+2019,Long+2022}, low surface brightness H\,II regions \citep{Micheva+2022,Lugo-Arlanda+2024}, emission line stars \citep{Roth+2018,Vaz+2023,Taibi+2024}, carbon stars \citep{Roth+2018,Taibi+2024}, and more.

\subsection{The role of (almost) ultra-faint dwarfs in galaxy evolution}
\label{subsec:respop-UFD}

\noindent
The very faintest galaxies near us are fascinating laboratories to study both dark matter and early star formation and feedback. Since 2005 we have seen a rapid growth in the number of know faint galaxies (see e.g. \citet{Simon2019}) and there are now more than 60 galaxies
known that are considered very faint dwarfs. These dwarfs are all exceptionally dark matter
rich with the faintest dwarfs being the most dark matter dominated sources known, though
even so it is thought that their dark matter halos are very low mass and thus the baryonic
content is likely to be very sensitive to feedback effects.

Understanding these systems in detail therefore has the potential to both place constraints
on dark matter models, as well as understanding star formation in the very smallest dark
matter halos. This is the focus of this science case idea.\\

 \begin{figure}[h!]
  \centering
  \includegraphics[width=1.0\textwidth]{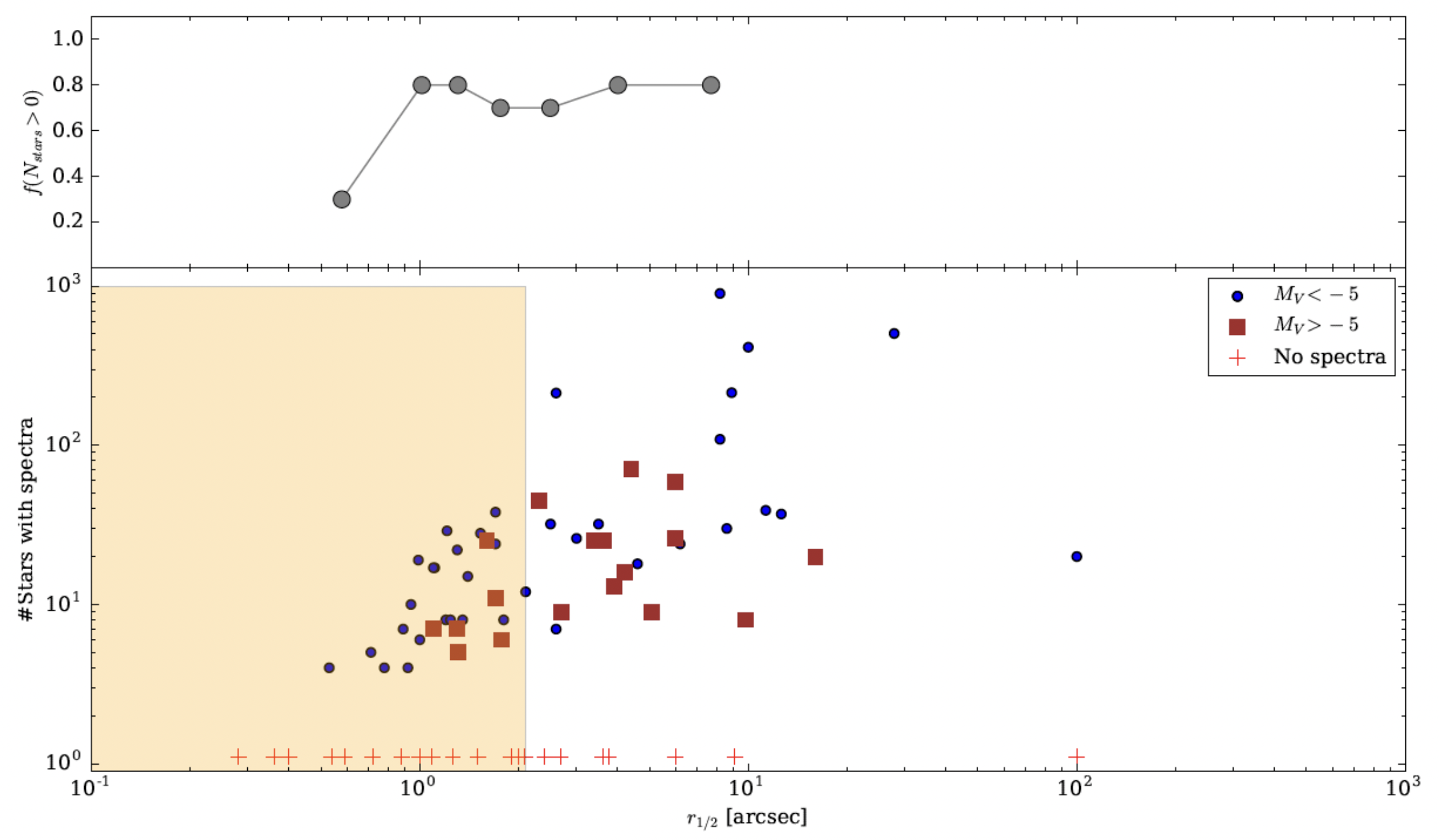}\\
  \caption{\small Top panel: The fraction of dwarf galaxies (M$_{V}$ $> -10$) with at least one spectroscopic observation as a function of halflight radius in arcseconds. The data are binned with 10 galaxies per bin. Bottom panel: The number of stars spectroscopically
confirmed to be members of the dwarf galaxies as a function of the half-light radius in arcminutes. The blue point shows
galaxies brighter than M$_{V}$ = -5 and the red squares those fainter than this. The objects with no spectroscopic observations in the
literature are shown as red plusses at their appropriate half-light radius. The shaded rectangle shows the region targeted by
this proposal. [Based on \citet{McConnachie2012}, figure not fully updated with latest data, so this is a lower limit].} 
 \label{fig:UDF1}
\end{figure}

\citet{Simon2019} argues that a magnitude of M$_{V}$=-7.7 separates the dwarf galaxies into faint
and ultra-faint, in that galaxies brighter than M$_{V}$=-7.7 appear to have some recent star
formation while fainter than that galaxies appear to be very old with no significant recent star
formation. Thus, by targeting galaxies spanning this range – fainter than, say, M$_{V}$=-10 we
would be able to survey a very wide range of star formation histories and understand what
conditions will stop star formation in a dark matter halo.\\

\noindent {\bf Main aims:} 
\begin{itemize}
\item Constrain the dark matter halo properties of faint dwarf galaxies.
\item Determine the star formation histories and metallicity distribution functions of faint
dwarf galaxies.
\item Quantify the fraction of stripped stars in faint dwarfs as a function of Galacto-centric
radius.
\end{itemize}

The state-of-the-art of constraining the dark matter profile in ultra-faint dwarfs is arguably the
MUSE-based study by \citet{Zoutendijk+2021a}, which also discusses stripping, but while this has a number of stars in the central regions, the coverage of stars in the outer regions is insufficient to strongly constrain the total dark matter
halo mass.\\
 
 \begin{figure}[h!]
  \centering
  \includegraphics[width=0.9\textwidth]{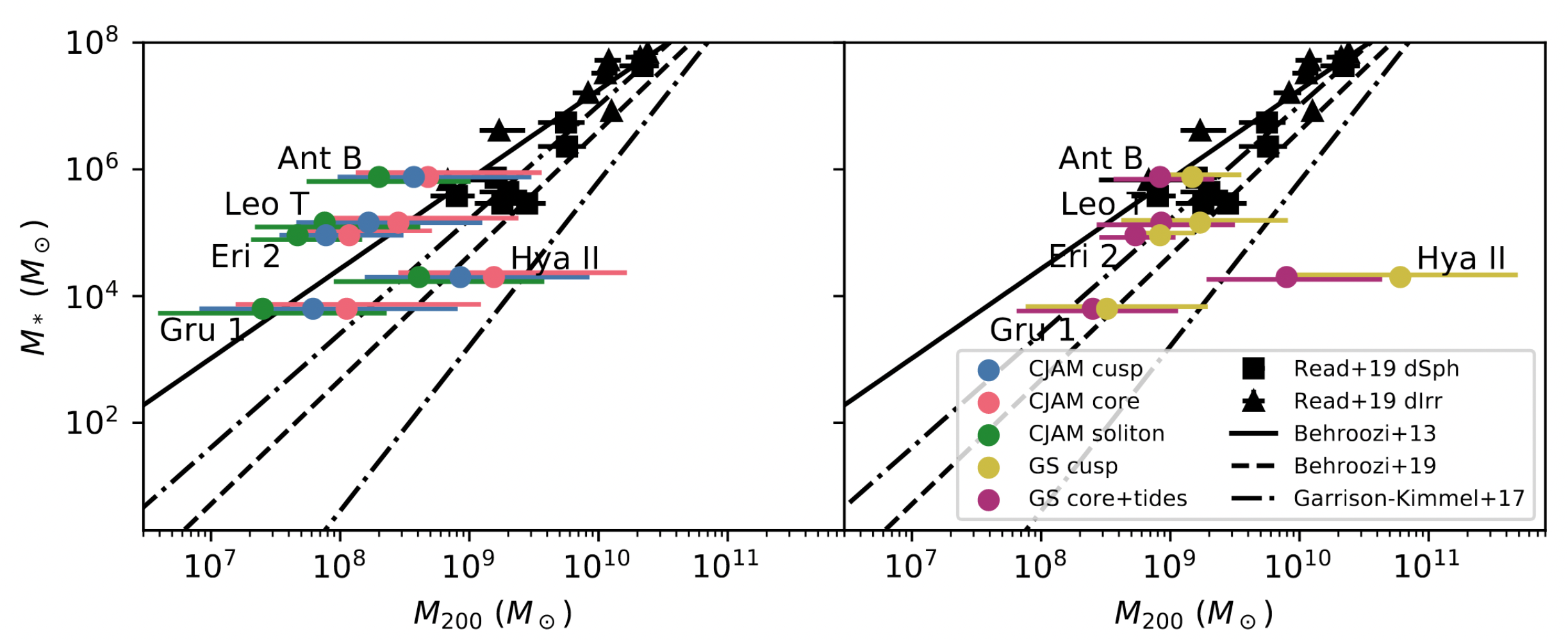}\\
  \caption{\small From \citet{Zoutendijk+2021b}.This shows a comparison of the galaxies’ stellar masses M and virial
masses M200, using two different methods (left and right), to expectations from models. The coloured symbols
are for ultra-faint dwarfs, while in black are brighter/classical dwarfs. With \wst\ we can populate this with up to an
order-of-magnitude more galaxies.} 
 \label{fig:UDF2}
\end{figure}
 
 \noindent
{\bf The procedure for doing this project would be three-fold:}
 \begin{itemize}
\item Use the central IFS to obtain a high density of stellar spectra in the densest part of
the dwarfs. This would use something like PampelMUSE (Kamann et al 2013) for
extraction of spectra using the high spatial resolution imaging which we will have
from Euclid by the time of \wst, and for many of the dwarfs known today from HST.
\item By using the multi-wavelength data, the fibres will target TBD numbers of stars
outside the location of the IFS to constrain the dark matter halo parameters and look
for stripped stars.
\item After memberships have been determined using the IFS and the MOS, we would go
back with the high resolution spectrograph on the brightest stars to determine
detailed abundances to constrain the enrichment histories.
\end{itemize}

\clearpage


\clearpage

\section{Extragalactic science}

\paragraph{Authors:}Jarle Brinchmann$^3$, Richard Ellis$^5$, Francesco Belfiore$^{14}$, Nicolas Bouché$^{13}$, Anne Verhamme$^{18}$, Sandro Tacchella$^{62,63}$, Stefano Zibetti$^{14}$, Darshan Kakkad$^{65}$, Tanya Urrutia$^{10}$, Mamta Pandey-Pommier$^{98}$, Mark Sargent$^{106}$, Jorryt Matthee$^{79}$, Themiya Nanayakkara$^{48}$, Filippo Mannucci$^{14}$, Floriane Leclercq$^{72}$, Anna Gallazzi$^{14}$, Clotilde Laigle$^{70}$, Igor Zinchenko$^{120}$, Oliver Mueller$^2$, Angela Adamo$^{15}$

\subsection{Introduction}

The upcoming decade will see an enormous investment in deep panoramic imaging.
Ambitious surveys driven primarily by cosmology will, starting in 2025, provide 0.2 arcsec image quality optical data over 15,000 deg$^2$ from \euclid\  and multi-band photometry to AB$\sim 27$ from \lsst. Associated low resolution spectroscopy from \desi, and shortly \pfs\, MOONS and \euclid, will characterise the demographics of the galaxy population (luminosities, masses, star formation rates), together with statistical properties of large-scale structure, out to relatively high redshift. In addition to massive catalogues of photometric redshifts and well-defined spectral energy distributions, there will also be spectroscopic redshifts for as many as 10$^8$ galaxies over $0<z<3$. \jwst\ and eventually ELT will complement this with higher quality spectra but over much smaller regions of sky. 

Now is the time to commence planning for \wst. By the late 2020s, deep multi-band imaging over very large areas will become available enabling efficient and reliable target selection with minimal contamination from interlopers and, as discussed earlier (Section \ref{Intro:radio}), radio surveys will be providing complementary, rich catalogues over much of the sky. \wst\ will exploit and complement this remarkable progress with a significant increase in the number and quality (in terms of spectral resolution and S/N) of spectra over unprecedented cosmic volumes with its multi-object spectrograph (MOS), while simultaneously sampling smaller scales with a high spatial density necessary for understanding physical process on galactic scales with its panoramic integral field spectrograph (IFS). This combination of high quality data on both large and small scales will represent a major advance in understanding how galaxies assemble over cosmic time. The spectroscopic resolution enabled by the large aperture and dedicated nature of \wst\ will extend our knowledge beyond merely charting redshifts from emission or absorption lines. Precision emission line ratios enabled by resolving doublets and measuring internal stellar velocity dispersions from high S/N absorption line profiles will yield crucial new physical information out to much higher redshifts than existing surveys or facilities will be able to achieve.

The IFS will also complement the MOS by providing spectroscopy of galaxies without the need for pre-imaging which will validate the selection function of the MOS, as well as provide physical information on small spatial scales where the MOS is inefficient due to fibre placement limitations --- within its FoV the IFS can be expected to get 50--100 times more redshifts than the MOS over a similar area. Most importantly, the IFS will provide spatially resolved data and hence internal kinematics for large samples of distant galaxies as well as gas-phase and stellar metallicity maps for brighter galaxies out to $z \sim 1$. This will be further strengthened by the unique combination of an IFS and a MOS working in concert. This will allow us to link the spatially resolved properties of galaxies to their location in the larger-scale structure in a way no current facility can accomplish.

\begin{figure}
    \centering
    \includegraphics[width=0.9\textwidth]{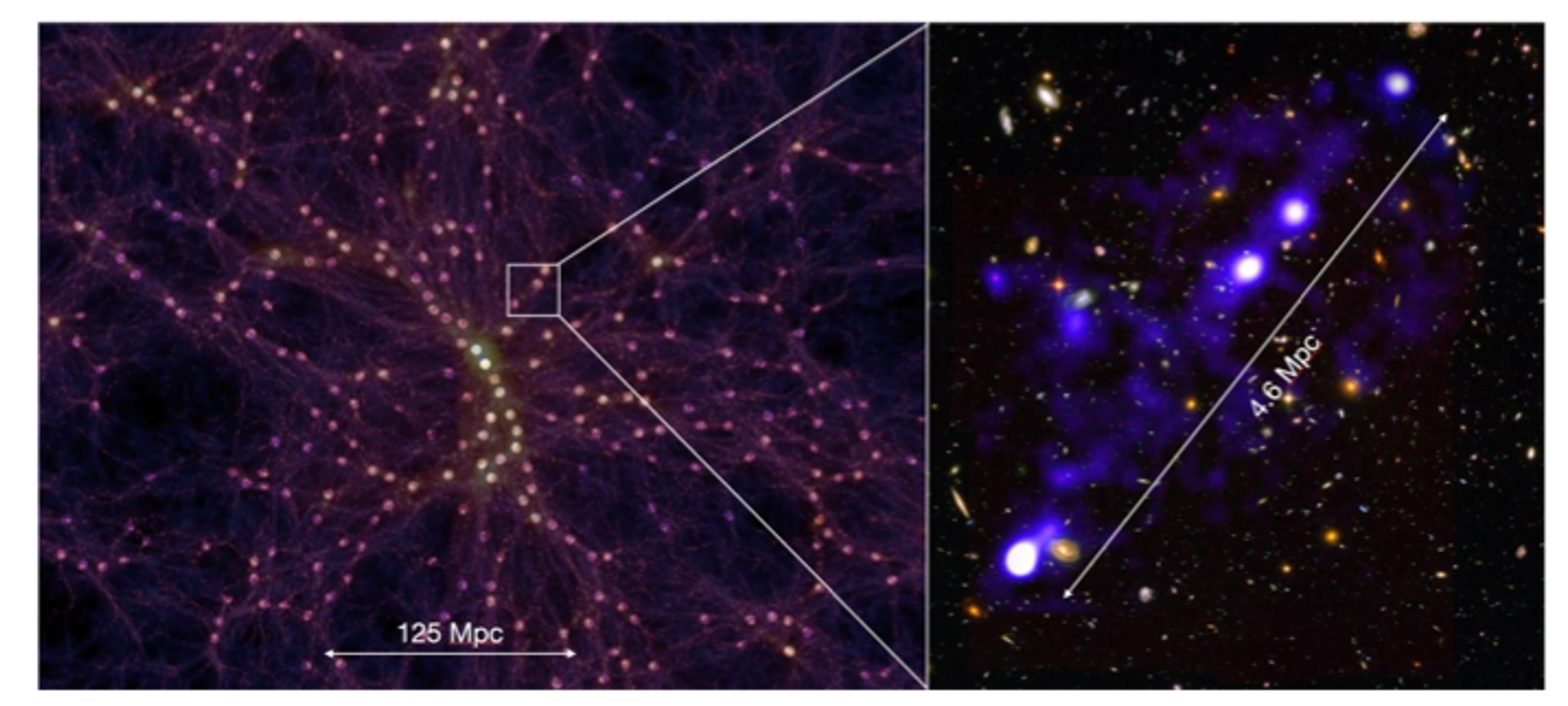}
    \caption{A major goal of extragalactic science with \wst\ is to understand the assembly history of galaxies in relation to the framework of dark matter in the so-called ``cosmic web''. This will be accomplished both on large cosmic scales (left) using the unparalleled field of view and multiplex gain of the MOS, and on finer scales (right) using its IFS where important baryonic feedback processes can be studied (Figure from~\citet{MUSE_CW})}
    \label{fig:web-overview}
\end{figure}

Panoramic IFS on 8-10m class telescopes such as \muse\ and \kcwi\ have already ably demonstrated their unparalleled sensitivity to emission lines in blind surveys, reaching an order of magnitude improvement in sensitivity relative to narrow-band searches \citep[e.g.][]{Wisotzki2016}. The IFS on \wst, with its much larger field of view, will provide an order of magnitude increase in the discovery space relative to the state of the art. Such massive blind emission line surveys will reach to much fainter limits than can be achieved through slitless spectroscopy or narrow-band imaging. The increased sensitivity will permit the direct detection of the circum- and inter-galactic gas in emission surrounding galaxies at cosmological distances. With present-day facilities such detections are rare \citep[c.f.][see also Figure~\ref{fig:web-overview}]{MUSE_CW} but the significantly increased survey volume enabled by \wst\ will transform such exploratory studies into a main-stream undertaking and provide rich information on the cycle of gas in and out of galaxies.

The above synergies between galaxy-scale processes and the larger cosmic structures can be accomplished at both high redshifts (z$\simeq2-3$) where star formation and AGN activities are at their peak, as well as at intermediate redshifts ($z<1.5$) where familiar optical spectroscopic diagnostics are available and finer detail and improved S/N is possible. In the nearby universe, \wst\ will allow for detailed mapping of a significant fraction of the Local Volume galaxies to understand the emergence of kpc-scale scaling laws. Simultaneously, the MOS can be used to sample outlying \hii\ regions and nearby dwarf galaxies. This synergistic sampling of small and large scales over a considerable range of look-back times will make \wst\ a uniquely efficient facility for understanding galaxy evolution.

The extragalactic science case for \wst\ is based, fundamentally, on understanding how the distribution function of galaxies in the present-day Universe was assembled, the detailed physical processes involved and how these are connected to and governed by the surrounding dark matter structure. Nearby datasets (e.g. SDSS, 2dFGRS, 2MASS, GALEX and WISE) have provided a comprehensive view of galaxies and associated structure in the local Universe ($z < 0.2$), as a function of stellar mass, stellar content, metallicity, star formation rate, environment, and activity level. Large ground-based telescopes have attempted to extend this survey strategy to $z\simeq$1, although as these are not normally dedicated facilities, progress has been limited. The crucial next step is to push these studies over large representative volumes beyond $z \sim 1$ where the star formation and AGN activity in the Universe was at its peak. 

The most fundamental feature of hierarchical structure formation is the complex distribution of matter in a structure referred to as the "cosmic web". Galaxies form and assemble within this evolving network of dark matter halos, filaments and voids, accreting material from it and ejecting processed gas into it. Landmark nearby galaxies surveys such as SDSS and 2dFGRS established important correlations between the properties of individual galaxies and their hosting structures. We now seek to map these correlations over a wide range of scales with high fidelity at epochs corresponding to the peak of star formation activity. This can only be done with a dedicated large aperture multi-object spectroscopic facility using optimised selection techniques and novel analysis methods.

The cosmic web presents the underlying framework for addressing important scientific questions that will still be highly relevant in the 2030s:

\begin{itemize}
\item How do small-, intermediate- and large-scale structures affect the accretion onto, star formation in and ejection of matter out of galaxies? 
\item How do outflows from galaxies chemically enrich the cosmic web and the intergalactic medium? Are there regions of the cosmic web that are chemically pristine at intermediate redshift which can provide a valuable glimpse of the earliest periods of galaxy formation?
\item How do the distribution functions of galaxy properties vary with redshift and environment at $z>1$? 
\item What are the properties of the progenitors of the most massive structures found in the present day Universe? 
\end{itemize}

While some of these questions will certainly be addressed by current or near-future spectroscopic campaigns, surveys with present facilities will generally lack the volume necessary for representative results, suffer from poor sampling on small scales and/or fundamentally will not produce spectroscopy of the required quality and resolution. These joint requirements are challenging: volumes of $\sim \mathrm{Gpc}^3$ are needed to sample the full range of web environments, yet galaxy number densities $\sim 10^{-2}\,\mathrm{h}^3\,\mathrm{Mpc}^{-3}$ are important to probe filamentary structures and highly non-linear knots. Spectra with high S/N and good resolution are essential to yield physical and kinematic parameters for each galaxy in order to provide the diagnostics required to address the questions above. With present surveys, this can only be achieved by the undesirable practice of stacking spectra.  Most importantly, diagnostic spectroscopy on small scales cannot be efficiently probed with multi-object spectographs, even on comparably large telescopes such as Subaru. Only by simultaneously harnessing a panoramic IFS can the above questions be properly addressed.

In the following we illustrate the remarkable scientific potential of \wst, both in terms of the primary goal of charting the distribution of galaxies in relation to the cosmic web over a wide range of physical scales as well as a selection of other important scientific opportunities. Note these additional science programmes represent only a selection of the extragalactic opportunities with such an impressive facility. In what follows  we assume \wst\ to be a 10--12m class facility with a 3 deg$^2$ field of view and a multiplex gain of 20,000. 

\subsection{Galaxy assembly and the cosmic web{\label{sec:extgal-cosmic-web}}}

\medskip
\noindent 
As briefly introduced above, the primary goal of this component of the case for \wst\ is to reconstruct the 3-D density distribution of galaxies with a fidelity comparable to the SDSS on large scales in
redshift bins of $\Delta z=0.5$ from $z=1$ to $z=4$ with $10^6$ galaxies per bin for six $\mathrm{Gpc}^3$ volumes using the MOS. On smaller scales the combination with the IFS will yield high numbers of
foreground-background pairs with small transverse separation leading to the characterisation of a full continuous distribution of transverse separations. Multiple tracers are possible for the cosmic web
(e.g. emission-line galaxies, massive galaxies, absorbing gas), and each have their own biases and sensitivities, some of which are poorly understood. In order to achieve a trade-off between systematic and
statistical uncertainties, the survey design will likely adopt a tiered approach whereby the unbiased, most sensitive, tracers will lie in the central IFS pointing where emission line studies can be undertaken without photometric selection. These deep IFS pointings prioritise a low systematic uncertainty and maximise discovery space, while the MOS observations prioritise low statistical uncertainty. The MOS component will drive the overall survey design and is based on several physical considerations.

Firstly, the survey volumes must be sufficiently large to contain all representative structures. Since the abundance of clusters similar to Coma (halo mass $\sim 10^{15}\,\mathrm{M}_\odot$) is several per
Gpc$^3$, this governs how we can probe the full range of densities, halo masses and environments in each redshift bin. Moreover, the number of galaxies per bin must also ensure we probe structures sufficiently finely on small scales to reliably detect proto-clusters and voids. Finally, on all scales the spectra must have sufficient S/N and resolution to measure key physical parameters such as continuum-based star-formation rates and stellar metallicities derived from UV absorption lines at $z > 2$. Such high quality data will then permit correlations of the basic physical properties of each galaxy in the context of their environment on both large and small scales.

\begin{figure}[ht!]
    \centering
    \includegraphics[width=0.7\textwidth]{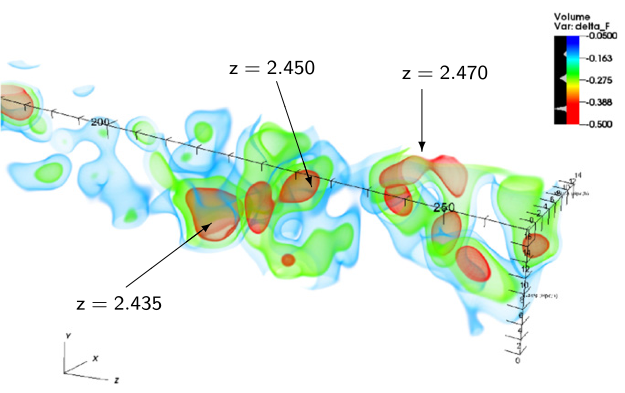} \\
    \includegraphics[width=0.7\textwidth]{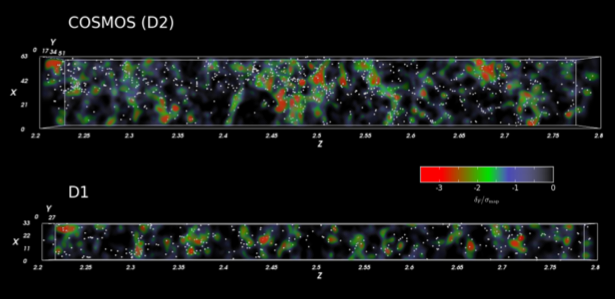} 
    \caption{Demonstration of 3-D tomography using the fluctuating transmission of neutral hydrogen seen via \lya\ absorption in the spectra of background $2<z<3$ galaxies. (Top) Density structures revealed using Keck/LRIS in the CLAMATO survey \citep{Lee-2018-CLAMATO}. (Bottom) Application of a similar technique in two survey fields from the LATIS survey using Magellan/IMACS \citep{Newman-2020-LATIS}. \wst\ has the capability to chart the cosmic web with unprecedented sensitivity over much larger cosmic volumes.}
    \label{fig:tomography}
\end{figure}

Traditionally, such surveys have used the 3-D galaxy distribution for delineating cosmic structures. However, over a restricted redshift range of $z\sim 2$ -- $3$, a more powerful technique has emerged based
on using 3-D topology of the \lya\ forest seen in absorption along the line of sight to background sources. The low density absorbing clouds of hydrogen responsible for the \lya\ forest trace the linear regime of density fluctuations and thus represent an invaluable tracer of the dark matter distribution. Long considered the realm of 30m-class telescopes \citep[e.g.][]{Evans2015-ELTMOS}, recent applications have shown adequate information for background galaxies is feasible with 8-10m-class telescopes \citep[e.g.][]{Lee-2016-CLAMATO,Lee-2018-CLAMATO,Newman-2020-LATIS} provided the continuum S/N and spectral resolution ($R\sim 1000$--$1500$) are adequate with reasonable blue sensitivity (Figure~\ref{fig:tomography}). Although \mosaic\ on the \elt\ has the potential to undertake such an application with finer sampling, its limited field of view makes it impractical to cover the required cosmic volumes. 4-metre based surveys such as \fourmost\ and \weave\ do not have the sensitivity to secure adequate signal/noise for absorption line spectroscopy at redshifts where the Ly-$\alpha$ forest can be observed.

We thus envisage two complementary ``Legacy" galaxy surveys with the MOS; one based on using galaxies as tracers over $0.3<z<7$, and another utilising the \lya\ forest between $z\simeq2$ -- $3$ using background
galaxies. The fluxes and surface densities of the sources required to chart evolution in the cosmic web using both galaxies and the \lya\ forest over the common redshift range $2<z<3$ are similar - indeed,
they are often the same sources.  In each $\Delta z=0.5$ volume bin, a $1\, \mathrm{Gpc}^3$ volume ($h=0.7$, hereafter) implies a sky area of $\sim 200\,\mathrm{deg}^2$. For $10^6$ galaxies per bin (i.e.\ a minimum density of $\sim 3\times 10^{-2}\,\mathrm{Mpc}^{-3}$ required to identify filaments and knots), necessitates a target density of $\sim 5000\,\mathrm{deg}^{-2}$ per bin or $\sim 30,000\,\mathrm{deg}^{-2}$ across the entire redshift range. To uniformly populate the full redshift range will require careful photometric selection in deep wide field catalogues but this will be
practical with multi-band data from \lsst\ and \euclid\ which will reach fainter than AB=25.5 in the optical \citep{LSST} and AB=24 in the near-IR \citep{Euclid-WideSurvey}. The limiting magnitude will typically vary from $\AB{i} \sim 23.1$ at $z\sim 1$ to $\AB{i} \sim 24.9$ at $z\sim 3$. Current 8-10m telescope
surveys spanning this redshift and magnitude range indicate exposure times of 2-3 hours for emission line redshifts and 5-7 hours for the most challenging $\AB{i} < 25.8$ sources to $z\sim 4$.

In the redshift range $2<z<3$, this target surface density is more than adequate to map the cosmic web using the \lya\ forest. \citet{Lee-2016-CLAMATO,Lee-2018-CLAMATO} achieved a spatial resolution of $\sim3.5\,\mathrm{h}^{-1}\,\mathrm{Mpc}$ targeting $z\sim 2.3$--$2.8$ galaxies with $\AB{g} \sim 24$--$24.8$ with a surface density of $\sim 1000\,\mathrm{deg}^{-2}$. Exposures with Keck LRIS at $R\sim 1000$ were typically 2 hours achieving a continuum S/N$\sim 3$ per \AA. To push to $z\sim 3$ with this S/N will require exposures of $\sim 7$ hours, comparable to that for the primary galaxy survey.

Simultaneously with the above MOS survey, the IFS can be employed to enhance the scientific goals by tracing the cosmic web in emission through \lya\ at $2<z<3$ \citep{MUSE_CW}. Studies of \lya\ in emission can also be extended to higher redshifts than the MOS survey in order to map the patchiness/coalescence of ionised bubbles during the final stages of cosmic reionisation ($z\sim 5$--$7$). By targetting regions containing known bright quasars, the statistics of \lya\ emitters will offer valuable insight into how these non-thermal sources contribute to concluding reionisation. In the same way, via judicious pointings IFS data can also be used to study intermediate redshift galaxies ($0.3<z<1.6$) and in Sections~\ref{sec:extgal-galaxies-cosmic-web}--\ref{sec:extgal-archeology} we discuss how we can use these data to relate the properties of galaxies to their location in the cosmic web with greater spatial fidelity.

As an illustration of the investment of time necessary for this ``Legacy" proposal, if we assume an unit exposure for the brightest targets of $\sim 2$ hours, with fainter targets studied in multiple
visits, to achieve the required surface density ($30,000\, \mathrm{deg}^{-2}$) over each $3\,\mathrm{deg}^2$ field would require 15--25 hours. A survey of $6\times10^6$ galaxies
would then take $\simeq$100 dark nights. Depending on other comparable \wst\ campaigns, this would represent an achievable goal in 3-4 years and is similar to long term investments being made in other surveys today (e.g. \pfs).

\subsection{The influence of the cosmic web on galaxy evolution at $z\sim 1$}
\label{sec:extgal-galaxies-cosmic-web}

\wst\ can also make significant advances in tracing evolving structures at lower redshifts ($z\sim1$) where most of the familiar rest-frame optical diagnostic lines of metallicity and star formation rate are available. Moreover, features such as filaments can be more easily resolved than at $z\sim2-3$ and, using the IFS, spatial sampling on scales $<1$ Mpc becomes practical. Extending studies of how the cosmic web influences galaxy evolution over a longer baseline in cosmic time can also assist in determining how galaxies quench \citep[e.g.][]{Aragon-Calvo-2019}, gain their angular momenta \citep[e.g.][]{Laigle-2015} and attain their present day morphological forms \citep[e.g.][]{Kuutma-2017}. 

Currently, there is little theoretical consensus on the role of the cosmic web in these topics. While specific star formation rates are reduced close to nodes and filaments in some numerical simulations \citep{Singh-2020-simulation,Xu-2020-simulation,Malavasi-2022-CosmicWeb}, others \citep{Kotecha-2022-filaments, Zheng-2022-filaments} report increased star formation (SF) activity and delayed quenching. Such inconsistencies likely reflect the wide range of scales involved ($~100$kpc to several Mpc). In particular, \citet{Song-2021-filament} find that gas regulation (and quenching) occur at the edge of a cosmic filament, a result that cannot be verified through observations if the sub-Mpc scale of the filament remains unresolved. Such issues illustrate the importance of mapping the cosmic web with a resolution substantially better than 1 cMpc. Recent $z<1$ surveys such as GAMA \citep{Kraljic-2018}, VIPERS \citep{Malavasi-2017} and 
COSMOS \citep{Laigle-2018} have resolutions $\sim5$ cMpc limited by MOS fibre densities of few 10$^3$deg$^{-2}$ corresponding to $<1$ arcmin$^{-2}$ (or $<\sim1$ cMpc$^{-2}$ at z=1-2). 

\begin{figure}[ht!]
    \centering
    \includegraphics[width=0.9\textwidth]{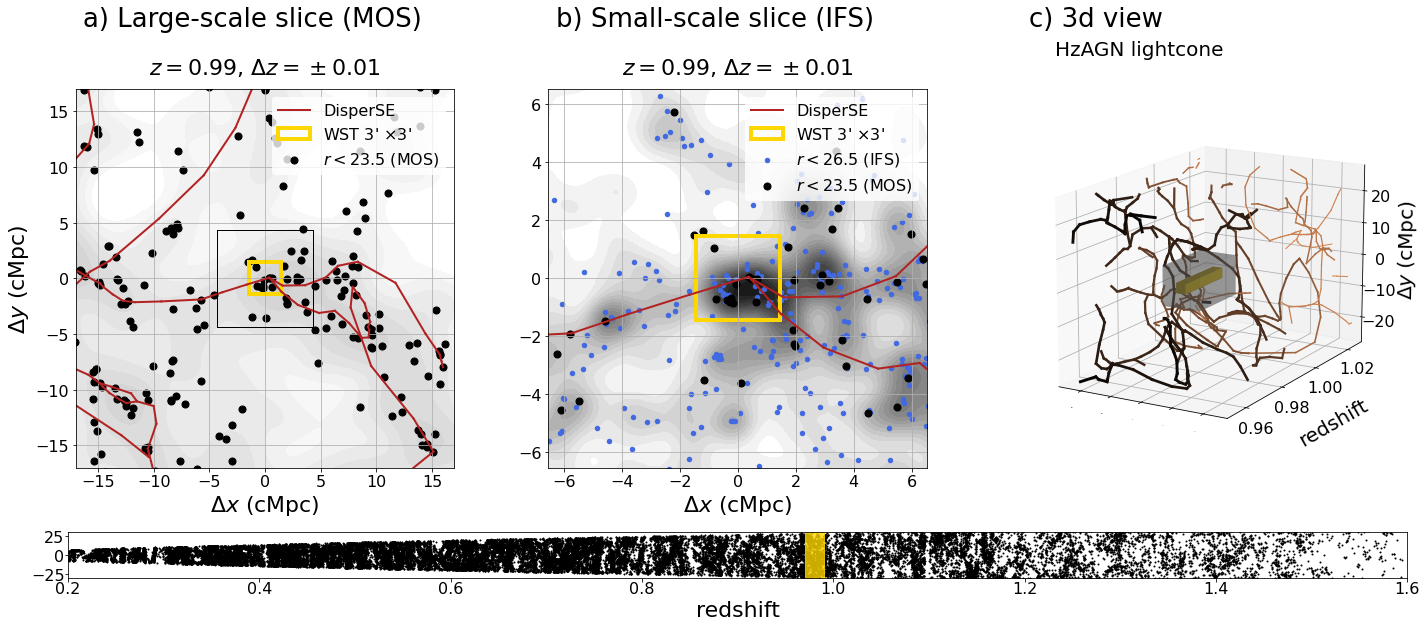}
    \caption{A realistic demonstration of the capacities of \wst\ for the study of galaxy properties in their large-scale environment.
a) Distribution of $r<23.5$ galaxies in a cube centered at z=1 drawn from the Horizon-AGN lightcone \citep{Dubois-2014,Laigle-2019}, as they could be observed with the MOS. Such galaxy sampling on a large field is required to locate sufficiently well the large-scale filaments (e.g. with DisPerSE \citealt{Sousbie-2011-DISPERSE}). b) Zoom around the central part of the field. The blue dots are $r<26.5$ galaxies distributed around the  cosmic filaments, which could be observed with the IFS ($3'\times 3'$ yellow square). This configuration would allow to investigate the connection between the large-scale environment of these galaxies and their properties and evolution, over a large range of redshift (bottom panel).}
    \label{fig:extgal-bouche_laigle}
\end{figure}

\wst's combination of a panoramic MOS and an on-axis IFS is ideally suited to making progress. As filaments are typically 3-20 cMpc long and $<1$ cMpc in width, the IFS can map fields of 3$\times$3 cMpc (3$\times$3 arcmin) and connect this information to large scale structures (5-100 cMpc) determined from the MOS. This will simultaneously map the cosmic web and determine the locations of all galaxies within a stellar mass range of 10$^8$ to 10$^{10}$ M$_{\odot}$ in the vicinity of the filaments. Fig.~\ref{fig:extgal-bouche_laigle} illustrates this complementary between the information on the large scales collected by the MOS and the small scales revealed by the IFS.

Such a $z\sim1$ MOS+IFS campaign could: 

\begin{itemize}
    \item map $\sim$10$^7$ emission line galaxies (ELGs) over $\sim$10$^3$ deg$^2$ with a target density 10$\times$ that of current MOS surveys
    \item secure a spectroscopic census of both active (ELG) and passive galaxies as demonstrated by \muse\ which, in a 2 hour exposure, can survey $\simeq$50 galaxies in 1 arcmin$^2$ to r=24.5 \citep[e.g.][]{MUSE_DRII}. 
\end{itemize}

The IFS is key as it (i) resolves cosmic filaments from the much higher galaxy density regions ($>$50 arcmin$^{-2}$) and (ii) enables studies of the properties of normal star-forming and passive galaxies in the vicinity (0.2 to 2 cMpc). Technically, several tools have been developed to map 3D structures from sparse datasets, e.g. DISPERSE \citep{Sousbie-2011-DISPERSE} or PolyPHORM (\citet{Elek-2020}, see also \citet{Libeskind-2018}). Leveraging line-of-sight information from multiple QSO surveys would be beneficial since metal lines at $z=1$ can trace the cool neutral gas. 

As an illustration of the investment of time necessary for this ``Legacy'' proposal, the cumulative number counts of galaxies to $r<23.5$ is $\sim 25 000\,\mathrm{deg}^{-2}$  \citep{Metcalfe-2001-numbercounts}. Sampling 30--50\% of these would fill the fibers over $2\,\mathrm{deg}^2$, and to map the large scales of the cosmic web would require at most 2 passes, or 4 hr per field assuming an unit exposure for the brightest targets of $\sim 2$ hours.  For illustration purposes, a legacy survey over $500\,\mathrm{deg}^2$ would cover $\sim 200$ fields/lines-of-sight,  provide around $5\times 10^6$ MOS/fiber spectra, yield 250,000 galaxies in the IFS and would then take $\approx 100$ dark nights.

\subsection{Gaseous flows and the baryon cycle}

The high quality spectra of $z\simeq2-3$ galaxies discussed in Section \ref{sec:extgal-cosmic-web} can be used to address key questions in the interface between galaxies and their local circumgalactic medium. Star-forming galaxies drive powerful, galactic-scale outflows and also accrete gas from the intergalactic medium (IGM). This flow of gas inwards and outwards of galaxies can be envisaged as the cycling of baryons which, we now understand, plays a major role in regulating galaxy growth and evolution. Specifically, gaseous inflows induce new star formation and black hole growth which, in turn, can create outflows that affect the thermal state of the surrounding gas via a self-regulating feedback loop. The importance of this self-regulation is supported by a wide range of theoretical models and observations on galactic and circumgalactic scales. Key pieces of evidence for the baryon cycle include the mass-metallicity relation and metal enrichment of the IGM  \citep{Tremonti-2004-MZ,Oppenheimer-Dave-2006,Lilly-2013-gasreg}, the quenching of star formation in massive galaxies \citep{Hopkins-2008-quenching}, and the mismatch between the galaxy stellar mass function and the dark matter halo mass function \citep{Benson-2003-LF,Keres-2009-hotcold,Oppenheimer-2010-MF}. 

Galactic winds from massive stars and AGN can be characterised by the incidence of blue-shifted interstellar absorption lines arising from cool, outflowing gas seen against the backlight of the stellar continuum, thereby testing momentum-driven versus energy-driven winds. To date, the signal to noise necessary to diagnose such outflows at $z\sim2$ has typically been achieved only through stacking spectra of selected Lyman break galaxies. The dedicated nature and larger aperture of \wst\ will make it possible to accomplish such measurement for individual galaxies and studied as a function of their location in the cosmic web. 

The interface between this outflowing gas and the inflowing component is the circumgalactic medium (CGM) which can be probed most effectively using QSO-galaxy or galaxy-galaxy pairs at different redshifts. Absorption lines seen in the spectrum of the background source arising from species at the redshift of the foreground galaxy provide important measures of how far the outflowing material extends and its kinematic and chemical properties. This extremely diffuse medium is multi-phase, and typically extends up to 300 kpc \citep{Tumlinson2017}. The dense sampling of background galaxies and QSOs required for the MOS-based tomography project discussed in Section \ref{sec:extgal-cosmic-web} will ensure a significant increase in the number of suitable pairs over earlier studies and hence lead to a major advance in understanding how galactic feedback regulates galaxy activity and chemical enrichment at z$\simeq$2-3. The panoramic IFS  will complement this absorption-based strategy by providing direct maps of the CGM and the outflowing emission into gaseous filaments thought to link galaxies. By the time of \wst, SKA will also be providing a complementary view of the distribution of neutral hydrogen (see Section \ref{Intro:radio}) ensuring a complete view of the cycle of gas at cosmic noon.

Studies of the baryon cycle are optimally performed at redshifts $z\simeq$2-3 for several reasons. Firstly, this is the formative period in galaxy evolution when star-forming and AGN activity is at its peak. Moreover many of the diagnostic spectroscopic features, indicative of both the chemical composition and kinematics of the gas, are redshifted into the optical region. UV resonant lines are ideal tracers of gas flows: their spatial distribution probes the extent of the gas, and their spectral distribution traces the  kinematics \citep{Prochaska2011,Dijkstra2017}. Lyman $\alpha$ is a further valuable rest-UV tracer of the CGM. Ubiquitous Lyman $\alpha$ haloes have been discovered around high redshift galaxies with MUSE \citep{Wisotzki2016, Leclercq2017}, but the diversity of their spatial and spectral distributions is not well understood and is likely multifactorial \citep[e.g.][]{Blaizot2023}, thus progress is likely to require much larger samples than is possible to assemble before \wst.

\begin{figure}[ht!]
\centering
\includegraphics[width=0.9\textwidth]{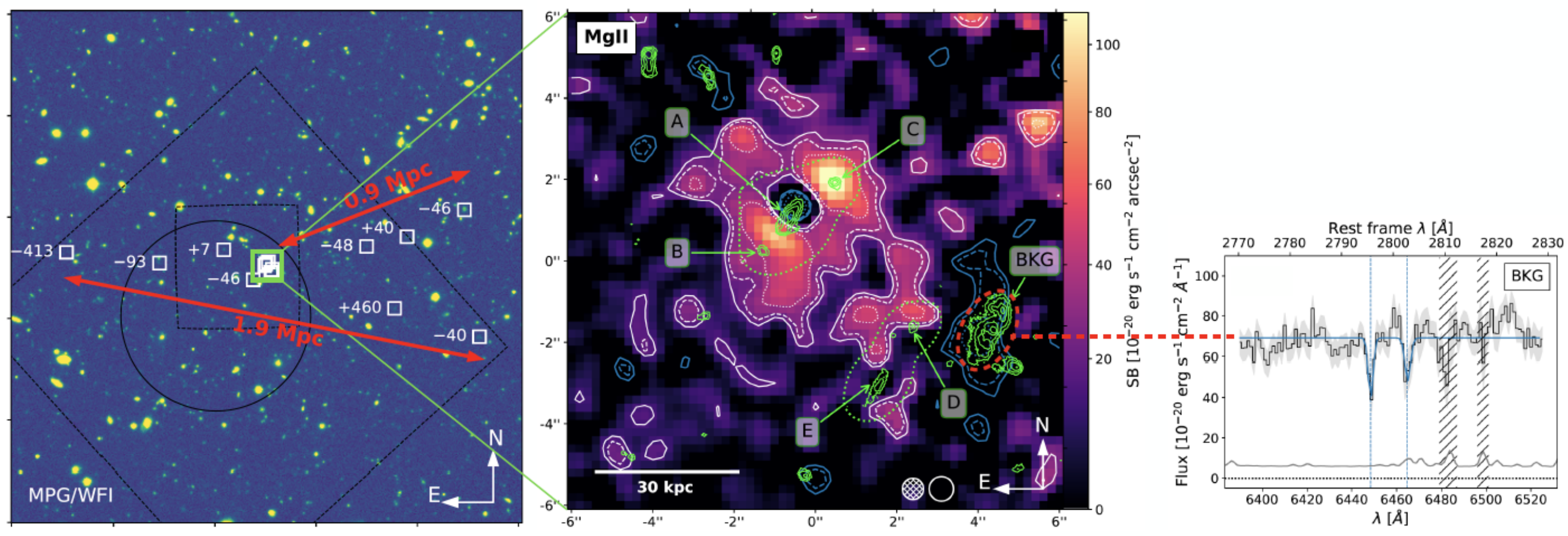}
\caption{Metal-enriched intragroup medium surrounding five galaxies at z$\simeq$1.3 embedded in a larger overdensity (left, 4'$\times$4') revealed by VLT/MUSE, both in Mg II emission (middle) and in MgII absorption (right) from \citet{Leclercq2022} \wst\ will provide a complete view of the gas surrounding galaxies by mapping the inner CGM in emission and by probing the diffuse CGM in absorption using rest-UV tracers redshifted into the optical. It will also enable us to connect the properties of the expelled gas to the larger-scale environment.}
\label{fig:extgal-MgII}
\end{figure}

Finally, for the redshift range $0.3 < z < 2.2$, the resonant MgII doublet is also observable with \wst.  
Challenging detections of MgII halos have been reported with \muse\ and \kcwi\ \citep{Burchett2021, Zabl2021}, thereby probing the chemical enrichment of the expelled gas into the CGM  \citep{Guo-2023-MgIIWinds}. \wst\ has the capability of simultaneously probing the denser CGM in emission with the IFS, and mapping the extent of the diffuse CGM in absorption with the MOS \citep[][see Figure~\ref{fig:extgal-MgII}]{Leclercq2022}. Within the redshift interval $2 < z < 2.2$, the CGM can be traced with both Lyman-$\alpha$ and MgII to study the relative distribution of enriched and pristine hydrogen gas. Collectively, therefore, there is great synergy between these detailed studies of the CGM using various UV tracers and the 'Legacy' surveys described in Section \ref{sec:extgal-cosmic-web}.

\medskip

\subsection{Archaeological studies of galaxies to $z\sim$1}
\label{sec:extgal-archeology}

\medskip

A systematic census of the physical properties of galaxies through
studies of their stellar populations and nature of their ISM can provide powerful constraints on models of galaxy
evolution. The SDSS survey clearly demonstrated the impact of
spectroscopic data for $>$10$^5$ galaxies by establishing a number of
fundamental trends relating to galaxy bimodality, archaeological
downsizing, the galaxy-SMBH connection and local environment. These
scaling relations link the masses, structural properties and
integrated star formation rates of galaxies to their stellar
populations and ISM through their chemical enrichment and cosmological
assembly histories. Such relations enable us to infer how various
physical mechanisms affect galaxy evolution over a range of scales and
mass regimes thereby offering a fundamental test bench for developing
and constraining galaxy evolution models.

These $z=0$ relations have in fact become cornerstones of the
calibration of state-of-the art models but they are not always well
reproduced by models \citep[e.g.][]{Garcia2024-MZ}. One important reason
for this is likely that the vast majority of our knowledge of these
scaling relations and their scatter is at $z=0$. Thus we lack solid
quantification of many, if not most, of how these trends change with
cosmic time. 

Furthermore, due to the complex nature of galaxy evolution, these
mean scaling relations reveal only part of the story. The scatter
associated with such relations offers additional information relating
to “stochastic” processes, such as mergers, gaseous inflows and
stellar and AGN-driven outflows \citep[e.g.][]{Garcia2024-MZ,Davies-2024,MattheeSchaye-2019}. However, as emphasised 
by \citet[][see also \citealt{Dutton2010-MZ}]{MattheeSchaye-2019}, in models the dark matter
halos play a central role in establishing the slope and scatter in at
least some scaling relations, thus linking large-scale structure to
the internal processes in galaxies. At present, however, our ability
to extract this information and testing these model predictions is
severely limited by the lack of a handle on the redshift evolution.

Further progress requires higher quality spectroscopic data and larger
statistical samples for which key physical parameters can be derived
individually. Taking the study of stellar populations as an example,
in addition to determining basic properties such as the star formation
rate, dust content and gas-phase metallicity, addressing the stellar
metallicity and past star formation history requires a major
investment in both signal to noise (S/N $\gtrsim 20$ per \AA\ and
resolution $\gtrsim 2500$). A large sample size is necessary to
place these higher level galaxy properties in the context of the
manifold of environmental and other parameters discussed above.

Over the past two decades, SDSS and related campaigns, e.g. GAMA 
\citep{Kraljic-2018}, have made significant progress in this regard 
for the $z<0.2$ universe, complementing the approach of inferring the 
earlier evolution of stellar populations from the "archaeological record" 
in nearby galaxies which, inevitably, becomes more uncertain with increasing
look-back time. However, applying these archaeological studies to high
quality observations of $0.2<z<1$ galaxies offers the prospect of
complementing the information obtained directly from surveys beyond a
redshift $z>1$ (Section \ref{sec:extgal-cosmic-web}). Although a few campaigns (e.g. LEGA-C, \citealt{vanderWel2016-LEGAC}) have
undertaken archaeological studies beyond $z\simeq$0.2 they are based
on samples of less than a few thousand galaxies over a relatively
limited range of stellar masses (a few times
$10^{10}$M$_{\odot}$). 

\wst\ has the capability to increase $z<1$
samples by two orders of magnitude with a mass completeness down to a
few $10^9$ M$_{\odot}$. Only with samples of this magnitude can we
expect to truly test the model predictions mentioned above. 

\begin{figure}[ht!]
\centering
\includegraphics[width=0.8\textwidth]{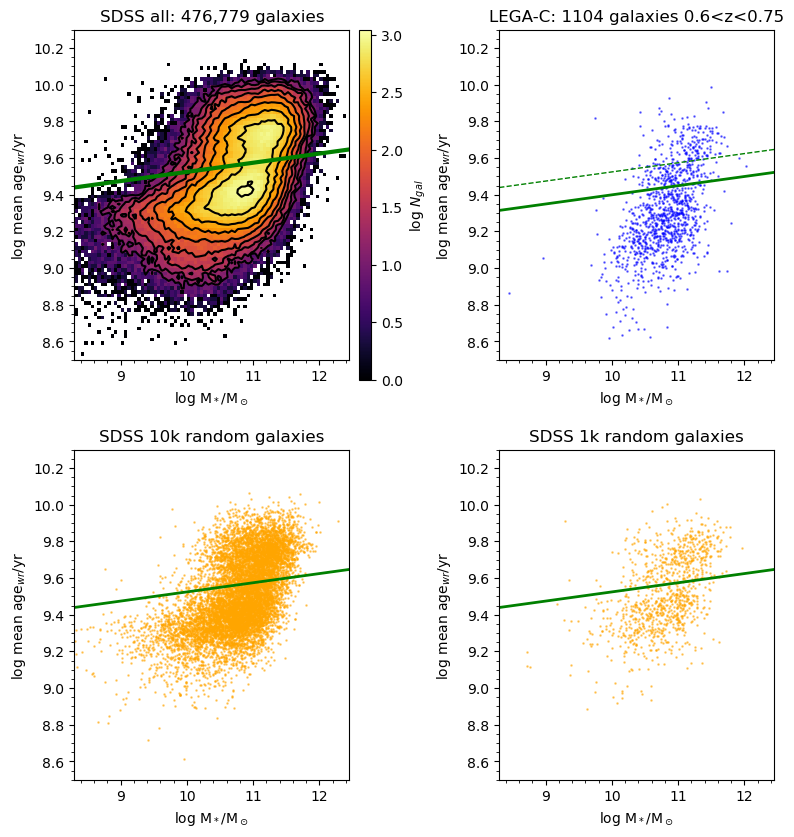}
    \caption{Galaxy distributions in mean stellar age vs stellar
mass. (Top left) Full SDSS sample for $z<$0.22 revealing a
mass-dependent age bimodality and a green valley (solid line). (Top
right) The smaller LEGA-C sample cannot reveal such distinctions
robustly as indicated by a tentative location of the green valley
(dashed line) with respect to SDSS. (Bottom panels) Subsets of the
SDSS sample comprising 10$^4$ (left) and 10$^3$ (right) galaxies
indicating the required sample size to locate key features. \wst\ has
the capability to generate samples as large as 10$^5$ galaxies which
can be studied in subsets according to various physical properties.}
    \label{fig:LEGAC}
\end{figure}

As an illustration based on LEGA-C observations taken at ESO's VLT
with VIMOS, exposure times ranging from a few 10s to $\sim 100$ hours would
deliver spectra at a resolution of R$\simeq$3000 with a S/N$>$10-20
per \AA\ in the continuum over the required mass range. The density of
available targets estimated for the well-studied COSMOS deep field
indicates $\simeq$7000 galaxies deg$^{-2}$ for $0.6<z<1$ down to
$10^{10}$ M$_{\odot}$ \citep[][LEGA-C DR3]{vanderWel-2021-LEGAC-DR3}. For a lower mass
limit of $10^{9}$ M$_{\odot}$, the target density would be
$\simeq$10-20,000 deg$^{-2}$. Thus a sample size of $10^5$ could be achieved with a moderate time investment of $\sim 250$ hours by adding $\sim 25$ hours in each of $\sim 10$ fields of the ``Legacy survey'' outlined in Section~\ref{sec:extgal-cosmic-web}.

In order to highlight the importance of a large ($>$10$^5$)
statistical sample, Figure~\ref{fig:LEGAC} shows the age-stellar mass distribution of galaxies in SDSS for $z< 0.22$ (Mattolini et al., in preparation) and LEGA-C (Gallazzi et al., in preparation) in the plane of mean stellar age vs stellar mass. Since ages are easier to constrain than metallicities/star formation histories, a S/N$>$10 suffices for
reliable ages, and for SDSS the large sample size means several hundred
galaxies are present in each bin and thus features such as the
mass-dependent age bimodality and the presence of a "green valley" can
be readily identified. The bottom panels show the distribution in the
same plane for two random subsamples of the SDSS, made of 10,000 and
1,000 galaxies, respectively. Only in the former case can any
bimodality be robustly quantified. Controlling for different
properties (e.g. redshift, environment, AGN activity, metallicity, gas
fraction, SFR etc.) is key to inferring the various physical
mechanisms involved. As \wst\ has the capability to survey $>10^5$
galaxies, subsamples of several thousand galaxies selected in various
way can be studied to make progress. The current state of the art at
intermediate redshift is represented by the LEGA-C survey where only a
tentative green valley can estimated from the limited data.

\subsection{A local volume galaxy IFS survey}
\label{sec:extgal-LV-IFS}

Galaxy evolution is driven by feedback processes occurring on a variety of scales. On the scales of giant molecular clouds, the star formation efficiency and multi-phase ISM is shaped by small-scale processes \citep{Krumholz2014, Dale2015, Krumholz2019, Chevance2023}. On galaxy-wide scales, outflows and secular evolution drive the interplay between galaxies, their dark matter halos, and the circum- and intergalactic media \citep{Tumlinson2017, Veilleux2020}. Understanding the distribution of baryonic matter and energy across these scales remains a key challenge for both theoretical and observational astrophysics in the next decade.

Cosmological hydrodynamical simulations can reproduce clustering, scaling relations, and the internal structures of galaxies \citep{Crain2023}. However, the injection of energy through feedback processes occurs on numerically unresolved scales, which requires introducing sub-grid models. On the other hand, small-scale ISM simulations are now relatively sophisticated, including a variety of small-scale processes (e.g. stellar winds, ionising radiation and radiative transfer, e.g. \citealt{Semenov2021, Grudic2021}). However, they still struggle to address galaxy-wide phenomena and are generally unable to take into account the full cosmological context.

Observationally, large galaxy surveys provide the statistics needed to study the assembly of galaxy subcomponents, and the link between galaxies and their large-scale environments. Addressing the physics of star formation, on the other hand, requires a statistical perspective on the individual components of the matter cycle (HII regions, star clusters) at a resolution comparable with the size and separation length characteristic of star-forming regions ($\sim$ 100 pc). The current generation of integral field spectrographs (e.g. \muse\ at the \vlt) has enabled mapping of nearby galaxies demonstrating the power of this approach \citep{McLeod+2020, Barnes2021, DellaBruna2022, Emsellem2022}. However, such samples are relatively modest and span a limited range of environments within galaxies (typically the inner few kpc) thus impeding our ability to match any insights drawn with the results obtained from much larger galaxy surveys probing kpc scales  (e.g. MaNGA, \citealt{sdss-manga} and SAMI \citealt{sami}). To make progress requires `cloud-scale' ($<$100 pc) mapping of a sufficiently large sample of galaxies ($>$10$^3$) across a range of morphological features (e.g. spiral arms, bars, bulges) and integrated properties (e.g. $M_{\star}$, SFR, and local environmental density). 

The momentum for an ambitious survey of resolved properties of nearby galaxies is provided by the bounty of high resolution imaging campaigns already available from ALMA and \jwst\ \citep{Leroy2021, Williams2024} and, in a few years, from \euclid, the \nancy\ and the \lsst. Collectively these will provide unprecedented details on the stellar components of galaxies in both the optical and near-IR. Future UV missions (e.g. UVEX \citealt{Kulkarni2021}, CASTOR \citealt{Cote2019}, LUVOIR \citealt{Tumlinson2019}) will also provide exquisite data on the sources of ionising radiation. Ultimately, SKA will provide comparable maps tracing star formation via radio continuum and atomic gas at 21cm (Section \ref{Intro:radio}). Optical integral field spectroscopy, on the other hand, represents a noticeable gap in this multi-wavelength landscape. 

\begin{figure}[t!]
\centering

\includegraphics[width=1\textwidth,trim=0 0 0 0, clip]{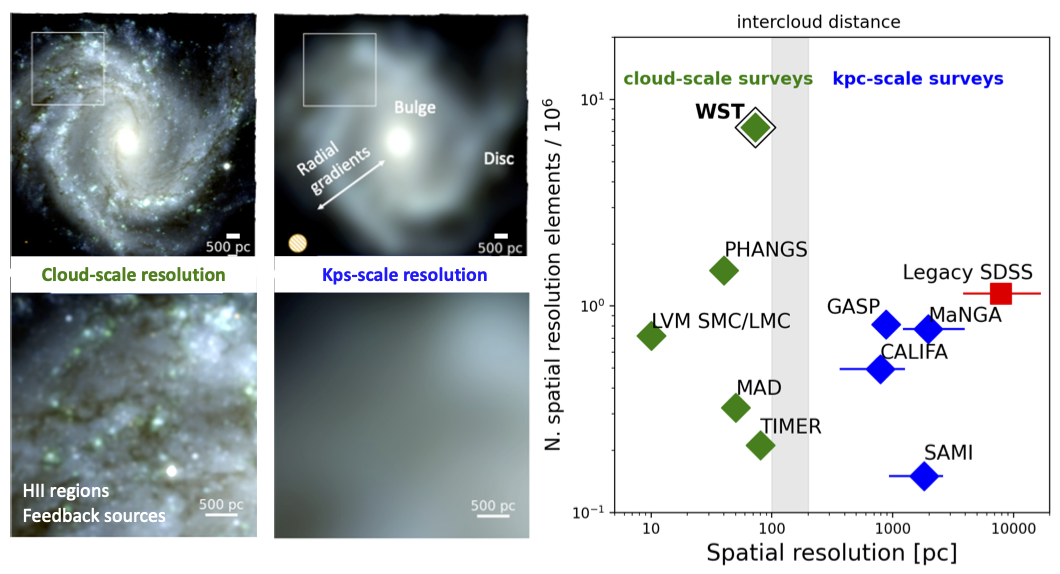}
\caption{\textbf{Left}: An illustration of the level of detail achievable in surveys that resolve the average distance between H\textsc{ii} regions (`cloud-scale')
 and those that observe galaxies at kpc resolution. \textbf{Right}: Comparison of the planned \wst\ Local Volume Legacy Survey with existing cloud-scale and kpc-scale survey. The Legacy SDSS (MOS survey) is also included for comparison. The \wst\ Local Volume Legacy survey would provide data for a statistical sample at cloud-scale resolution. Figures adapted from \cite{Emsellem2022}. \label{fig:IFS_survey_local} }

\end{figure}

With its large IFS, \wst\ is uniquely placed to perform an ambitious spectroscopic survey in the Local Volume (D$<$ 25 Mpc) at better than 100 pc resolution (Fig. \ref{fig:IFS_survey_local}). Key questions such a survey would address include:

\medskip
\noindent{\bf (i) The small-scale matter cycle:} Extensive multi-wavelength campaigns (e.g. \citealt{Kennicutt2003, Dale2009}) have revealed relations between gas, dust, and stars on kpc scales. However, much of the relevant physics requires observational resolutions better than $\sim$100 pc. For example, the well-studied relation between the surface density of gas and SFR (the Schmidt-Kennicutt law, \citealt{Kennicutt2012}) begins to break down at resolution better than $\sim$ 100 pc. Yet observations in this regime can probe the discrete phases along an evolutionary sequence from molecular clouds to HII regions, massive stars, and star clusters. The positions and number density of these different tracers are sensitive to the timescales associated with stellar feedback \citep{Corbelli2017, Kruijssen2019, Chevance2022}. In the closest galaxies (D $\lesssim$ 4 Mpc) individual massive stars can be spectroscopically characterise (Section \ref{subsec:respop-intro}). Relating nebulae to their ionising sources will permit precise measurement of feedback pressure and the escape fraction of ionising photons both as functions of local and global galaxy properties \citep{McLeod2021, DellaBruna2022, Teh2023}. These are fundamental measurements which can be compared and input into small-scale feedback simulations. 

Resolving individual nebulae will allow statistical studies of the metallicity of HII regions, in combination with the stellar population, uncovering the small-scale production and flow of metals in galaxies \citep{Maiolino2019, Kewley2019}. \wst's large aperture will ensure the detection of weak auroral lines yielding “direct” metallicity estimates beyond the inner gas-rich regions in poorly explored environments for star formation, where conversion from neutral to molecular hydrogen becomes highly inefficient. A sample of $\sim$10$^4$ regions would adequately span approximately 2 dex in the metallicity, N/O ratio, and ionisation parameter. Such a sample would be two orders of magnitude larger than the current state-of-the-art (e.g. CHAOS, \citealt{Berg2020}). Such high spatial resolution measures would also be useful to tracing gas-phase metallicity variations in nearby  (D$<25$ Mpc) galaxies hosting supernovae (Section 4).  Noting the detection rate of auroral lines in current \muse\ data and the average number density of HII regions, realising this program is consistent with a sample of $\sim$10$^3$ galaxies. Gas phase metallicity measurements will be combined with those from integrated-light spectroscopy and, where available, of resolved stars (Section \ref{subsec:respop-intro}) to chart the local history of chemical enrichment, the impact of metal diffusion, and the role of morphological features (e.g. spiral arms, bars) in mixing.

\medskip
\noindent{\bf (ii) Assembly history of galaxy components:}
\wst\ data will resolve the formation and chemical enrichment histories of local galaxies based on integrated-light spectroscopy \citep{Conroy2013}. Via non-parametric SFH modelling \citep{Leja2019}, novel stellar population models \citep{Eldridge2017}, and joint modelling of dynamics and stellar populations \citep{Poci2021}, the proposed survey will provide a detailed atlas of the assembly of galaxy sub-components. This would allow us to study the fraction of stars accreted by merger, the inside-out versus outside-in growth of galaxy discs as a function of large-scale environment, and the impact of secular evolution, e.g. the role of bars and bulges, the mixing of stellar populations over time.

\medskip
\noindent{\bf (iii) The galaxy-halo interface:} \nancy\ and \lsst\ will provide an unprecedented mapping of the low-surface-brightness universe and the halo population surrounding massive nearby galaxies. IFS and MOS spectroscopy of these faint structures and their associated globular clusters and planetary nebulae can address their assembly histories. A MOS survey would provide radial velocities for tens to hundreds of globular clusters around galaxies (especially the most massive ones), allowing an estimation of the dynamical mass, and therefore dark matter fraction, independently of rotation curves, also for passive, gas-poor galaxies (e.g. \citealt{Schuberth2010, Chaturvedi2022, Napolitano2022}).  

While recent efforts have uncovered a plethora of nearby dwarf satellite galaxies, spectroscopic follow programmes are still conducted largely on a case-by-case basis \citep{Fahrion2020}. The relative velocities in some systems in the Local Volume have revealed a tension with predictions from cosmological simulations (Cen A, \citealt{Muller2018}, the MW and M31, \citealt{Pawlowski2021}, NGC4490 \citealt{Karachentsev2024}). The MOS is well-suited to undertake a systematic study of dwarf satellites and improve upon the current state-of-the-art. The IFS will also allow detailed studies of the disc-halo interface of $\sim$ 100 highly inclined star-forming galaxies. Combined with HI data from MeerKat/SKA, \wst\ will diagnose multi-phase gas properties crucial for understanding the mutual interactions between gaseous inflow, outflow, and circumgalactic gas \citep{Fraternali2002, Levy2019} for systems of different stellar masses and distances from the main sequence. 

The above science cases collectively exploit the unique capabilities of \wst's large format IFS to study the resolved properties of the gaseous phase in nearby galaxies. However, the applications are highly synergistic with similar probes of the individual stars where they can be spatially resolved as is the case for the nearer galaxies (e.g. $D<3$ Mpc). In such cases, the MOS capability can additionally be brought into action. For convenience, these science cases based on \wst\ studies of resolved stellar populations are discussed separately in Section \ref{sec:respop}.

 \textbf{Observing strategy and targets}: \wst\ IFS, with its large $3' \times 3'$ field of view, is ideally suited for this Nearby Galaxies Reference Survey. For the largest closest ($D< 5$ Mpc) galaxies, mosaics will be necessary. \wst\ MOS can be used simultaneously to observe associated targets at larger galactocentric distances, including outlying HII regions, globular clusters, low-surface-brightness streams and satellites, thereby completing the full environmental census (see e.g. Fig. \ref{fig:NGC6397}). Source imaging data and catalogues from LSST ($0.36-1.0 ~\rm \mu m$) and Roman ($0.7- 3 ~ \rm \mu m$) will be publicly accessible several years before \wst\ starts operation.
 
The survey would map the inner discs ($r < \rm R_{25}$) of massive [$\log(M_\star/M_\odot)>9$] nearby galaxies with D$<$ 25 Mpc, where the average Paranal seeing of $0.8''$ corresponds to a resolution better than 100 pc, as required to resolve the average distance between HII regions. Using the z $=$ 0 Multiwavelength Galaxy Synthesis catalogue of nearby galaxies \citep{Leroy2019} and applying basic selection criteria (e.g. D$>$2 Mpc, Dec $< 20^\circ$, $|b_{gal}|>10^\circ$) leads to a sample of 1020 galaxies, requiring $\sim$ 1200 IFS pointings. This, effectively volume-limited, sample would have an enormous legacy value and could be carried out effectively in dark and/or grey nights.

\medskip
\subsection{Active Galactic Nuclei and their host galaxies}

\medskip
Ever since the discovery of quasars over 60 years ago \citep{3c273}, our understanding of active galactic nuclei, the growth of Super-massive Black Holes and their influence on their hosts' star formation have advanced significantly. We have now identified luminous quasars to the earliest epochs at redshift, z$>$7.5 \citep[e.g.,][]{banados18,yang20} with a new census of lower luminosity candidates emerging from surveys conducted with JWST \citep[e.g., ][]{larson23,maiolino23}. 

These advances arose through spectroscopic campaigns following up promising AGN candidates, from deep pencil-beam surveys \cite[e.g., Chandra Deep Field South, CDFS;][]{cdfs} to wide, all-sky studies of radio \cite[e.g., First Bright QSO Survey, FBQS;][]{fbqs}, near-infrared (e.g., UKIDSS, VHS), mid-infrared \cite[e.g., Spitzer Mid-infrared Active Galactic nucleus Survey, SMAGS;][]{lacy13}, X-ray \cite[e.g. XMM-XXL;][]{xmm-xxl} and optical surveys \cite[e.g., SDSS;][]{sdss-qsos}. To date, such campaigns have delivered over a million AGN spectra. Scientific highlights include (i) the discovery of broad absorption line quasars (BALQSOs), showing velocity-shifted absorption troughs \citep[e.g.,][]{trump06, dai08, 2022Natur.605..244B}, (ii) obscured quasars up to high $z$ across a wide luminosity range \citep[e.g.,][]{viitanen23}, (iii) a surprising lack of metallicity evolution in the broad line region at z$>$5 \citep[e.g.,][]{juarez09, 2020ApJ...905...51S, 2020ApJ...898..105O}, and (iv) characterisation of the evolving AGN luminosity function \citep[e.g.,][]{rankine24} and black hole mass function \citep[e.g.,][]{vestergaard09, 2022ApJ...941..106F}. 

Integral Field Spectroscopy has further revolutionised this field by mapping the host galaxy and AGN properties from the nearby universe out to the epoch of re-ionisation \citep[e.g.,][]{2019ApJ...887..196F}. In fact, IFU spectrographs have become workhorse instruments for studies at high $z$ on 8-10 m class telescopes (e.g., VLT/MUSE: \citet{muse}, Gemini/NIFS: \citet{nifs}, Keck/OSIRIS: \citet{osiris}) and for $z<1$ using SAMI \citep{sami}, SDSS/MANGA \citep{sdss-manga} and CALIFA \citep{Sanchez+2012}. By modelling the pixel-by-pixel continuum and emission line spectra, IFS surveys have mapped AGN outflows, determined how AGN affect interstellar medium (ISM) properties such as density, temperature and metallicity,and characterised the distribution of stellar populations in the vicinity of AGN. Together with state-of-the-art simulations, models have provided critical constraints on how AGN are fueled, how they assemble over time and how AGN feedback in the form of radiation pressure driven outflows or radio jets govern star formation in galaxies. Nonetheless, conflicting results remain and numerous questions about how AGN affect galaxy evolution remain unanswered.

In the next few years, new surveys will shape our understanding of how supermassive black holes form from initial seeds, how they grow and govern feedback in various types of galaxies. Many results based on \jwst\ data are claiming a larger fraction of AGN at high $z$ than earlier work suggested which raises questions about their role in cosmic reionisation. \euclid\ recently started its WIDE and DEEP surveys providing data for $\sim10^{9}$ galaxies with photometric redshifts of which at least $10^{6}$ are expected to be AGN. Ground-based surveys such as SDSS-V and DESI (and shortly Rubin/LSST) have detected numerous metal-poor AGN hosts \citep[e.g.,][]{zou24}. In the future, high resolution follow-up with extremely long-based (e.g. space) interferometry of SKA detected AGN up to the epoch of reionisation will allow us to resolve the torus at parsec or sub-parsec scales. Finally, in the era of \wst, ATHENA, the largest X-ray observatory to be launched in 2035, will unveil SMBH growth in heavily obscured environments by detecting thousands of Compton-thick AGN out to high redshift \citep[e.g.,][]{2022MNRAS.509.3015H}. Clearly alongside this voluminous multi-wavelength data, there will be a major requirement for rest-frame optical spectroscopy given its abundant number of diagnostic features.

Although DESI, SDSS-V and 4MOST will provide one-dimensional spectroscopic follow-up of LSST or SKA targets, strategic follow-up programmes with \wst\ IFS will characterise the host galaxy gas kinematics, stellar populations and ISM conditions. As an illustration rather than a complete inventory, we indicate science cases where \wst\ can transform our understanding of AGN and their role in regulating galaxy assembly. Of particular importance is the fact that \wst\ is a dedicated survey facility with a high multiplex advantage
and wide area IFS. This contrasts with the more powerful ELT where a more targeted strategy is appropriate. Indeed, we can envisage \wst\ being a natural feeder for more detailed studies of individual AGN with ELT.

We now turn to specific examples of questions that can only be address effectively with \wst.

\medskip
\noindent{\bf (i) \wst-SKA synergy: Jet-mode AGN feedback at low and high-redshift:} \wst\ spectroscopy will be crucial in the wholesale determination of redshifts for sources detected by SKA. The radio-based spectral energy distribution does not contain useful features for determining accurate redshifts. There is thus an obvious opportunity for a specialised SKA-WST survey aimed at establishing optical counterparts for a significant portion of SKA detections, including radio sources detected by instruments like MeerKAT. Leveraging the IFS capabilities of \wst, a SKA-\wst\ survey would not only provide redshifts but also yield valuable insights into the host galaxy properties including the ionisation radiation field, chemical abundances, stellar populations and star formation rate (SFR).

\begin{figure}
\centering
\includegraphics[scale=0.5]{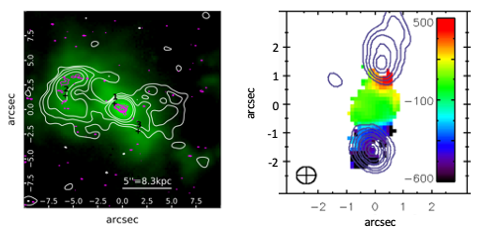}
\caption{Examples of the synergy between radio imaging and rest-frame optical IFS observations to unveil the interaction between radio jets and the host galaxy ISM. {\it Left:} [OIII]$\lambda$5007-based ionised gas morphology in the Teacup galaxy \citep[e.g.,][]{venturi23} where white contours show radio emission \citep[e.g.,][]{harrison15} carving out the ionised gas structure. {\it Right:} [OIII]$\lambda$5007 velocity field with GHz radio continuum contours (in purple) from VLA \citep{nesvadba17} in a z$\sim$2.4 radio AGN. The North radio lobe appears to have broken out of a cocoon of hot gas. SKA will identify several such jetted AGN at high $z$ and \wst\ follow-up can decipher jet-mode feedback for a large statistical.}
\label{fig:jetISM}
\end{figure}

Examining the interplay between radio jets and the ISM of jetted AGNs, and the role of jets in governing star formation is currently restricted to studies of low redshift galaxies (Figure \ref{fig:jetISM}). However, radiative feedback from growing black holes is expected to have its dominant impact at z$\sim$1--3, an epoch characterised by the peak of volume-averaged black hole growth rate. The proposed SKA-WST survey, exploiting the high sensitivity and multiplexing capabilities of \wst, will allow us to extend these studies to the more formative period in cosmic history. Even at low redshift, the \wst's unique IFS field-of-view will play a pivotal role in understanding the impact of radio hot spots on larger spatial scales. 

\medskip
\noindent{\bf (ii) New discovery space with \wst\ studies of a statistical sample of AGN:} \wst\ has the capability to obtain spectra of millions of AGN host galaxies which has the potential to reveal new scientific opportunities as summarised below:

\medskip
\noindent {\it Low-luminosity AGN}: The increased sensitivity of \wst\ will allow us to search for signatures of active black holes in fainter galaxies detectable using AGN variability with \lsst, and via non-thermal signatures with SKA and ATHENA. Traditionally, due to sensitivity limitations, low-luminosity AGN have not been studied in as much detail as quasars or moderate-to-high luminosity AGN.
Yet early \jwst\ results have indicated a likely substantial population of low luminosity AGN. Spatially-resolved spectroscopy with IFS will allow us to characterise the gas kinematics, stellar populations and ISM conditions in these low luminosity systems. Furthermore, resolved diagnostic diagrams will allow us to potentially detect recoil AGN, expected in off-nuclear locations in dwarf galaxies \citep[e.g.,][]{bellovary19}.  Complementary radio and X-ray observations with facilities described above will further constrain the interplay between jet-mode and radiative-mode feedback in these populations out to higher redshift. 

\medskip
\noindent {\it Binary or Dual AGN}: Supermassive black hole pairs or dual AGN systems provide crucial insight into merger-driven black hole growth scenarios. As such, these systems are prime candidates ESA's Laser Interferometric Space Antenna (LISA) mission, for which one of the major objective is to determine how supermassive black holes form and grow over cosmic time. \wst\ will play a key role in the detection of binary AGN at close separations via spatially-resolved spectroscopy \citep{mannucci23}. Furthermore, long-term temporal variability in spectra will be valuable to detect binary systems \citep[e.g.,][]{minev21}, further explained in the \wst\ Time-domain Section \ref{sec:timedomain}.

\medskip
\noindent {\it Demographic studies of AGN across cosmic time:} Through statistical analyses of extensive galaxy samples from SKA, ATHENA and LSST, \wst\ can effectively constrain the black hole mass function as a function of redshift. Moreover, constructing the rest-frame quasar UV luminosity functions offers crucial information on the concurrent evolution of black hole mass, galaxy luminosity and dark matter halo mass. Finally, the IFS can follow-up numerous Compton-thick (CT) AGN detectable with ATHENA up to z$\sim$3. CT-AGN are believed to represent a crucial phase during the growth of AGN and therefore, their resolved spectroscopic follow-up would indicate the presence and the associated energy of outflows driven by growing black holes.

\medskip
\noindent{\bf WST-AGN Time-domain science:}

\noindent As a dedicated facility \wst\ has the potential to conduct time-variability campaigns which will directly benefit the studies of AGN.  These opportunities are described in more detail in the Time-domain Section \ref{sec:timedomain} of but are briefly summarised here for convenience.

\begin{enumerate}
\item Due to its large aperture and advanced multiplexing capability, \wst\ will extend the reach of reverberation mapping experiments to fainter populations. This will yield black hole masses in a regime inaccessible with current spectroscopic facilities.
\item Puzzling rare events, such as Changing-look AGN (CL-AGN), are often discovered by means of long-term spectroscopic monitoring in the rest-frame optical wavelengths. These monitoring campaigns have largely relied on long-slit spectroscopy. Leveraging the IFS will enable comprehensive characterisation of the host galaxies associated with such rare examples, as well as AGN selected on their variability.
\item Broad absorption lines (BALs) in quasar spectra are thought to originate from turbulent accretion discs. Such quasars host extremely high velocity outflows, which are expected to introduce significant feedback that affects star formation in their host galaxies. Short to long-term spectroscopic monitoring (a few days to years) of a representative sample of BAL-QSOs will reveal diagnostic variability and provide valuable insight into the geometry and dynamics of these outflows and ultimately the accretion physics.  
\end{enumerate}

\medskip

\clearpage


\clearpage

\section{Cosmology}

\paragraph{Authors:} 
Jean-Paul Kneib$^2$, Andrea Cimatti$^4$, Richard I.\ Anderson$^2$, William d'Assignies D.$^{49}$, Emilio Bellini$^{23}$, Stefano Camera$^{36,37,38}$, Carmelita Carbone$^{27}$, Sofia Contarini$^{47}$, Stephanie Escoffier$^{56}$, Daniel Forero-S\'anchez$^2$, Vid Ir\v{s}i\v{c}$^{62,63}$, Benjamin Joachimi$^5$, Katarina Kraljic$^{66}$, Ofer Lahav$^5$, Khee-Gan Lee$^{73}$, Guillaume Mahler$^{76,7,8}$, Nicola Malavasi$^{10}$, Federico Marulli$^{4,16,78}$, Federico Montano$^{36,37}$, Michele Moresco$^{4,16}$, Chiara Moretti$^{23,92}$, Lauro Moscardini$^4$, Alice Pisani$^{101}$, Mamta Pandey-Pommier$^{98}$U, Johan Richard$^{13}$, Mickael Rigault$^{123}$, Antoine Rocher$^2$, Piero Rosati$^{105}$, Elena Sarpa$^{23}$, Carlo Schimd$^{29}$, Emiliano Sefusatti$^{107}$, Oem Trivedi$^{124}$, Maria Tsedrik$^9$, Aurelien Verdier$^2$, Francesco Verdiani$^{23}$, Giovanni Verza$^{111,112}$,\\ Matteo Viel$^{23,113,114,107}$, Pauline Vielzeuf$^{56}$, Christophe Y\`eche$^{119}$, Jiaxi Yu$^2$

\subsection{Introduction}

\begin{figure}[h]
    \centering
    \includegraphics[height=0.44\textwidth]{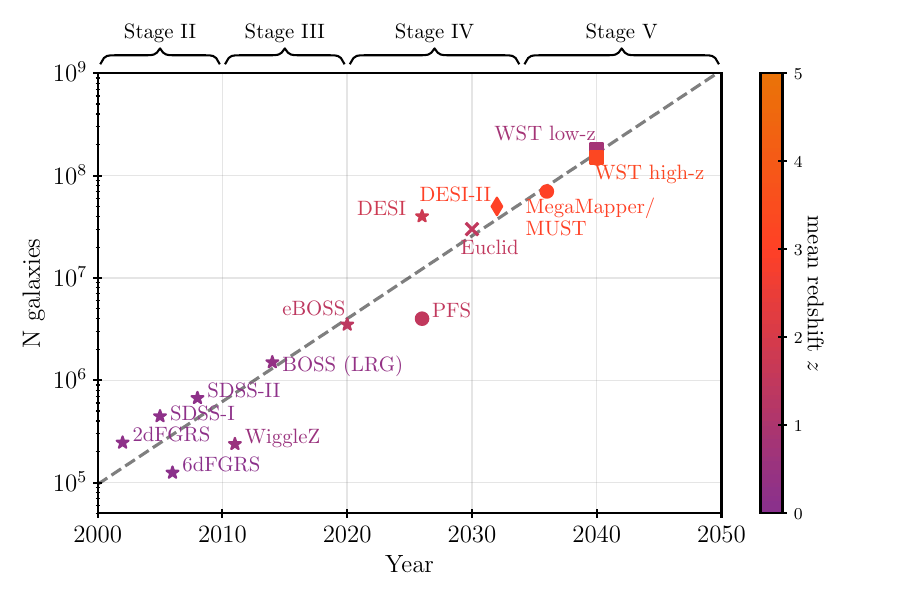}
        \includegraphics[height=0.44\textwidth]{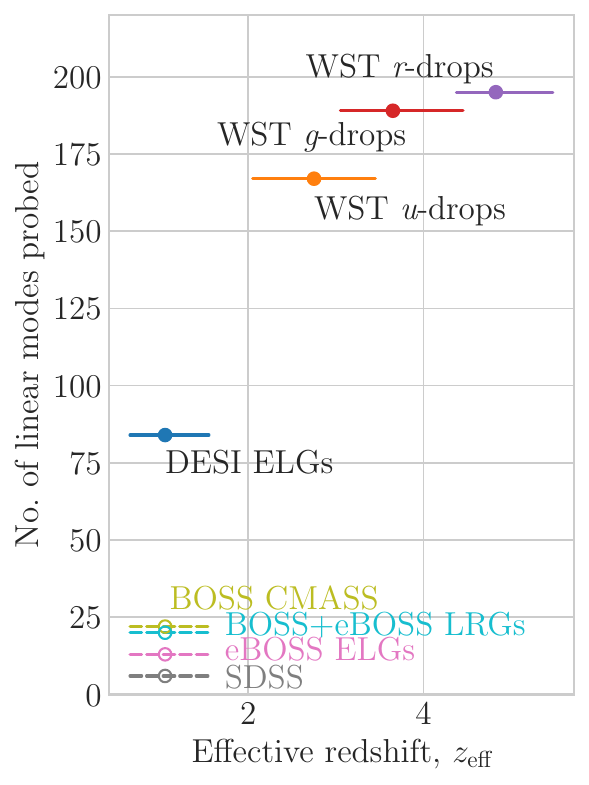} 
    \caption{\textit{Left panel:} Progress in wide cosmology spectroscopic redshift surveys from year 2000 to 2050. The dashed-line is qualitative and shows the improvement of the number of galaxy spectra over the years, the slope is fitted to the points and corresponds to a gain of a factor $\sim 8$ in the number of redshifts every 10 years. The colour scale indicates the mean redshift of each survey.
    Note that the different stages indicated at the top of the figure refers to the nomenclature of the Dark Energy Task Force report \citep{DEtaskforce2006}.
    \textit{Right panel:} Comparison of the number of linear modes probed by different surveys. The width of the horizontal bars gives the redshift range of the targets, with the abscissa of the circle marker corresponding to the effective redshift, and the ordinate being the number of linear modes probed.
    }
    \label{fig:cosmology-progress}
\end{figure}

Cosmology is experiencing a fruitful period since the confirmation of the accelerated expansion of the universe in 1998 \citep{SNIa1998}. This discovery inspired the exploration of the origin of this acceleration: whether due by the existence of a cosmological constant $\Lambda$, a dynamical dark energy that evolves with time, or a deviation from General Relativity (GR) \citep{DEtaskforce2006,BOSS2013,eBOSS2015}. The Cosmic Microwave Background (CMB) measurements and the Supernovae-Ia have established the $\Lambda$-Cold Dark Matter ($\Lambda$CDM) as the standard cosmological model, enlightening the dominant component of dark matter and dark energy in the energy content of the universe \citep{planck2016,Planck2020}. Meanwhile, imaging and spectroscopic surveys provide further constraints of dark energy and dark matter using weak lensing, galaxy clusters, galaxy clustering, and SNe Ia probes. 
In the early 2000s, pioneering multi-fibre experiments such as the  2-degree field galaxy redshift survey (2dFGRS) \citep{2005MNRAS.362..505C, 2001MNRAS.328.1039C} and the Sloan Digital Sky Survey 
(SDSS) \citep{Eisenstein:2005} 
measured hundred of thousands spectra of galaxies constraining for the first time the Baryonic Acoustic Oscillations to constrain cosmology with large-scale structures.
Further dedicated Baryonic Oscillation Spectroscopic  Surveys  (Stage-III surveys: SDSS-III/BOSS and SDSS-IV/eBOSS) have provided powerful probes encompassing  BAO studies, and gravity tests with redshift space distortion (RSD) \citep{BOSS2013, eBOSS2015} (Figure \ref{fig:cosmology-progress}).

The two decade-long ground-based spectroscopic survey, Sloan Digital Sky Surveys \citep[SDSS;][]{SDSS_I,SDSS_II,SDSS_III,SDSS_IV}, initiated in a systematic way the 3D galaxy mapping and clustering measurements. SDSS has measured the expansion history of the universe up to redshift 2.2 with luminous red galaxies (LRGs), emission line galaxies (ELGs), and quasi-stellar objects (QSOs), achieving a percentage-level precision in the angular-diameter distances and 6-percent precision in the structure growth rate measurements \citep{eBOSS_LRG_cosmology_2PCF,eBOSS_LRG_cosmology_Pk,eBOSS_ELG_cosmology_2PCF,eBOSS_ELG_cosmology_Pk,eBOSS_QSO_cosmology_2PCF,eBOSS_QSO_cosmology_Pk}. Combined with CMB, SNIa and weak lensing results from the Dark Energy Survey (DES), SDSS has enhanced the validity of GR and $\Lambda$CDM while opening a window for the time-evolving equation-of-state for dark energy \citep{SDSS_cosmology}. 
With similar galaxy and quasar probes, the largest ongoing spectroscopic survey, the Dark Energy Spectroscopic Instrument \citep[DESI: the first Stage-IV survey][]{DESI2016a} observes a 40\% larger footprint with at least 2.5 times higher object surface density than SDSS. \desi\ aims to achieve sub-percent-precision cosmological measurements. The other forthcoming Stage-IV redshift surveys: \fourmost, \pfs, and \textit{Euclid} will complement the \desi\ observations in redshift range and sky-coverage. The 4-meter Multi-Object Spectroscopic Telescope \citep[4MOST;][]{4MOST,4MOST_cosmology} aims to cover the Southern-Hemisphere (not covered by \desi) targeting Bright galaxies (BG), LRGs, and QSOs. The Subaru Prime Focus Spectrograph \citep[PFS;][]{PFS} focuses on a smaller footprint of 1'400 
 deg$^2$ mapping galaxies and quasars at $0.8<z<2.4$. 
 The \textit{Euclid} space mission \citep[][]{Euclid} is about to start its survey including measuring grism-spectra of $0.9<z<2$ galaxies.
 In addition to the attempt to differentiate dark energy models, spectroscopic surveys in the current stage also explore new questions in cosmology: Is the Gaussianity assumption of the primordial universe a valid assumption? What is the hierarchy of neutrinos?

To further address these unknowns, the cosmology redshift survey community is advocating Stage-V spectroscopic surveys such as MUltiplexed Survey Telescope (MUST\footnote{\url{https://must.astro.tsinghua.edu.cn}}), MegaMapper \citep{megamapper}, MaunaKea Spectroscopic Explorer \citep[MSE;][]{MSE_Instrument,MSE_Science} and WST. Using large aperture ($>4$meter) telescope(s) those surveys plan to use Lyman Break Galaxies (LBGs) and Lyman-alpha emitters (LAEs) to extend the precise large-scale universe mapping up to redshift 5 in the next decade(s), enabling the observation of the linear modes in the primordial universe \citep{survey_roadmap}. 
\desi\ collaboration also prepares for an extension of \desi\ {(DESI-II; P5 report\footnote{\url{https://www.usparticlephysics.org/2023-p5-report/assets/pdf/P5Report2023_121023-DRAFT_single-pages.pdf}})} as a pilot survey for future projects. \cref{fig:cosmology-progress} displays a representation of past, current, and planned spectroscopic surveys over the years along with the number of linear modes probed by each survey. 

Better understanding of the cosmological model and related fundamental physics questions will come from improved spectroscopic surveys and cross-analysis with other cosmological probes \citep[e.g., ][]{SDSS_cosmology}. The imaging surveys of the \textit{Euclid} space mission \citep{Euclid} and the LSST of the Rubin Observatory \citep{LSST} will provide the targets for future spectroscopic surveys. 

The combined analysis of Stage-V surveys and the next-generation CMB experiment, CMB-S4 \citep{CMB_S4} and LiteBIRD \citep{litebird}, will also put tighter constraints on parameters. Similarly, 21-cm intensity mapping surveys such as the Hydrogen Intensity and Real-time Analysis experiment \citep[HIRAX][]{HIRAX} will achieve BAO measurement at $0.8<z<2.5$ in the coming years, that will be enhanced by cross-analysis with the Stage-V surveys.

\begin{figure}[h]
    \centering
    \includegraphics[width=0.8\textwidth]{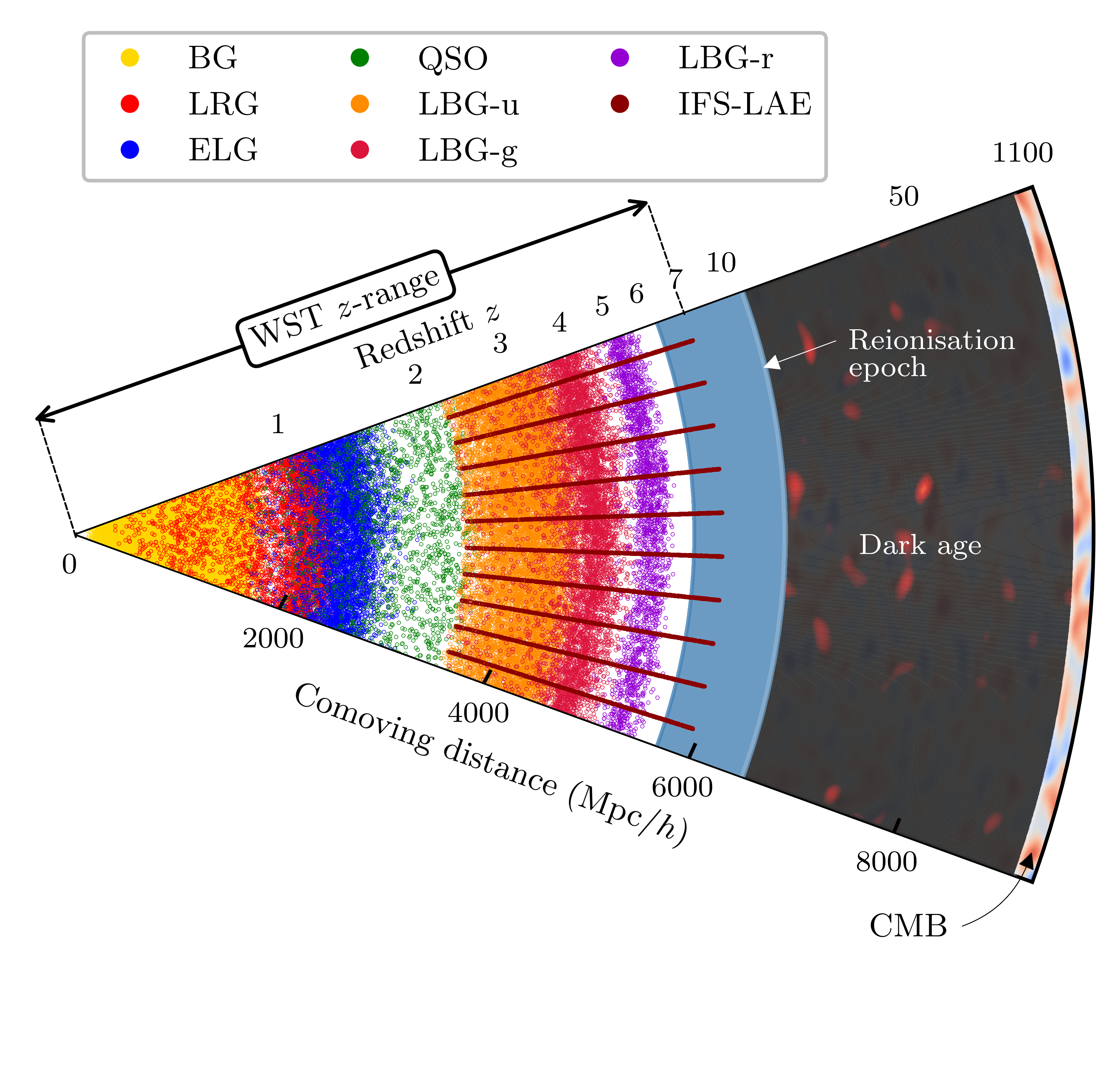}
    \caption{Schematic light-cone representation of the \wst\ Cosmology Surveys. \wst\ unique feature is its capacity to probe large-scale structures with galaxies in the redshift range $2<z<7$. MOS-LR targets will probe the range $0<z<5.5$ while IFS will probe thousands of pencil beams with LAEs up to $z\sim7$. }
    \label{fig:wst-lightcone}
\end{figure}

The Wide-field Spectroscopic Telescope with its large telescope aperture (12m or $9\times$ the collecting surface of \desi) and its high density of fiber ($20\,000$ fibers over $3.1\,\deg^2$ or $\sim10\times$ the fiber density of \desi) has all the characteristics of a competitive Stage-V survey. \wst\ will be capable to map the large-scale structure of the Universe with galaxies up to redshift $z\sim 5$ measuring precisely the local primordial non-Gaussianity parameter, $f_{NL}$, the sum of neutrino masses, as well as probing early dark energy models. \wst\ will also provide a huge legacy survey at $z<\sim 2$ enabling new cosmology science in particular on non-linear scales. The IFS will be also a unique opportunity to perform BAO measurement up to $z \sim 7$. A representation of the cosmological surveys is shown in \cref{fig:wst-lightcone} and the survey design is detailed in \cref{sec:survey_design}.

\begin{figure}[ht]
  \centering
\includegraphics[width=\textwidth]{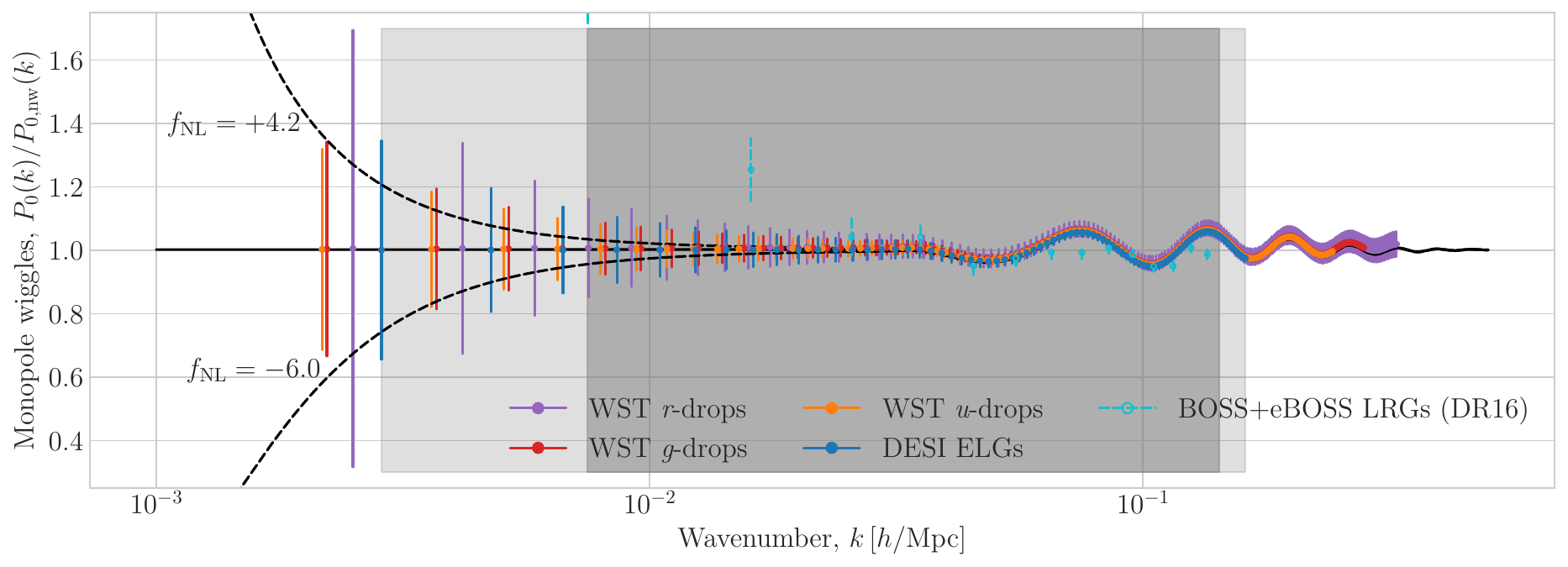}
   \caption{
   Comparison of \wst\ to state-of-the-art and Stage-IV ground-based spectroscopic galaxy surveys for the ratio of the monopole of the galaxy power spectrum to its smooth, `no-wiggle' broadband shape. The black, solid curve is the theoretical prediction in \(\Lambda\)CDM; empty cyan markers and dashed error bars are from current data \citep[BOSS+eBOSS LRGs DR16][]{eBOSS_LRG_cosmology_Pk} and forecasts for Stage-IV surveys \citep[][]{DESI2016a}, whist filled markers and solid error bars are forecasts for the various \wst\ samples (see legend). The grey-shaded areas mark the range of linear scales accessible currently (eBOSS, dark grey) and in the coming future (\desi, light grey). Finally, black dotted curves bracket the \(68\%\) credible intervals on the value of the local primordial non-Gaussianity parameter, \(f_{\rm NL}\), as measured by the \textit{Planck} satellite. By reaching smaller (higher $k$) and larger (smaller $k$) scales, \wst\ will bring stronger constraints on neutrino mass and $f_{NL}$ measurement compare to previous experiments.}
   \label{fig:p_k}
\end{figure}

\subsection{Opening-up high-redshift cosmology}

After the successful investigation of the \desi\ and (soon) \fourmost\ surveys in probing Dark Energy thanks to BAO measurements, the \wst\ Cosmology survey will conduct sub-percent galaxy clustering measurement beyond redshift two. Through clustering analysis using the power spectrum, $P(k)$, and the bispectrum, $B(k)$, we will conduct precise measurement of the large-scale structure of the universe in a broad $k$ range from $k\sim 0.002.h$/Mpc to $k\sim 0.3.h$/Mpc (equivalently from $\sim 20$ Mpc/$h$ to $\sim 3$ Gpc/$h$ comoving separation)  (see \cref{fig:p_k}). Thanks to these new measurements, sensitive constraints can be placed on the level of non-gaussianity, the sum of neutrino mass---expecting to discriminate between the normal and inverse hierarchy (see \cref{section:neutrino}), but also inflationary model, precise $H_0$ value, modified gravity and the early expansion. 

\begin{figure}
    \centering
    \includegraphics[width=0.7\textwidth]{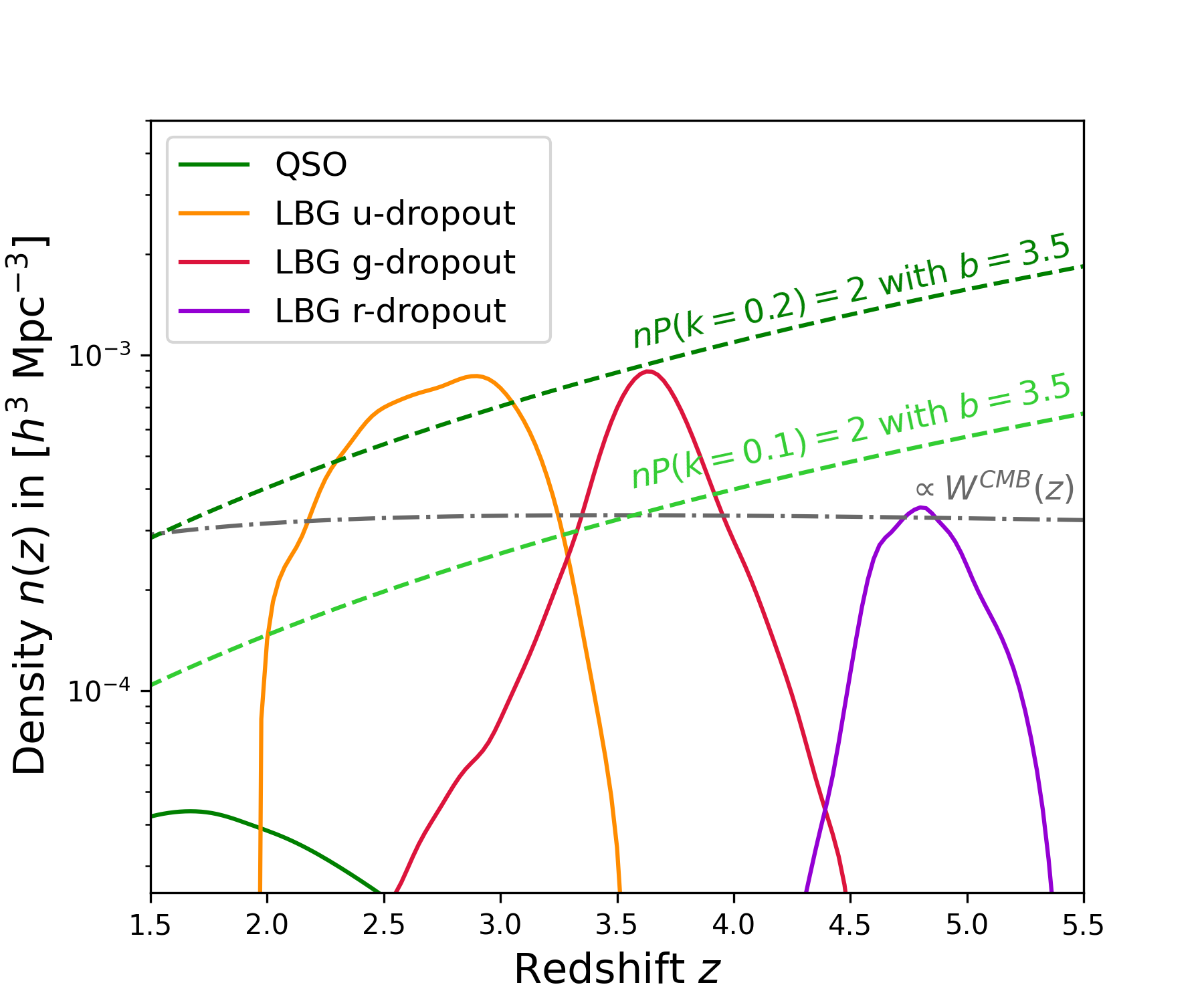}
    \caption{Expected spectroscopic redshift distribution of galaxies for the high-$z$ survey. Green dashed-lines show the qualitative parameter $nP(k=0.1, k=0.2)=2$ and the grey dashed-line shows the CMB lensing kernel as a function of redshift.
    }
    \label{fig:Nz_zhigh2}
\end{figure}

To achieve this, we will take advantage of the efficient Lyman-Break Galaxies (LBG) selection techniques at $2<z<5$ that will be available with the deep $u$-, $g$-, and $r$-band \lsst\ photometry.
The high redshift cosmology survey will also include quasars at $z>2$ selected by their photometry or by variability criteria to complement the LBG tracers and allow cross-correlation studies with the quasar itself and the Lyman-$\alpha$ forest. \cref{fig:Nz_zhigh2} shows the expected spectroscopic distribution of objects for the high-$z$ survey.

\subsubsection{Dark energy, growth and BAO}

\begin{figure}
    \centering
    \includegraphics[width=0.9\textwidth]{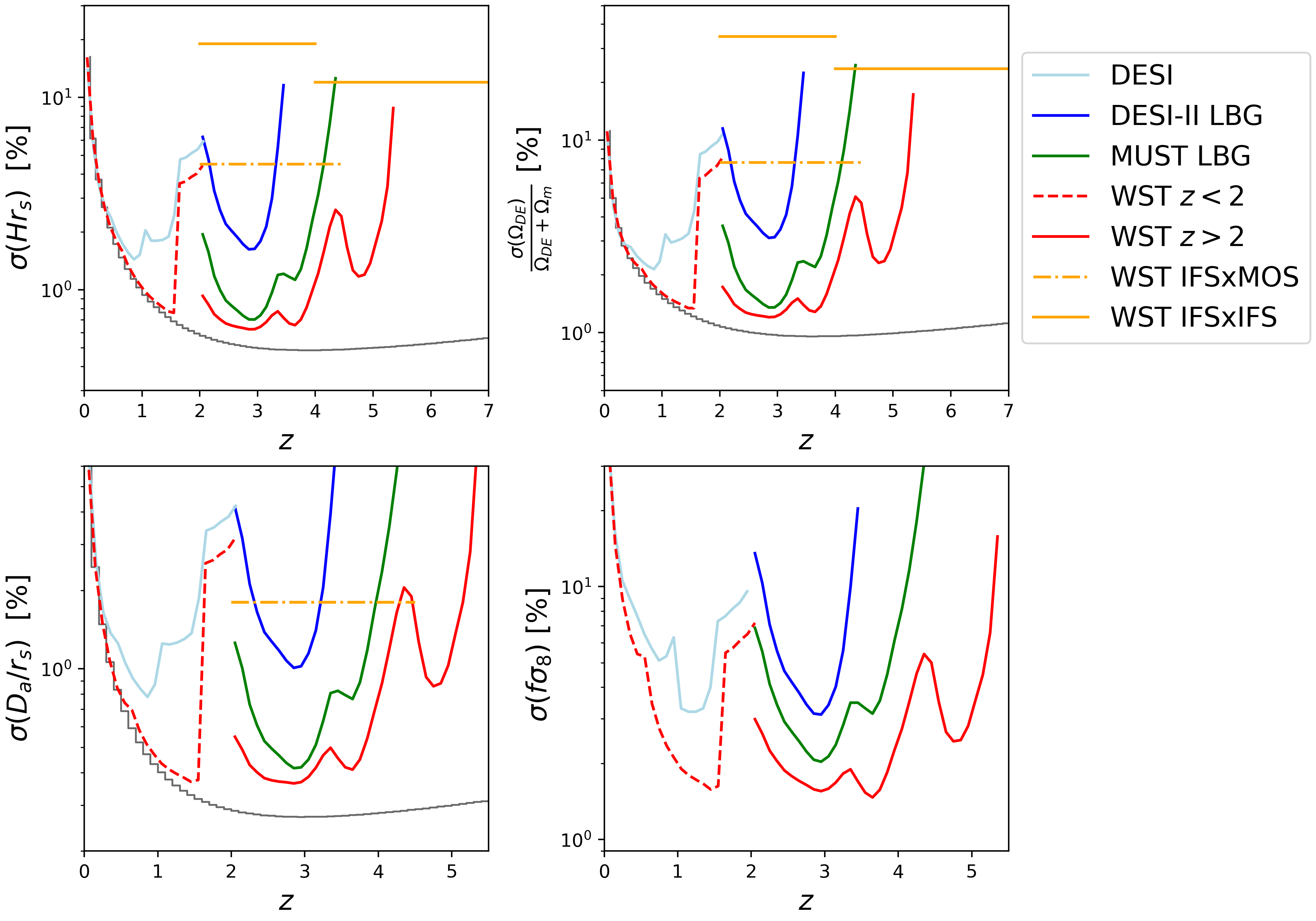}
    \caption{
    Forecast sensitivity of \wst\ on BAO parameters $D_A(z)/r_{\rm s}$ (\textit{bottom left}), $H(z)r_{\rm s}$ (\textit{top left}), and on structure growth $f\sigma_8$ (\textit{bottom right}) compare to \desi\ (\textit{light-blue lines}), DESI-II (\textit{darkblue lines}), and Megamapper/MUST (\textit{green lines}). The sensitivity of BAO measurements is also propagated to a dynamical dark energy prediction $\sigma(\Omega_{\rm DE})$ (\textit{top right}). Contribution from the legacy survey (and $z<2$ QSO) is shown in \textit{red dashed-lines}, from the high-$z$ survey ($2<z<5.5$) in \textit{red solid lines}. We also include the additional measurements IFS, either from the cross-correlation with the MOS-LR in \textit{yellow dashed lines}, and from its auto-correlation in \textit{yellow solid lines} (up to $z\sim7$). The grey lines represent the cosmic variance limit.}
    \label{fig:forecast_BAO_growth}
\end{figure}

 Galaxy clustering is, to date, a key probe in cosmology to constraint Dark Energy with the measurement of the scale of baryon acoustic oscillations (BAO) and gravity by measuring the growth rate of structure, $f$, using redshift space distortions induced by peculiar velocity of galaxies. The BAO scale was measured with first detection by Sloan Digital Sky Survey (SDSS) \citep{Eisenstein:2005} and 2dFGRS \citep{cole_2df_2005} and further investigated with SDSS/BOSS and eBOSS \citep{SDSS_cosmology}. \desi\  will provide precise measurement of the BAO scale and the growth rate of structure $f$ over a broad range of galaxy redshifts from $0<z<1.6$ \citep{DESI_validation}. 

\wst\ will similarly be able to constrain the scale of the BAO and the growth of structure with unprecedented precision in a new redshift window: $2 < z < 5.5$. Galaxies such as LBGs will be used to map the universe then dominated by matter, measure the expansion that is decelerating and probe the structures growth with high efficiency (assuming a standard flat-$\Lambda$CDM model). \wst\ will also improve significantly the $z<1.6$ constraints from Stage IV spectroscopic surveys, by reaching the cosmic variance limit.

Using the number densities described in \cref{sec:high_z_survey} and \cref{fig:Nz_zhigh2},  over a footprint of $15\,000$ deg$^2$, we can predict the sensitivity of $D_A(z)/r_{\rm s}$, $H(z)r_{\rm s}$, $\Omega_{\rm DE}$ and $f\sigma_8$ for individual redshift bins. $D_A$ is the angular diameter distance, $r_{\rm s}$ is the sound horizon at the drag epoch, $H$ is the Hubble parameter, $f$ is the growth factor, $\sigma_8$ is the amplitude of density fluctuations on the scale of $8\,h^{-1}\mathrm{Mpc}$, and $\Omega_{\rm DE}$ is the dark energy density parameter. Anticipating over the \cref{sec:legacy_survey,sec:Lyman-alpha blind survey}, we also report the forecasts for the Legacy low-redshift mapping $z<1.6$ (\cref{sec:legacy_survey}) and the Lyman-$\alpha$ parallel (\cref{sec:Lyman-alpha blind survey}) surveys. The forecasts were conducted using a Fisher matrix approach, detailed in \citet{eboss_forecast} and \citet{fisher_forecast_dassignies} and were run for $\Delta z=0.1$ redshift bins, except for the IFS-LAE, where only one or two large redshift bins have been used. The results are presented in \cref{fig:forecast_BAO_growth}. Due to the lack of current measurements of galaxy bias at high redshift, we consider a single galaxy bias value for all LBG samples  $b_{\rm LBG}=3.5$, and a galaxy bias  $b_{\rm LAE}=2.5$ for the LAE-IFS sample. The bias model used for QSO is taken from \cite{bias_QSO} and for low-$z$ samples, we used $b_{\rm BGS}=1.34/D(z)$, $b_{\rm LRG}=1.7/D(z)$, $b_{\rm ELG}=0.84/D(z)$, with $D(z)$ the growth function normalised at present time \citep{desicollaboration2016desi}. We assume a sky coverage of $5000\,\deg^2$ for DESI-II, and $15\,000\,\deg^2$ for Megamapper/MUST. \desi\ forecast results are taken from \cite{DESI_validation}.

Overall, \wst\ will provide constraints on dark energy and gravity at redshifts that are not yet probed with current LSS surveys, and will be highly competitive with respect to similar future surveys at that time. In addition, LAEs observed with the IFS will provide the first opportunity to constrain cosmology up to redshift $\sim$7 (see \cref{sec:Lyman-alpha blind survey}). 
The \wst\ MOS-LR constraints are represented in red, with dashed lines corresponding to the legacy low-$z$ survey, and solid lines to the high redshift one. For $D_A(z)/r_{\rm s}$, $H(z)r_{\rm s}$, $\Omega_{DE}$ we predict that \wst\ measurement will reach the cosmic variance limit at redshift $z<1.5$, improving significantly \desi\ predicted results (solid light blue line). \wst\ will also significantly improve the  $f\sigma_8$ measurements at low-$z$ compared to \desi\ (in particular with the $0.6<z<1.6$ ELG sample). At higher redshifts, the constraints from the LBG sample of \wst\ cover a much larger redshift range ($2<z<5.5$) compared to DESI-II (solid blue line) and Megamapper/MUST (solid green line) but also perform significantly better on the overlapping redshift range, with improvement between 30 to 300 $\%$. 

The precision on the growth rate of structure measurement, $f\sigma_8$, is improved by $\sim 35\%$ compared to Megamapper/MUST over the entire redshift range $2<z<4.5$ reaching a precision of $<2\%$ around $z\sim 3$. 
For the BAO parameters, $D_A(z)/r_{\rm s}$ and $H(z)r_{\rm s}$, the precision is significantly improved compared to \desi\ and DESI-II and will be $15-20\%$ better compare to Megamapper/MUST. The precision will reach up to $\sim 0.4\%$ and $\sim 0.7\%$ at redshift $2<z<3.5$ for $D_A(z)/r_{\rm s}$ and $H(z)r_{\rm s}$ respectively. The Lyman-$\alpha$ parallel survey will be able to constrain radial BAO measurements to a precision of $\sim 10\%$ for the first time in a redshift range $4<z<7$. All the radial BAO results are propagated to a dynamical dark energy forecast as in \cite{Sailer_forecasts}.
We use the  measurement of $h(z)$ from BAO and a CMB prior for $\sigma (\omega_{\rm m})=2.4\times 10^{-4}$. The constraints on dark energy can be written as $\sigma_{\Omega_{\rm DE}}^2=\Omega_{\rm m}^2(z)(4\sigma(\log h(z))^2+\sigma(\log \omega_{\rm m})^2)$ (assuming a flat-$\Lambda$CDM universe matter dominated, i.e.\ $\Omega_{\rm DE}(z)+\Omega_{\rm m}(z)=1$). \wst\ will also outperform the other foreseen spectroscopic experiment (Megamapper/MUST), with better precision, measuring the fraction of DE at 1.5 $\%$ at redshifts 0.8--1.5 and 2--4  and over unexplored redshift range: 3 $\%$ at redshifts 4.2--5.

\subsubsection{Probing fundamental physics: Neutrinos, Non-Gaussianities and Relativistic Doppler} 
\label{section:neutrino}

\paragraph{Neutrinos mass\\}

\begin{figure}[h!]
\centering
\includegraphics[width=\textwidth]{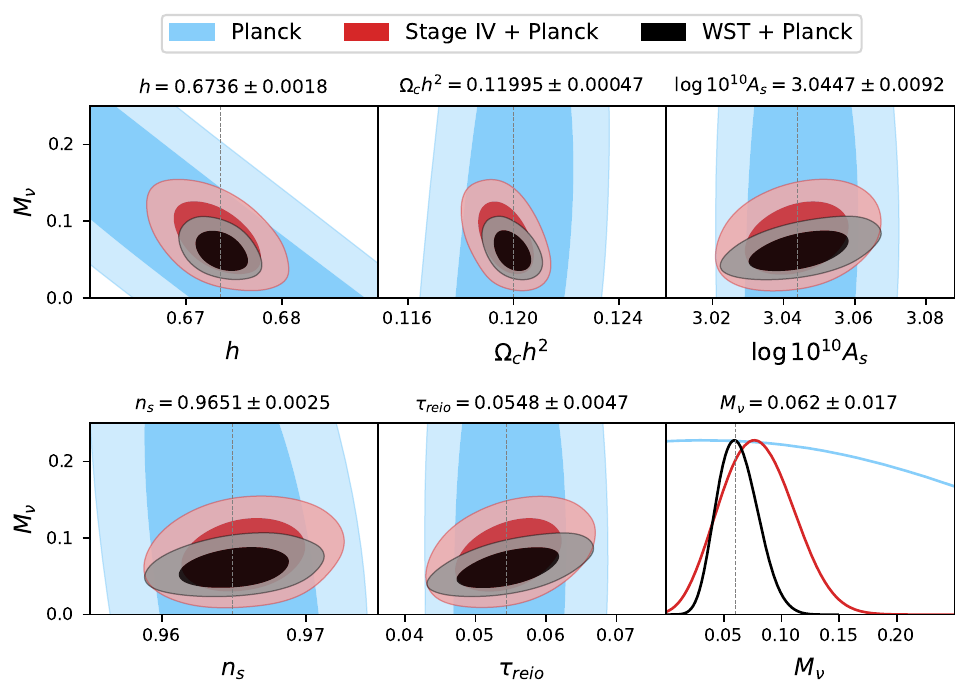}
\caption{Results of the MCMC forecast, in black, combining \wst\ total LBG sample power spectrum and Planck likelihoods. The power spectrum has been fitted employing the fully nonlinear model up to $k_{\mathrm{max}}=0.3 \, h\mathrm{Mpc}^{-1}$. Focusing on $M_\nu$, a comparison of the \wst\ survey constraints to those of all-sky space mission dataset is reported as well. Vertical lines represent the fiducial values chosen for the dataset. 
}
\label{fig:mnu-eft}
\end{figure}

The observation of flavor oscillations of atmospheric and solar neutrinos \citep{ahmad02,fukuda98} as well as oscillation studies at reactors and accelerators unequivocally prove neutrinos to possess non-zero rest masses \citep[e.g.][]{esteban19}, contradicting the Standard Model expectation of them being massless. These experiments return a lower limit on the total neutrino mass of  0.056 eV, while Tritium beta-decay experiments provide an upper limit of about 1.1 eV on the electron neutrino \citep{katrin19}.

In the context of the cosmological model, massive neutrinos contribute at low redshift to the matter content and their free-streaming is responsible for a characteristic small-scale suppression in the power spectrum of matter density perturbations. This feature has been studied extensively both at a linear and non-linear level, on several observables like galaxy clustering, weak lensing, voids, or the Lyman-$\alpha$ forest \citep[see e.g.][]{lesgourgues13}. It is interesting to notice that the effect on the matter power spectrum becomes relevant for wavenumbers $k \gtrsim 0.1\,h\,{\rm Mpc}^{-1}$, well within the quasi-linear regime where we can take advantage of analytical modelling based on perturbation theory. In linear theory, the suppression with respect to the massless neutrino case saturates at $\Delta\,P(k)/P(k) \sim -8\, f_{\nu}$, a value directly proportional to the neutrino density fraction $f_{\nu}=\Omega_{\nu}/\Omega_{\rm m}$, while in the non-linear case it increases to $-10\, f_{\nu}$. The fraction $f_{\nu}$ is in turn proportional to the sum of neutrino masses $M_{\nu}=\Sigma_{i=1,3} m_i$. This means that a detection of the effect can lead to the determination of the neutrino mass scale, independent (and complementary) to laboratory experiments. 

At present a variety of cosmological observations already provide upper limits to $M_\nu$ a factor 10 more stringent than beta-decay experiments. These limits, in the ballpark of $0.12$--$0.3\,\mathrm{eV}$ at $95\%$ C.L., have been obtained from CMB observations combined with BAO and SNe \citep{Planck2020},  the Lyman-$\alpha$ forest \citep{Palanque2016}, galaxy clustering  \citep{ivanov,tanseri} and cluster abundance \citep{eRosita_2024arXiv240208458G}. These results show the strong potential for discovery from cosmological surveys in the upcoming years.

To assess the sensitivity of the \wst\ survey to neutrino mass, we produce forecasts by running Monte-Carlo Markov chains on a synthetic dataset. In particular, we perform a full-shape analysis considering the three multipoles of the redshift-space galaxy power spectrum, assuming a minimal $\Lambda$CDM with massive neutrinos cosmology.
The theoretical model adopted is based on the Effective Field Theory of the Large Scale Structure \citep{baumann2012, carrasco2012}, as implemented in the \texttt{PBJ} code \citep{oddo2020, oddo2021, moretti2023}.

The results of this analysis, performed in combination with a mock Planck likelihood, are reported in \cref{fig:mnu-eft}.
In the conservative case of a fiducial $M_\nu = 0.06 \, \mathrm{eV}$ (the lower limit from neutrino oscillations) we forecast a precision of $\sigma_{M_\nu}=0.017 \, \mathrm{eV}$ for the \wst\ LBGs survey, a substantial improvement with respect to e.g.\ a Stage IV survey where the same analysis yields $\sigma_{M_\nu}=0.031 \, \mathrm{eV}$. Given the lower limit from laboratory experiments, this forecast for \wst\ can be interpreted as a piece of evidence for non-null neutrino masses at $3.5 \sigma$ significance, while a larger value can be expected from larger fiducial $M_\nu$. \\

\paragraph{Primordial non-Gaussianities \\}

\begin{figure}
    \centering
    \includegraphics[scale=0.8]{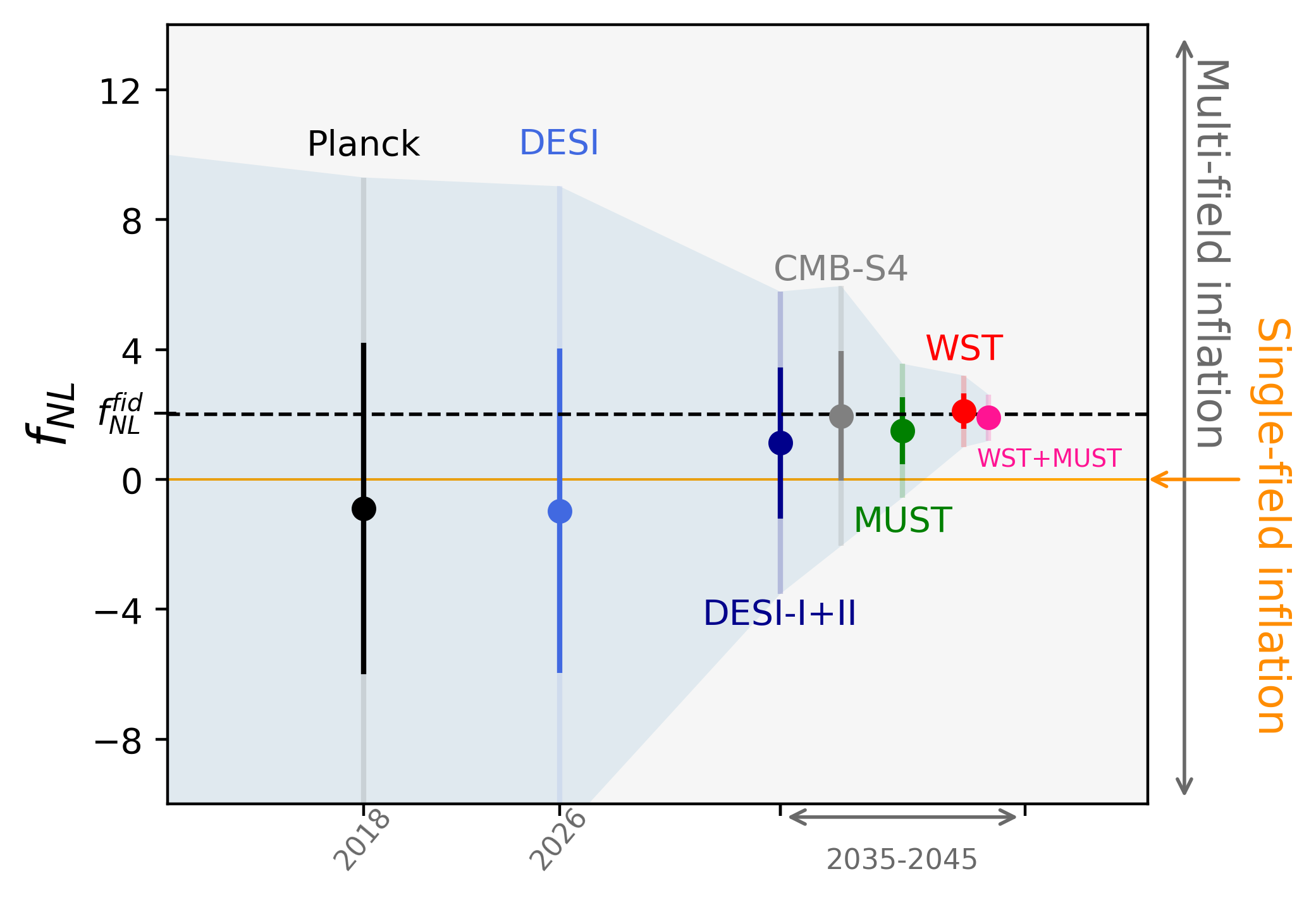}
    \caption{Forecast on local non-Gaussianity (errorbars) for future LSS surveys (power-spectrum only), CMB-S4, and Planck's best constraint.  We highlight the predicted domain of $f_{\rm NL}$ for multi-field (in grey) and single-field inflation (in orange). We illustrate potential detection with a fiducial $f_{\rm NL}=2$ (dashed line).}
    \label{fig:fnl_inflation}
\end{figure}

As galaxy redshift surveys cover increasingly large volumes, they become more and more competitive with CMB observations \citep{Planck2020} as a probe of primordial perturbations and, therefore, an important test for inflationary models beyond the standard single-field, slow-roll scenario. Two effects are relevant in this respect. The first is a possible primordial component to the galaxy bispectrum \citep{ScoccimarroSefusattiZaldarriaga2004, SefusattiKomatsu2007}. This is indeed the only potentially measurable effect of “non-local” primordial non-Gaussianity (described by the equilateral and orthogonal shape-dependences), resulting typically from single-field models. The second effect is a scale-dependent correction to linear galaxy bias, $\Delta b (k)\propto f_{\rm NL}(b-1)/k^2$, particularly significant at large scales \citep{NG_Dalal2008, Matarrese_Verde_2008, Carbone_etal2008, Grossi_etal2009}. Such correction is induced by “local” primordial non-Gaussianity, characterised by a large signal in squeezed bispectrum configurations, a specific prediction of multi-field inflation. Recent galaxy clustering observations are indeed providing interesting constraints on both non-local and local non-Gaussianity \citep{CabassEtal2022, CabassEtal2022B, DAmicoEtal2022A}. 

Given the very large volume probed and thanks to the large expected bias of the LBG samples, \wst\ will be a breakthrough experiment to constrain primordial non-Gaussianities and potentially detect the signature of multi-field inflation. In addition, a high galaxy density will allow to beat down shot noise at small scales, where most of the bispectrum signal is distributed over a large number of triangular configurations. This has the potential to significantly improve current limits on orthogonal and equilateral primordial non-Gaussianity.   

Here we limit ourselves to forecasts the potential of \wst\ to constrain local primordial non-Gaussianity, the model where large-scale structures are closer to CMB ones. The current best constraint from the Planck mission is $f_{\rm NL}=-0.9\pm 5.1$ \citep{Planck_fnl}. The large-scale structure data provide $-4<f_{\rm NL}<27$ at 68\% C.L. instead \citep{Cagliari2023}.

We show the results of our forecast for the combination of \wst\ high-$z$ LBGs and low-$z$ objects, and assuming the maximal mode $k_{\rm max}=0.1\,h\,\mathrm{Mpc}^{-1}$, in \cref{fig:fnl_inflation}. We compare it with the expected uncertainties for \desi, DESI-II, MUST, \citep[see, e.g.][]{NG_bispectrum, Barreira2022, fisher_forecast_dassignies} and the best constraints from the Planck Collaboration, along with future projections from the Stage-IV experiments \citep[S4, ][]{PNG_S4}. Our findings suggest a precision of $\sigma(f_{\rm NL})\approx 0.55$ with \wst,  a factor 2  better than Megamapper/MUST, and a factor 5 better than \desi (I+II)  or S4. The better performance of \wst\ with respect to Megamapper/MUST is mainly due to the larger volume, with a marginal contribution from a multi-tracer analysis \citep{Seljak_2009} allowed by the \wst\ samples. The \wst\ measurement would be the first reaching sub-unity precision, a critical milestone to constrain inflationary models, potentially able to differentiate between single and multi-field inflation. To illustrate this point, in \cref{fig:fnl_inflation} we assume a fiducial value $f_{\rm NL}=2$ as one could expect in a multi-field inflation model. \wst\ could then exclude a single-field inflation scenario at least at the 3$\sigma$ level. The constraints can be further improved to $\sigma(f_{\rm NL})\approx 0.36$  by joining constraints with Megamapper/MUST, since the latter covers a different volume at high-$z$. Stage-V spectroscopic surveys, and \wst\ in particular, appear to be among the most promising projects for advancing our understanding of the high-energy physics of inflation. 

\paragraph{Relativistic Doppler effect \\}

\begin{figure}[h!]
    \centering
    \includegraphics[width=0.49\textwidth]{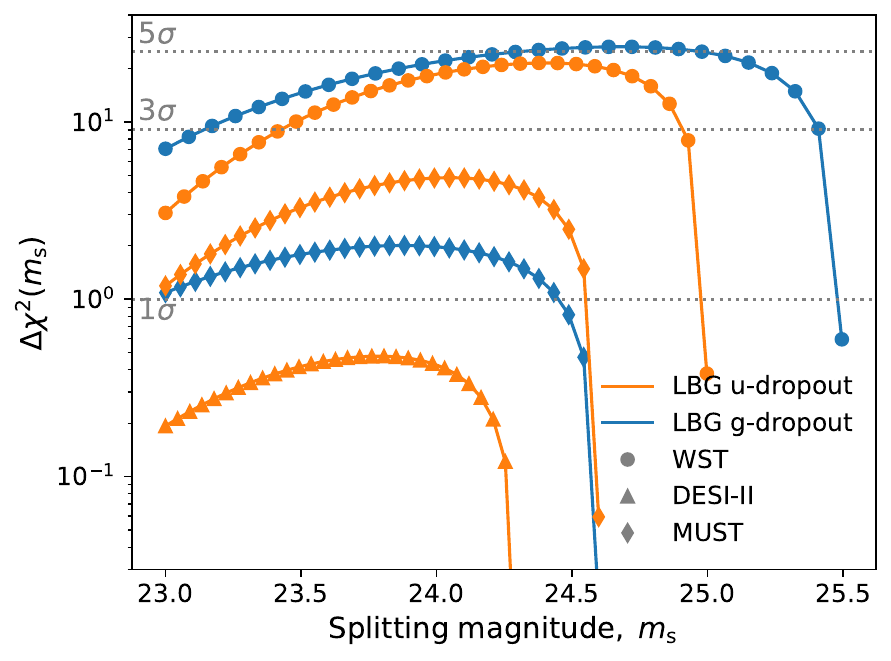}
    \includegraphics[width=0.49\textwidth]{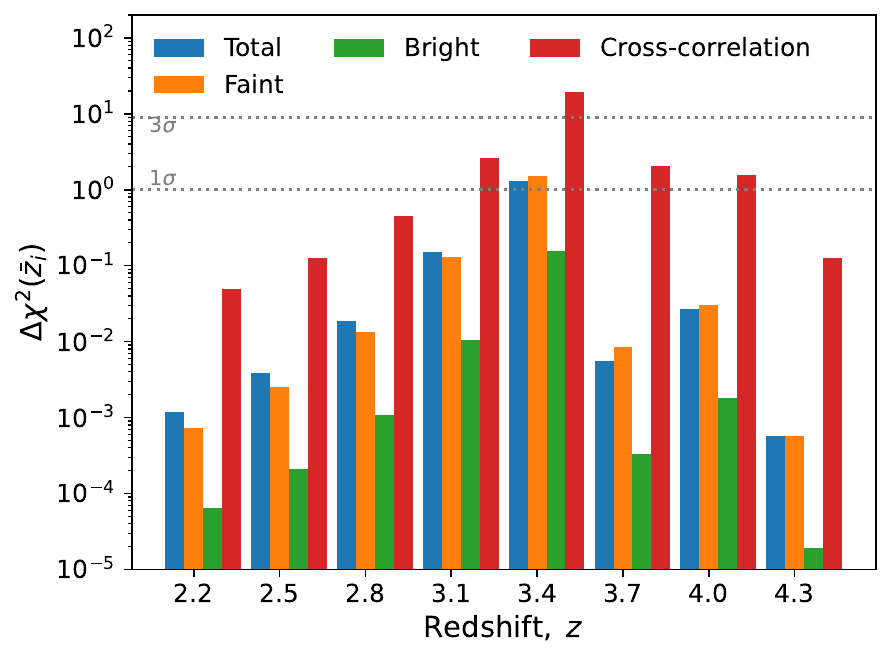}
    \caption{Left: cumulative statistical significance for the Doppler contribution in a faint-bright cross-power spectrum measurement as a function of $m_{\rm s}$. Orange curves are expressed in terms of $i$ magnitude and refer to LBG $g$-dropout; whilst blue $r$ magnitude-based lines are for LBG-u samples. DESI-II ($5000 \; {\rm deg^2}$), Megamapper/MUST  ($15000 \; {\rm deg^2}$), and \wst\  ($15000 \; {\rm deg^2}$) values are reported with triangles, diamonds and circles, respectively. 
    Right: Detection significance of relativistic Doppler contribution for \wst\ LBG-g dropout sample. The $\varDelta\chi^2$ variable is evaluated against a null hypothesis of no Doppler contribution in the galaxy power spectrum. Colour code: blue, orange, and green respectively for the auto-correlation power spectrum of total, faint, and bright samples; red is for the faint-bright cross-correlation power spectrum. Being the maximum $i$ magnitude of $25.5$, we set the splitting magnitude to $m_{\rm s}=24.5$.}
    \label{fig:Doppler_chi2}
\end{figure}

The cosmic LSS is an invaluable test bench to study gravity. As such, it is continuously probed to assess the reliability of our description of the gravitational interaction provided by GR on both large and small scales. Fluctuations in number counts of galaxies---and, in general, any other biased tracers of the LSS---are customarily described at the level of linear perturbations in terms of the density contrast of matter fluctuations, modulated by the linear galaxy bias, and redshift-space distortions. To these, the weak-lensing effect of cosmic magnification should also be taken into account, especially for high-redshift samples. However, the proper relativistic treatment of galaxy clustering features a number of additional terms that come from the de-projection of number counts on the past light-cone \citep{2010PhRvD..82h3508Y,2011PhRvD..84f3505B,2011PhRvD..84d3516C}. Such terms are mostly subdominant but become important on very large scales. Also, a probe to test GR is gravitational redshift around galaxy clusters, which has been already measured in current surveys \citep{2011Natur.477..567W,2015PhRvL.114g1103S,2023A&A...669A..29R}. In this context, the detection of a relativistic effect on the largest cosmic scales would therefore represent a confirmation of the validity of Einstein's theory of gravity in a regime far from that of strong fields where it has tested to exquisite accuracy. 

Among those large-scale effects, the most significant is a relativistic Doppler, whose detection has not been achieved so far by ongoing observational campaigns because of the limited cosmic volume. In Fourier space, where we often study galaxy clustering thanks to its capability to separate small and large scales, the Doppler term acts as a subdominant correction in the relation between the galaxy density contrast and that of matter. In particular, it is inversely proportional to the modulus of the wavenumber $k$, therefore dominant on the very large scales plagued by cosmic variance, and sample-dependent. 
Cross-correlation analyses on galaxy power spectra look more promising than auto-correlation ones, due to the presence of the relativistic effect in a non-vanishing imaginary term that might be relevant even at intermediate scales \citep{2009JCAP...11..026M}. In order to study cross-correlation using a single dataset we can divide a galaxy population into two sub-samples: a bright and a faint one \citep{2014PhRvD..89h3535B,2016JCAP...08..021B,2017JCAP...01..032G}. Then, by carefully selecting the flux split between the sub-samples we can maximise the probability of detecting the Doppler signal \citep{2023MNRAS.525.4611B}. 

To quantify the \wst\ sensitivity to this kind of signature in the data we rely on the \(\varDelta\chi^2\) test statistics and thereby obtain the following forecasts. Here, we use the theoretical predictions of the cross-power spectrum of the bright and the faint samples to produce synthetic data. The \(\varDelta\chi^2\) thus corresponds to the chi-square for a null-hypothesis of no Doppler term, against the synthetic data set that includes Doppler.
The left panel of \cref{fig:Doppler_chi2} depicts the cumulative detection significance associated with a measurement of the faint-bright cross-correlation power spectrum as a function of the splitting magnitude \(m_{\rm s}\). We use \( i \) magnitude for LBG $g$-dropout sample and \( r \) magnitude for LBG $u$-dropout. Thanks to its increased sensitivity and wide sky coverage, the \wst\ survey will be able to achieve a high-significance detection of the relativistic contribution. Considering the entire redshift interval, \wst\ will reach a \(5\, \sigma\) detection with the LBG $g$-dropout sample, whereas the \( 4\, \sigma\) level will be attained by LBG $u$-dropout (either standard or extended) samples.

We also set up a comparison with other forthcoming surveys by reporting estimations of \(\varDelta\chi^2\) of DESI-II and Megamapper/MUST. The enhanced number of sources that will be observed by \wst\ (e.g.\ w.r.t.\ Megamapper/MUST, which will have a similar sky coverage) plays a crucial role in the detection of relativistic contributions on the largest scales of the universe. We also show in the right panel of \cref{fig:Doppler_chi2} the relativistic Doppler differential detection significance for the LBG $g$-dropout sample in the case of an optimal splitting \(i\) magnitude of \(24.5 \), reported for a redshift-bin width \(\varDelta z \sim 0.3 \). Red bars refer to the cross-correlation between the faint and the bright samples, while blue, orange, and green bars illustrate the auto-correlation of the total, faint and bright populations, respectively. As expected, the cross-power spectrum leads to at least an order of magnitude improvement over auto-correlations. This demonstrates the constraining power of the bright/faint split technique when applied to a survey that can provide us with two dense sub-samples, as \wst\ will be.

\subsubsection{Dark Matter}

\begin{figure}
\centering
\includegraphics[width=0.6\textwidth]{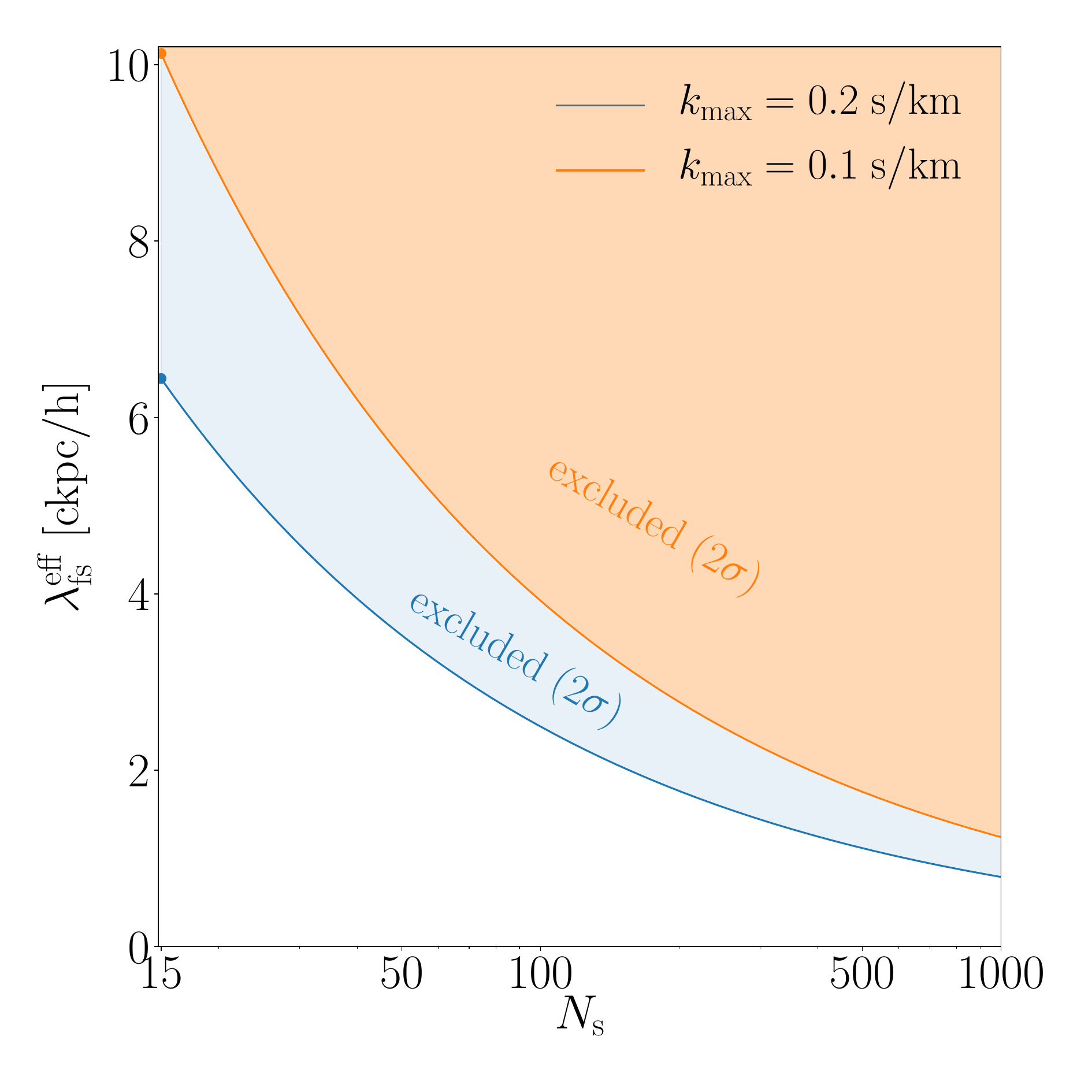}
\caption{A forecast on the constraining power of a \wst-like survey on the dark matter free-streaming scale as a function of number of spectra at $z>4$, which can be extracted from a 1D \lya flux power spectrum analysis. The number of HR spectra in a \wst-like survey may exceed $1,000$, and depends on the exact observing strategy and exposure times. Two exclusion regions are shown for different scale cuts ($k_{\rm max}$). At fixed resolution of the MOS-HR mode ($R\sim 40,000$) the value $k_{\rm max}$ in the survey will depend on the S/N distribution. The two points (blue and orange) at 15 spectra correspond to state-of-the-art constraints from UVES and HIRES data. With about 100 QSO spectra the current limits could improve by a factor $\sim$ 3. The relation between free streaming scale and DM properties is model dependent and can be exploited once a given DM model is chosen like thermal warm DM, Fuzzy DM or sterile neutrinos. The current state-of-the-art, for high-z high resolution QSO spectra, is represented by the two circles at $N_{\rm s}=15$.}
\label{fig:dm_fs}
\end{figure}

The nature of dark matter (DM) can be investigated by exploiting the impact of DM free streaming on the transmitted 1D \lya flux power (see e.g.\ \cite{baur16,irsic23}). At present, very tight constraints can be placed on the mass of a warm DM thermal relic and these are presented as lower limits in the range between 3 and 5 keV (2$\sigma$ C.L.), depending on data sets used, priors on the intergalactic medium thermal history, marginalization over nuisance (astrophysical and instrumental) parameters (e.g.\ \cite{villasenor23,irsic23}).
These constraints can be placed by using both low resolution low signal-to-noise QSO spectra (as provided by eBOSS or \desi) and high resolution high signal-to-noise ones (obtained from UVES/VLT, Keck/HIRES etc.). The standard practice is to analyse these two data sets separately, mainly because the dedicated set of CPU-time expensive hydro-dynamical simulations needed to interpret the data would have a very different set of requirements. An ideal situation could be offered by a new \wst\ survey which could combine the advantages of a large statistical sample of QSO spectra (as typically offered by low-resolution data sets) and high resolution and high signal-to-noise (from a smaller sample). In this way, the 1D flux power will probe the long-lever arm of 1D flux power from large scales to small scales from a unique and homogeneous set of QSO spectra, with statistical and systematic errors under control. This in turn will allow us to break much more effectively, than by using present data, the strong degeneracies present between astrophysics (gas pressure and thermal state) and cosmology (like DM free-streaming). 

The constraining power of the dark matter free-streaming depends both on the number of observed spectra (and the total path-length covered by the spectra) as well as the maximum wavenumber the theoretical modelling of the power spectrum can safely interpret. The value of $k_{\rm max}$ is defined as the value where $n_{\rm eff}P_F = 1$ for the 1D flux power spectrum, with the effective number density given by the power spectrum of the spectral noise ($P_N = n_{\rm eff}^{-1}$). In the regime of faint objects the the noise power $P_N$ scales inversely proportionally with signal-to-noise ratios of the spectrum ($P_N\propto \mathrm{S/N}^{-2}$). If all the spectra have the same S/N, then $nP(k=0.2\;\mathrm{s/km})=0.76$ and $nP(k=0.16\;\mathrm{s/km})=1.69$ for $g=21$ at $z>4$. Using the predicted quasar number density distributions for \wst-like survey, $nP(k=0.2\;\mathrm{s/km}))=5.06$ for $g<21$ at $z>4$ and $nP(k=0.2\;\mathrm{s/km})=0.35$ for $g<23$ at $z>4$. 

From \cref{fig:dm_fs}, it is clear that the larger number of QSOs provided by \wst\ will be extremely important for reaching smaller scales in terms of DM free streaming constraints and could potentially reach the $\sim$ 1 comoving kpc$/h$ regime, providing very tight constraints also on the physics of small scale structures of the IGM. 

It is also expected that a \wst\ \lya\ survey could constrain other DM scenarios like fuzzy dark matter, sterile neutrinos and mixed (Cold and Warm) dark matter models.

\subsubsection{Synergies}
\paragraph{Photometric calibration\\}
The dark energy probes to be employed by \lsst\ and \textit{Euclid} will depend on estimates of photometric redshifts, either for individual galaxies or through the moments of the tomographic bins distribution $n_{\rm p}(z)$. The lack of complete spectroscopic data at redshift higher than 2 poses challenges to the redshift calibration of \lsst\ and \textit{Euclid} \citep{clust_z_euclid}.
The cross-correlation of photometric objects with extensive high-redshift spectroscopic samples measured by \wst\ and reported in \cref{fig:Nz_zhigh2} would tightly constrain the high-redshift distributions. Indeed the relative uncertainty on $n_{\rm p}(z)$ scales as the square root of the number of spectra \citep{McQuinn_clust_z}. 

\paragraph{CMB lensing\\}
Weak lensing of the CMB integrated along the line of sight is a tracer of the matter distribution, and one of the central science cases for upcoming CMB surveys \cite{Simons_obs,CMB_S4, litebird}.
The \wst\ survey being in the southern hemisphere, it is ideally situated for cross-correlations with SO or CMB-S4 experiments.  
CMB lensing as a matter tracer will be included in the \lsst\ and \textit{Euclid} analysis \citep{LSST_CMB, Ilic_etal2022}.
Nonetheless most of the cosmological constrain will come from $z<2$ bins whereas a \wst-like survey will extend the domain to higher redshift, where the CMB lensing kernel peaks \citep{LEWIS_2006}, as illustrated in \cref{fig:Nz_zhigh2}. Thus the high-redshift \wst\ spectroscopic survey is highly complementary to other upcoming and planned photometric experiments. For example, the high-redshift LSS\texttimes CMB lensing can improve constraints on massive neutrino $M_\nu$, gravitational slip parameter $\gamma=\Phi/\Psi$ by a $30$--$50\%$ factor, and up to $90\%$ for the amplitude of early dark energy $f_{\rm EDE}$ with respects to LSS  only \citep{Sailer_forecasts,cosmology_noon}. 

\subsection{Cosmology Lyman-$\alpha$ parallel survey}
\label{sec:Lyman-alpha blind survey}

The IFS can be used in parallel to the MOS-LR observations to identify Lyman-$\alpha$ Emitters (LAEs) within the redshift range $2<z<7$. LAEs are believed to be young, low-mass galaxies at high redshifts, exhibiting a high star-forming rate \citep{Ly_alpha_gal,Ly_universe}. Despite ongoing efforts, a comprehensive understanding of the factors influencing the  Ly-$\alpha$ emission strength in LAE spectra remains elusive, thus introducing complexity to our comprehension of this galactic population. Yet, in this section, we will focus on the possibility of constraining cosmological parameters using this LAE sample as tracers of the underlying mass density.

Due to the faint and limited spectrum of these targets, the typical procedure for target selection involves narrow or medium-band imaging, presenting challenges in using these targets for massive spectroscopy. However, the use of the IFS eliminates the need for the target selection step, and allows an efficient measurement of LAEs.
The concept of the cosmology Lyman-$\alpha$ parallel survey is to take advantage of any telescope pointing in the extragalactic sky not targeting specifically a bright object such as a nearby galaxy. The parallel observations can encompass several thousands independent tiny patches (9 arcmin$^2$ each). The accessible redshift range of LAEs covers $2<z<7$. With multiple thousands of lines of sight, the extraction of cosmological information with LAEs can be derived for the first time through the 1D power spectrum. 

We derive our survey properties estimates based on the MUSE 8.3 arcmin$^2$ MOSAIC observations \cite{MUSE_DRII}. 
Assuming 900 LAEs observed for every IFS patch and a total of about $7000$ pointed observations (considering only the Dark time survey see Section \ref{sec:high_z_survey}), this would yield 6 million LAEs over a cumulative $16\,\deg^2$ area, boasting a density of $400\,000$ LAEs per square degree.

\begin{figure}
    \centering
    \includegraphics[scale=0.5]{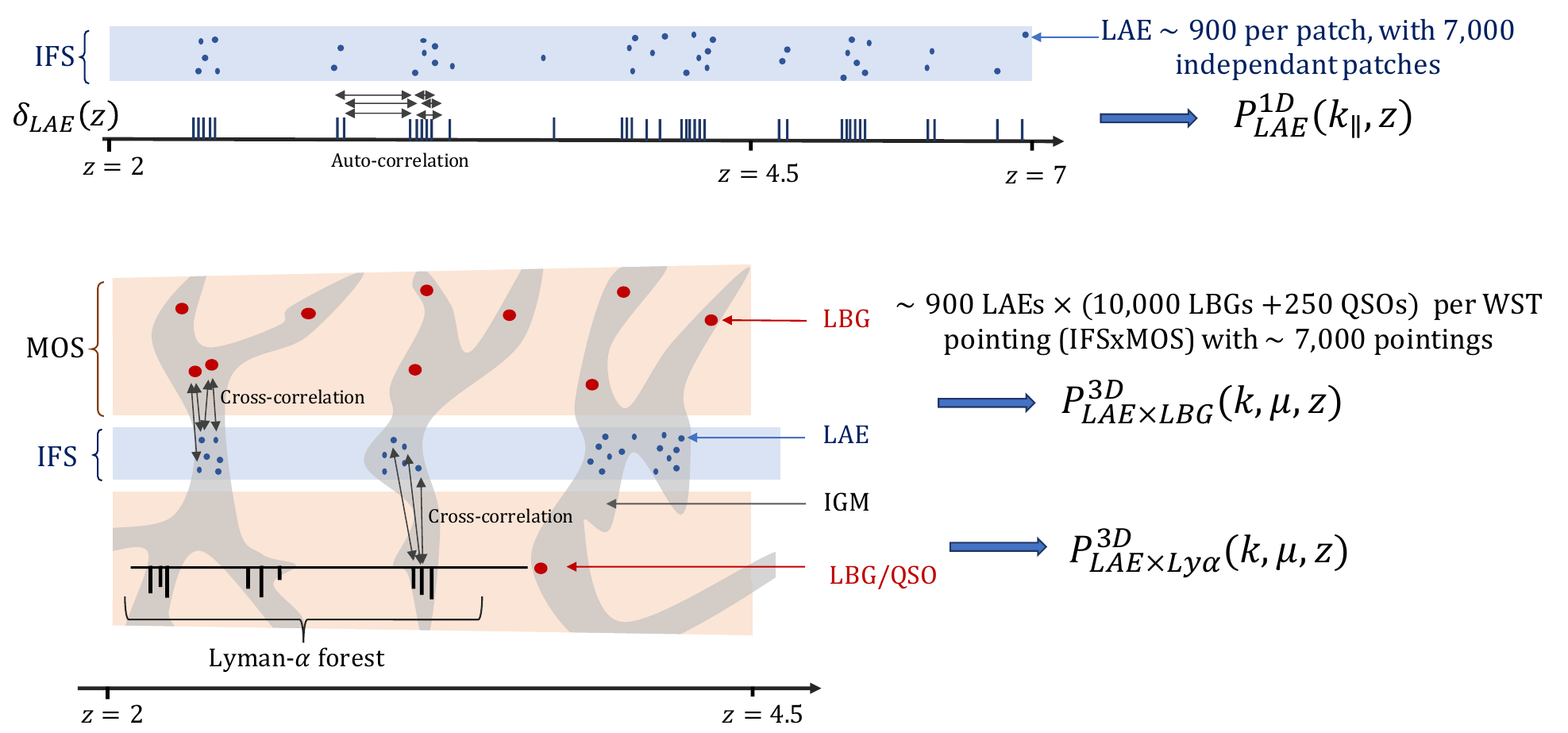}
    \caption{The two clustering analysis using the LAEs from the IFS. For the auto-correlation (\textit{top}), we consider only line of sight overdensities, giving a 1-dimensional power-spectrum. For the cross-correlations (\textit{bottom}), we have a 3 dimensional one, excluding $\mu\sim 1$ (horizontal arrows) because of the null intersection of the tracer-volumes.}
    \label{fig:LAExLAE_LAExLBG}
\end{figure}

Two primary cosmological analyses can be conducted with such samples: the 1D LAE power spectrum derived from LAE auto-correlation, and the 3D power spectrum obtained through cross-correlations with Lyman Break Galaxies (LBG) and Quasi-Stellar Objects (QSO) MOS-LR observations (either in position or using the Ly-$\alpha$ forest in front of the QSOs). These various analyses are illustrated in \cref{fig:LAExLAE_LAExLBG}. 

\subsubsection{IFS-LAE auto-correlation}

We investigate the constraint of 1D correlation along the line of sight, for (parallel) BAO as a proof of concept. Any parameter that impacts the dark matter power spectrum can in theory be (weakly) constrained on this large redshift range. 

Given \wst\ survey strategy, we predict about $7000$ pointed observations, and we stack their auto-correlations in Fourier space to derive the power spectrum $P^{1D}_{\rm LAE}(k_\parallel,z)$. The BAO results are reported in \cref{fig:forecast_BAO_growth}. We get an $18\%$ measurement of $sH$ at $2<z<4$ which is an order of magnitude larger than the constraint from the MOS-LR LSS, and a $10 \%$ measurement over the higher redshift range $4<z<7$ where no other measurement has ever been done. The improvement of the measurement at higher redshift is mainly due to the reduction of the non-linearities affecting the BAO peak, since in both cases the measurement is highly cosmic variance limited. We propagate the uncertainties to $\Omega_{\rm DE}$ and predict a 20$\%$ measurement. Thus even if not competitive with the LSS-MOS-LR measurement, this would be  the first cosmological probe from this early epoch.

\subsubsection{LAEs cross-correlation with LBGs}

A second possibility is the cross-correlation of LAEs from the IFS with the LBGs from the MOS. We investigate the potential constraint on the BAO parameters, from the clustering of LAEs and LBGs. 

Both tracers do not cover the same volume, and have different densities. For each tracer we model its  covariance by $C=P+1/n$, with P the power-spectrum, and $1/n$ the poissonian shot-noise, and we use the product of the two covariances for the Fisher matrix calculation.  

The survey volume is still limited by the IFS volume, which in the case of the cross-correlation is further restricted to $2<z<4.5$. In comparison with the LBG-autocorrelation, here for LAE $nP\gg  1$ and the (LBG) shot noise contaminates linearly instead of quadratically the measurement. Because of the geometry of the IFSxMOS-LR, we do not consider all the $\mu \in \left[ -1,1\right]$, where $\mu$ is the cosine of the angle to the line of sight. One can think of this angle as the inclination of the arrows in Figure \ref{fig:LAExLAE_LAExLBG}, which should not be too horizontal in order to cover both the IFS and MOS-LR regions. 

We predict a measurement of the angular distance $D_A/r_{\rm s}$ with 2 $\%$ precision, and $r_{\rm s}H$ at $5\%$ as reported in \cref{fig:forecast_BAO_growth}. Thus the constraint is still competitive with the $<1\%$ constraint from the MOS-LR. Nonetheless, it is an interesting complementary (but not independent) measurement at the percent level, with a different noise contamination.

A second probe from the IFS cross MOS-LR -correlation is the cross-correlations with the Ly-$\alpha$ forest. Indeed LAEs are a very good tracer of the IGM \citep{MUSE_CW}, so we expect the correlation with the forests to be high.  We do not produce quantitative forecast, since this would  be at the level of the 3D Ly-$\alpha$ power spectrum, and would involve complex simulation/modelling. Indeed the cross-correlation of Lyman-$\alpha$ forest with QSO is predicted to improve by roughly a factor 2 the QSO-clustering-only measurement, for \desi\ for example \cite{DESI_validation}. Here the situation is a bit more complex because we are dealing with LBGs whose Lyman-$\alpha$ forest is difficult to use, and QSO, whose densities is order of magnitude less than the LBG-one. Nonetheless, it is likely that a percent level constrain as in the previous case can be achieved, but with associated challenges.

\subsection{A combined MOS-LR+IFS survey of the growth of galaxy clusters}

\begin{figure}[t]
  \centering
  \includegraphics[width=0.8\textwidth]{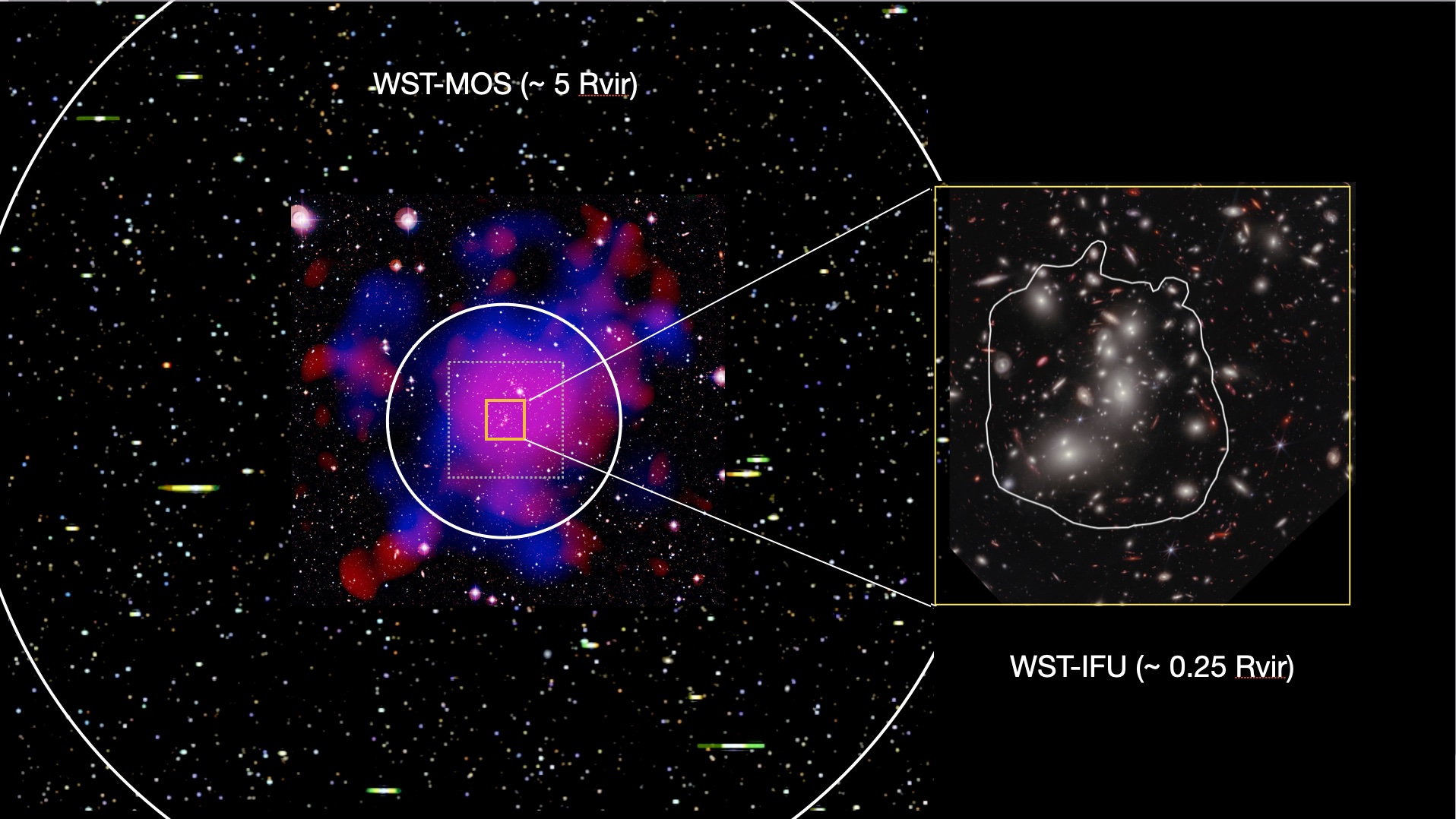}
  \caption{\small A schematic of the \wst\ footprint using MOS-LR (outer and inner circles) and IFU (central square) observing the massive clusters Abell 2744 ($z=0.31$). Mass measurements of the clusters can be obtained using lensing (\textit{Euclid}, JWST), X-ray (e.g.\ eROSITA) and dynamics (\wst). A dedicated cluster survey can probe how clusters grow with time and how they connect to the filaments and large scale structures. 
  }
  \label{fig:cluster_pointing}
\end{figure}

\begin{figure}[h]
  \centering
  \includegraphics[width=0.7\textwidth]{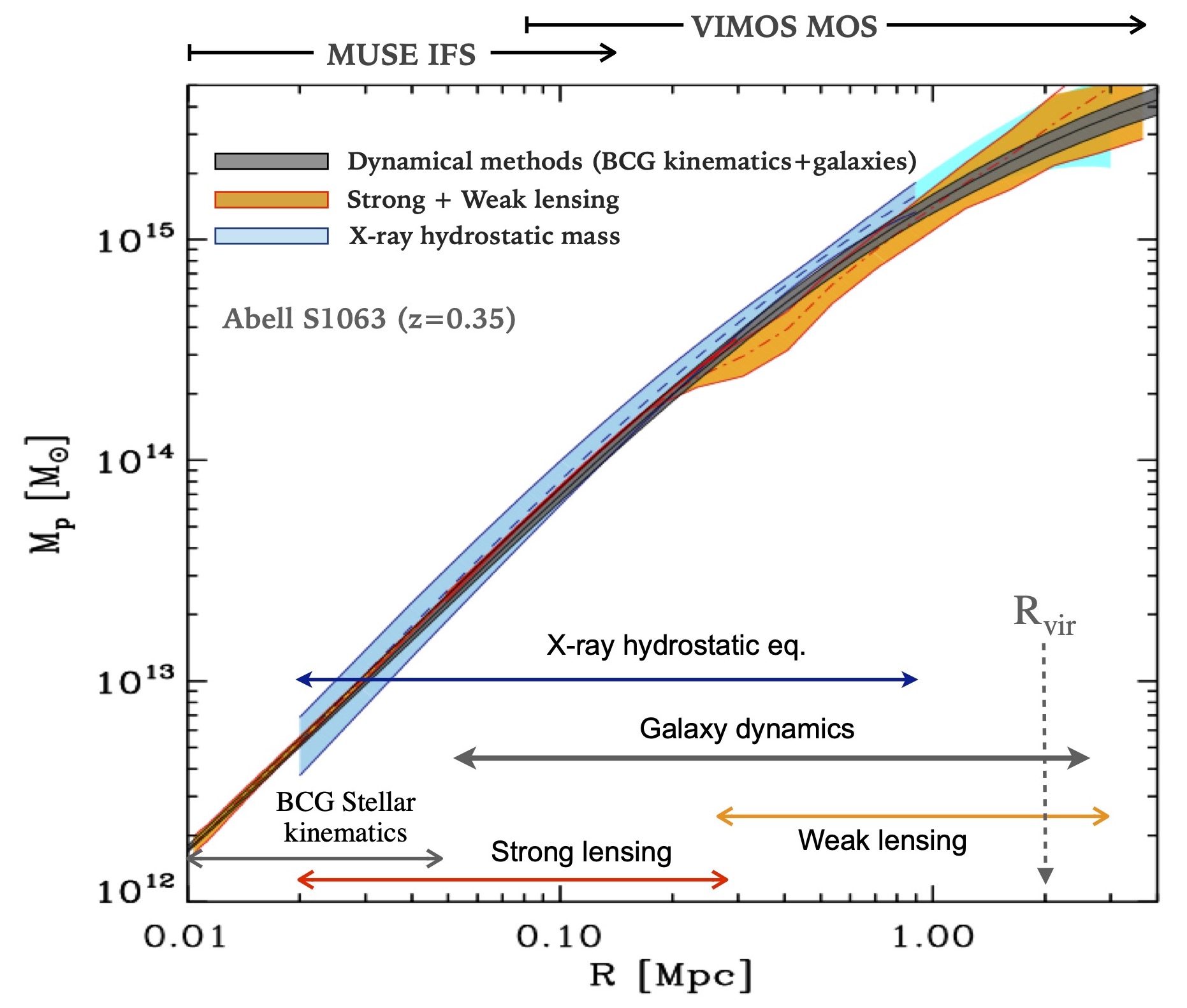}
  \caption{\small Total projected mass density profile of the Hubble Frontier Fields cluster Abell S1063 at $z=0.35$, obtained from different mass probes \citep{Sartoris2020, Biviano2023}. The dynamical mass profile (black/grey curve) from the innermost regions out to $\sim\!3$ Mpc is obtained with VLT/MUSE IFU (1' FoV) and VIMOS-MOS ($\sim\! 20'$ FoV) observations, by combining the spatially resolved internal kinematics of the BCG (MUSE) with over 1000 galaxy velocities  (from MUSE$+$VIMOS redshifts). The  lensing profile is obtained by joining a strong lensing model of the cluster core (exploiting $\sim\! 60$ multiple images discovered with MUSE) with a weak lensing analysis from wide-field ground-based imaging. The X-ray hydrostatic mass is derived from Chandra observations.
  }
  \label{fig:cluster_masses}
\end{figure}

\begin{figure}[t]
  \centering
    \includegraphics[width=0.65\textwidth]{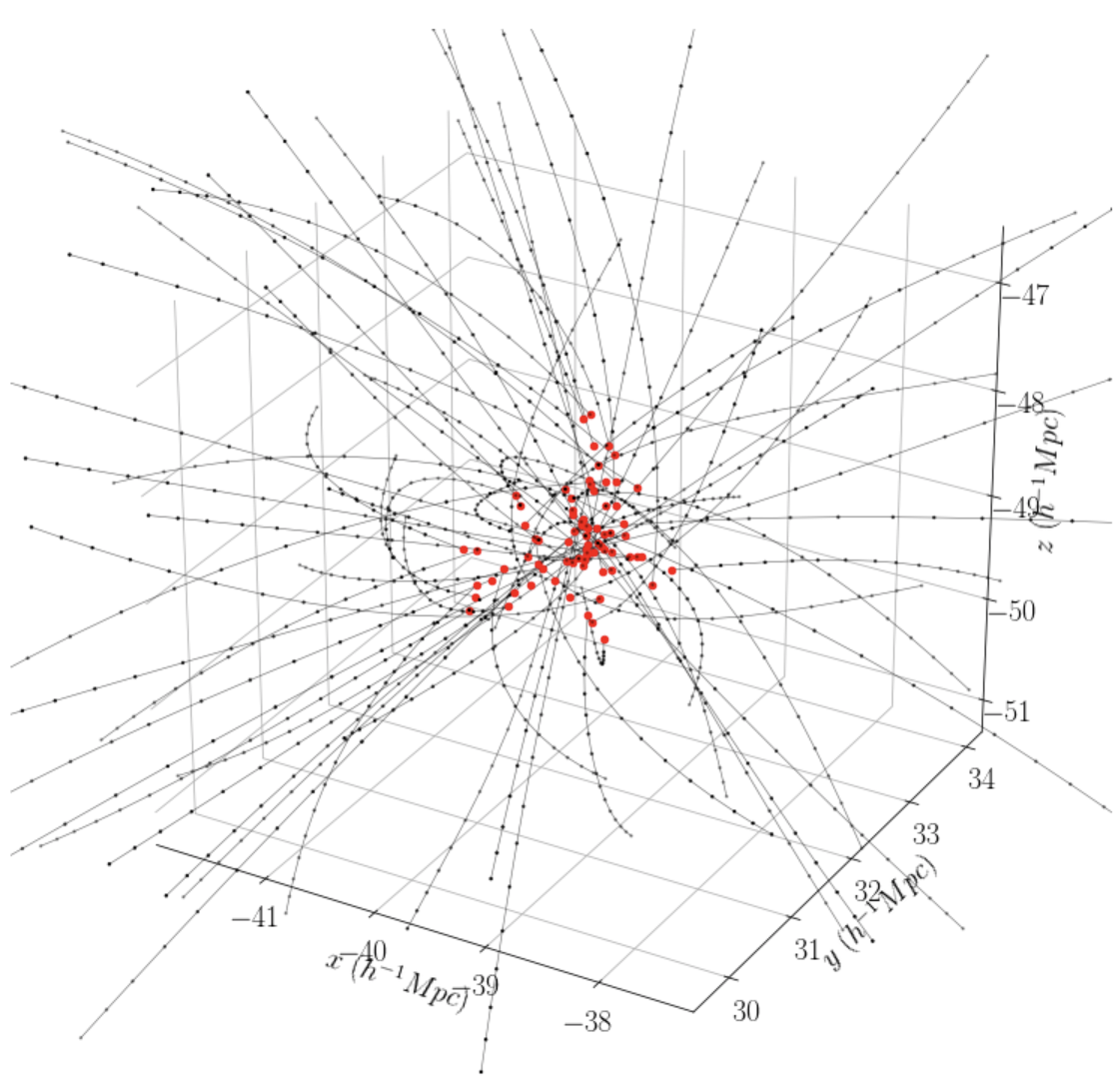}
   \caption{\small Orbits of galaxies in a simulated cluster of mass similar to Virgo cluster, reconstructed backward-in-time from their observed positions at $z=0.07$ (red dots) up to $z = 1.1$ over 14 time-steps (black points) using the eFAM method \citep[credit:][]{Sarpa+2022}.}
  \label{fig:cluster_dynamics}
\end{figure}

\begin{figure}[t]
  \centering
    \includegraphics[trim={0 0 0.1cm 0},clip,width=0.85\textwidth]{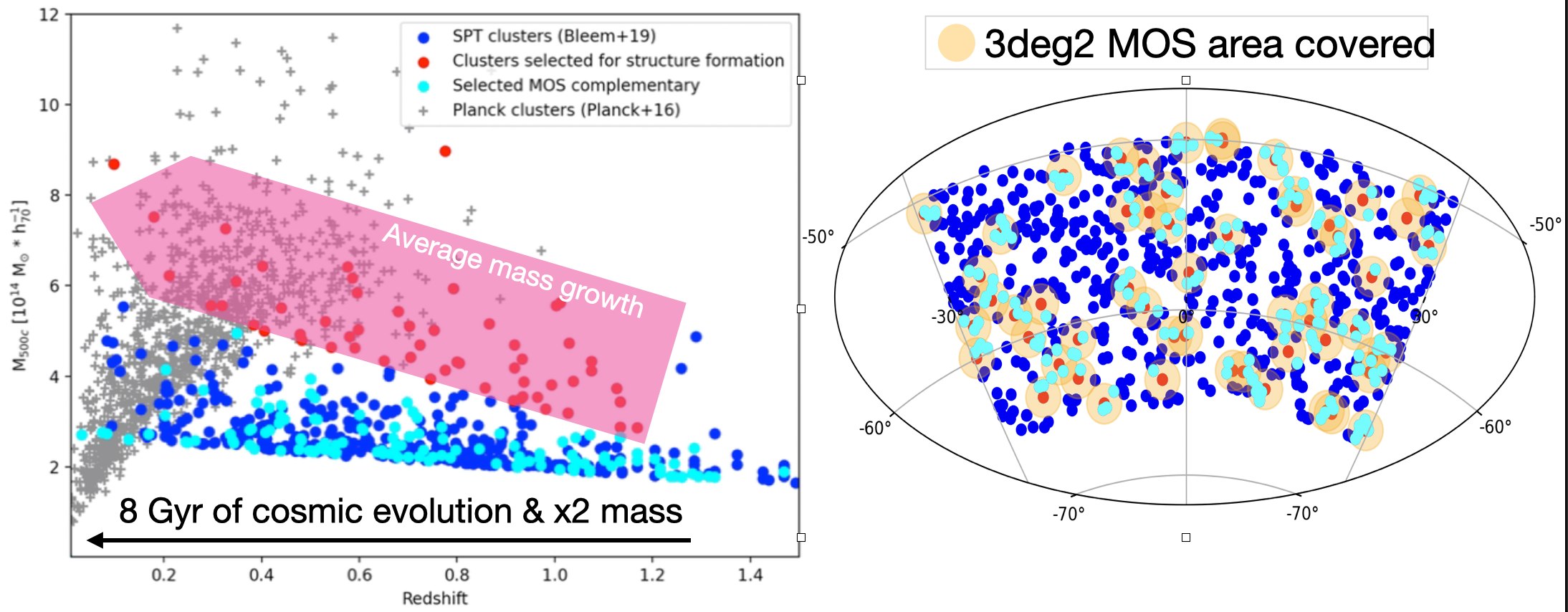}
   \caption{\small \emph{Left:} Selection of clusters according to their mass evolutionary track with respect to redshift. \emph{Right:} Footprint of the different \wst\ MOS-LR pointings over a few thousands deg$^2$ showing the targeted cluster distribution. Each \wst\ IFU pointing can reach on average of 3.4 clusters within the MOS FoV. In both figure, red points are cluster targets for the IFS, cyan points show the nearby \wst\ MOS-LR selected clusters. Dark blue circles are the non-targeted lower mass and isolated clusters. The large pink arrow line represent the average mass growth \citep{Fakhouri2008} for massive clusters.}
  \label{fig:cluster_survey}
\end{figure}

Clusters of galaxies are the largest, more massive, gravitationally bound structures observable in the Universe emerging after photon decoupling, containing tens of galaxy groups and hundreds of galaxies. According to the hierarchical structure formation typical of a universe dominated by CDM, structures grow by accretion and merging with small halos (galaxies) assembling first, followed by medium halos (galaxy groups) and then large halos (galaxy clusters) forming around the denser nodes of the cosmic web. As latest virialized structures, galaxy clusters keep memory of the primordial density field (Gaussian or mildly non-Gaussian) while hosting galaxies at the latest evolutionary stage \citep[see review by ][]{Kravtsov2012}. Moreover, the most energetic phenomena take place in them, including complex interplay between gravity-induced dynamics of collapse runaway cooling of the hot ICM (e.g., \citealt{Fabian1994,McDonald2012}) and outbursts of AGN outflow activities and radio mode feedback that heat the surrounding intergalactic medium via shocks (e.g., \citealt{Rafferty2008,Pommier2013,Hlavacek2022}).

Galaxy clusters are therefore at the cross-roads of cosmology and astrophysics, constituting unique laboratories for testing models of gravitational structure formation, galaxy evolution, thermodynamics of the intergalactic medium, and plasma physics.

Building a comprehensive view of massive clusters over a wide redshift range, including the dynamics and morphology of their environment down to several viral radii incorporating the filamentary structures that feed them, not only allows the study of cosmological scenarios, but also enables a comprehensive study of structure formation on multiple scales.

The synergy between the IFS and MOS modes of the \wst, combined with their large field of view, will be particularly powerful in improving our understanding of the mass and galaxy assembly history of clusters (\cref{fig:cluster_pointing}) and enabling an accurate measurement of mass density profiles using different methodologies. Recent progress in this field, using such an approach, is highlighted in \cref{fig:cluster_masses}, where the VLT/MUSE IFS and VIMOS-MOS observations were combined to obtain a mass profile over a wide radial range from dynamical and a gravitational lensing methods.  With \wst\ observations of intermediate redshift massive clusters, thousands of cluster member redshifts will be measured out to $\sim \! 10$ Mpc, by combining IFS observations in high galaxy density inner regions with panoramic MOS observations in the outskirts. The wide-field IFS, will be critical to develop high-precision strong lensing models out to radii $\gtrsim 0.5\,{\rm  Mpc}$, by unveiling hundreds of multiply lensed images per cluster. The extensive MOS spectroscopic coverage will also be critical to select the background galaxies needed for weak lensing analysis out and beyond the virial radius.

Moreover, the \wst\ spectroscopic data set on massive clusters will provide high-quality (projected) phase space diagrams, including both member and infalling galaxies, as well as orbital information on subsets of cluster galaxies (e.g.\ \citealt{Biviano2013, Mercurio2021}). These observations can be compared with similar quantities extracted from cosmological simulations, thus testing $\Lambda$CDM predictions on cluster assembly history.

The dynamics of galaxies and, more in general, the velocity field in the periphery of the cluster can be reconstructed by non-linear methods such as NAM \citep{Shaya+2017}, eFAM \citep{Sarpa+2022}, or others techniques using the angular position and redshift (phase space) of galaxies as constraints; see Fig.~\ref{fig:cluster_dynamics}. Accurate 3D positions for a large sample of galaxies over a volume encompassing the cluster outskirts, as measured by the high-multiplex MOS, are crucial to reconstruct galaxy orbits. This information can be further employed to assess the role of environmental effects at high-redshift on galaxy quenching in the outskirts of clusters, extending the low-redshift results obtained with WINGS and OmegaWINGS clusters \citep{Salerno+2020}. The reconstructed dynamics of core and outskirts galaxies can then be used to relate their physical properties in dense environments (SFR, metallicity, etc.) to the topological type of the shallower environments they went through their journey (voids, sheets, filaments; see \citealt{Sarpa+2022}).
Moreover, reconstructed orbits can be used to investigate the origin of the ionized gaseous queues observed in VESTIGE \citep{Boselli+2014,Boselli+2018} and GASP \citep{Poggianti+2017} surveys for Virgo and nearby clusters but for larger statistical samples. 

Hierarchical models of structure growth show that when entering a cluster, galaxies start spiralling down the potential well and lose 10\% to 50\% of their mass after a few Gyr \citep{Han2016,vdBosch2018}, mainly through tidal stripping \citep{vdBosch2018}. On average, a cluster at $z=0$ would have increased its mass by a factor of 5 from $z=1$ \citep[][see \cref{fig:cluster_survey}]{Fakhouri2008}, mainly due to infalling sub-halos \citep{Wu2013}. The importance of sub-halos infalling into clusters compared to other mechanisms (pure accretion, cluster mergers) has been estimated from simulations \citep{Bahe2019} but remain difficult to observed due to the different physical scale involved from the outskirts of the clusters (at few times its virial radius) to the inner part of its core (within less than hundred kpc and violent merging with the central galaxy). Previous studies investigated the sub-halo fraction (e.g.\ \citealt{Richard2010,Golden-Marx2019,Mahler2019,Mahler2020,Meneghetti2020}), however they did not connect it to the cluster growth itself. It is clear that much larger number statistics and a wide redshift range are needed to compensate the intrinsic scatter of such processes.

\begin{figure}
  \centering
  \includegraphics[width=0.49\textwidth]{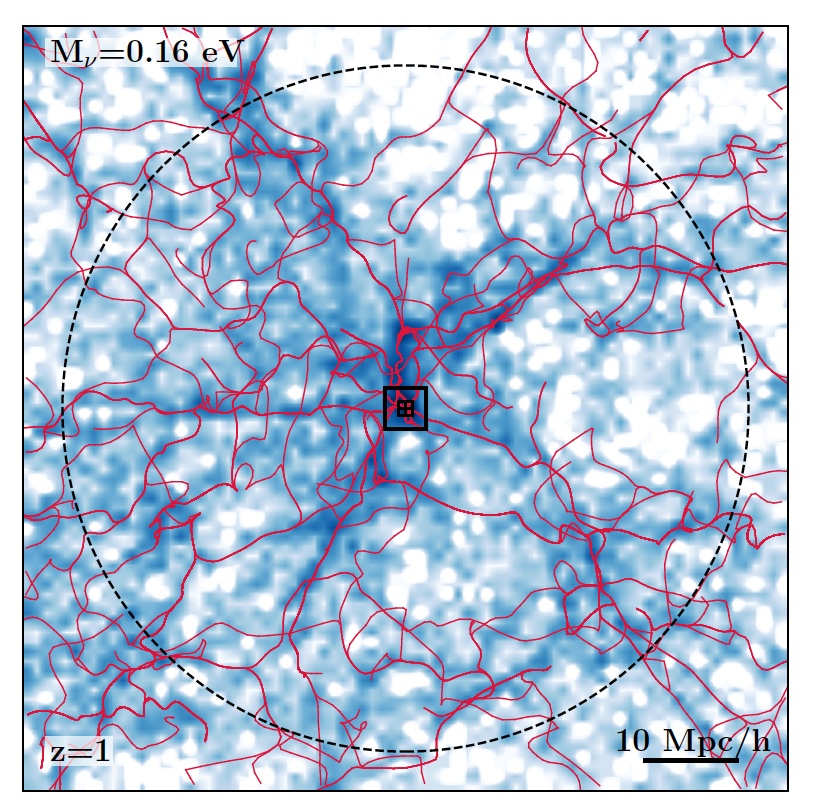}
   \includegraphics[width=0.49\textwidth]{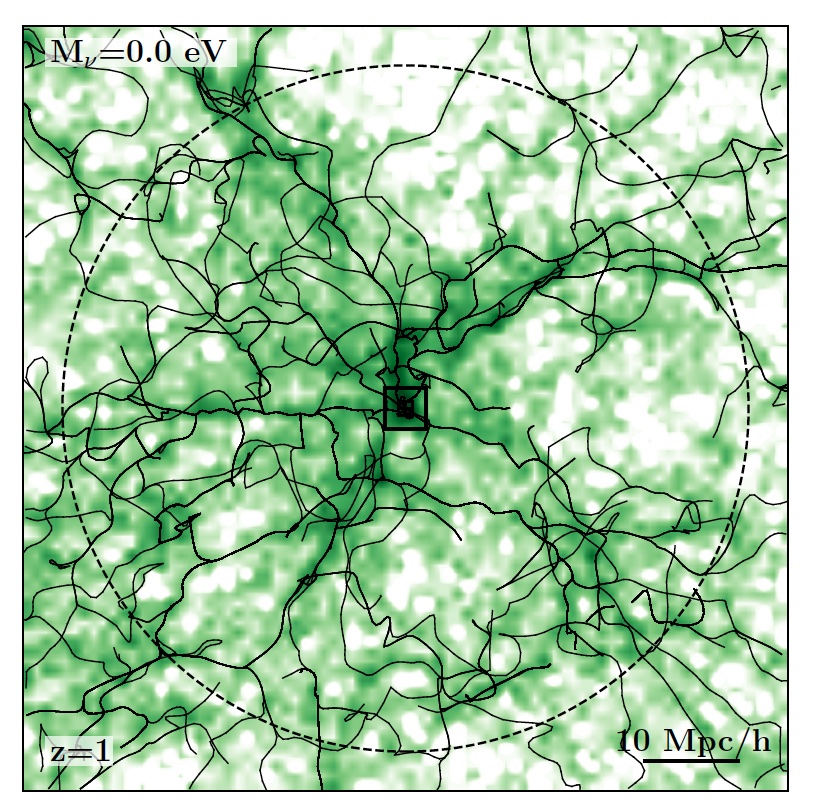}
     \caption{Example of the most massive cluster in the DEMNUni simulations \citep{Carbone_etal2016, Castorina_Carbone_2015} with M$_\nu$ = 0.16 eV (left) and M$_\nu$ = 0.0 eV (right) at $z=1$ together with the reconstructed filamentary network in a 40 $h^{-1}$Mpc thick slice. Dashed circle centered on the cluster indicates the \wst\ MOS-LR footprint while the innermost squares represent the IFU and $3\times3$ IFU (moisaicking) footprints.}
\label{fig:demnuni_cluster_skel}     
\end{figure}

\begin{figure}
  \centering
  \includegraphics[width=0.49\textwidth]{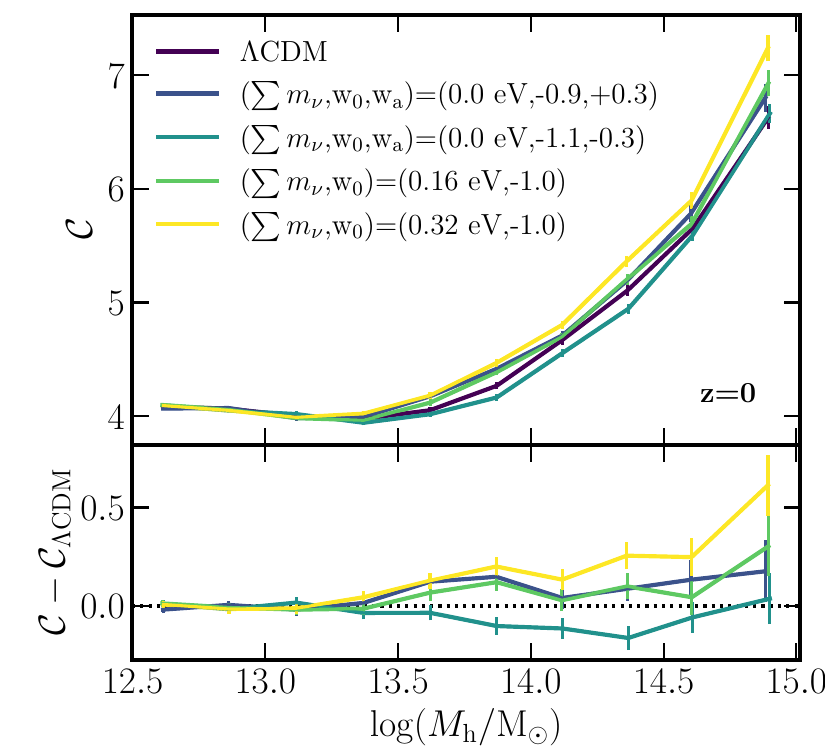}
   \includegraphics[width=0.49\textwidth]{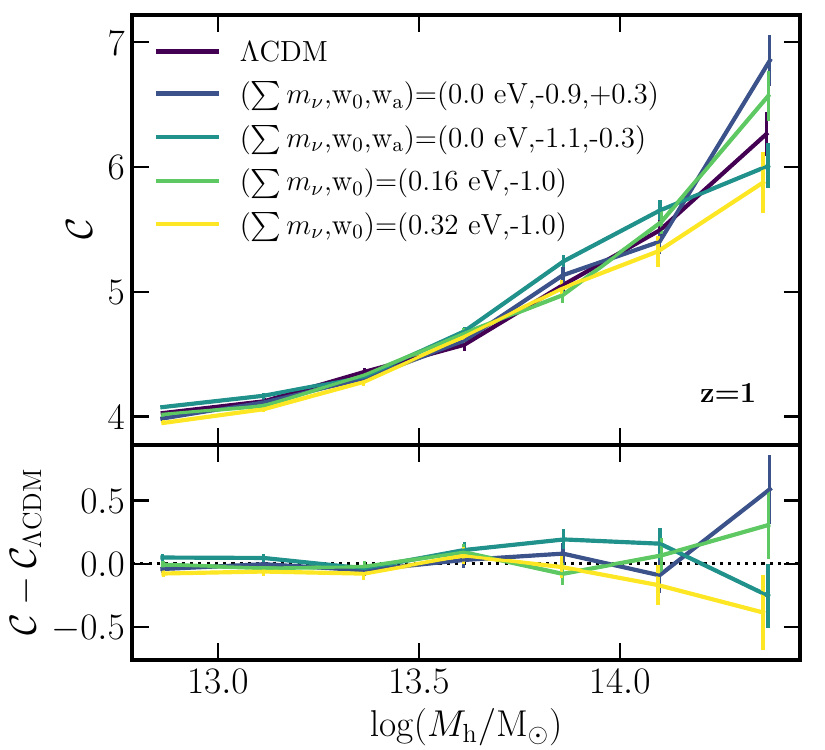}
     \caption{Mean connectivity of halos in DEMNUni simulations \citep{Carbone_etal2016, Parimbelli_etal2022, Hernandez_Molinero_2024} (top panels) and its difference wrt the $\Lambda$CDM model (bottom panels) as a function of halo mass at $z=0$ (left) and $z=1$ (right). As expected, the connectivity increases with increasing halo mass regardless of cosmological model. The redshift evolution of the connectivity of the cosmic web can in principle be used to probe the underlying cosmological model.}
\label{fig:demnuni_connect}     
\end{figure}

Besides dynamical probes, the morphology of galaxy clusters and their outskirts offers new probes for cosmology. In particular the connectivity $\mathcal{C}$, i.e.\ the number of filaments (see e.g.\ \cref{fig:demnuni_cluster_skel}) feeding the clusters’ core, is a topological measure not only probing the (non) Gaussianity of the initial density field \citep{CodisEtal2018}, but also depending on the later cosmological dynamics, i.e.\ on cosmic environment with effect on the SFR and morphology of galaxies \citep{DarraghFordEtal2019,KraljicEtal2020}. On large and intermediate scales, this is regulated by the cold/warm/hot nature of dark matter \citep{HyeongHan_2024NatAs} (including massive neutrinos) and dark energy (including modified theories of gravity) in a still unexplored way (see Fig.~\ref{fig:demnuni_connect}). Using a high-multiplex MOS-LR covering large area, one can probe $\mathcal{C}$ out of tens of Mpc. Besides, simultaneous observations with a high-resolution wide-field IFS can (i) accurately map the internal morphology of clusters, which is directly sensitive to the cosmological model \citep[e.g.][]{BonnetEtal2022}, (ii) locate the endpoints of the filaments in the crowded internal regions of clusters, and (iii) estimate the spin of the halo and subhaloes probing a possible relation with galaxy morphology \citep{CodisEtal2012}.

A challenging use of the wide-field high-resolution IFS is the measurement of kinematic lensing, i.e.\ the deformation of the 2D velocity map of galaxies induced by lensing that offers an independent, pointwise, not statistical measurement of gravitational lensing convergence and shear of galaxies \citep{Blain2002}. Kinematic lensing is expected to improve the mass modelling of the large-scale structure and clusters, with a smaller number of galaxies than required by traditional (photometric) methods, provided they are sufficiently isolated in order for the rotational kinematics being not very disturbed. The technique has been proposed for radio high-resolution HI velocity maps \citep{Morales2006}, further investigated and tested on NGC~3621 and NGC~5236 from The HI Nearby Galaxy Survey (THINGS) data \citep{deBurgh-Day+2015}, and proposed in NIR domain with Roman Space Telescope
\citep{Xu+2023}, where the potential limiting factors are the spectral and spatial resolution.

The unique capabilities of the \wst\ offer the most efficient way to uniquely bring together all the requirements needed for these studies and probe clusters' growth from its outskirts to the inner regions, over more than 8 Gyr of cosmic evolution. Clusters will be selected out to $z\sim 1.5$ from large large sky surveys, such as those in the millimetre based on the SZ effect (for a nearly mass-selected sample), and in the X-ray band (for a X-ray luminosity selection), particularly from eROSITA \citep{Bulbul2024}, as well as in the optical/near-IR with \textit{Euclid} (WL) and \lsst. As shown in the left panel of \cref{fig:cluster_survey}, based on the South pole cluster detection \citep{Bleem2023}, the \wst\ cluster survey will be designed over an area with about 0.112 cluster per square degree, for a total of 2016 clusters, considering the full \lsst\ footprint. Thanks to the large number of fibres available in the MOS-LR instruments, we can further probe on average 3.4 clusters per pointing for a total of 6876 clusters, ultimately mapping for the first time the large scale intra-cluster network of galaxies (\cref{fig:demnuni_cluster_skel}). 

\subsection{The legacy low-redshift ($z<1.6$) mapping
\label{sec:legacy_survey} }

Going beyond the galaxy cluster that will cover about 1/3 of the sky fraction, a legacy galaxy survey of the low redshift Universe ($z<1.6$) will have many applications supporting strong and weak lensing measurement, producing a legacy catalogue for extragalactic transient identification (including GW, GRB, FRB, SNe …) and radio surveys. This survey will also enable the study of the geometry and topology of the LSS and its components (nodes, sheets, filaments, and voids) and cross-correlation with other surveys (HI maps, CMB maps, weak lensing), supporting the estimation of cosmological parameters to better precision than the Stage-IV surveys.

This legacy survey would extend the cluster survey in the sky coverage, as well as providing complementary targets to the cluster survey strategy. It could also accomodate any transient TOO telescope pointing, or run in parallel of any specific IFS only pointings.

Reaching a high target density allows to probe precisely small scales in the galaxy power spectrum (high k) and will provide information on neutrino mass,  on the  galaxy-halo connections, constrain baryonic effects (in particular for weak lensing measurement), and probe dark matter models.

\begin{figure}[t]
  \centering
  \includegraphics[width=\textwidth]{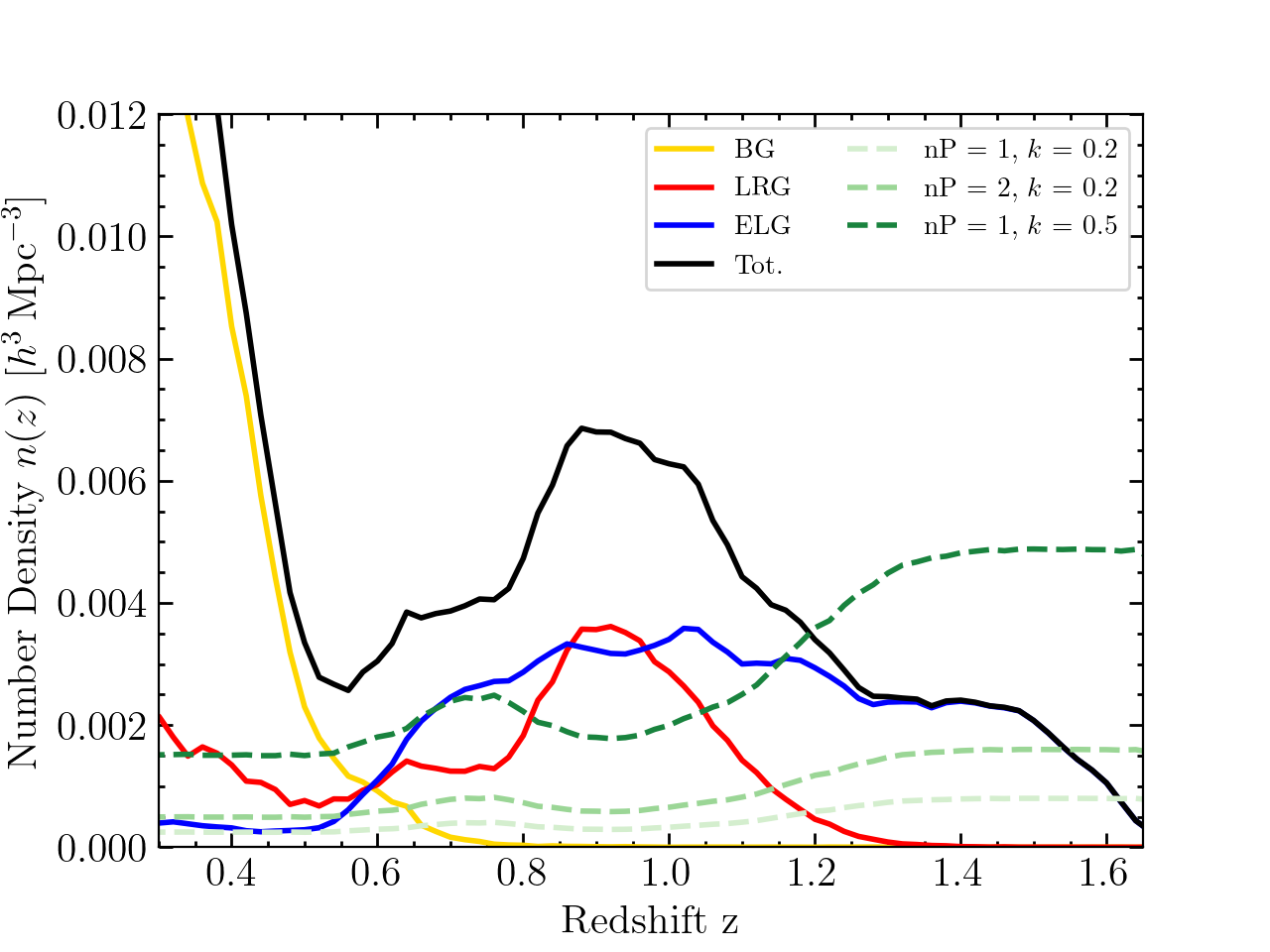}
   \caption{Target density using modified \desi\ cuts for \wst\ low-redshift legacy survey. BGs are showed in yellow, LRGs in red and ELGs in blue. The black line corresponds to the total density of the three type of objects. In green are shown the required density to reach $\bar{n}P(k)=1,2$ for 2 different $k$ scales: 0.2 and 0.5 $h\cdot $Mpc$^{-1}$. }
   \label{fig:lowz_dens}
\end{figure}

The target selection for the legacy survey shown in Figure \ref{fig:lowz_dens} has been extrapolated from the DECaLS DR10.1 photometric data \citep{2016AAS...22831701B} by extending the \desi\ recipes to fainter magnitude (typically 1 magnitude) for the Bright galaxies \citep{Hahn_2023}, Luminous Red Galaxies \citep{Zhou_2023} and the Emission Line Galaxies \citep{Raichoor_2023}. As DECaLS is not going deep enough in the g band, we had to use the COSMOS2020 data \citep{WangJerabkova2021} and modify slightly the colour cuts done in \desi\ for the ELG. 
Such selection can easily reach the target density of $\sim10\,000\,\deg^{-2}$, which is more than the \wst\ MOS-LR fiber density ($\sim 6666$ fibres/deg$^2$).

The corresponding target distribution is shown in \cref{fig:lowz_dens} with the number of targets needed to reach $\bar{n}P(k)=1,2$, where $\bar{n}$ is the average number density and $P(k)$ is the amplitude of the observed power spectrum at scale $k$ (set to 0.2 and 0.5).
The higher $\bar{n}P(k)$, the better we can reconstruct the BAO features damped by non-linear structure formation \citep{Font_Ribera_2014}, reaching a 50$\%$ acoustic peak reconstruction for $\bar{n}P(k)=2$ with $\mu = 0$ and $k=0.2$.
Bias for BG and LRG are the expected one from \desi\ \citep{desicollaboration2016desi} while for ELG we used the one derived from the ELG Halo Occupation Distribution (HOD) \citep{rocher2024desi}.
With the high density provided by \wst, we can easily reach $\bar{n}P(k)=2$ but we can also probe small scales ($k=0.5$).

\subsubsection{The galaxy-dark matter connection}

\begin{figure}[t]
  \centering
  \includegraphics[width=0.95\textwidth]{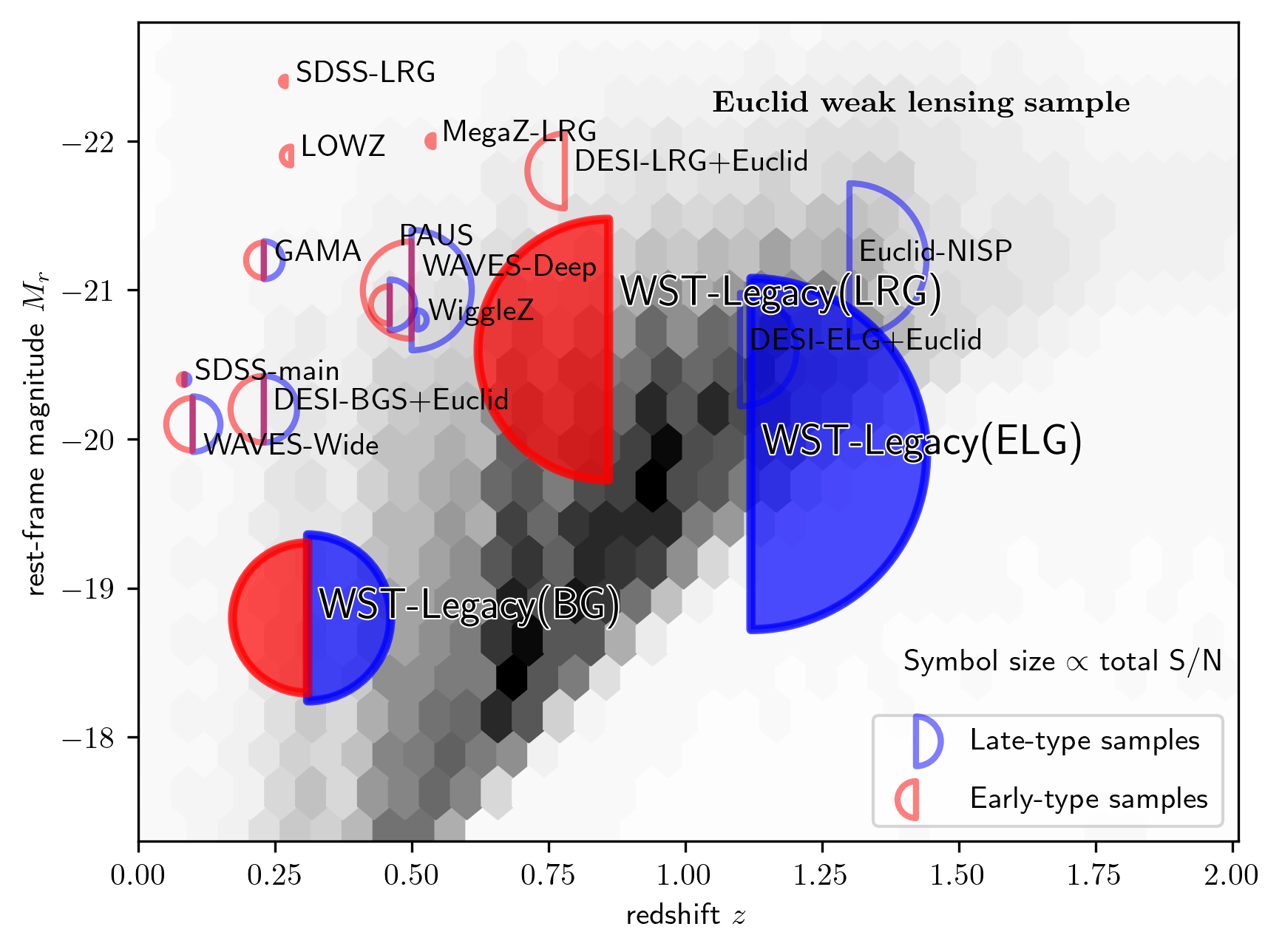}
   \caption{Intrinsic alignment constraining power of the \wst\ Legacy Survey galaxy samples in comparison to a selection of current and forthcoming measurements. The symbol area is proportional to the expected signal-to-noise of a galaxy position-shape cross-correlation assuming a uniform signal amplitude amongst the samples. The symbols are centred on the typical redshifts and rest-frame $r$-band magnitude of the sample. Blue (red) symbols indicate late-type (early-type) galaxy samples. The grey-scale density plot in the background illustrates the expected distribution of the \textit{Euclid} weak lensing galaxy sample.}
   \label{fig:ia}
\end{figure}

Large galaxy surveys are the key supplier of cosmological information in the low-redshift Universe. Projects like \desi\ \citep{DESI2016a}, the \textit{Euclid} mission \citep{Euclid}, and \lsst\ \citep{LSST} will deliver sub-percent precision cosmological measurements by the mid-2030s. These will only be achieved with a much-improved understanding of how galaxies as the carriers of the signals relate to the underlying dark matter distribution of the cosmic large-scale structure. Moreover, this link is crucial to complete our understanding of how galaxies form and assemble into the diverse populations observed in the low-redshift Universe \citep{wechsler18}. The \wst\ Legacy Survey will enable order-of-magnitude improvements in the sensitivity of direct probes of the galaxy-dark matter connection.

As a prominent example we consider the intrinsic alignment of galaxies, i.e.\ the tendency of galaxy shapes (usually in projection) to align with each other and with the surrounding large-scale structure (see \citealt{joachimi15,kiessling15,kirk15,troxel15} for reviews). Intrinsic alignments play a dual role of a limiting astrophysical systematic in cosmological weak lensing measurements and a probe of how galaxies interact with the tidal fields of the surrounding dark matter environment. Our understanding of galaxy alignments is currently poor and limited by the availability of observations, with cosmological simulations currently failing to yield a fully consistent picture \citep[e.g.][]{samuroff21}. Of the highest priority are an as-yet missing clear detection of alignments amongst late-type (rotationally supported) galaxies, establishing the currently highly uncertain evolution with redshift of the effect, and pushing constraints for early-type (pressure-supported) galaxies to the lower luminosities typically targeted by galaxy imaging surveys.

In \cref{fig:ia} we show simple forecasts for the signal-to-noise of intrinsic alignment measurements in \wst\ Legacy Survey samples combined with galaxy shapes from the \lsst. These are contrasted against analogous predictions for analyses of galaxy samples from a variety of existing and planned surveys, assuming a standardised alignment signal that does not depend on luminosity, redshift, and galaxy type. We show the constraining power for early- and late-type samples separately via the area of the red and blue semi-circles, respectively, but note that the details of the sample selection vary considerably from dataset to dataset.

All three \wst-Legacy samples will enable a step change in intrinsic alignment science over the state of the art: compared to the equivalent samples from the completed \desi\ and \textit{Euclid} surveys, the constraining power will increase by at least a factor 4, and by an order of magnitude for the LRG sample. \wst\ facilitates the study of intrinsically fainter galaxies, covering the bulk of the redshift and luminosity range of galaxies used in weak lensing cosmology for the first time (see the grey \textit{Euclid} density distribution in \cref{fig:ia}; note that the \lsst\ Gold Sample used for weak lensing analysis will be qualitatively similar). These specs will allow for unprecedented insights into the dependence of elliptical galaxy alignments on host halo mass \citep[see][]{piras18}, and into satellite and central galaxy alignments with the BG sample (compare to \citealt{georgiou19,johnston19} for studies on samples with similar number density but over 180$\,{\rm deg}^2$ only). 

Weak lensing itself is a powerful tool to directly map the dark matter distribution around samples of \lq lenses\rq. In the \wst\ era the completed datasets from \textit{Euclid}, the \lsst, and the \nancy\ \citep{RomanTelescope} will provide background screens of source galaxies that are highly complementary in terms of sky coverage, depth, and spatial resolution. This will enable high-precision studies of the dark matter distribution around \wst\ galaxies in the new redshift-luminosity regime illustrated in \cref{fig:ia}, including dark matter halo masses and radial profiles \citep[e.g.][]{mandelbaum06}, halo ellipticity \citep[e.g.][]{schrabback21,robison23}, as well as group, satellite, and filament environments \citep[e.g.][]{clampitt16,sifon18}.

Moreover, by combining galaxy weak lensing with clustering and e.g.\ stellar mass function measurements, it is possible to extract the dark matter halo occupation statistics of galaxies jointly with cosmological information from small physical scales that is complementary to traditional linear-scale cosmological analysis (e.g., \citealt{leauthaud17,dvornik23}; see also \citealt{amon23} and references therein). Taking the largest sample considered in \citet{amon23}, i.e.\ SDSS-BOSS lenses combined with Dark Energy Survey Year-3 lensing, as reference, both the clustering and weak lensing measurements with \wst\ Legacy Sample lenses will reduce their noise by a factor of 30. The combination of great depth and high surface density allows \wst\ to deeply probe into galaxy-size dark matter haloes. With these unique specs, \wst\ will tightly constrain the stellar-to-halo mass relation over an extremely wide mass range, determine the statistics of satellite galaxies occupying their host halo, and make decisive statements on the significance of assembly bias and the feedback through baryonic processes on the dark matter distribution, key information to ultimately resolve the $\Lambda$CDM structure growth tension \citep{abdalla22}.

\subsubsection{Cosmic voids}

Cosmic voids, the under-dense regions of the cosmic web, are powerful tools to constrain cosmology \citep{Pisani19}, both as a stand-alone probe and in combination with traditional techniques \citep{pelliciari}.
Their number counts and shapes, as well as the dynamics of galaxies on their boundaries, offer well-established new means to probe cosmology \citep[see e.g.][]{Lavaux2012, Pisani2015, Hamaus2016, Nadathur16, Nadathur2020, Hamaus2020, Zhao2020, Zhao2022, Woodfinden2022, Contarini2023}. In particular the stacked void-galaxy correlation function can be used for the Alcock-Paczynski test \citep{Lavaux2012, Ceccarelli13, Paz13, Hamaus14}, or to extract information by modelling the RSD \citep[see e.g.][]{Hamaus2015, Chuang2017, Correa21, Correa22}. The sizes of voids depend on the nature of the dark energy component \citep{Verza19,Verza22, Verza23}, are of same order of magnitude as the screening length of some modified theories of gravity \citep{Perico19, Contarini21}, and comparable to the free-streaming scale of massive neutrinos \citep{Massara2015, Kreisch19, Schuster_etal2019, Contarini21, Verza23}.

The void clustering is also a powerful tool to extract cosmology \citep{Aubert2022, Kreisch2022, Zhao2022, Tamone2023}, as well as void lensing, where tests of gravity are based on the weak lensing signal of cosmic voids \citep{Sanchez2017, Baker2018, Boschetti2023}. Finally, cosmological constraints from voids can also be improved by including cross-correlations between voids and galaxy lensing \citep{Bonici2023}, or by considering the imprint of voids on the CMB \citep{Kovacs2019, Vielzeuf2021, Vielzeuf_etal2023}.

In the following paragraphs, we present the cosmological forecasts derived from the size function of cosmic voids in the \wst\ survey\footnote{The analyses presented in this section have been performed with the \texttt{CosmoBolognaLib} \citep{Marulli2016}, available at \href{https://gitlab.com/federicomarulli/CosmoBolognaLib}{gitlab.com/federicomarulli/CosmoBolognaLib}.}. The void size function describes the comoving number density of cosmic voids as a function of their effective radius. Its theoretical model is based on the excursion-set formalism, first developed by \cite{SvdW04} and \cite{Jennings13}. More recently, \citet{Verza24} developed a theoretical model that merges the excursion-set formalism with the statistics of the Lagrangian density peak \citep{Bardeen86}, allowing for a robust treatment at all scales, particularly including the smallest ones.

The void size function model we employ in this study for the \wst\ forecasts is the one presented in \citet{Contarini2022}, already used for \textit{Euclid} spectroscopic survey forecasts and to analyze current BOSS data \citep{Contarini2023, Contarini2024}.

In spectroscopic galaxy surveys, voids are detected within the redshift-space galaxy distribution. A void finding algorithm is run on the 3D distribution of tracers \citep[e.g.][]{Sutter15, Nadathur16, Paz23}, followed, when necessary, by a cleaning algorithm \citep{Ronconi17} to align the definition of voids used in the identification process with the one assumed in the theoretical model of the void size function. Since it relies on observed galaxies, the analysis requires modeling both RSD and galaxy bias with nuisance parameters, typically calibrated through N-body simulations. Here, we adopt the same calibration as in \cite{Contarini2022} and marginalize over the two nuisance parameters.

For this forecast we consider the \wst\ Legacy and Synergy survey, assuming access to the positions of $180$ million galaxies in the redshift range $z<1.5$, covering a sky area of $15,000 \ \mathrm{deg}^2$. In this analysis, we do not consider voids that could potentially be detected in the high-redshift survey ($3<z<6$) since they are expected to be shallow and therefore harder to identify and model. We generate synthetic data from the theoretical model of the void size function. Specifically, we theoretically estimate the average number counts expected from the \wst\ galaxy catalog, along with the corresponding statistical uncertainty.

We assume that the \wst\ voids are traced by galaxies with properties similar to those analyzed in \cite{Contarini2022}. We consider that the linear bias, $b_\mathrm{eff}(z)$, and the number density of galaxies, $n(z)$, follow the same trends as sampled from the Flagship simulation, the official \textit{Euclid} mock light-cone \citep{Potter17, EC20}, spanning the redshift range expected for the \textit{Euclid} spectroscopic survey, i.e.\ $0.9<z<1.8$. We extend the predictions of $b_\mathrm{eff}(z)$ down to the redshift range covered by the low-redshift survey of \wst\ and then rescale $n(z)$ consistently with the expected galaxy density.

\begin{figure}
    \centering
    \includegraphics[scale=0.32]{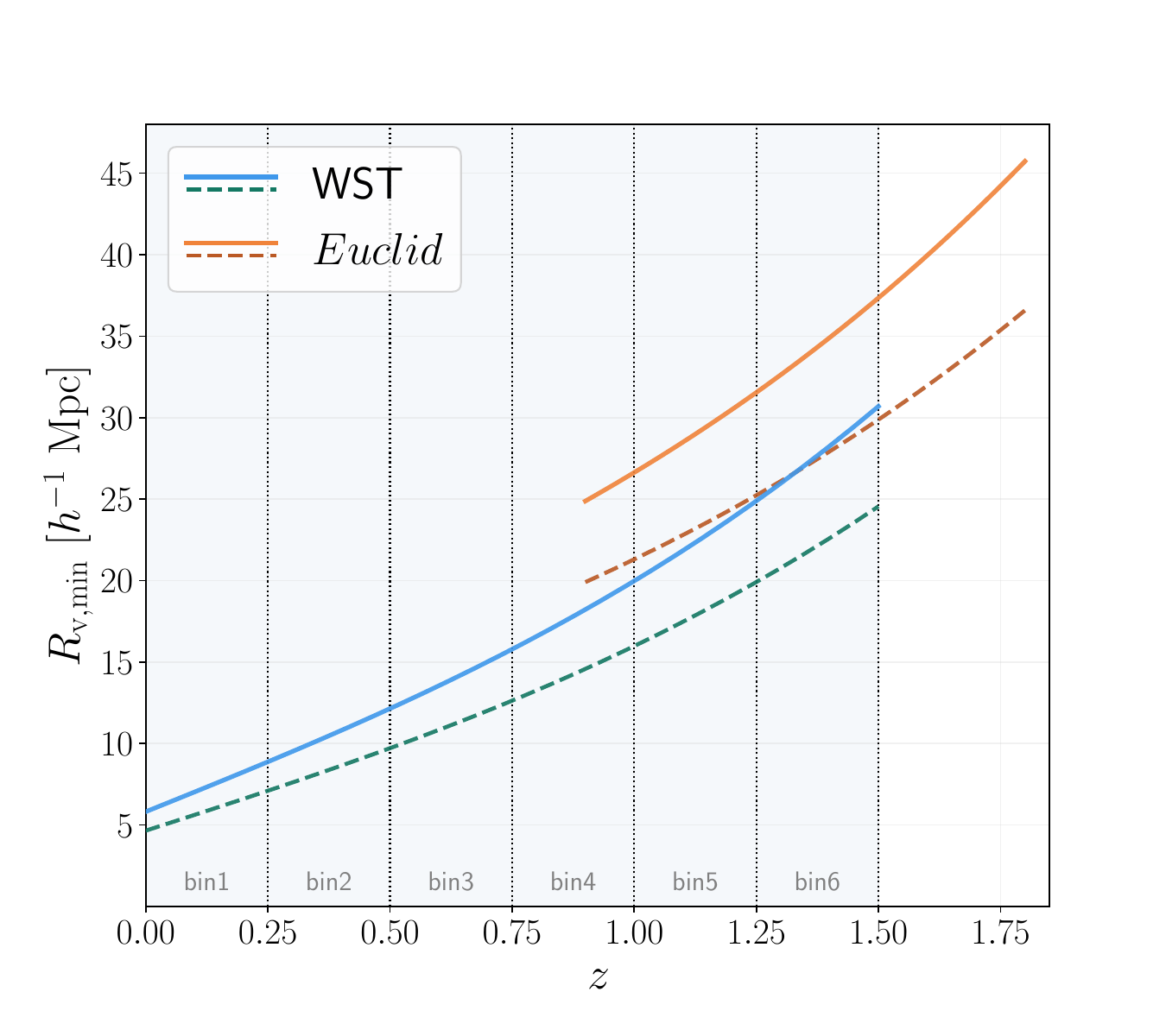}
    \includegraphics[scale=0.32]{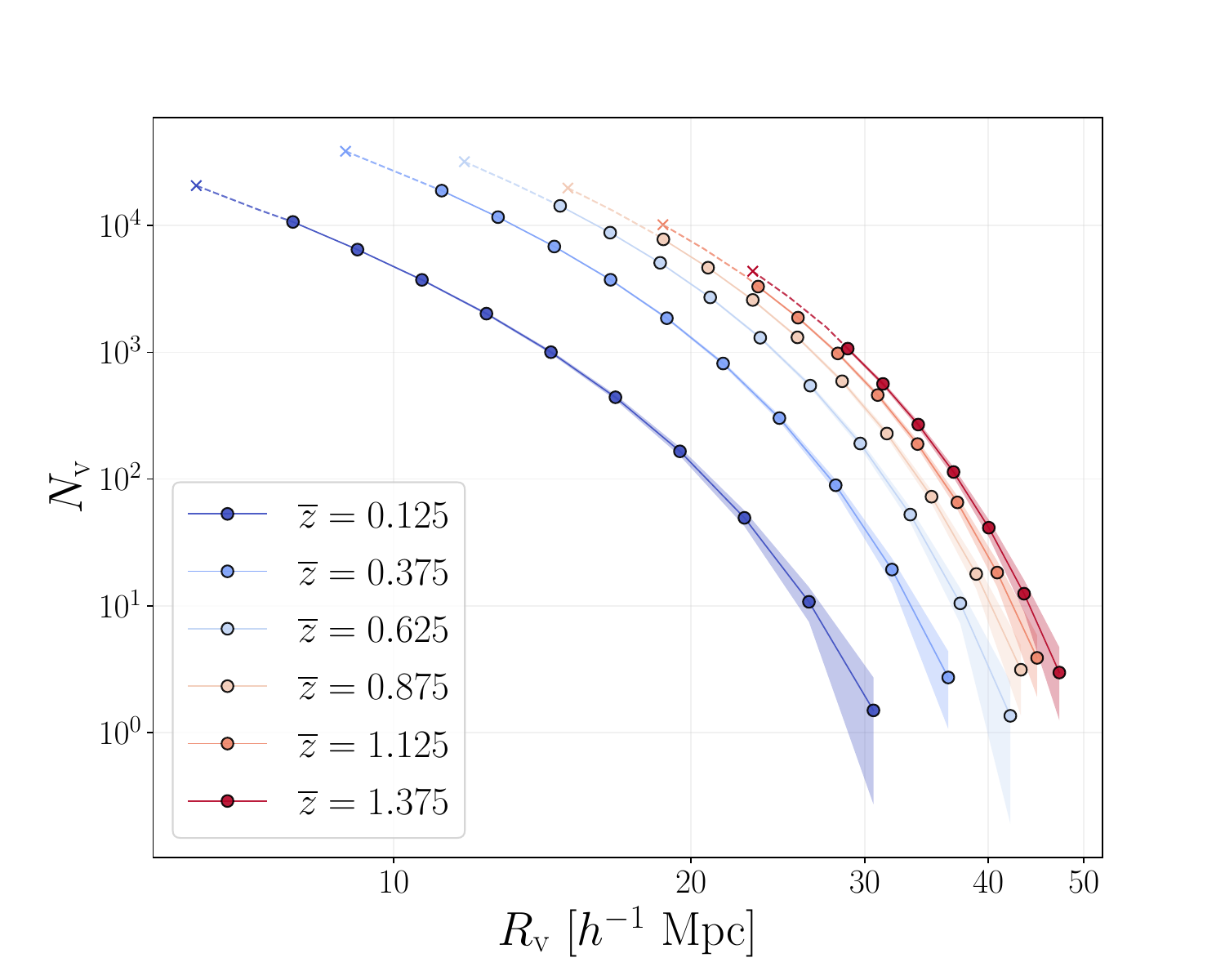}
    \caption{\textit{Left}: Minimum void radii considered to derive cosmological forecasts for the \wst\ (blue) and \textit{Euclid} (orange) surveys. Solid lines are computed with a pessimistic (thus more conservative) approach, dashed lines represent an optimistic setting. \textit{Right}: Number counts of cosmic voids as a function of their effective radii, expected for the \wst\ survey. Different colours depict the predictions for different mean redshifts, extracted from six equi-spaced redshift bins at $z < 1.5$. Shaded areas indicate the Poisson errors associated with the void number counts, while dashed lines ending with a cross illustrate an extension of the predicted void number counts down to the minimum spatial scale that could optimistically be modeled.}
    \label{fig:minR_VSF+predicted}
\end{figure}

The $n(z)$ function is crucial for determining the scales at which the void size function model is applicable when considering voids observed in a discrete distribution of tracers. Below a certain scale, the measured void size function is affected by incompleteness, i.e.\ loss of void counts due to the limited spatial resolution of the tracer catalog. To address this issue, the standard approach is to consider only those voids with radii greater than the mean separation of the tracers (galaxies, in our case), $\mathrm{MGS}(z)=n(z)^{-1/3}$, multiplied by a factor, $f_\mathrm{cut}$. Common choices for $f_\mathrm{cut}$ range between $2$ and $2.5$ \citep[see, e.g.][]{Pisani2015b, Ronconi19, Verza19, Contarini19, Contarini21, Contarini2022, Contarini2023, Paz23}. 
Here, we derive cosmological forecasts considering two settings: a pessimistic one, where the simulated void size functions are computed assuming a minimum void radius, $R_\mathrm{v,min}=f_\mathrm{cut} , \mathrm{MGS}(z)$ with $f_\mathrm{cut}=2.5$, and a less conservative, optimistic one, with $f_\mathrm{cut}=2$. We illustrate the predictions for $R_\mathrm{v,min}$ as a function of redshift for \wst\ and \textit{Euclid} in the left panel of \cref{fig:minR_VSF+predicted}, obtaining these values from the expected $n(z)$ for the two surveys.

The expected void counts are computed in six equi-spaced bins of redshift, also depicted in \cref{fig:minR_VSF+predicted} (left panel) for \wst. We assume a fiducial cosmology consistent with \citet{Planck2020} for both \wst\ and \textit{Euclid}. 
We use the synthetic void data to compute cosmological forecasts, fixing all cosmological parameters except for the cold dark matter density, $\Omega_\mathrm{cdm}$, and the neutrino mass, $M_\nu$. We assume an effective number of neutrino species $N_\mathrm{eff}=2.04$ and relate the neutrino mass to the neutrino density parameters as $\Omega_\nu=M_\nu / (93.14 \ h^2 , \mathrm{eV})$ \citep{Mangano05}. Finally, we assume a flat universe geometry by rescaling the dark energy density as $\Omega_\mathrm{de}=1-\Omega_\mathrm{m}$, where $\Omega_\mathrm{m}=\Omega_\mathrm{cdm}+\Omega_\mathrm{b}+\Omega_\nu$. Synthetic data are then extracted from the void size function model for void radii values logarithmically spaced. We assign wide uniform priors to the free parameters, except for the two nuisance parameters of the void size function model, which are assumed to be correlated and follow a 2D Gaussian prior based on the calibration performed in \cite{Contarini2022}.
We assume a Poissonian likelihood \citep{Sahlen16} and run Markov Chain Monte Carlo (MCMC) chains to sample the posterior distribution of $\Omega_\mathrm{cdm}$ and $M_\nu$, treating $\Omega_\mathrm{de}$ and $\sigma_8$ as derived parameters.

The right panel of \cref{fig:minR_VSF+predicted} shows the predicted void size function, along with the difference in the scale cut values between the two considered settings.
The total number of expected voids for \wst\ is approximately 45,000 [1 million] in the pessimistic [optimistic] configuration. For comparison, the total number of expected cleaned voids for \textit{Euclid} is about 2.5 thousand [15 thousand] in the pessimistic [optimistic] configuration. Higher numbers are expected for the void-galaxy cross-correlation since it requires a less severe cleaning procedure. 

\begin{figure}
    \centering
    \includegraphics[scale=0.3]{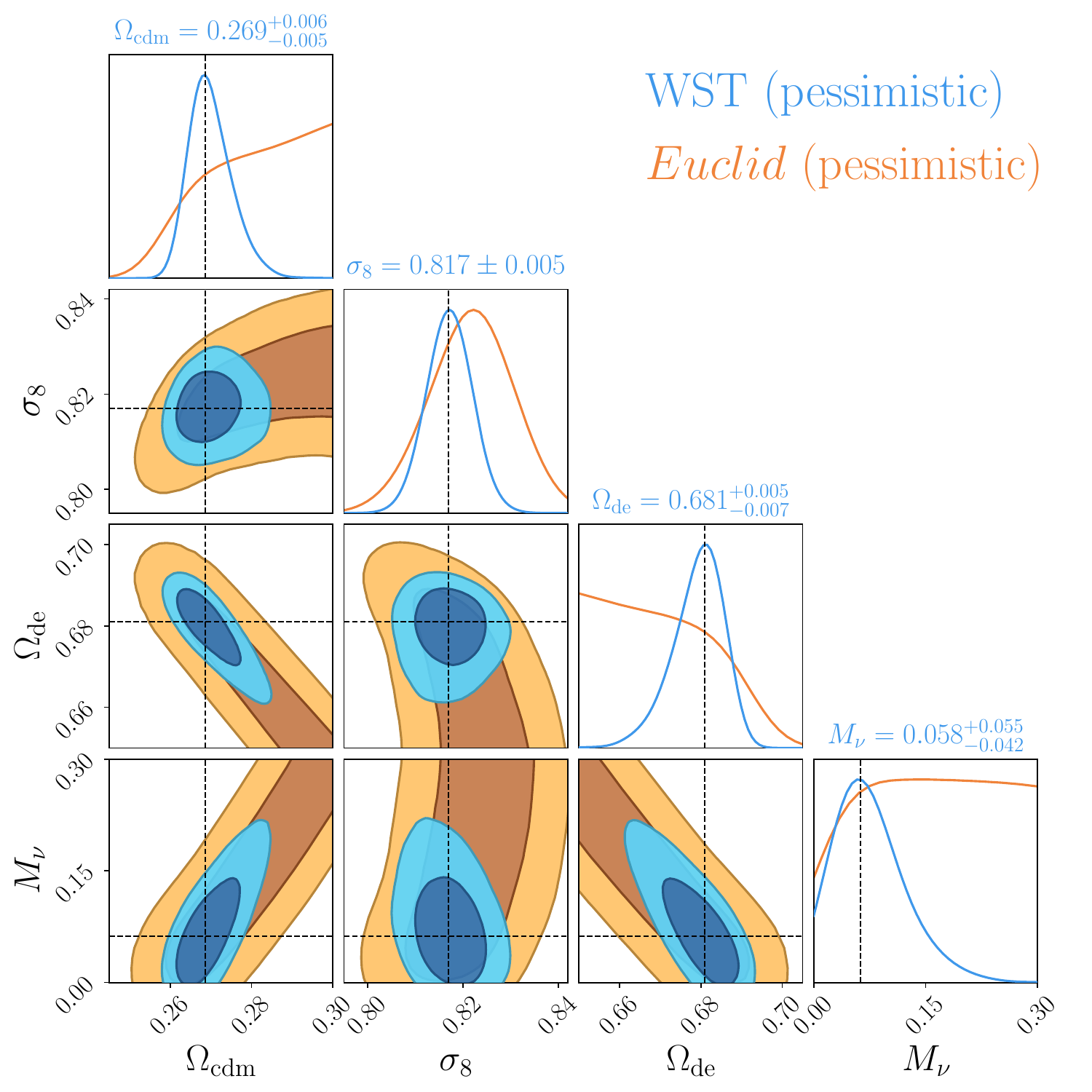}
    \includegraphics[scale=0.3]{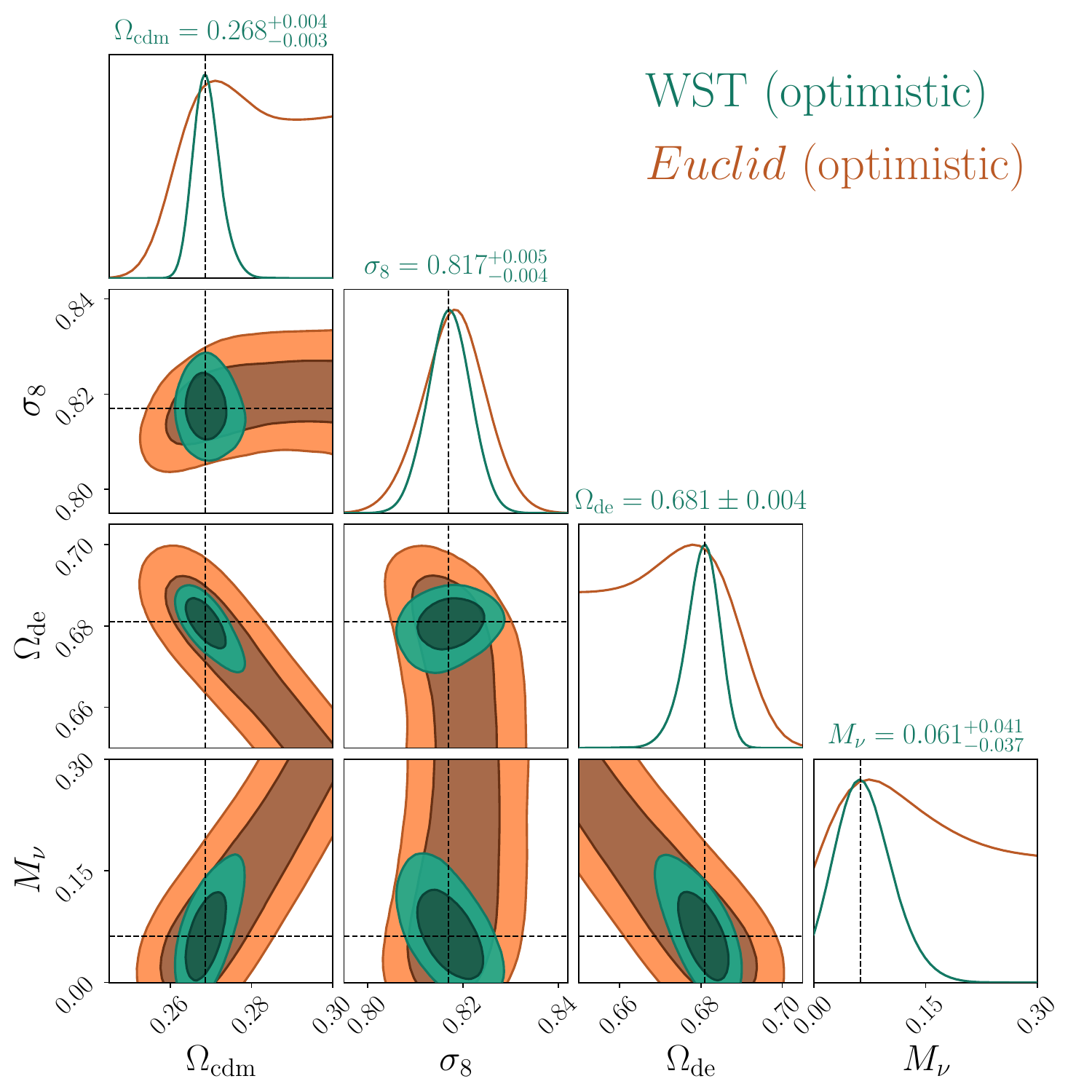}
    \caption{Cosmological forecasts from void number counts computed for \wst\ (in blue), compared to those expected for a \textit{Euclid}-like survey (in orange). In the case of \wst, the constraints are derived from the predictions on void size functions reported in the right panel of \cref{fig:minR_VSF+predicted}. \textit{Left}: Contours representing the $68\%$ and $95\%$ confidence levels obtained by simulating the expected cosmic void counts with a pessimistic approach, i.e.\ assuming $R_\mathrm{v,min}=2.5 \, \mathrm{MGS}(z)$. 
    \textit{Right}: Similar contours but for the optimistic setting, i.e.\ assuming $R_\mathrm{v,min}=2 \, \mathrm{MGS}(z)$.
    We indicate with black dashed lines the value of the fiducial cosmological parameters and we report the parameter values derived for \wst\ above the projected 1D constraints, together with the corresponding uncertainty.}
    \label{fig:forecasts_VSF}
\end{figure}

We present the results computed in the pessimistic [optimistic] configuration in the left [right] panel of \cref{fig:forecasts_VSF}. 

While the constraining power does not improve significantly when moving from the pessimistic to the optimistic configuration, it changes dramatically between \wst\ and \textit{Euclid}. For example, \wst\ will allow to reduce the relative error on $M_\nu$ by a factor $\sim 6$.

From the comparison of their confidence contours, it clearly emerges that for \wst, the degeneracy between $\Omega_\mathrm{cdm}$ and $M_\nu$ is milder, and all the considered cosmological parameters are effectively constrained. We underline that, in this preliminary analysis, the only differences between \wst\ and \textit{Euclid} impacting our forecasts concern the $n(z)$ (resulting in different minimum scale cuts) and the redshift range. In conclusion, even though \textit{Euclid} will cover a larger volume of the Universe compared to \wst, the significantly larger number of voids expected for the latter at lower redshifts, due to the high tracer density, will lead to remarkably tighter constraints on the analysed cosmological parameters. The high tracer density
of \wst\ will require an extremely robust modelling of the low-end side of the void size function, recently explored in \citep{Verza24}. While this could, in principle, impact the posterior distribution of the explored cosmological parameters, the overall constraining power is expected to be similar.

Aside from the void size function, the stacked void-galaxy cross-correlation function \citep{Lavaux2012, Hamaus14, Pisani14, Hamaus2020, Nadathur2020, Aubert2022} provides tight constraints on both the growth of structures, via RSD effects, and the expansion history of the universe, through the Alcock-Paczynski test \citep{Lavaux2012, Cai2016, Hamaus17, Hamaus2020, Hamaus22}.

The observed void-galaxy cross-correlation function can be modeled using parameter-free de-projection methods, and jointly account for geometrical and dynamical distortions, as well as tracer bias \citep{Pisani14, Hamaus2020}. To reliably forecast the constraining power of the void-galaxy cross-correlation function, it is essential to consider the underdense nature of voids and their spatial extent \citep{Schuster2022, Schuster2023}, requiring simulated light-cones that match the survey specifications, so far not available for \wst. Nevertheless, the expected constraining power is anticipated to increase significantly with the number of voids. For a survey similar to \textit{Euclid}, analyzing this statistic can provide constraints on the expansion history and relevant cosmological parameters (such as dark energy density) at a level comparable to or even better than more traditional techniques like baryon acoustic oscillations and other galaxy clustering analyses \citep{Hamaus22}. Consequently, a high level of constraining power is expected from the \wst\ void-galaxy cross-correlation function. 

\subsubsection{Gravitational waves as standard sirens}
\label{cosmo:GW}

\begin{figure}[t]
  \centering
  \includegraphics[width=0.465\textwidth]{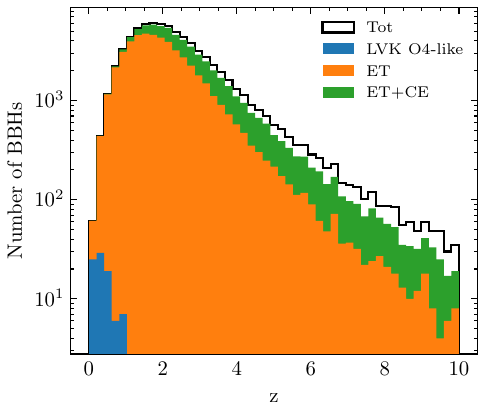}
  \includegraphics[width=0.475\textwidth]{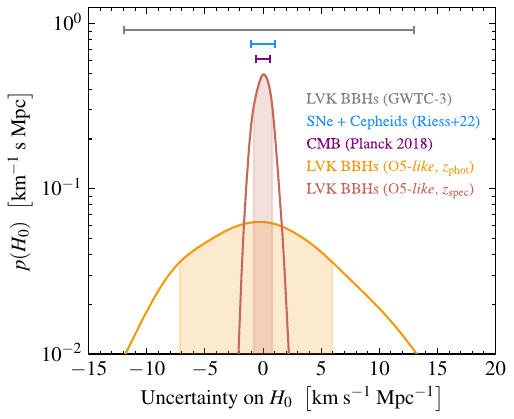}
   \caption{Forecasts on the improvement achievable with \wst\ on cosmological analyses with gravitational waves as dark sirens.
   Left panel: redshift distribution of the expected black hole binaries observable for the observing Run 4 (O4) in the LIGO-Virgo-KAGRA network (in blue), for Einstein Telescope (ET, in orange) and Einstein Telescope combined with Cosmic Explorer (CE, in green, see \citealt{Iacovelli2022}). Right panel: uncertainty on the Hubble constant achievable for the Observing Run 5 with an LVK network configuration considering a complete photometric catalog compared to a complete spectroscopic catalog as would be provided by \wst\ (respectively in yellow and in red, adapted from \citealt{Borghi2024}); for comparison, the gray error bar represents the current uncertainty obtained from the analysis of LVK BBHs with the Gravitational Wave Transient Catalog 3 \citep[GWTC3][]{GWTC3-2023}, the violet one the accuracy on $H_0$ reached by CMB analysis \citep{Planck2020} and the blue one from SNe+Cepheids analysis \citep{Riess2022}.}
   \label{fig:GW}
\end{figure}

Gravitational waves (GW) are extremely promising cosmological probes, especially considering the new foreseen \citep[Advanced LIGO-Virgo,][]{AdVLIGO2015,AdVVirgo2015} and 3rd generation (3G) GW observatories (Einstein Telescope, \citealt{ET}; Cosmic Explorer, \citealt{CE}; LISA, \citealt{LISA2}). The coalescence of compact objects (such as binary black holes, BBH, and binary neutron stars, BNS) generates GW that can be used as ``standard sirens'', providing a direct determination of the luminosity distance $d_L$ without relying on any further calibration. However, the downside of this approach is the degeneracy of the signal with redshift, which has to be broken to derive cosmological constraints. Different methods have been proposed in the literature \citep{Moresco2022}, either based on the discovery of the electromagnetic counterpart of the GW event ({\it bright siren}), or on the statistical association of the redshift of the host through the use of a galaxy catalog ({\it dark siren}), or by taking advantage of features in the distribution of astrophysical properties of the merging binaries to be used as scale indicators, like the peak in the binary BBH function ({\it spectral siren}). \wst\ will be an extremely powerful and versatile instrument to follow-up and search the electromagnetic counterparts of bright sirens events, and this science case is discussed in Sect. \ref{td:GW}.
Concerning the dark siren science case, GWs have been used to obtain new constraints on the Hubble constant $H_0$ by cross-correlating at the moment mostly with the GLADE+ \citep{Dalya2022} and Dark Energy Survey \citep[DES][]{DES2016} catalogs. An example of a first analysis is given by the DES paper \citep{SoaresSantos2019,Palmese2020}.
However, it has been shown that it is crucial to obtain a more homogeneous, complete, and deep galaxy catalog to fully apply this method, especially in view of the new GW observation campaigns with current interferometers (LIGO-Virgo-KAGRA, LVK) and with the expected 3G network (ET, CE). 
The deep \wst\ spectroscopic samples over very wide sky areas will represent a crucial piece in future GW analyses by improving dramatically the results in an unexplored redshift range, allowing us to reach the percent accuracy on the Hubble constant, and significantly improving the determination of several other parameters \citep[e.g., see][]{Borghi2024}. The plan is to provide a spectroscopic catalog as deep and complete as possible from follow-up spectroscopic observations of the wide photometric catalogs that will be obtained in the near future by the incoming \textit{Euclid} and \lsst. 

As an example, in \cite{Borghi2024} (but see also Ciancarella et al in prep) it has been shown that already with the sensitivity of the Observing Run 5 of LVK the analysis of the 100 BBH events with signal-to-noise ratio larger than 25 can provide a determination of the Hubble constant to percent level if complemented with the information of a complete galaxy catalog with spectroscopic redshift. On the other hand, if only photometric redshifts are available the signal degrades by a factor $\sim$10, as shown in \cref{fig:GW}. This demonstrates how fundamental is such a complete and deep spectroscopic redshift catalog as \wst\ could provide, especially since 3G GW observatories will increase by a factor $\sim$10$^3$ the number of detected GW events, opening also new window of detectability as a function of redshift (see \cref{fig:GW}).

Moreover, this will allow us to measure not only $H_0$ but also to test dark energy and early dark energy models. To this aim, it will be necessary to reconstruct the $z$-$d_L$ relation for GW sources without an electromagnetic counterpart. The cross-correlation of GW detection with \wst\ galaxy catalogs will be extremely important not only for cosmological but also for astrophysical reasons, since it will allow us to measure the bias of binary mergers at high redshifts and give insight into their host galaxies evolution. Finally, another important probe will be the cross-correlation of unresolved GW events (stochastic gravitational background SGWB) with high redshift galaxy catalogs as the one provided by \wst. Inflation predicts a cosmological background of gravitational waves. A detection, in the proper frequency range, of its cross-correlation with galaxy survey, especially at high redshifts, could shed light on the early universe and its initial conditions. In the pre-run phase, forecasts will also be very useful to assess the \wst\ performance in combination with GW events.

\subsubsection{Cosmic chronometers}

\begin{figure}[t!]
  \centering
  \includegraphics[width=0.95\textwidth]{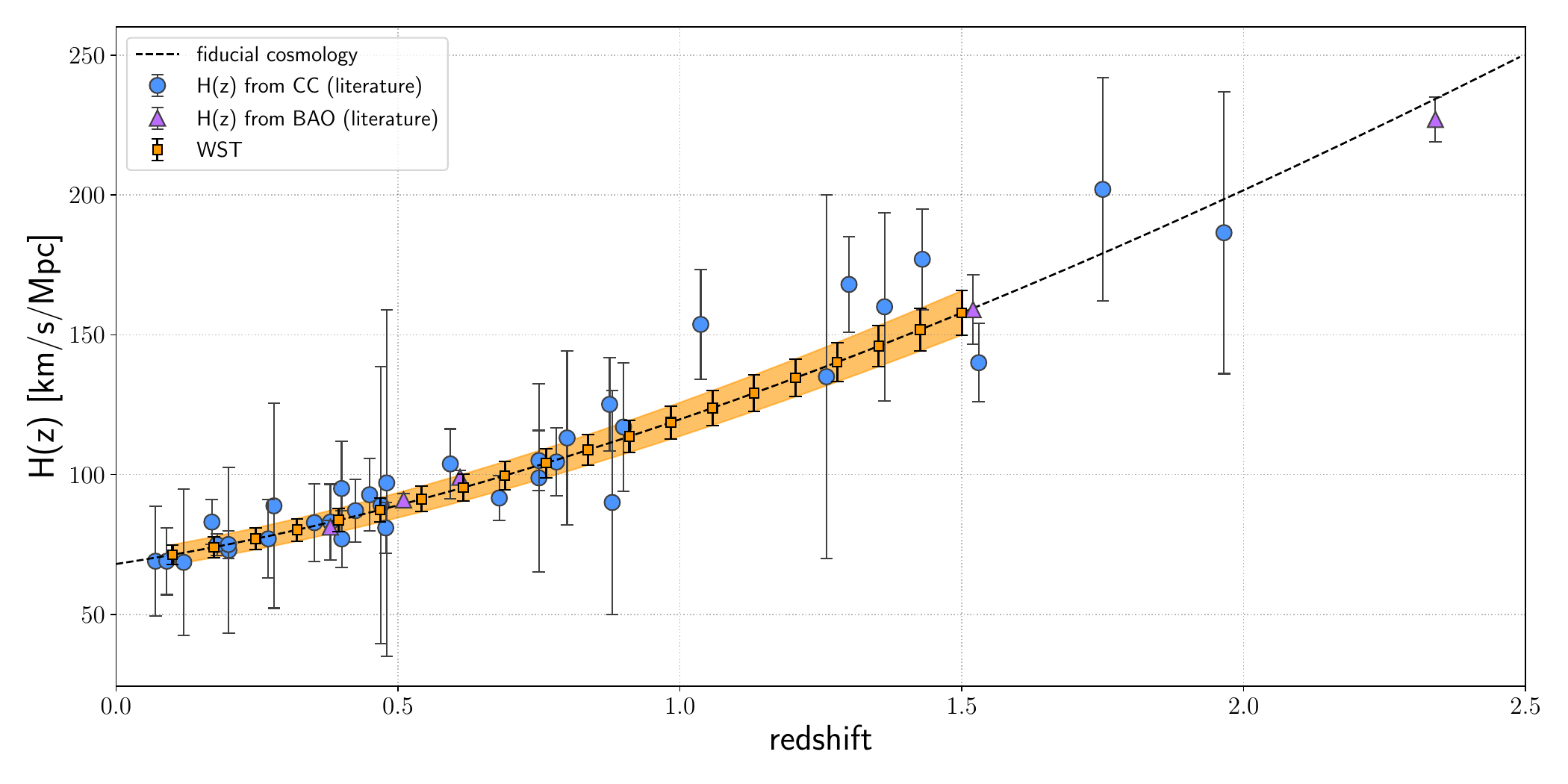}
   \caption{Forecast on the constraints on the Hubble parameter $H(z)$ that can be obtained with \wst\ considering the cosmic chronometers approach. The gray circle points represent the current $H(z)$ measurements from the literature with cosmic chronometers, while the triangle points show the constraints from Baryon Acoustic Oscillation analyses; the orange shaded area is the accuracy that can be reached from the analysis of very massive and passively evolving galaxies in \wst\ in the range $0<z<1.5$.}
   \label{fig:CC}
\end{figure}

In parallel with more standard approaches, in the last decades new and alternative techniques to derive constraints on the expansion history of the Universe have emerged. These novel cosmological probes are more and more fundamental since they can bypass the limitations of standard methods, provide independent measurements, enhance the accuracy and precision of the results, and give information to check and mitigate systematic effects \citep[for a review, see][]{Moresco2022}. In this context, 
as first introduced by \cite{Jimenez2002}, a measurement of the differential age evolution of the Universe over a redshift range provides a direct estimate of the expansion rate of the Universe through the simple equation $H(z)=-\frac{1}{(1+z)}\frac{dz}{dt}$, without any cosmological assumption. It has been demonstrated that very massive and passively evolving galaxies can play a significant role in this context, since they can be exploited as ``cosmic chronometers'', being a population capable of tracing the oldest objects in the Universe in a very homogeneous way; therefore, by measuring their redshift and spectra, their ageing (dt) over a redshift interval (dz) can be constrained, providing a direct measurement of the Hubble parameter $H(z)$. This method has proven to be very promising so far, providing constraints on the Hubble parameter in the redshift range $0<z<2$ with an accuracy ranging between 5 and 15\% (see \citealt{Moresco2022,Moresco2023} for reviews on the topic). So far, one of the limitations of this technique has been the fact that no dedicated spectroscopic survey exists to target specifically massive and passively evolving galaxies, which remains a population elusive and hard to reveal. The characteristic of \wst, both in terms of survey area, spectral resolution, and spectroscopic signal-to-noise ratio achievable will provide a unique and homogeneous way to sample this population across a very wide range of cosmic times, as shown in \cref{fig:CC}, allowing us to sample with unprecedented accuracy the expansion history of the Universe. In particular, where previously this method was capable of reaching an accuracy on $H(z)$ of $\sim5$\% only on a few points up to $z\sim0.5$, and up to $\sim15$\% up to $z\sim1.5-2$, with \wst\ this approach is expected to reach an accuracy of the order of $\sim5$\% over the entire redshift range $0<z<1.5$.

\subsubsection{Probing the cosmic baryon distribution with FRB foreground mapping}

Baryons constitute only $\Omega_b \approx 5\%$ of the Universe, but represent all ``normal'' visible matter. 
All the observable stars and galaxies in the Universe, however, account for $<10\%$ of the cosmic baryon budget as measured by Big Bang Nucleosynthesis (BBN) and cosmic microwave background (CMB) anisotropies (e.g., \citealt{fukugita:2004}). 
Instead, most cosmic baryons are expected to reside in the diffuse gas well outside the visible stellar components of galaxies, either in the circumgalactic or intergalactic medium (CGM and IGM) \citep{cen:2006}.
Until recently, heroic observational efforts using, e.g., X-rays, quasar absorption lines, and the Sunyaev-Zel'dovich effects have failed to account for the full cosmic baryon budget \citep[e.g.][]{de-graaff:2019}, 
primarily because interpreting these probes requires model assumptions on variables such as gas temperature and metallicity, and can only detect baryons in specific regimes.
This `missing baryon problem', however, can be addressed by fast radio bursts (FRBs).

\begin{figure*}
    \centering
    \includegraphics[width=0.9\textwidth]{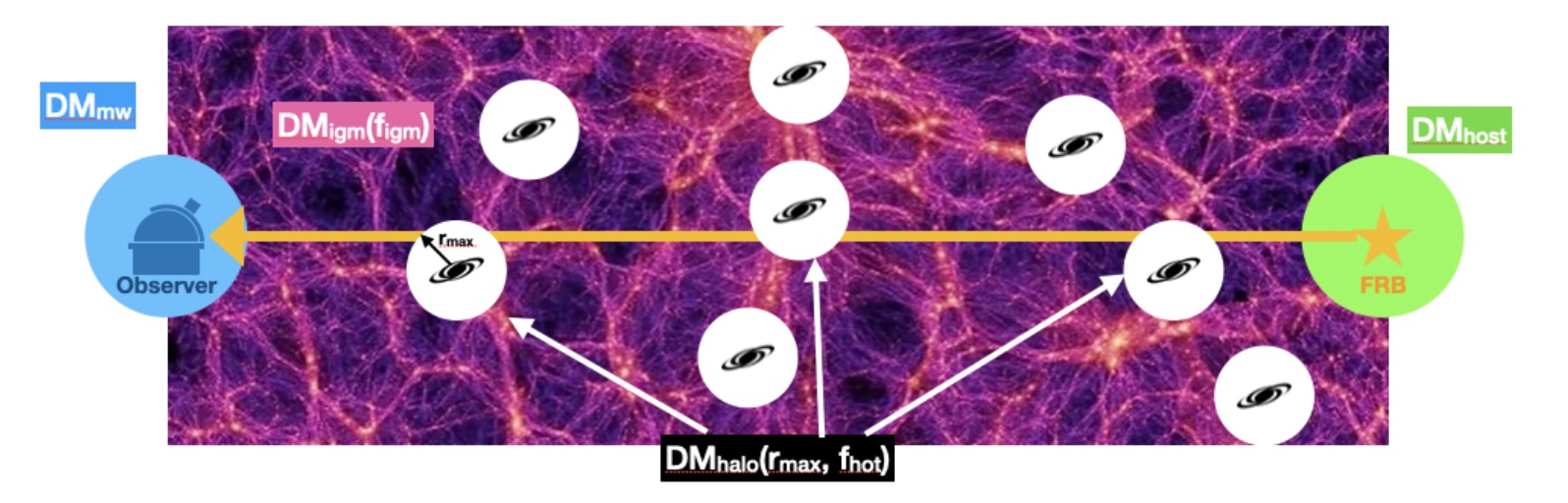}
    \caption{Schematic diagram showing the various components in a typical FRB dispersion measure, $\mathrm{DM_{frb}}$. While the Milky Way contribution, $\mathrm{DM_{mw}}$, is well-constrained by Galactic pulsars, the other components cannot \textit{a priori} be separated using the FRB measurement itself. Detailed spectroscopic foreground galaxies, however, allow the possibility of decomposing these components (e.g., \citealt{simha:2020,lee:2023}): $\mathrm{DM_{halo}}$ can be modelled from directly-intersected galactic halos, while $\mathrm{DM_{igm}}$ is modelled through cosmic web maps reconstructed from the large-scale galaxy distribution.}
    \label{fig:frb_schematic}
\end{figure*}

First discovered by \citet{lorimer:2007}, FRBs are millisecond-duration radio signals whose exact origins are still unclear, but are likely related to magnetars (see \citealt{cordes:2019} for a review). 
As these transient radio signals traverse the Universe, electromagnetic interactions with free electrons lead to a frequency-dependent time delay. This is usually quantified by the dispersion measure, or $\mathrm{DM_{frb}} = \int n_e(l) dl$, i.e.\ the path integral of the free electron density $n_e$ over the path $l$.
Since the vast majority of IGM and CGM gas is expected to be ionized, $\mathrm{DM_{frb}}$ thus offers a clean probe of the cosmic baryons, although there are also contributions from the Milky Way and host galaxy as well as possibly the FRB engine (see \cref{fig:frb_schematic}).
The constraining power of FRBs is enhanced if they can be \textit{localized} to specific host galaxies with measurable redshifts, thus revealing the limits of the $\mathrm{DM_{frb}}$ integral. 

In a landmark paper, \citealt{macquart:2020} used a sample of 7 localized FRBs to demonstrate that the $\mathrm{DM_{frb}}$--$z$ relationship is consistent with $\Omega_b$ as determined by BBN and CMB observations.
This result technically solved the missing baryon problem, but the next step is to ask \textit{where} in the cosmos are the baryons distributed? For example, what fraction resides in the IGM vs the CGM? \citet{khrykin:2023} recently showed that the relative partition between the low-density IGM and CGM within $r<r_{200}$ of galactic halos is sensitive to the nature of galaxy and AGN feedback.  
Large-scale redistribution of baryons by feedback smooths the overall matter distribution \citep[e.g.][]{chisari:2018}, and needs to be taken into account in e.g., weak lensing, analyses.

However, at fixed redshift $z$, there is considerable variance in the $\mathrm{DM_{frb}}(z)$ values due to cosmic web fluctuations along the path (e.g., \citealt{jaroszynski:2019,takahashi:2021}), as well as stochastic intersections with gas halos of individual foreground galaxies \citep{zhu:2021,walker:2023}.
It has been estimated that $\sim 10^3$ localized FRBs would be required to beat down this variance and place constraints on feedback models \citep{batten:2021}. 
However, \citet{lee:2022} showed that adding foreground spectroscopic data dramatically improves the constraining power of FRBs toward the cosmic baryon distribution. 
This generally arises from two components: (i) a wide and shallow galaxy redshift survey to map cosmic web fluctuations traversed by the FRB path, and (ii) a deeper census of individual galaxies whose halo gas might be intersected directly by the FRB sightline.
This concept has recently enabled the first constraints on the IGM and CGM cosmic baryon fractions using just 8 FRB sightlines from the FLIMFLAM Survey \citep{khrykin:2024}, a joint observational program primarily using the AAOmega fibre spectrograph on the 3.9m AAT, as well as deeper observations with 8-10m facilities.

\begin{figure*}
\begin{centering}
    \includegraphics[width=0.55\textwidth]{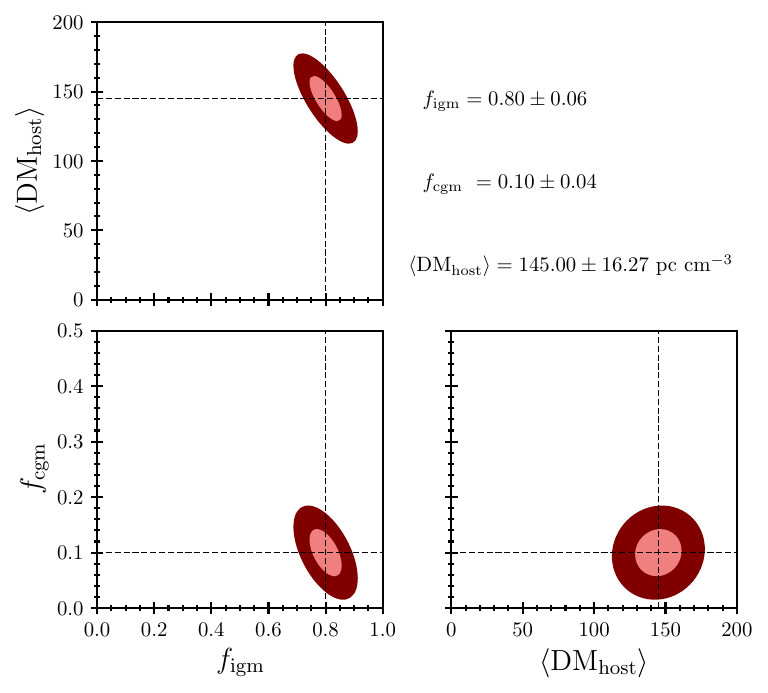}
    \caption{A Fisher matrix forecast of the measurements on the cosmic baryon distribution at $\langle z\rangle=0.2$ from a sample of $N=200$ FRBs at $0.1<z_\mathrm{frb} <0.4$ if complete spectroscopic data can be obtained from their foreground. 
    The forecasting methodology is outlined in \citet{lee:2022}.
    Such precision will enable detailed constraints on the nature of galaxy feedback \citep{khrykin:2023}.     
    \label{fig:frb_forecast}}
    \end{centering}
\end{figure*}

\cref{fig:frb_forecast} shows a Fisher matrix forecast on the cosmic baryon constraints enabled by $N=200$ low-$z$ FRBs localized at $0.1<z<0.4$. These forecasts are for a simple model based on the scenario shown in \cref{fig:frb_schematic}, in which a fraction $f_{igm}$ of all cosmic baryons reside in the diffuse IGM, $f_{cgm}$ is the fraction residing collectively in the CGM of galaxy halos, and $\langle \mathrm{DM_{host}}\rangle$ is the mean host dispersion of the FRB sample.

The number of FRB discoveries will increase exponentially over the next decade. By the time \wst\ comes online, SKA will be discovering $\sim 10^2$ fast radio bursts (FRBs) per day that are at $z<2$ \citep{hashimoto:2020}. 
\wst\ would be an excellent facility to follow up the most promising of these FRBs, by simultaneously localising the host galaxy as well as targeting foreground galaxies for cosmic baryon analysis.
With the IFS, the host galaxy redshift can be targeted including intervening galaxies (with impact parameters of several arcmins) in the foreground, while the MOS-LR will target large-scale structure galaxy tracers in the entire foreground light cone. 

Up to FRB redshifts of $z\sim 1.3$, the selection of MOS-LR large-scale tracer sample will be the same as the Grey Time Legacy Survey (\cref{sec:legacy_survey}). 
The primary difference would be that the pointings would be centred on the FRBs in order to stack multiple IFS exposures across the $\sim 3$ fibre configurations of the MOS-LR, which would improve the confidence of the host galaxy spectrum and facilitate a complete census of closely-intersected galaxies.
While leading to increased complexity in the tiling strategy and angular selection function of the Grey Time Legacy Survey, 
just one FRB pointing per Grey Time night would lead to samples of $N>1000$ FRBs within 3 years.

Such a sample would allow detailed modelling of the cosmic baryons beyond the simple model shown in \cref{fig:frb_forecast}. 
For example, we could constrain the redshift evolution of $f_{igm}$ to study the growth of the WHIM, while $f_{cgm}$ could be studied as a function of galaxy halo mass \citep{ayromlou:2023,khrykin:2023} to probe the nature of feedback across different mass regimes.
Simultaneously, detailed information from the host galaxy spectrum and FRB signal (radio properties such as scattering and scintillation) could be combined with the $\mathrm{DM_{host}}$ constraints to shed light into the nature of the FRBs.

With these results, cosmologists will finally achieve a full census of the ``easiest"  5\% of the cosmic mass-energy content that have thus far eluded characterisation.

\subsection{Cosmology survey design}
\label{sec:survey_design}

\wst\ will provide an order of magnitude more galaxy redshifts than have ever been observed before, overpassing the milestone of 100 million galaxy redshifts. It will both open a new galaxy redshift window at $2<z<7$ as well as consolidate the current lower redshift ($z<1.6$) surveys for cosmology and legacy science. The cosmology survey will be run during both Dark Time and Grey Time. The number density of cosmology targets is larger than the planned fibre density of the MOS-LR instrument (sse below) demonstrating that the current instrumental concept is adequate. However, because of the broad range of magnitudes in the targets and hence their required exposure time, it will  be possible to accommodate ancillary targets in parallel to the cosmology targets such as transient objects or science projects with low density of targets (typically with a number density level below 100 targets per square degree).

As mentioned above the prime imaging data to conduct the target selection will be the \lsst\ which will have released their 10-year survey data before the start of any \wst\ observation. Of course, other multi-wavelength survey data will be useful for specific target selection (such as radio data from SKAO, X-ray data from eROSITA, or CMB data from CMB-S4). However, the numbers given here are a first estimate limited to \lsst\ data and will be revisited in more detail in the future.
The extragalactic \lsst\ footprint covers $\sim18\,000\,\deg^2$ and thus will be the prime region to be covered by the \wst\ Cosmology survey.

\begin{figure}[t]
    \centering
    \includegraphics[width=0.9\textwidth]{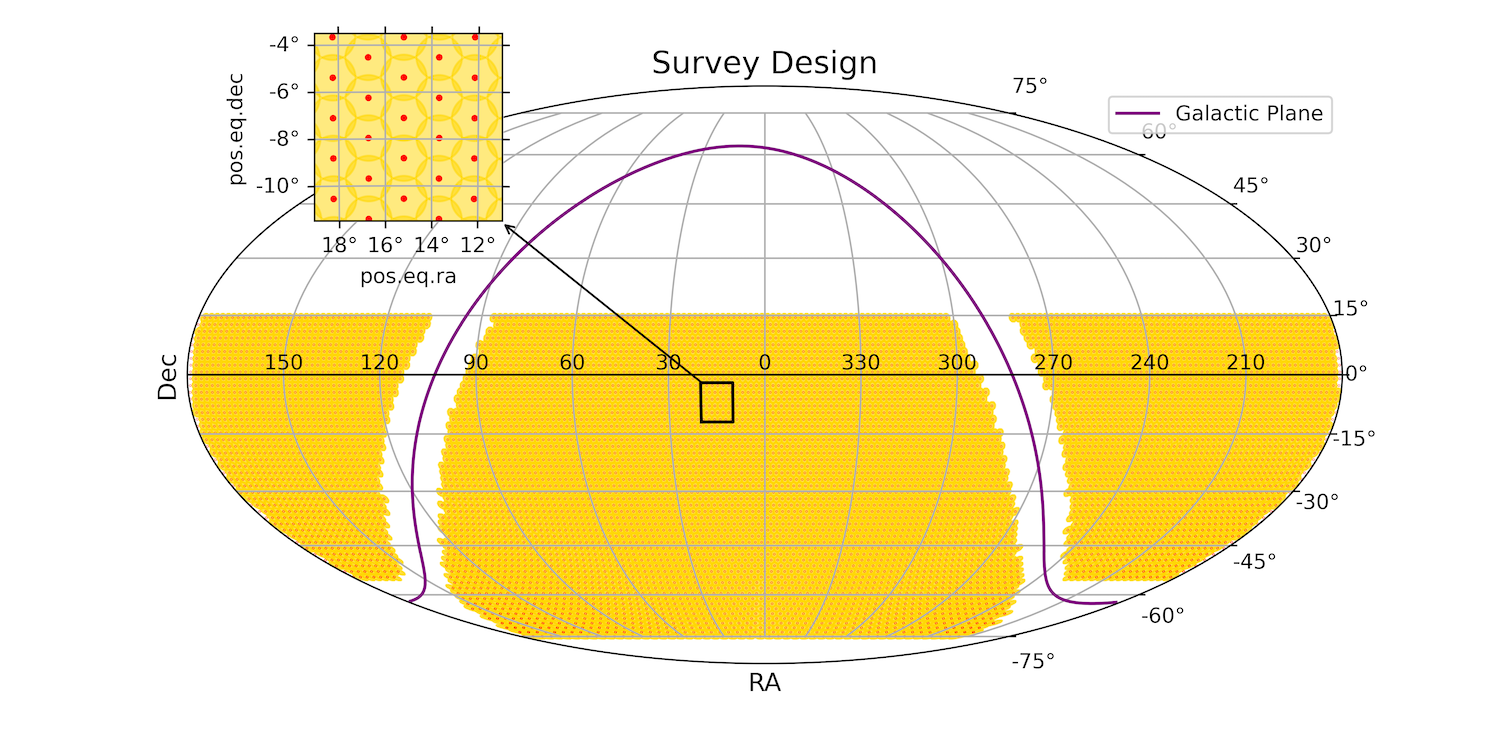}
    \caption{The Extra-galactic Sky footprint covered with an example of MOS-LR (yellow circles) and IFS (red points) footprint coverage for the high redshift survey during dark time. About $7000$ pointings are needed to have an homogeneous coverage of the $18\,000\,\deg^2$ of the survey, taking into account of some overlap between fields.
    }
    \label{fig:footprint_design_dark_time}
\end{figure}

\subsubsection{High-redshift galaxy and quasar targets}

Lyman break galaxies (LBGs) are actively star-forming galaxies at high redshift \citep{Steidel:1996}. LBGs are the most promising tracers to carry out large-scale clustering measurements in the redshift range $2 < z <5$, a region only probed for now by Lyman-$\alpha$ forest measurements from high redshift quasars (QSO).  Their rest-frame spectra are characterised by a complete absorption at the Lyman-limit ($912\,\r{A}$), due to surrounding neutral Hydrogen absorption, and a soft decrement shorter of the Lyman-$\alpha$ line ($1216\,\r{A}$) from Ly-$\alpha$ forest absorbers \citep{Dropout_Wilson_White}. 

LBGs are faint systems and require long effective exposure times to confidently assess their type and measure their redshifts. Thus, with its 12m aperture and its high density of fiber, \wst\ is a perfect telescope to measure high LBGs densities over the full accessible extra-galactic sky.

In practice, a $u$-dropout galaxy lack detection in the $u$-band, while having significant flux in redder bands. $u$-dropout galaxies have a mean redshift of $z\sim3$. Similarly, the $g$ and $r$-dropout galaxy lack detection in the $g$ and $r$-band and probe LBGs at $z\sim4$ and $z\sim 5$. In terms of depth at 5$\sigma$ detection limit, 10-year \lsst\ photometry will give respectively for the $ugriz$ bands, $\sim(25.5,27,27,26.5,25.5)$ AB magnitude. This will allow a confident selection in $u$-dropout with $r<25$, in $g$-dropout with $i<25.5$ and in $r$-dropout with $z<25.5$.

In addition, \desi\ has demonstrated its ability to observe QSOs up to $r<23$ \citep{DESI_QSO_TS}. With the \wst, we can reasonably expect to reach a magnitude limit of $r<24$ with a density of QSOs at the order of 400 deg$^{-2}$ (170 deg$^{-2}$ at $z>2.0$), following the quasar luminosity function of \cite{Palanque2016}.

This allows us to study the auto and cross-correlation function of their Ly-$\alpha$ forest as well as the cross-correlation function between the Ly-$\alpha$ forest of QSOs and LBG tracers.

\subsubsection{Dark Time Fiducial Survey}\label{sec:high_z_survey}

\begin{figure}[t]
\centering
    \includegraphics[width=1.\textwidth]{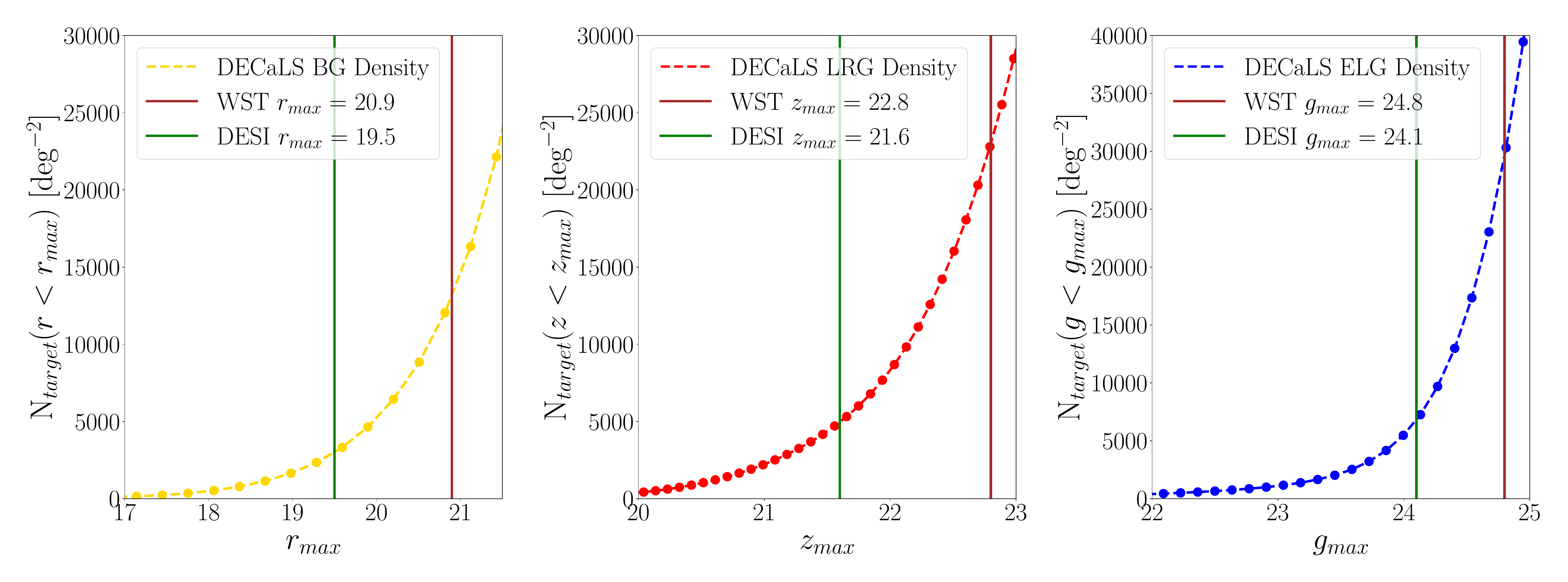} \\
    \includegraphics[width=1.\textwidth]{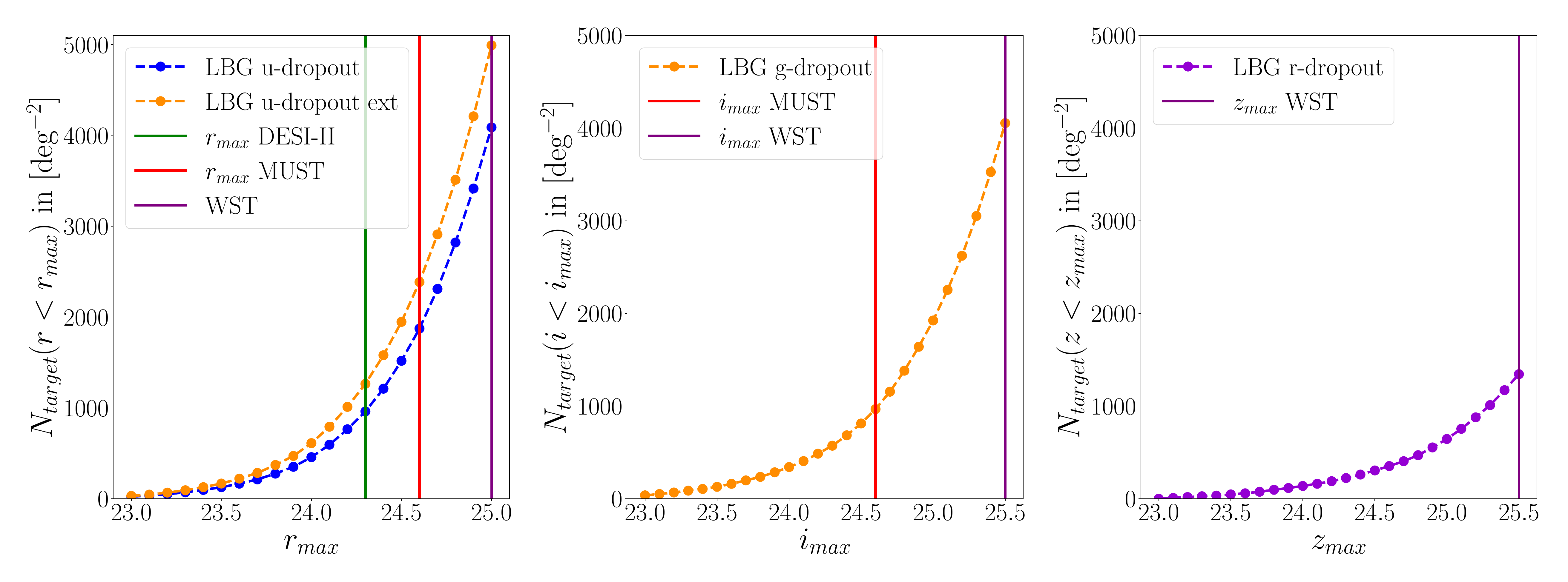}
\caption{\textit{Top panels}: Luminosity functions for the BG (\textit{left}), LRG (\textit{center}) and ELG (\textit{right}) samples for the Legacy survey (grey time). Vertical green lines correspond to the maximal depth used for \desi\ target selection and can be compared with that predicted for \wst\ (in purple).
\textit{Bottom panels}: Luminosity functions for the LBG $u-$dropout (\textit{left}), $g-$dropout (\textit{center}), and $r-$dropout  (\textit{right}) samples for the high-$z$ survey (dark time). Vertical green lines correspond to the maximal depth used for \desi\ target selection and can be compared with that predicted for Megamapper/MUST (in red) and \wst\ (in purple).}
\label{fig:lumin_func}
\end{figure}

To obtain the anticipated tracer and spectroscopic densities for the fiducial \wst\ survey, we extrapolate data from \desi\ observations carried out using CLAUDS and HSC imaging \citep{CLAUDS_imaging_2019, HSC_imaging_2022}.
The Signal-to-Noise Ratio (S/N) describes how well a source is measured by an instrument.
We model its time evolution using the CCD equation below, assuming a sky noise dominated regime for faint objects, which is a relevant limit for high-redshift cosmology, 
 $   {\rm S/N}\propto I(m) \sqrt{S_{\rm tel} \cdot t},$
with $I(m)$ the flux of a magnitude $m$ object. Thus, we assume that the observing time for an object with magnitude $m$ will be statistically reduced by a factor $S_{\rm WST}/S_{\rm DESI}=9$ compared with \desi\footnote{This argument is based only on the mirror surface. The time of observing is also expected to be reduced for WST thanks to better atmospheric conditions [Chile vs. Arizona].}.

Furthermore, two galaxies with a magnitude difference $\Delta m$ have a flux difference of $10^{\Delta m/2.5}$. Therefore, assuming a fixed S/N, the observation time difference between the two galaxies are given by the ratio of their fluxes, leading to $t_2=10^{\frac{2\,\Delta m}{2.5}}t_1$.

Consequently, for the $u$-dropout sample, assuming two-hours exposure with \desi\ to reach objects at $r_{\rm max}=24.2$, this would correspond to 15 minutes of exposure with WST for the same magnitude.
In practice there is no efficiency gain with \desi\ increasing this two-hours exposure, and we expect the same for the corresponding WST 15 min exposure. We assume the efficiency to be independent of the $r$-magnitude, but we implement additional 15-minutes exposures following  the $t_2/t_1$ equation above. The efficiency saturates at 80$\%$ for $z_{\rm photo}>2.8$, but linearly drop to 20$\%$ at $z_{\rm photo}>2$. This behaviour is mainly due to two effects. First, the throughput decreases in the UV part of the spectrum due losses in mirror reflectivity and fibre transmission. Then, due to evolution of the LBG population, for $z<2.8$, the Ly-$\alpha$ emission is less pronounced, making it more difficult to determine the redshift through the Ly-$\alpha$ line.
Thus for \desi\, the target selection was optimised to avoid targets at $z<2.8$,
maximising the effective efficiency of observation.
An extension of the standard selection was define to increase the target density at $2<z<2.8$. We assume DESI-II and Megamapper/MUST u-dropout will target u-dropout through the standard selection, while WST will use the extended one with a maximal $r$-magnitude $r_{\rm max}=25$. We further assume that we can improve the UV throughput and  the target selection to favour LBGs with stronger Ly-$\alpha$ emission, increasing the efficiency by 25 $\%$ for the $2<z<2.8$ range, with respect to \desi.

For the $g$-dropout ($r$-dropout) sample, we also extrapolate the \desi\ preliminary results, although they are statistically less significant. Given the fainter magnitudes of these LBGs, which are mainly around $z\sim 4$ ($5$), we consider that an exposure of one hour corresponds to an efficiency of 50\%, for $i_{\rm max}=25.5$ ($z_{\rm max}=25.5$), with WST. These assumptions are based on simple approximations and do not take into account any redshift or magnitude dependence on the efficiency. 

The different luminosity functions for the $ugr$-dropout samples are presented in the bottom panel of \cref{fig:lumin_func} as well as the magnitude cut assumed for DESI-II, $u$-dropout; Megamapper/MUST, $ug$-dropout; and WST, $ugr$-dropout samples. The WST target density is $\sim 10$ (3) times larger than that of DESI-II (Megamapper/MUST). \Cref{fig:cumul_sky_dens}  display the cumulative target density as a function of redshift for both legacy and high-$z$ surveys. In the end, we expect to reach a total target density of $\sim 10\,000$ targets per square.
Such high densities which are typically 30\% more than the fiber density demonstrates that we could either accommodate target with different exposure time, or refine the target strategy to optimise the target densities (this can be the scope of future analysis).

 Assuming a fibre assignment of 90$\%$ (based on the fibre density, and the different tracer exposure time) and convolving the observation efficiency with the photometric redshift distribution of targets, we derive the expected spectroscopic distribution of objects, shown in \cref{fig:Nz_zhigh2}.

Observing $18\,000\,\deg^2$ or $7000$ pointed observations with one hour exposure per pointing represent a total of 700 nights of dark-time with 10h/nights. Assuming 35\% of Dark Time during the year and accounting for 80\% survey efficiency, this translate into a 7-year observing program. Altogether a total of about 150 million redshift measurement should be obtained during this Dark Time survey.
 
These survey characteristics are derived through simple assumptions, and a proper Exposure-Time-Calculator (ETC) would need to be implemented in the future to refine these estimates.

\subsubsection{Galaxy clusters and Legacy Survey targets}

 For the galaxy clusters survey the goal is to measure with the MOS-LR instruments a significant (80-90\%) fraction of the cluster members (whether they are red or blue galaxies) as well as the galaxies in nearby filaments. Target selesction will be conducted using \lsst\ photometry and applying a wide redshift photometric cut around the cluster redshift. The IFS instrument will map the core of the cluster. Depending of the redshift of the cluster we will conduct a single IFS pointing (for high redshift cluster $z>0.3$) or a 2x2 or 3x3 mosaic pointings for lower redshift clusters. IFS will cover the region of strong gravitational lensing (arcs and multiple images) as well as enabling measurement of the velocity dispersion of the central galaxy.

\begin{figure}
    \centering
    \includegraphics[width=0.495\textwidth]{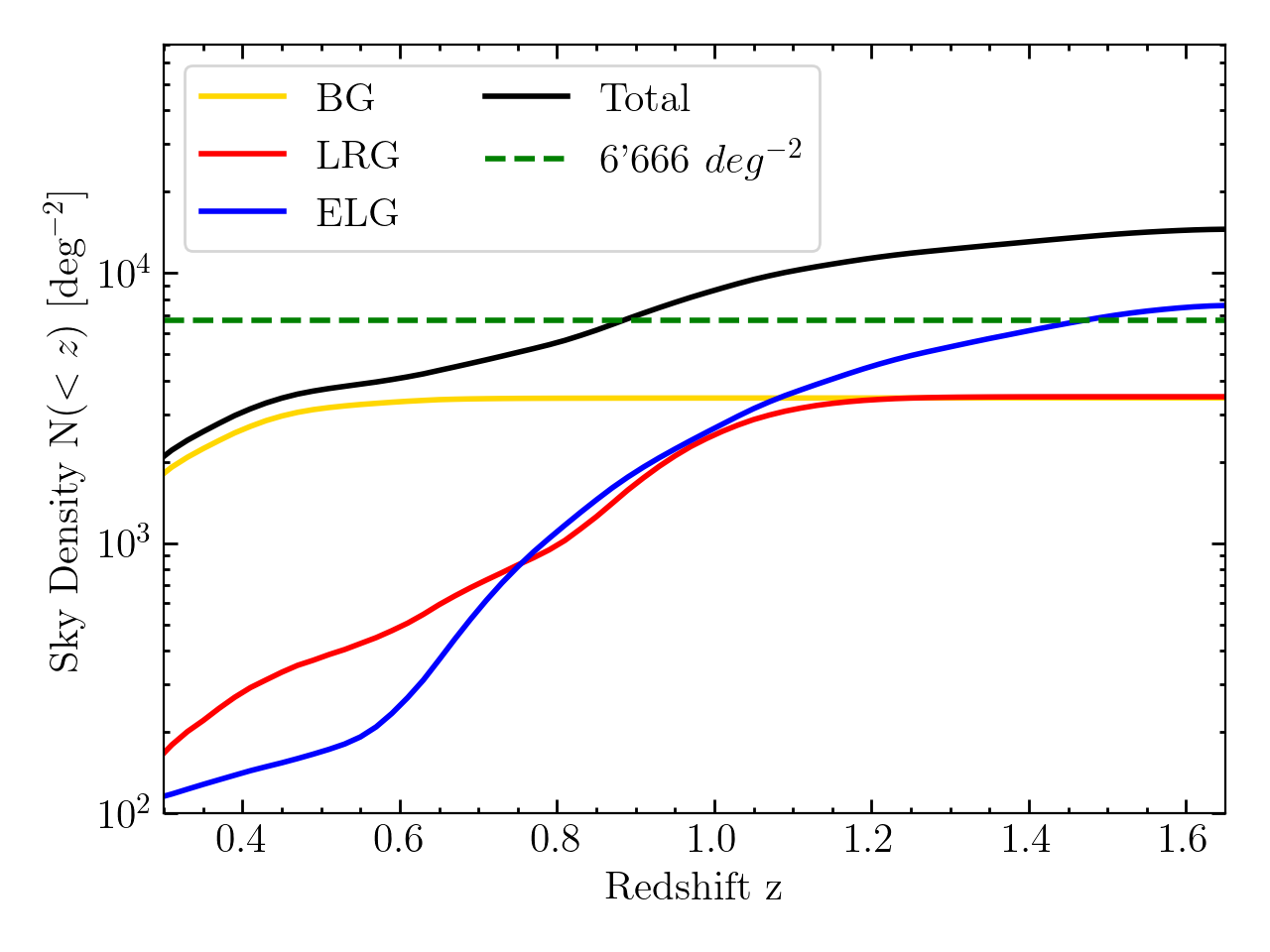}
    \includegraphics[width=0.495\textwidth]{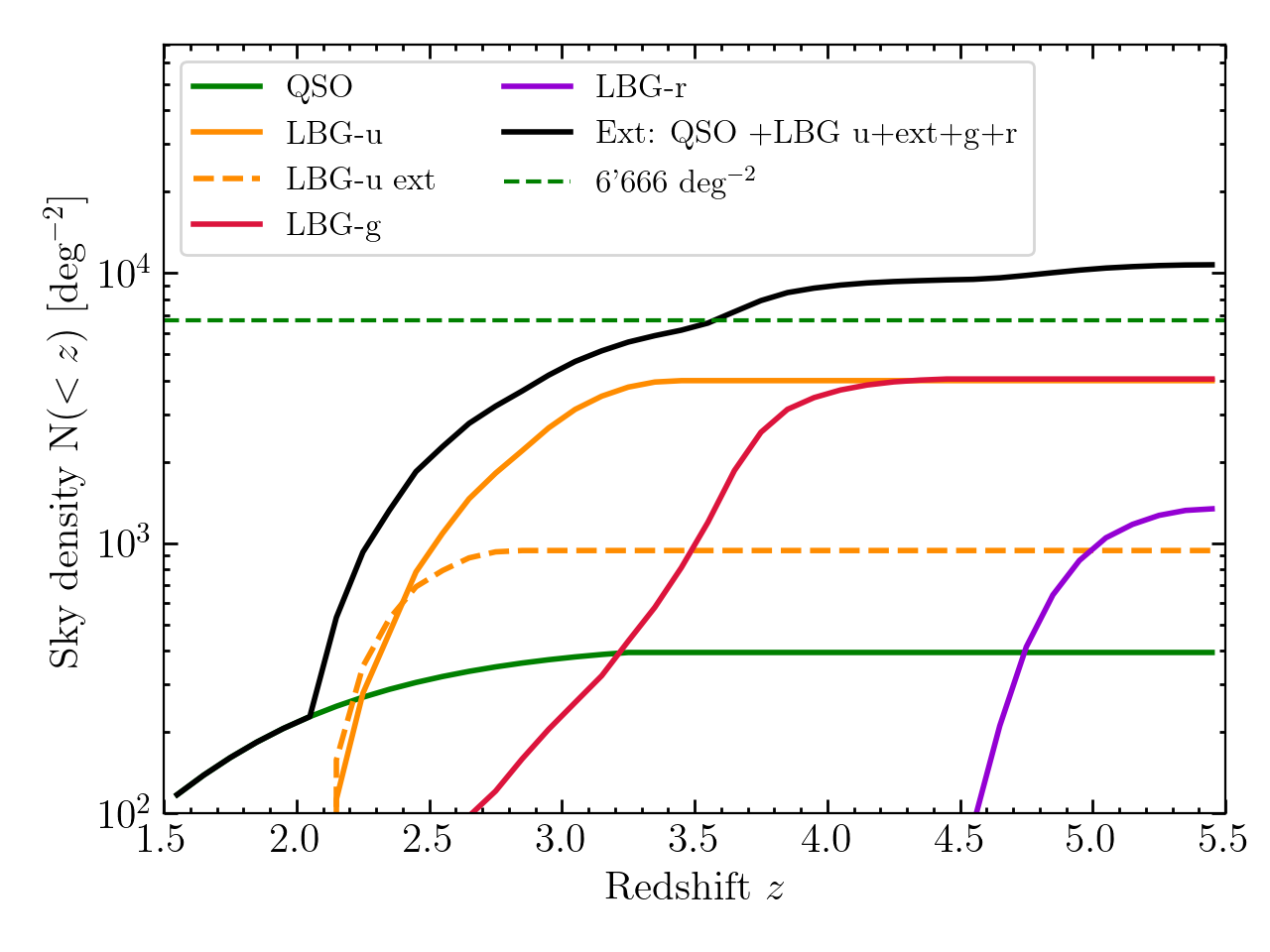}
    \caption{Cumulative number of targets as a function of redshift for the legacy survey (\textit{left}) and the high-$z$ survey (\textit{right}). Each galaxy tracer is represented by a coloured line as labelled and the total number of object is represented by the solid black line. The green dashed-line represent the number of fibre per square degree on the sky.}
    \label{fig:cumul_sky_dens}
\end{figure}

 For the Legacy Survey strategy we will use only the MOS-LR instrument. Although, there is ample time to define the targets to be selected, we explore a possible target selection based on the \desi\ targets but extending the magnitude limit to typically one magnitude fainter as shown in the top part of Figure \ref{fig:lumin_func}. This selection leads to a cumulative target density of about $11\,000\,\deg^{-2}$ as shown on the left panel of Figure  \ref{fig:cumul_sky_dens}. There are clearly no shortage of targets to be used by the WST fibers.

\subsubsection{Grey Time Fiducial Survey}

 The Cluster  and the Low-redshift Legacy Surveys will be conducted in synergy during Grey time.
 Also we can easily accommodate the exact central pointing to follow a telescope-TOO that would point the IFS to a transient target.

 Assuming a survey footprint of $18\,000\,\deg^2$ for the extra-galactic sky, a minimum of $7000$ MOS-LR pointing will be needed (see Figure \ref{fig:footprint_design_dark_time}). Yet because the IFS constraint to point to a cluster galaxy center or to a transient event, we estimate that a minimum of $\sim$40\% supplementary MOS-LR pointings will be necessary to have an homogeneous coverage of the $18\,000\,\deg^2$ of the extra-galactic sky, meaning a total of $10\,000$ pointings. Assuming an average 45min exposure per pointing this represent about 500 nights of grey time. Assuming $40\%$ of Grey nights per year and accounting for a $80\%$ efficiency, this translate into a 7-year program.
 As many targets will get their redshift measurement in less than 15min, we estimate that this survey will measure $\sim200$ million redshifts.

\clearpage

\section*{Affiliations}
\addcontentsline{toc}{section}{\protect\numberline{}Affiliations}

$^1$European Southern Observatory, Karl-Schwarzschild-Strasse 2, 85748 Garching bei M\"unchen, Germany\\
$^2$Institute of Physics, Laboratory of Astrophysics, \'Ecole Polytechnique F\'ed\'erale de Lausanne (EPFL)\\
$^3$Instituto de Astrofísica e Ciências do Espaço, Universidade do Porto\\
$^4$Dipartimento di Fisica e Astronomia ``Augusto Righi'' - Alma Mater Studiorum Universit\`{a} di Bologna\\
$^5$Department of Physics and Astronomy, University College London\\
$^6$Observatoire de la Côte d'Azur, Université Côte d'Azur, CNRS, Laboratoire Lagrange\\
$^7$Centre for Extragalactic Astronomy, Department of Physics, Durham University\\
$^8$Institute for Computational Cosmology, Department of Physics, University of Durham\\
$^9$Institute for Astronomy, University of Edinburgh\\
$^{10}$Leibniz-Institut f\"ur Astrophysik Potsdam (AIP)\\
$^{11}$Nicolaus Copernicus Astronomical Center, Polish Academy of Sciences\\
$^{12}$Kapteyn Astronomical Institute, University of Groningen\\
$^{13}$Univ Lyon, Univ Lyon1, Ens de Lyon, CNRS, Centre de Recherche Astrophysique de Lyon (CRAL)\\
$^{14}$INAF - Osservatorio Astrofisico di Arcetri\\
$^{15}$Stockholm University\\
$^{16}$INAF - Osservatorio di Astrofisica e Scienza dello Spazio di Bologna\\
$^{17}$Universidad de Valparaíso\\
$^{18}$University of Geneva\\
$^{19}$Research School of Astronomy and Astrophysics, Australian National University, Canberra, ACT 2611, Australia\\
$^{20}$ARC Centre of Excellence for All Sky Astrophysics in 3 Dimensions (ASTRO 3D), Australia\\
$^{21}$Instituto de Astrof\'isica de Canarias\\
$^{22}$Universidad de La Laguna\\
$^{23}$SISSA - Scuola Internazionale Superiore di Studi Avanzati\\
$^{24}$University of Florence\\
$^{25}$INAF - Osservatorio Astronomico di Roma\\
$^{26}$GEPI, Observatoire de Paris, Universite PSL, CNRS\\
$^{27}$INAF – Istituto di Astrofisica Spaziale e Fisica cosmica Milano\\
$^{28}$Sydney Institute for Astronomy, School of Physics, The University of Sydney\\
$^{29}$Aix Marseille Univ, CNRS, CNES, LAM\\
$^{30}$INAF - Osservatorio Astronomico di Palermo\\
$^{31}$University of Rome Tor Vergata\\
$^{32}$Newcastle University\\
$^{33}$Gran Sasso Science Institute (GSSI)\\
$^{34}$UK Astronomy Technology Ctr., Royal Observatory, Edinburgh, United Kingdom\\
$^{35}$INAF - OACT\\
$^{36}$Dipartimento di Fisica, Universit\`a degli Studi di Torino\\
$^{37}$Istituto Nazionale di Fisica Nucleare -- INFN, Sezione di Torino\\
$^{38}$INAF - Osservatorio Astrofisico di Torino\\
$^{39}$School of Mathematical and Physical Sciences, Macquarie University\\
$^{40}$INAF – Istituto di Radioastronomia\\
$^{41}$Institut f\"ur Astrophysik, Georg-August-Universit\"at\\
$^{42}$Pontificia Universidad Catolica de Chile, Instituto de Astrofisica\\
$^{43}$Millennium Institute of Astrophysics, Santiago, Chile\\
$^{44}$Pontificia Universidad Catolica de Chile, Centro de Astroingenieria\\
$^{45}$University of Padova - Department of Physics and Astronomy "G. Galilei"\\
$^{46}$Centro de Astrobiolog\'ia (CAB), CSIC-INTA, Madrid, Spain\\
$^{47}$Max-Planck-Institut f\"ur Extraterrestrische Physik\\
$^{48}$Centre for Astrophysics and Supercomputing, Swinburne University\\
$^{49}$Institut de Física d’Altes Energies (IFAE), The Barcelona Institute of Science and Technology\\
$^{50}$INAF - Osservatorio Astronomico di Padova\\
$^{51}$ASI - Space Science Data Center\\
$^{52}$National Astronomical Observatory of Japan\\
$^{53}$INAF - Osservatorio Astronomico di Palermo\\
$^{54}$Institute of Astronomy, University of Cambridge\\
$^{55}$ICRAR, The University of Western Australia\\
$^{56}$Aix Marseille Univ, CNRS/IN2P3, CPPM\\
$^{57}$Centro de Estudios de F\'isica del Cosmos de Aragón (CEFCA)\\
$^{58}$Astrophysics Research Institute, Liverpool John Moores University\\
$^{59}$University of Leeds\\
$^{60}$University of Belgrade - Faculty of Mathematics, Department of astronomy\\
$^{61}$Hamburger Sternwarte, Universit\"at Hamburg\\
$^{62}$Kavli Institute for Cosmology, University of Cambridge\\
$^{63}$Cavendish Laboratory, University of Cambridge\\
$^{64}$INAF - Osservatorio Astronomico di Capodimonte\\
$^{65}$Space Telescope Science Institute\\
$^{66}$Observatoire Astronomique de Strasbourg, Universite de Strasbourg, CNRS\\
$^{67}$Tuorla Observatory, Department of Physics and Astronomy\\
$^{68}$Finnish Centre for Astronomy with ESO (FINCA)\\
$^{69}$Caltech/IPAC\\
$^{70}$Institut d'Astrophysique de Paris\\
$^{71}$Department of Astronomy, University of Vienna\\
$^{72}$Department of Astronomy, The University of Texas at Austin\\
$^{73}$Kavli Institute for the Physics and Mathematics of the Universe (WPI), UTIAS, The University of Tokyo\\
$^{74}$Universidad de Chile\\
$^{75}$School of Physics, Trinity College Dublin, The University of Dublin\\
$^{76}$STAR Institute, Liege, Belgium\\
$^{77}$Blue Skies Space, Italia SRL\\
$^{78}$INFN - Sezione di Bologna\\
$^{79}$Institute of Science and Technology Austria (ISTA)\\
$^{80}$School of Physics and Astronomy, University of Birmingham\\
$^{81}$Astronomical Institute, Faculty of Mathematics and Physics, Charles University\\
$^{82}$Institut d'Astronomie et d'Astrophysique, Universit\'e Libre de Bruxelles (ULB)\\
$^{83}$Royal Observatory of Belgium\\
$^{84}$Max Planck Institute for Astronomy\\
$^{85}$INAF-IAPS\\
$^{86}$Instituto de Astrofísica, Dep. de Ciencias Físicas, Facultad de Ciencias Exactas, Universidad Andres Bello\\
$^{87}$Vatican Observatory, V00120 Vatican City State, Italy\\
$^{88}$Departamento de Fisica, Universidade Federal de Santa Catarina, Trinidade 88040-900, Florianopolis, Brazil\\
$^{89}$Leiden Observatory, Leiden University\\
$^{90}$School of Physics, University of New South Wales\\
$^{91}$UNSW Data Science Hub, University of New South Wales\\
$^{92}$ICSC - Centro Nazionale di Ricerca in High Performance Computing, Big Data e Quantum Computing\\
$^{93}$University of Valencia\\
$^{94}$Astrophysics Research Centre, School of Mathematics and Physics, Queens University Belfast\\
$^{95}$INAF-Osservatorio Astronomico d'Abruzzo\\
$^{96}$European Space Agency, European Space Astronomy Centre\\
$^{97}$Laborat\'orio Nacional de Astrof\'isica, MCTI, Brazil\\
$^{98}$University Catholic of Lyon, University of Lyon\\
$^{99}$Lund Observatory, Division of Astrophysics, Department of Physics, Lund University\\
$^{100}$The Oskar Klein Centre, Department of Astronomy, Stockholm University\\
$^{101}$CPPM/CCA/Cooper/Princeton\\
$^{102}$Astronomical Observatory Belgrade\\
$^{103}$Departament de F\'isica, Universitat Polit\`ecnica de Catalunya\\
$^{104}$INAF - Osservatorio Astronomico di Brera\\
$^{105}$University of Ferrara\\
$^{106}$International Space Science Institute (ISSI)\\
$^{107}$INAF - Osservatorio Astronomico di Trieste\\
$^{108}$Institute of Theoretical Physics and Astronomy, Vilnius University\\
$^{109}$Department of Astronomy \& Astrophysics, University of California\\
$^{110}$College of Science, Australian National University, ACT, Australia\\
$^{111}$Center for Cosmology and Particle Physics, Department of Physics, New York University\\
$^{112}$Center for Computational Astrophysics, Flatiron Institute\\
$^{113}$INFN – National Institute for Nuclear Physics\\
$^{114}$IFPU, Institute for Fundamental Physics of the Universe\\
$^{115}$Armagh Observatory and Planetarium\\
$^{116}$Keele University\\
$^{117}$University of Maryland\\
$^{118}$ Boston University\\
$^{119}$IRFU, CEA, Université Paris-Saclay\\
$^{120}$LMU-Munich\\
$^{121}$University of Potsdam, Institute of Physics and Astronomy\\
$^{122}$Zentrum für Astronomie der Universit{\"a}t Heidelberg, Astronomisches Rechen-Institut, M{\"o}nchhofstr. 12-14, 69120 Heidelberg, Germany\\
$^{123}$Univ Lyon, Univ Claude Bernard Lyon 1, CNRS, IP2I Lyon / IN2P3\\
$^{124}$International Centre for Space Sciences and Cosmology, Ahmedabad University\\

 \clearpage

\section*{Acknowledgments}
\addcontentsline{toc}{section}{\protect\numberline{}Acknowledgments}

Front-page image: credit to Rossella Spiga (INAF).
\medskip

\noindent Richard I. Anderson acknowledges support from the European Research Council (ERC) under the European Union's Horizon 2020 research and innovation programme (Grant Agreement No. 947660). Richard I. Anderson is funded by the Swiss National Science Foundation through an Eccellenza Professorial Fellowship (award PCEFP2\_194638).

\noindent Patricia Arevalo acknowledges support from $ANID NCN2023\_002$
\smallskip

\noindent Giuseppina Battaglia acknowledges support from the Agencia Estatal de Investigacion del Ministerio de Ciencia en Innovacion (AEI-MICIN) and the European Regional Development Fund (ERDF) under grant number PID2020-118778GB-I00/10.13039/501100011033 and the AEI under grant number CEX2019-000920-S.
\smallskip

\noindent Francesco Belfiore acknowledges funding from the INAF Fundamental Astrophysics program. 
\smallskip

\noindent Sofia Bisero acknowledges support from the {\it Erasmus+} programme.
\smallskip

\noindent Susanna Bisogni acknowledges support from the INAF 2022 Minigrant project, "A novel machine learning approach to delve into the AGN central engine", Ob.Fun. 1.05.12.04.01
\smallskip

\noindent St\'ephane Blondin was supported by the `Programme National de Physique Stellaire' (PNPS) of CNRS/INSU cofunded by CEA and CNES.
\smallskip

\noindent Rosaria Bonito acknowledges support by the INAF Mini-Grant “Physical properties of Accreting young stellar objects: exploration of their light Curves and Emission (PACE)” and by the Large Grant INAF 2022 YODA (YSOs Outflows, Disks and Accretion: towards a global framework for the evolution of planet forming systems).
\smallskip

\noindent Jarle Brichmann was supported by Fundação para a Ciência e a Tecnologia (FCT) through the research grants DOI: 10.54499/UIDB/04434/2020 and DOI: 10.54499/UIDP/04434/2020 and through national funds with DOI: 2020.03379.CEECIND/CP1631/CT0003
\smallskip

\noindent Stefano Camera acknowledges support from the Italian Ministry of University and Research (\textsc{mur}), PRIN 2022 `EXSKALIBUR – Euclid-Cross-SKA: Likelihood Inference Building for Universe's Research' and from the European Union -- Next Generation EU.
\smallskip

\noindent Support for M. Catelan is provided by ANID's Millennium Science Initiative through grant ICN12\textunderscore 009, awarded to the Millennium Institute of Astrophysics (MAS); by ANID's Basal project FB210003; and by FONDECYT grant \#1171273.
\smallskip

\noindent Valentina D'Orazi acknowledges the financial contribution from PRIN-MUR 2022YP5ACE.
\smallskip

\noindent Camilla Danielski acknowledges financial support from the INAF initiative ``IAF Astronomy Fellowships in Italy'', grant name GExoLife.
\smallskip

\noindent Suhail Dhawan acknowledges support from the Marie Curie Individual Fellowship under grant ID 890695 and a Junior Research Fellowship at Lucy Cavendish College. 
\smallskip

\noindent Ana Escorza thanks "La Caixa" Foundation (ID 100010434) for their support under fellowship code LCF/BQ/PI23/11970031
\smallskip

\noindent Miriam Garcia gratefully acknowledges support by grants PID2022-137779OB-C41, PID2022-140483NB-C22 and PID2019-105552RB-C41, funded by the Spanish Ministry of Science and Innovation/State Agency of Research MCIN/AEI/10.13039/501100011033.
\smallskip

\noindent Robert Grand is supported by an STFC Ernest Rutherford Fellowship (ST/W003643/1).
\smallskip

\noindent Amina Helmi acknowledges financial support from a Spinoza prize.
\smallskip

\noindent Artemio Herrero  acknowledges support by the Spanish Ministrios de Ciencia, Innovación y Universidades through grants PID2021-122397NB-C21 and SEV2015-0548. 
\smallskip

\noindent Vanessa Hill was supported by the Programme National Cosmology et Galaxies (PNCG) of CNRS/INSU with INP and IN2P3, co- funded by CEA and CNES
\smallskip

\noindent Dragana Ilic acknowledges funding provided by the University of Belgrade -- Faculty of Mathematics (contract 451-03-66/2024-03/200104), Astronomical Observatory Belgrade (contract 451-03-66/2024-03/200002), through grants by the Ministry of Education, Science, and Technological Development of the Republic of Serbia. 
\smallskip

\noindent Vid Irsic is supported by the Kavli Foundation.
\smallskip

\noindent Luca Izzo is supported by an INAF Data Grant (F.O.1.05.23.05.15)
\smallskip

\noindent Benjamin Joachimi acknowledges support by the ERC-selected UKRI Frontier Research Grant EP/Y03015X/1 and by STFC Consolidated Grant ST/V000780/1.
\smallskip

\noindent Sebastian Kamann acknowledges funding from UKRI in the form of a Future Leaders Fellowship (grant no. MR/T022868/1).
\smallskip

\noindent Jean-Paul Kneib  acknowledge support from the Swiss National Science Foundation(SNF) "Cosmology with 3D Maps of the Universe" research grant: $200020\_207379$
\smallskip

\noindent Georges Kordopatis was supported by the Programme National Cosmology et Galaxies (PNCG) of CNRS/INSU with INP and IN2P3, co-funded by CEA and CNES.
\smallskip

\noindent Andjelka B. Kova{\v c}evi{\' c}  is supported by the Ministry of Science, Technological Development, and Innovation of R. Serbia through projects the University of Belgrade - Faculty of Mathematics (contract 451-03-66/2024-03/200104).
\smallskip

\noindent Hanindyo Kuncarayakti  was funded by the Research Council of Finland projects 324504, 328898, and 353019.
\smallskip

\noindent Sara Lucatello acknowledges the support from PRIN INAF grant ObFu 1.05.01.85.14 (“Building up the halo: chemo-dynamical tagging in the age of large surveys”, PI. S. Lucatello)
Laura Magrini acknowledges INAF for funding through the Checs minigrant and the Large Grant EPOCH, and the Ministero dell'Universita' e la Ricerca (MUR) for the grant PRIN project n.2022X4TM3H "Cosmic POT". 
\smallskip

\noindent Kate Maguire is funded by the EU H2020 ERC grant no. 758638
\smallskip

\noindent Nicola Malavasi acknowledges funding by the European Union through a Marie Sk{\l}odowska-Curie Action Postdoctoral Fellowship (Grant Agreement: 101061448, project: MEMORY).
\smallskip

\noindent Davide Massari acknowledges support from PRIN MIUR2022 Progetto "CHRONOS" (PI: S. Cassisi) funded by European Union - Next Generation EU
\smallskip

\noindent Thibault Merle is granted by the BELSPO Belgian federal research program FED-tWIN under the research profile Prf-2020-033\_BISTRO.
\smallskip

\noindent Dante Minniti thanks the support from the ANID BASAL Center for Astrophysics and Associated Technologies (CATA) projects ACE210002 and FB210003, from Fondecyt Regular No. 1220724, and from CNPq Brasil Project 350104/2022-0.
\smallskip

\noindent Themiya  Nanayakkara acknowledges support from the Australian Research Council Laureate Fellowship FL180100060.
\smallskip

\noindent Matt Nicholl is supported by the European Research Council (ERC) under the European Union’s Horizon 2020 research and innovation programme (grant agreement No.~948381) and by UK Space Agency Grant No.~ST/Y000692/1.
\smallskip

\noindent Francesca Onori acknowledges support from MIUR, PRIN 2020 (grant 2020KB33TP) “Multimessenger astronomy in the Einstein Telescope Era (METE)”
\smallskip

\noindent Swayamtrupta Panda acknowledges the financial support of the Conselho Nacional de Desenvolvimento Científico e Tecnológico (CNPq) Fellowships 300936/2023-0 and 301628/2024-6.
\smallskip

\noindent Mamta Pandey-Pommier acknowledges the support of the Indo-French Centre for the Promotion of Advanced Research (Centre Franco-Indien pour la Promotion de la Recherche Avancée) under project no. 6504-3.
\smallskip

\noindent Michal Pawlak is supported by the BEKKER fellowship BPN/BEK/2022/1/00106 from the Polish National Agency for Academic Exchange
\smallskip

\noindent Alice Pisani acknowledge support from the Simons Foundation to the Center for Computational Astrophysics at the Flatiron Institute. 
\smallskip

\noindent Luka \v C. Popovi\'c acknowledge funding provided by the Astronomical Observatory Belgrade (contract 451-03-66/2024-03/200002), through grants by the Ministry of Education, Science, and Technological Development of the Republic of Serbia.
\smallskip

\noindent Roberto Raddi acknowledges support from Grant RYC2021-030837-I funded by MCIN/AEI/ 10.13039/501100011033 and by “European Union NextGeneration EU/PRTR”
\smallskip

\noindent Alberto Rebassa-Mansergas acknowledges support from the Spanish MINECO grant PID2020-117252GB-I00 and by the AGAUR/Generalitat de Catalunya grant SGR-386/2021.
\smallskip

\noindent Antoine Rocher acknowledges support from the Swiss National Science Foundation (SNF) "Cosmology with 3D Maps of the Universe" research grant $200020\_207379$
\smallskip

\noindent Andreas A. C. Sander is supported by the Deutsche Forschungsgemeinschaft (DFG - German Research Foundation) in the form of an Emmy Noether Research Group – Project-ID 445674056 (SA4064/1-1, PI Sander). AACS further acknowledges support by funding from the Federal Ministry of Education and Research (BMBF) and the Baden-W{\"u}rttemberg Ministry of Science as part of the Excellence Strategy of the German Federal and State Governments.
\smallskip

\noindent Jason L. Sanders acknowledges the support of the Royal Society.
\smallskip

\noindent Rodolfo Smiljanic acknowledges support from the National Science Centre, Poland, project 2019/34/E/ST9/00133.
\smallskip

\noindent Graham P. Smith acknowledges support from The Royal Society, the Leverhulme Trust, and from STFC grant ST/S006206/1 for UK participation in LSST.
\smallskip

\noindent Guillaume Thomas acknowledge support from the Agencia Estatal de Investigación (AEI) under grant Ayudas a centros de excelen- cia Severo Ochoa convocatoria 2019 with reference CEX2019- 000920-S, from the Agencia Estatal de Investigación (AEI) of the Ministerio de Ciencia e Innovación (MCINN) under grant with reference FJC2018-037323-I, of the Agencia Estatal de Investigación del Ministerio de Ciencia e Innovación (AEI- MCINN) under grant En la frontera de la arqueología galacác- tica: evolucíon de la materia luminoso y obscura de la vía Láctea y las galaxias enenas del Grupo Local with reference PID2020-118778GB-I00.
\smallskip

\noindent Tanya Urrutia acknowledges funding from the ERC-AdG grant SPECMAP-CGM-101020943.
\smallskip

\noindent Aurélien Verdier acknowledges support from the Swiss National Science Foundation (SNF) "Cosmology with 3D Maps of the Universe" research grant $200020\_207379$
\smallskip

\noindent Sofia Bisero and Susanna Diana Vergani acknowledge support by the {\it API Gravitational waves and compact objects} of Paris Observatory and by the {\it Programme National des Hautes Énergies (PNHE)} of CNRS/INSU co-funded by CNRS/IN2P3, CNRS/INP, CEA and CNES.
\smallskip

\noindent Giovanni Verza acknowledges NASA grant RSA 1692291 and support from the Simons Foundation to the Center for Computational Astrophysics at the Flatiron Institute.
\smallskip

\noindent Giustina Vietri acknowledges financial support from the INAF 2022 Minigrant project "Searching for UV ultra-fast outflow in AGN by exploiting widearea public spectroscopic surveys", Ob.Fun. 1.05.12.04.01
\smallskip

\noindent Jorick S. Vink acknowledges support from STFC grant ST/V000233/1
\smallskip

\noindent Carlos Viscasillas Vázquez acknowledge support from Lithuanian Science Council (LMTLT, grant No. P-MIP-23-24). INAF MiniGrant Checs.
\smallskip

\phantomsection
\addcontentsline{toc}{section}{References}
\small{
\bibliography{main}}

\appendix

\section{Membership of the WST Science Team\label{appendix}}

{\it As of February 2024.}

\begin{multicols}{2}

\noindent Erin Abraham \\
Angela	 Adamo	\\
Ana	 Afonso	\\
Jose	 Afonso	\\
David	 Aguado	\\
Mohammad	 Akhlaghi	\\
Evelyne	 Alecian	\\
David	 Alexander	\\
Sinan	 Alis	\\
Anish	 Amarsi	\\
Joseph	 Anderson	\\
Richard~I.	 Anderson	\\
Igor	 Andreoni	\\
James	 Angthopo	\\
Francesca	 Annibali	\\
James	 Annis	\\
Anke	 Arden-Arentsen	\\
Patricia Ar\'evalo \\
Magda	 Arnaboldi	\\
Hakim	 Atek	\\
Katie	 Auchettl	\\
Marc	 Audard	\\
Roland	 Bacon	\\
William M.	 Baker	\\
Eduardo	 Ba\~{n}ados	\\
Paramita	 Barai	\\
Sandro	 Bardelli	\\
Sydney A.	 Barnes	\\
Stefania	 Barsanti	\\
Giuseppina	 Battaglia	\\
Andrew	 Battisti	\\
Amelia	 Bayo	\\
Francesco	 Belfiore	\\
Michele	 Bellazzini	\\
Sirio	 Belli	\\
Emilio	 Bellini	\\
Vardha N.	 Bennert	\\
Maria	 Bergemann	\\
Santiago	 Bernal	\\
Leda Berni \\
Anupam	 Bhardwaj	\\
Matteo	 Biagetti	\\
Simone	 Bianchi	\\
Katia	 Biazzo	\\
Arjan	 Bik	\\
Maciej	 Bilicki	\\
Sofia	 Bisero	\\
Susanna	 Bisogni	\\
Jeremy	 Blaizot	\\
Joss	 Bland-Hawthorn	\\
St\'ephane	 Blondin	\\
Julia	 Bodensteiner	\\
Henri	 Boffin	\\
Andrea	 Bolamperti	\\
Micol	 Bolzonella	\\
Hemanth	 Bommireddy	\\
Piercarlo	 Bonifacio	\\
Rosaria	 Bonito	\\
Giuseppe	 Bono	\\
Leindert	 Boogaard	\\
Maria Teresa	 Botticella	\\
Connor	 Bottrell	\\
Nicolas	 Bouch\'e	\\
Dominic	 Bowman	\\
Alexey	 Boyarsky	\\
Vittorio Francesco	 Braga	\\
Marica	 Branchesi	\\
Jarle	 Brinchmann	\\
Marcella	 Brusa	\\
Fernando	 Buitrago	\\
Innocenza	 Busa	\\
Stefano	 Camera	\\
Alex	 Cameron	\\
Karina	 Caputi	\\
Julio A.	 Carballo-Bello	\\
Carmelita	 Carbone	\\
Ricardo	 Carrera	\\
Joseph	 Caruana	\\
Giada	 Casali	\\
Letizia	 Cassar\'a	\\
Norberto	 Castro Rodriguez	\\
M\'arcio	 Catelan	\\
Lorenzo	 Cavallo	\\
Pierluigi	 Cerulo	\\
Alan	 Chan	\\
Cristina	 Chiappini	\\
Norbert	 Christlieb	\\
Benedetta	 Ciardi	\\
Andrea	 Cimatti	\\
Bianca-Iulia	 Ciocan	\\
Maria-Rosa L.	 Cioni	\\
Michele	 Cirasuolo	\\
Michelle	 Cluver	\\
Matthew	Colless	\\
Thomas Edward	 Collett	\\
Michelle	 Collins	\\
Alice	 Concas	\\
Sofia	 Contarini	\\
Thierry	 Contini	\\
Yannick	 Copin	\\
Edvige	 Corbelli	\\
Luis	 Corral	\\
Jesus	 Corral-Santana	\\
Warrick	Couch	\\
Giovanni	 Cresci	\\
Stefano	 Cristiani	\\
Olga	 Cucciati	\\
Andrei	 Cuceu	\\
Mark	 Cunningham	\\
Emma	 Curtis Lake	\\
Filippo	 D'Ammando	\\
William	 d'Assignies Doumerg	\\
Valerio	 D'Elia	\\
Francesco	 D'Eugenio	\\
Valentina	 D'Orazi	\\
Andr\'e Rodrigo	 da Silva	\\
Maria Giovanna	 Dainotti	\\
Emanuele	 Dalessandro	\\
Francesco	Damiani	\\
Maria Luiza L.	 Dantas	\\
Alexandre	 David-Uraz	\\
Luke	 Davies	\\
Annalisa	 De Cia	\\
Richard	 de Grijs 	\\
Roelof S. de Jong	\\
Gabriella	 De Lucia	\\
Gayandhi	 De Silva	\\
Rafael S.	 de Souza	\\
Antonio	 de Ugarte Postigo	\\
Elisa	 Delgado-Mena	\\
Ricardo	 Demarco	\\
Camilla Danielski \\
Miroslava	 Dessauges-Zavadsky	\\
Suhail	 Dhawan	\\
Paola	 Di Matteo	\\
Bruno	 Dias	\\
Anastasio	 D\'iaz-S\'anchez	\\
Fabio Rosario	 Ditrani	\\
Mariano	 Dominguez	\\
Darko	 Donevski	\\
Andris	 Dorozsmai	\\
Alyssa	 Drake	\\
Arnas	 Drazdauskas	\\
Simon	 Driver	\\
Ulyana Dupletsa \\
Rajeshwari	 Dutta	\\
Richard	 Ellis	\\
Eric	 Emsellem	\\
Andrea	 Enia	\\
Benoit	 Epinat	\\
Denis	 Erka	\\
St\'ephanie	 Escoffier	\\
Ana	 Escorza	\\
Michele	 Fabrizio	\\
Anja	 Feldmeier-Krause	\\
Jeremy	 Fensh	\\
Eleonora	 Fiorellino	\\
Giuliana	 Fiorentino	\\
Adriano	 Fontana	\\
Francesco	 Fontani	\\
Daniel	 Forero Sanchez	\\
Jaime	 Forero-Romero	\\
Matteo	 Fossati	\\
Sotiria	 Fotopoulou	\\
Elena	 Franciosini	\\
Patrick	 Fran\c{c}ois	\\
Alexander	 Fritz	\\
Michele	 Fumagalli	\\
Dimitri	Gadotti	\\
Anna	 Gallazzi	\\
Francisco Jose	 Galindo Guil	\\
Poshak	 Gandhi	\\
Miriam	 Garcia	\\
Jorge Garcia-Rojas \\
Thibault	 Garel	\\
Adriana	 Gargiulo	\\
Bianca	 Garilli	\\
Francisco	 Garz\'on	\\
Marisa	 Girardi	\\
Karl	 Glazebrook	\\
Satya	 Gontcho A Gontcho	\\
Jonay	 Gonzalez	\\
Oscar	 Gonzalez	\\
Ylva	G\"{o}tberg	\\
Celine	 Gouin	\\
Matthew	 Graham	\\
Robert	 Grand	\\
Eva	 Grebel	\\
Valeria	 Grisoni	\\
Brent	 Groves	\\
Carlotta	 Gruppioni	\\
Mario Giusepppe	 Guarcello	\\
Marco	 Gullieuszik	\\
Yucheng	 Guo	\\
Boris	 Haeussler	\\
Christopher	 Haines	\\
Christopher	 Harrison	\\
Johanna	 Hartke	\\
Will	 Hartley	\\
Matthew	 Hayes	\\
Misha	 Haywood	\\
Nandini  Hazra \\
Wojciech	 Hellwing	\\
Amina	 Helmi 	\\
Artemio	 Herrero	\\
Pascale	 Hibon	\\
Michael	 Hilker	\\
Vanessa	 Hill	\\
Michaela	 Hirschmann	\\
Benedict	 Hofmann	\\
Carrie	 Holt	\\
Andrew Hopkins \\
Song	 Huang	\\
Daniela	 Iglesias	\\
Olivier	 Ilbert	\\
Dragana Ili\'c	\\
Enrichetta	 Iodice	\\
Angela	 Iovino 	\\
Vid Ir\v{s}i\v{c} \\
Valentin	 Ivanov	\\
Luca	 Izzo	\\
Pascale	 Jablonka	\\
Maja	 Jablonska	\\
Mathilde	 Jauzac	\\
Rob	 Jeffries	\\
Tereza	 Jerabkova	\\
Benjamin	 Joachimi	\\
Sean	 Johnson	\\
David	 Jones	\\
Peter Jonker \\
\'Eric	 Jullo	\\
Koki	 Kakiichi	\\
Darshan	 Kakkad	\\
Sebastian Kamann \\
Vanshika	 Kansal	\\
Cenk	 Kayhan	\\
Laura	 Keating	\\
Nandita	 Khetan	\\
Robert	 Kincaid 	\\
Rain	 Kipper	\\
Jakub	 Klencki	\\
Jean-Paul	 Kneib	\\
Christian	 Knigge	\\
Chiaki	 Kobayashi	\\
Tadayuki	 Kodama	\\
Piotr	 Kolaczek-Szymanski	\\
Dakshesh	 Kololgi	\\
Georges	 Kordopatis	\\
Rubina	 Kotak	\\
Andjelka B. Kova\v cevi\'c	\\
Davor	 Krajnovi\'c	\\
Katarina	 Kraljic	\\
Jens-Kristian	 Krogager	\\
Pratyush	 Kumar Das	\\
Hanindyo	 Kuncarayakti	\\
Yuna Kwon \\
Fiorangela	 La Forgia	\\
Fabio	 La Franca	\\
Ofer	 Lahav	\\
Clotilde	 Laigle	\\
Carmela	 Lardo	\\
S\o{}ren	 Larsen	\\
Monica	 Lazzarin	\\
Zoe	 Le Conte	\\
Ryan	 Leama	\\
Floriane	 Leclercq	\\
Khee-Gan	 Lee	\\
Matt	 Lehnert	\\
Ellen	 Leitinger	\\
Seunghwan	 Lim	\\
Paulina  Lira \\
Eleonora Loffredo \\
Marcella	 Longhetti	\\
Tobias Jakob	 Looser	\\
Angel Rafael	 Lopez Sanchez	\\
Sara	 Lucatello	\\
Junais	 M.	\\
Filippo	 Maccagni	\\
Manuela	 Magliocchetti	\\
Laura	 Magrini	\\
Kate Maguire \\
Guillaume	 Mahler	\\
Vincenzo	 Mainieri	\\
Fatemeh	 Zahra Majidi	\\
Nicola	 Malavasi	\\
Kasia	 Malek	\\
Chiara	 Mancini	\\
Catherine	 Manea	\\
Filippo	 Mannucci	\\
Antonino	 Marasco	\\
Marcella	 Marconi	\\
Rui	 Marques-Chaves	\\
Valerio	 Marra	\\
Sarah	 Martell	\\
Nicolas	 Martin	\\
John Eduard	 Mart\'inez Fern\'andez	\\
Mary Loli	 Mart\'inez-Aldama	\\
Lucimara	 Martins	\\
Silvia	 Martocchia	\\
Federico	 Marulli	\\
Michael	Maseda	\\
Davide	 Massari	\\
Tadafumi	 Matsuno	\\
Jorryt	 Matthee	\\
Sophie	 Maurogordato	\\
Alan McConnachie \\
Richard	 McDermid	\\
Rebecca	 McElroy	\\
Sean	 McGee	\\
Anna	 McLeod	\\
Richard	 McMahon	\\
Stefan	 Meingast	\\
Andrea	 Melandri	\\
Jens	 Melinder	\\
Jaroslav	 Merc	\\
Amata	 Mercurio	\\
Thibault	 Merle	\\
Leo	 Michel-Dansac	\\
Alessandra	 Migliorini	\\
Marco	 Mignoli	\\
Sarunas	 Mikolaitis	\\
Bryan	 Miller	\\
Ivan	 Minchev	\\
Dante	 Minniti	\\
Abhisek	 Mohapatra	\\
Marta	 Molero	\\
Chayan	 Mondal	\\
Maria	 Mongui\'o Montells	\\
Ana	 Monreal Ibero	\\
Federico	 Montano	\\
Ben	 Montet	\\
Michele	 Moresco	\\
Alessia	 Moretti	\\
Chiara	 Moretti	\\
Alice Mori \\
Lauro	 Moscardini	\\
Andres	 Moya	\\
Alessio	 Mucciarelli	\\
Tamal	 Mukherjee	\\
Oliver	M\"{u}ller	\\
David	 Murphy	\\
Michael	 Murphy	\\
Themiya	 Nanayakkara	\\
Peter	 Nemeth	\\
Samir Nepal \\
Nadine	 Neumayer	\\
Jeffrey	 Newman	\\
Matt	 Nicholl	\\
Brunella	 Nisini	\\
Peder	 Norberg	\\
Thomas	 Nordlander	\\
Kieran	 O'Brien	\\
Daniela	 Olave-Rojas	\\
Javier	 Olivares	\\
Ronaldo	 Oliveira da Silva	\\
Gulnara	 Omarova	\\
Francesca	 Onori	\\
Cyrielle	 Opitom	\\
Matt	 Owers	\\
Sergen	 Ozdemir	\\
Eric	 Paic	\\
Anna Francesca	 Pala	\\
Lovro	 Palaversa	\\
Swayamtrupta	 Panda	\\
Giulia	 Papini	\\
Ciro	 Pappalardo	\\
Luca	 Pasquini	\\
Lee	 Patrick	\\
Michal	 Pawlak	\\
Rahna	 Payyasseri	\\
Greco	 Pe\~{n}a	\\
Harold	 Pena Herazo	\\
Laura	 Pentericci	\\
Ismael	 Perez-Fournon	\\
Jose Manuel	 Perez-Martinez	\\
Priscila	 Pessi	\\
Gabriele	 Pezzulli	\\
Matthew	 Pieri	\\
Grzegorz	 Pietrzynski	\\
Andrzej	 Pigulski	\\
Silvia	 Piranomonte	\\
Alice	 Pisani	\\
Joanne	 Pledger	\\
Adriano	 Poci	\\
Bianca	 Poggianti	\\
Agnieszka	 Pollo	\\
Mamta	 Pommier	\\
Emanuela	 Pompei	\\
Paola	 Popesso	\\
Luka \v C. Popovi\'c	\\
Katja	Poppenh\"{a}ger	\\
Lucia	 Pozzetti	\\
Loredana	 Prisinzano	\\
Annagrazia	 Puglisi	\\
Salvatore	 Quai	\\
Roberto	 Raddi	\\
Fabio	 Ragosta	\\
Monica	 Rainer	\\
Sofia	 Randich	\\
Milena	 Ratajczak	\\
Bridget Ratcliffe \\
Alberto	 Rebassa-Mansergas	\\
Andrea	 Reguitti	\\
Umberto	 Rescigno	\\
Yves Revaz \\
Johan	 Richard	\\
Mickael	 Rigault	\\
Vincenzo	 Ripepi	\\
Aaron	 Robotham	\\
Antoine	 Rocher	\\
Gonzalo	 Rojas Garcia	\\
Donatella	 Romano	\\
Michael	 Romano	\\
David	 Rosario	\\
Piero	 Rosati	\\
Joakim	 Rosdahl	\\
Andrea	 Rossi	\\
Martin	 M. Roth	\\
Tomasz	 R\'ozanski	\\
Florian	 Ruppin	\\
Giuseppe	 Sacco	\\
Kanak	 Saha	\\
Amelie	 Saintonge	\\
Nick	 Samaras	\\
Hugues	 Sana	\\
Paula	 Sanchez Saez	\\
Ruben	 Sanchez-Janssen	\\
Andreas	 A.C. Sander 	\\
Jason	 Sanders	\\
Chandra Shekhar	 Saraf	\\
Mark	 Sargent	\\
Elena Sarpa \\
Sarath	 Satheesh Sheeba	\\
Ricardo	 Schiavon	\\
Carlo	 Schimd	\\
Ilane	 Schroetter	\\
Steve	 Schulze	\\
Marco	 Scodeggio	\\
Emiliano	 Sefusatti	\\
Ho	 Seong Hwang	\\
Sanjib	 Sharma	\\
Shreeya	 Shetye	\\
Charlotte	 Simmonds	\\
Sergio	 Simon-Diaz	\\
Rajeev	 Singh Rathour	\\
Malgorzata	 Siudek	\\
Thirupathi	 Sivarani	\\
Asa	 Skuladottir	\\
Rodolfo	 Smiljanic	\\
Dan	 Smith	\\
Graham P.	 Smith	\\
Jennifer	 Sobeck	\\
Azlizan Adhyaqsa	 Soemitro	\\
Jubee	 Sohn	\\
Roberto	 Soria	\\
Lorenzo	 Spina	\\
Else	 Starkenburg	\\
Matthias	 Steinmetz	\\
Edita	 Stonkute	\\
Mark	 Sullivan	\\
Will	 Sutherland	\\
Mark	 Swinbank	\\
Robert	 Szabo	\\
Sandro	 Tacchella	\\
Salvatore	 Taibi	\\
Margherita	 Talia	\\
Grazina	 Tautvaisiene	\\
Edward N Taylor	\\
Peter	 Taylor	\\
Elmo	 Tempel	\\
Matthew	 Temple	\\
Aishwarya Linesh	 Thakur	\\
Sabine	 Thater	\\
Christopher	 Theissen	\\
Guillaume	 Thomas	\\
Christina	Th\"{o}ne	\\
Yuan-Sen	 Ting	\\
Eline	 Tolstoy	\\
Crescenzo	 Tortora	\\
Paolo	 Tozzi	\\
Michele	 Trabucchi	\\
Scott	 Trager	\\
Gregor Traven \\
Grant	 Tremblay	\\
Laurence	 Tresse	\\
Oem	 Trivedi	\\
Maria	 Tsantaki	\\
Maria	 Tsedrik	\\
Hannah	 Turner	\\
Tanya	 Urrutia	\\
Christopher	 Usher	\\
Elena	 Valenti	\\
Francesco	 Valentino	\\
Mathieu	 Van der Swaelmen	\\
Arjen	 van der Wel	\\
Sophie	 Van Eck	\\
Irene	 Vanni	\\
Giacomo	 Venturi	\\
Alfonso Veropalumbo \\
Francesco	 Verdiani	\\
Aur\'elien	 Verdier	\\
Daniela	 Vergani	\\
Susanna Diana	 Vergani	\\
Anne	 Verhamme	\\
Aprajita	 Verma	\\
Joel	 Vernet	\\
Giovanni	 Verza	\\
Matteo	 Viel	\\
Giustina	 Vietri	\\
Cristian	 Vignali	\\
Jorge Andr\'es	 Villa V\'elez	\\
Jorick   S. Vink \\
Miguel	 Vioque	\\
Carlos	 Viscasillas V\'azquez	\\
Benedetta	 Vulcani	\\
Jakob	 Walcher	\\
Roland	 Walter	\\
Michael	 Walther	\\
Nicholas	 Walton	\\
Hai-Feng	 Wang	\\
Peter M. Weilbacher \\
Martin	 Wendt	\\
Gillard	 William	\\
Emily	 Wisnioski	\\
Clare	 Worley	\\
Nick Wright \\
Mengyuan	 Xiao	\\
Quanzhi	 Ye	\\
Christophe	 Yeche	\\
Ilsang	 Yoon	\\
Jiaxi	 Yu	\\
Tayyaba	 Zafar	\\
Simone	 Zaggia	\\
Stefano	 Zarattini	\\
Werner	 Zeilinger	\\
Cheng	 Zhao	\\
Stefano	 Zibetti	\\
Bodo	 Ziegler	\\
Igor	 Zinchenko	\\
Elena	 Zucca	\\

\end{multicols}

\end{document}